%% file: main.tex
\documentclass[11pt]{article}
\usepackage{setspace}
\usepackage{graphicx}
\usepackage[font=small,labelfont=bf]{caption}
\usepackage[font=small,labelfont=bf]{subcaption}
\usepackage{amsmath,amsthm,amssymb,cancel}
\usepackage{textcomp}
\usepackage{float}
\usepackage{mathrsfs}
\usepackage{physics}
\usepackage{hyperref}
\usepackage{cite}
\usepackage{url}
\usepackage{ifthen}
\usepackage[top=15mm, bottom=15mm, left=15mm, right=15mm]{geometry}
\usepackage[table]{xcolor}
\usepackage[utf8x]{inputenc}
\usepackage{xspace}
\usepackage[normalem]{ulem}
\usepackage{listings}
\usepackage{multirow}

  \hypersetup{%
  pdftitle={Impact of g-2},
  pdfauthor={Peter Athron, Csaba Balazs, Douglas HJ Jacob, Wojciech Kotlarski, Dominik Stoeckinger, Hyejung Stoeckinger-Kim},
  pdfkeywords={anomalous,magnetic,moment,FlexibleSUSY,supersymmetry,MSSM,simplified,models,EFT},
  colorlinks=true,urlcolor=blue,linkcolor=blue,citecolor=blue,filecolor=blue}
\newcommand\code[1]{{\lstinline!#1!}}

\newcommand{\update}[1]{\textcolor{black}{#1}}

\newcommand{\oldtext}[1]{\textcolor{red}{Old text removed here}}

\newcommand{\fs}{\texttt{FlexibleSUSY}\@\xspace}
\newcommand{\sarah}{\texttt{SARAH}\@\xspace}
\newcommand{\softsusy}{\texttt{SOFTSUSY}\@\xspace}
\newcommand{\ddcalc}{\texttt{DDCalc}\@\xspace}
\newcommand{\DDCalc}{\ddcalc}
\newcommand{\micromegas}{\texttt{micrOMEGAs}\@\xspace}

\newcommand{\pythia}{\texttt{PYTHIA}\@\xspace}
\newcommand{\checkmate}{\texttt{CheckMate}\@\xspace}
\newcommand{\madanalysis}{\texttt{MadAnalysis}\@\xspace}
\newcommand{\smodels}{\texttt{SModelS}\@\xspace}
\newcommand{\colliderbit}{\texttt{ColliderBit}\@\xspace}
\newcommand{\darkbit}{\texttt{DarkBit}\@\xspace}

\newcommand{\GAMBIT}{\texttt{GAMBIT}\@\xspace}
\newcommand{\gambit}{\GAMBIT}
\newcommand{\GMTCalc}{\texttt{GM2Calc}\@\xspace}
\newcommand{\multinest}{\texttt{MultiNest}\@\xspace}

\newcommand{\amu}{a_\mu}
\newcommand{\damu}{\Delta a_\mu}
\newcommand{\amuNEW}{\amu^{\text{2021}}}
\newcommand{\damuNEW}{\damu^{\text{2021}}}
\newcommand{\damuBNL}{\damu^{\text{BNL}}}

\newcommand{\FieldP}{$({\bf 1},{\bf 1}, 1)$}
\newcommand{\FieldN}{$({\bf 1},{\bf 1}, 0)$}
\newcommand{\FieldC}{$({\bf 1},{\bf 1},-1)$}
\newcommand{\FieldD}{$({\bf 1},{\bf 2},-1/2)$}

\newcommand{\FieldY}{$({\bf 1},{\bf 2},-3/2)$}
\newcommand{\FieldA}{$({\bf 1},{\bf 3}, 0)$}
\newcommand{\FieldT}{$({\bf 1},{\bf 3},-1)$}

\newcommand{\OneFieldN}{\FieldN\@\xspace}
\newcommand{\OneFieldC}{\FieldC\@\xspace}
\newcommand{\OneFieldD}{\FieldD\@\xspace}

\newcommand{\OneFieldA}{\FieldA\@\xspace}
\newcommand{\OneFieldT}{\FieldT\@\xspace}

\newcommand{\CombinedFieldsSNFC}{\multirow{2}{*}{\FieldN$_0$} &\multirow{2}{*}{-- \FieldC$_{1/2}$}\@\xspace}
\newcommand{\CombinedFieldsSCFN}{\FieldC$_0$ &-- \FieldN$_{1/2}$\@\xspace}
\newcommand{\CombinedFieldsSDFN}{\FieldD$_0$ &-- \FieldN$_{1/2}$\@\xspace}
\newcommand{\CombinedFieldsSNFD}{\multirow{2}{*}{\FieldN$_{0}$} &\multirow{2}{*}{-- \FieldD$_{1/2}$}\@\xspace}
\newcommand{\CombinedFieldsSDFC}{\multirow{2}{*}{\FieldD$_{0}$} &\multirow{2}{*}{-- \FieldC$_{1/2}$}\@\xspace}
\newcommand{\CombinedFieldsSCFD}{\FieldC$_0$ &-- \FieldD$_{1/2}$\@\xspace}
\newcommand{\CombinedFieldsSDFD}{\FieldD$_0$ &-- \FieldD$_{1/2}$\@\xspace}
\newcommand{\CombinedFieldsSDFA}{\multirow{2}{*}{\FieldD$_{0}$} &\multirow{2}{*}{-- \FieldA$_{1/2}$}\@\xspace}
\newcommand{\CombinedFieldsSDFT}{\multirow{2}{*}{\FieldD$_{0}$} &\multirow{2}{*}{-- \FieldT$_{1/2}$}\@\xspace}
\newcommand{\CombinedFieldsSAFD}{\FieldA$_{0}$ &-- \FieldD$_{1/2}$\@\xspace}
\newcommand{\CombinedFieldsSAFT}{\multirow{2}{*}{\FieldA$_{0}$} &\multirow{2}{*}{-- \FieldT$_{1/2}$}\@\xspace}
\newcommand{\CombinedFieldsSTFD}{\FieldT$_{0}$ &-- \FieldD$_{1/2}$\@\xspace}
\newcommand{\CombinedFieldsSTFA}{\FieldT$_{0}$ &-- \FieldA$_{1/2}$\@\xspace}
\newcommand{\CombinedFieldsFCVN}{\FieldC$_{1/2}$ &-- \FieldN$_{1}$\@\xspace}
\newcommand{\CombinedFieldsFDVN}{\FieldD$_{1/2}$ &-- \FieldN$_{1}$\@\xspace}
\newcommand{\CombinedFieldsFDVA}{\FieldD$_{1/2}$ &-- \FieldA$_{1}$\@\xspace}
\newcommand{\CombinedFieldsFNVP}{\FieldN$_{1/2}$ &-- \FieldP$_{1}$\@\xspace}
\newcommand{\CombinedFieldsFDVC}{\FieldD$_{1/2}$ &-- \FieldC$_{1}$\@\xspace}
\newcommand{\CombinedFieldsFTVA}{\FieldT$_{1/2}$ &-- \FieldA$_{1}$\@\xspace}

\newcommand{\TwoFieldsSNFC}{\FieldN$_0$ -- \FieldC$_{1/2}$\@\xspace}
\newcommand{\TwoFieldsFNVC}{\FieldN$_{1/2}$ -- \FieldC$_{1}$\@\xspace}

\newcommand{\TwoFieldsSNFD}{\FieldN$_{0}$ -- \FieldD$_{1/2}$\@\xspace}

\newcommand{\TwoFieldsFYFC}{\FieldY$_{1/2}$ -- \FieldC$_{1/2}$\@\xspace}

\newcommand{\TwoFieldsFYFT}{\FieldY$_{1/2}$ -- \FieldT$_{1/2}$\@\xspace}

\newcommand{\TwoFieldsSDFC}{\FieldD$_{0}$ -- \FieldC$_{1/2}$\@\xspace}

\newcommand{\TwoFieldsFDVC}{\FieldD$_{1/2}$ -- \FieldC$_{1}$\@\xspace}

\newcommand{\TwoFieldsSDFD}{\FieldD$_{0}$ -- \FieldD$_{1/2}$\@\xspace}

\newcommand{\TwoFieldsSDFA}{\FieldD$_{0}$ -- \FieldA$_{1/2}$\@\xspace}

\newcommand{\TwoFieldsFDVA}{\FieldD$_{1/2}$ -- \FieldA$_{1}$\@\xspace}
    
\newcommand{\TwoFieldsSDFT}{\FieldD$_{0}$ -- \FieldT$_{1/2}$\@\xspace}

\newcommand{\TwoFieldsSAFT}{\FieldA$_{0}$ -- \FieldT$_{1/2}$\@\xspace}

\title{New physics explanations of $\amu$ in light of the FNAL muon $g-2$ measurement}
\author{Peter Athron$^{a,b}$, Csaba Bal\'azs$^b$, Douglas HJ Jacob$^b$, Wojciech Kotlarski$^c$, \\Dominik St\"ockinger$^c$, Hyejung St\"ockinger-Kim$^c$}
\date{}
\begin{document}
\maketitle
\noindent$^a$ Department of Physics and Institute of Theoretical Physics, Nanjing Normal University, Nanjing, Jiangsu 210023, China \\
$^b$ School of Physics and Astronomy, Monash University, Melbourne, Victoria 3800, Australia\\
$^c$ Institut f\"ur Kern- und Teilchenphysik, TU Dresden, Zellescher Weg 19, 01069 Dresden, Germany\\
E-mail:  \href{mailto:peter.athron@coepp.org.au}{peter.athron@coepp.org.au},
\href{mailto:csaba.balazs@monash.edu}{csaba.balazs@monash.edu},
\href{mailto:douglas.jacob@monash.edu}{douglas.jacob@monash.edu}, \\
\href{mailto:wojciech.kotlarski@tu-dresden.de}{wojciech.kotlarski@tu-dresden.de},
\href{mailto:dominik.stoeckinger@tu-dresden.de}{dominik.stoeckinger@tu-dresden.de},
\href{mailto:hyejung.stoeckinger-kim@tu-dresden.de}{hyejung.stoeckinger-kim@tu-dresden.de}


\begin{abstract}
The Fermilab Muon $g-2$ experiment recently reported its first
  measurement of the anomalous magnetic moment \update{$\amu^{\textrm{FNAL}}$,
    which is in full agreement with the previous BNL measurement
    and pushes the world average deviation $\damuNEW$ from the Standard
    Model to a significance of $4.2\sigma$. }
  Here we provide an
  extensive survey of its impact on beyond
  the Standard Model physics.  We use state-of-the-art calculations
  and a sophisticated set of tools to make predictions for $\amu$,
  dark matter and LHC searches in a wide range of simple models with
  up to three new fields, that represent some of the few ways that
  large $\damu$ can be explained.  In addition for \update{the particularly
  well motivated Minimal
  Supersymmetric Standard Model}, we exhaustively cover  the
  scenarios where large $\damu$ can be explained while simultaneously
  satisfying all relevant data from other experiments.
  Generally, the $\amu$
  result can only be explained by rather small masses and/or large
  couplings and enhanced chirality flips, which can lead to conflicts
  with limits from LHC and dark matter experiments.  Our results show
  that the new measurement excludes a large number of models and
  provides crucial constraints on others.  Two-Higgs doublet and
  leptoquark models provide viable explanations of $\amu$ only in
  specific versions and in specific parameter ranges. Among all models
  with up to three fields, only models with chirality enhancements can
  accommodate $\amu$ and dark matter simultaneously. The MSSM \update{can
  simultaneously explain} $\amu$ and dark matter for Bino-like LSP
  in several coannihilation regions. \update{Allowing under abundance of the dark matter relic density,} the Higgsino- and particularly Wino-like LSP
  scenarios become promising explanations of the $\amu$ result.  
\end{abstract}

\newpage
\tableofcontents
\section{Introduction} \label{sec:Introduction}
Precision measurements of the anomalous magnetic moment of the muon,
$\amu$, provide excellent tests of physics beyond the Standard Model
(BSM), and the results can give hints at what form it might
take. Recently the E989 experiment \cite{Grange:2015fou} at the Fermi
National Laboratory (FNAL) published the \update{most precise
  measurement} of the anomalous magnetic moment of the muon\update{~\cite{PhysRevLett.126.141801}}.  This result, and the previous result from
Brookhaven National Laboratory (BNL) \cite{Bennett:2006fi} (adjusted
according to the latest value of $\mu_\mu/\mu_p$ as in
Ref.\ \cite{aoyama2020anomalous}) and \update{the new world average \cite{PhysRevLett.126.141801}}
are
\begin{align}
  \amu^{\textrm{FNAL}} &=    \update{(116\,592\,040\pm54)\times10^{-11}} ,\label{Eq:FNALresult}\\
  \amu^{\textrm{BNL}} &=    (116\,592\,089\pm63)\times10^{-11}
  ,\\
  \amuNEW &= \update{(116\,592\,061\pm41)\times10^{-11}}.
  \label{Eq:amuNEW}
\end{align}
The FNAL measurement \update{is fully compatible with } the previous
best measurement and \update{has a smaller uncertainty}.
Compared to the BNL result, the new world average $\amuNEW$ has a
\update{slightly decreased} central value 
and a \update{30\%} reduced statistics-dominated uncertainty.
In parallel to the FNAL measurement, a worldwide theory initiative
provided the White Paper
\cite{aoyama2020anomalous} with the best estimate for the central
theory prediction in the Standard Model (SM). Its value and
uncertainty are
\begin{align}
  \amu^{\textrm{SM}} &= (116\,591\,810\pm43)\times10^{-11} .
\label{amuSMWP}
\end{align}
This SM prediction is based
on up-to-date predictions of QED \cite{Aoyama:2012wk,Aoyama:2019ryr},
electroweak \cite{Czarnecki:2002nt,Gnendiger:2013pva}, hadronic vacuum
polarization
\cite{Davier:2017zfy,Keshavarzi:2018mgv,Colangelo:2018mtw,Hoferichter:2019gzf,Davier:2019can,Keshavarzi:2019abf,Kurz:2014wya}
and hadronic light-by-light contributions
\cite{Melnikov:2003xd,Masjuan:2017tvw,Colangelo:2017fiz,Hoferichter:2018kwz,Gerardin:2019vio,Bijnens:2019ghy,Colangelo:2019uex,Pauk:2014rta,Danilkin:2016hnh,Jegerlehner:2017gek,Knecht:2018sci,Eichmann:2019bqf,Roig:2019reh,Blum:2019ugy,Colangelo:2014qya}. For
further discussion of recent progress we refer to
Ref.\ \cite{aoyama2020anomalous}.\footnote{%
  The White Paper also contains an extensive discussion of promising
  progress of lattice QCD calculations for the hadronic vacuum
  polarization. The lattice world average evaluated in
  Ref.\ \cite{aoyama2020anomalous}, based on \cite{Chakraborty:2017tqp,Borsanyi:2017zdw,
Blum:2018mom,Giusti:2019xct,Shintani:2019wai,Davies:2019efs,Gerardin:2019rua,Aubin:2019usy,Giusti:2019hkz}, is compatible
  with the data-based result 
\cite{Davier:2017zfy,Keshavarzi:2018mgv,Colangelo:2018mtw,Hoferichter:2019gzf,Davier:2019can,Keshavarzi:2019abf,Kurz:2014wya}, has a higher
  central value and larger uncertainty. More recent lattice results
  are obtained in Refs.\ \cite{Borsanyi:2020mff,Lehner:2020crt}. Scrutiny of these results is
  ongoing (see e.g.\ Ref.\ \cite{Colangelo:2020lcg}) and further progress can be expected.\label{footnoteLQCD}
  }
The experimental measurements show the following deviations from the
updated theoretical SM prediction: 
	\begin{align} \label{eqn:MuonGm2Discrepancy}
           \Delta \amu^{\textrm{FNAL}} &= \update{(23.0 \pm 6.9) \times 10^{-10}} ,\\
           \Delta \amu^{\textrm{BNL}} &= (27.9 \pm 7.6) \times 10^{-10} ,\label{eqn:BNLDiscrepancy}\\
           \damuNEW &= \update{(25.1 \pm 5.9) \times 10^{-10}} .\label{eqn:avgDiscrepancy}
\end{align}
In each case the uncertainties are combined
by summing them in quadrature.  \update{In the last line $\damuNEW$ is the new, updated
deviation based on the experimental world average and the
SM White Paper result.}
The long standing
discrepancy between the BNL measurement and the SM theory prediction
\update{is confirmed and sharpened}. Its significance is increased
from \update{$3.7\sigma$} to  
 to \update{$4.2 \sigma$} by the combination with FNAL data.

This \update{improvement} has a significant impact on our understanding of BSM physics as it \update{strengthens} a major constraint on a variety of otherwise plausible SM extensions.
In this paper we provide a comprehensive overview of this impact the FNAL measurement has on BSM physics.
We examine
the impact in minimal 1-, 2- and 3-field extensions of the SM, and in the well-motivated Minimal Supersymmetric Standard Model
(MSSM).  Within this theoretical framework we provide a thorough overview of the impact the FNAL
measurement has and highlight promising scenarios that can
explain it.  In our investigation we use state-of-the-art $\amu$ calculations.
For the simple SM extensions we use FlexibleSUSY \cite{Athron:2014yba,Athron:2017fvs}, which includes the universal leading logarithmic two-loop QED contributions in addition to the full one-loop calculation.
For the MSSM we use \GMTCalc \cite{Athron:2015rva}, which
implements a dedicated high-precision MSSM calculation including two-loop and higher-order contributions based
on the on-shell scheme.

Reviews and general discussions of BSM contributions to $\amu$  have been given in
Refs.\ \cite{Melnikov:2006sr,Jegerlehner:2017gek,Jegerlehner:2009ry,Czarnecki2001,Stockinger:2006zn,Stockinger:1900zz,Lindner:2016bgg}. 
Previously the deviation from BNL has been studied extensively in the
literature.  There was intensive activity proposing BSM
explanations of the BNL result after its first discovery and in
following years 
\cite{Arnowitt:2001be,Czarnecki:2001pv,Baltz:2001ts,Everett:2001tq,Feng:2001tr,Chattopadhyay:2001vx,Choudhury:2001ad,Komine:2001fz,Mahanta:2001gg,Das:2001it,Cheung:2001ip,Kephart:2001iu,Ma:2001mr,Hisano:2001qz,Xing:2001qn,Ibrahim:2001ym,Ellis:2001yu,Dedes:2001hh,Einhorn:2001mf,Choi:2001pz,Kang:2001sq,Kim:2001se,Martin:2001st,Rajpoot:2001tz,deS.Pires:2001da,Komine:2001hy,Cheung:2001hz,Baek:2001nz,Raidal:2001pf,Carvalho:2001ex,Baer:2001kn,Chacko:2001xd,Wu:2001vq,Baek:2001kh,Chen:2001kn,Arhrib:2001xx,Enqvist:2001qz,Cerdeno:2001aj,Kim:2001eg,Blazek:2001zm,Cho:2001nfa,Barshay:2001eq,Arnowitt:2001pm,
 Belanger:2001am,deBoer:2001in,Roszkowski:2001sb,Daikoku:2001wm,Adhikari:2001sf,Wang:2001ig,deBoer:2001nu,Kersting:2001zz,Zhou:2001ew,Endo:2001ym,Cacciapaglia:2001pa,Ma:2001tb,
  Cho:2001hx,
  Park:2001uc,Kim:2001rc,Agashe:2001ra,Calmet:2001si,Appelquist:2001jz,
  Das:2004ay,
  Xiong:2001rt,Calmet:2001dc,Dai:2001vv,Yue:2001db,
  Tabbakh:2006zy,Blanke:2007db,
  Iltan:2001nk,Krawczyk:2001pe,Larios:2001ma,Krawczyk:2002df,
  Chakraverty:2001yg,Mahanta:2001yc,
  Huang:2001zx,Gninenko:2001hx,Lynch:2001zr,Baek:2001kca,Ma:2001md,Murakami:2001cs,
  Pospelov:2008zw,Heo:2008dq,
  Huang:2002dj,Kiritsis:2002aj,Das:2002en,Baek:2002cc,Chattopadhyay:2002jx,Byrne:2002cw,Kim:2002cy,Baek:2002wm,Martin:2002eu,Boyarkina:2003mr,Chavez:2004nr,Sawa:2005py,Cembranos:2005sr,Hambye:2006zn,Boyarkin:2008zz,Aguilar:2008qj,Hektor:2008xu,Domingo:2008bb,Biggio:2008in,
  Adachi:2009fw, Cheung:2009fc, Hofer:2009xb,Fukuyama:2009xk,Ho:2010yp,Crivellin:2010ty,
  Heinemeyer:2003dq,Heinemeyer:2004yq,Feng:2006ei,Marchetti:2008hw,vonWeitershausen:2010zr
  }.
Many ideas came under pressure from results at the LHC, and scenarios
were proposed which could resolve tensions between $\amu$, LHC results
and other constraints
\cite{
  Fargnoli:2013zda,Fargnoli:2013zia,
  Cho:2011rk,Endo:2011gy,Endo:2011mc,Endo:2011xq,Ibe:2012qu,Kim:2012fc,Endo:2013bba,Ibe:2013oha,Akula:2013ioa,Zhang:2013hva,Endo:2013lva,Bhattacharyya:2013xma,Evans:2013uza,Iwamoto:2014ywa,Kersten:2014xaa,Gogoladze:2014cha,Badziak:2014kea,Kowalska:2015zja,Chakrabortty:2015ika,
  Padley:2015uma,
  Bach:2015doa,Harigaya:2015jba,Chowdhury:2015rja,Khalil:2015wua,Ajaib:2015yma,Harigaya:2015kfa,Gogoladze:2015jua,Baez:2015sqj,Belyaev:2016oxy,Li:2016ucz,Okada:2016wlm,Kobakhidze:2016mdx,Belanger:2017vpq,Fukuyama:2016mqb
  ,Choudhury:2017acn,Hagiwara:2017lse,Endo:2020mqz,Chakraborti:2020vjp,Horigome:2021qof
  ,Yin:2016shg,Zhu:2016ncq,Hussain:2017fbp,Ning:2017dng,Frank:2017ohg,Bagnaschi:2017tru,Li:2017fbg,Pozzo:2018anw,Altin:2017sxx,Chakraborti:2017dpu,Yanagida:2017dao,Choudhury:2017fuu,Endo:2017zrj,Wang:2018vxp,Bhattacharyya:2018inr,Cox:2018qyi,Cox:2018vsv,Yang:2018guw,Tran:2018kxv,Wang:2018vrr, Abdughani:2019wai,Kotlarski:2019muo,Dong:2019iaf,Ibe:2019jbx,Yang:2020bmh,Yanagida:2020jzy,Han:2020exx,Cao:2019evo,Yamaguchi:2016oqz,Yanagida:2018eho,Shimizu:2015ara,Yin:2016pkz,Su:2020lrv
  ,Hundi:2012uf,Megias:2017dzd,
  Doff:2015nru,Hong:2016uou,
  Broggio:2014mna,Han:2015yys,Wang:2014sda,Ilisie:2015tra,Abe:2015oca,Chun:2015hsa,Wang:2016rvz,Chun:2016hzs,Abe:2017jqo,Cherchiglia:2016eui,Cherchiglia:2017uwv,Wang:2018hnw,Han:2018znu,Chun:2019oix,Iguro:2019sly,Wang:2019ngf,Sabatta:2019nfg,Jana:2020pxx,Botella:2020xzf,Rose:2020nxm,Li:2020dbg,Ghosh:2021jeg,Mondal:2021vou,
  Das:2016vkr,ColuccioLeskow:2016dox,Kowalska:2018ulj,Dorsner:2019itg,Crivellin:2020tsz,Gherardi:2020qhc,Babu:2020hun,Crivellin:2020mjs,
  Heeck:2011wj,Altmannshofer:2014pba,Altmannshofer:2016oaq,Belanger:2015nma,Altmannshofer:2016brv,Huang:2021nkl,Gninenko:2014pea,Araki:2015mya,Biswas:2016yjr,Kamada:2018zxi,Gninenko:2018tlp,Patra:2016shz,Amaral:2020tga,Bodas:2021fsy,Daikoku:2020nhr,
  Davoudiasl:2012qa,Davoudiasl:2012ig,Davoudiasl:2014kua,Lee:2014tba,Mohlabeng:2019vrz,
  Kannike2012,Dermisek:2013gta,Raby:2017igl,Poh:2017tfo,Kawamura:2020qxo,Frank:2020smf,deJesus:2020upp,Endo:2020tkb,Huang:2020ris,Chun:2020uzw,Dermisek:2020cod,Dermisek:2021ajd,
  Chen:2015vqy,Marciano:2016yhf,Bauer:2017nlg,Liu:2018xkx,Bauer:2019gfk,Cornella:2019uxs,Abdallah:2020biq,Escribano:2020wua,Buttazzo:2020vfs,
  Hundi:2011si,
  Huh:2013hga,Okada:2013ija,Kelso:2013zfa,
  Babu:2014lwa,Cogollo:2014tra,
  Ajaib:2015ika,Binh:2015jfz,Hektor:2015zba,Cogollo:2015fpa,Allanach:2015gkd,Chakrabortty:2015zpm,
  JinChun:2016onh, 
  Belanger:2016ywb, Nishida:2016lyk,
  Hong:2017tel,Crivellin:2018qmi,Li:2018aov,
  Liu:2018pdq, 
    Chen:2019nud,
  Badziak:2019gaf,Datta:2019bzu,
  Endo:2019bcj,CarcamoHernandez:2019ydc, 
  Liu:2020nsm,
  Calibbi:2020emz,Chen:2020jvl,Chua:2020dya,
  Saad:2020ihm,Acuna:2020ccz,Frank:2020kvp,Dutta:2020scq,
  Chen:2020tfr,
  Liu:2020ser,Dorsner:2020aaz,
  Nagai:2020xbq,Arbelaez:2020rbq,Abdullahi:2020nyr,
  Jana:2020joi,Chakrabarty:2020jro,Banerjee:2020zvi,Dinh:2020inx,Baryshevsky:2020aez,Ramsey-Musolf:2020ndm,
  Chen:2021rnl,
  Liu:2020sim,Aebischer:2021uvt,Cao:2021lmj,
  Fajfer:2021cxa,Yin:2021yqy,Greljo:2021xmg,
  Abdullah:2019ofw,Baker:2021yli,
  Freitas:2014pua,Calibbi:2014yha,Queiroz:2014zfa, 
Biggio:2016wyy, 
Biggio:2014ela,Kowalska:2017iqv,Athron:2017drj,
  Banerjee:2018eaf,Chiang:2018bnu,
Calibbi:2018rzv  
}. 
Many of
these constructions use supersymmetry in some way and will be
discussed in our Section \ref{sec:MSSM}, but this list also includes
solutions motivated by extra dimensions
\cite{Park:2001uc,Kim:2001rc,Agashe:2001ra,Calmet:2001si,Das:2004ay,Hundi:2012uf,Megias:2017dzd,Appelquist:2001jz}
and technicolor or compositeness
\cite{Xiong:2001rt,Calmet:2001dc,Dai:2001vv,Yue:2001db,Doff:2015nru,Tabbakh:2006zy,Hong:2016uou,Blanke:2007db},
or even introducing unparticle physics \cite{Hektor:2008xu}, as well
as just extending models with new states like the two-Higgs doublet
models
\cite{Iltan:2001nk,Krawczyk:2001pe,Krawczyk:2002df,Larios:2001ma,Broggio:2014mna,Wang:2014sda,Han:2015yys,Abe:2015oca,Ilisie:2015tra,Chun:2015hsa,Wang:2016rvz,Chun:2016hzs,Abe:2017jqo,Cherchiglia:2016eui,Cherchiglia:2017uwv,Wang:2018hnw,Han:2018znu,Iguro:2019sly,Wang:2019ngf,Chun:2019oix,Sabatta:2019nfg,Jana:2020pxx,Botella:2020xzf,Rose:2020nxm,Li:2020dbg,Ghosh:2021jeg,Mondal:2021vou}
or adding leptoquarks
\cite{Chakraverty:2001yg,Mahanta:2001yc,Das:2016vkr,ColuccioLeskow:2016dox,Kowalska:2018ulj,Dorsner:2019itg,Crivellin:2020tsz,Gherardi:2020qhc,Babu:2020hun,Crivellin:2020mjs},
new gauge bosons (including sub-GeV gauge bosons, dark photons and generalizations)
\cite{Huang:2001zx,Gninenko:2001hx,Lynch:2001zr,Murakami:2001cs,Pospelov:2008zw,Heo:2008dq,Heeck:2011wj,Davoudiasl:2012qa,Davoudiasl:2012ig,Davoudiasl:2014kua,Lee:2014tba,Gninenko:2014pea,Araki:2015mya,Belanger:2015nma,Altmannshofer:2016brv,Baek:2001kca,Ma:2001md,Altmannshofer:2016oaq,Patra:2016shz,Biswas:2016yjr,Kamada:2018zxi,Gninenko:2018tlp,Altmannshofer:2014pba,Mohlabeng:2019vrz,Amaral:2020tga,Bodas:2021fsy,Daikoku:2020nhr,Huang:2021nkl},
Higgs triplets \cite{Fukuyama:2009xk,Ho:2010yp} and vector-like
leptons
\cite{Davoudiasl:2012ig,Kannike2012,Dermisek:2013gta,Raby:2017igl,Poh:2017tfo,Kawamura:2020qxo,Frank:2020smf,deJesus:2020upp,Endo:2020tkb,Huang:2020ris,Chun:2020uzw,Dermisek:2020cod,Dermisek:2021ajd},
or very light,
neutral and weakly interacting scalar particles \cite{Chen:2015vqy,Marciano:2016yhf,Bauer:2017nlg,Liu:2018xkx,Bauer:2019gfk,Cornella:2019uxs,Abdallah:2020biq,Escribano:2020wua,Buttazzo:2020vfs}.
Some works have taken
a systematic approach, classifying states according to
representations and investigating a large set of them
\cite{Freitas:2014pua,Calibbi:2014yha,Queiroz:2014zfa,Biggio:2014ela,Biggio:2016wyy,Kowalska:2017iqv,Athron:2017drj,
  Banerjee:2018eaf,Chiang:2018bnu,Calibbi:2018rzv,Biggio:2008in}
\footnote{\update{Finally on the same day as the release of the FNAL result a very large number of papers were already released interpreting it \cite{das2021fimpwimp, buenabad2021challenges,wang2021gutscale,abdughani2021common,chen2021muon,ge2021probing,cadeddu2021muon,brdar2021semisecretly,cao2021imporved,chakraborti2021new,ibe2021muon,cox2021muon, babu2021muon,han2021muon, heinemeyer2021new,calibbi2021implications,amaral2021distinguishing,bai2021muon,baum2021tiny, yin2021muon,anselmi2021fake, nomura2021explanations, vanbeekveld2021dark,wang2021revisiting,gu2021heavy,zhu2021probing,criado2021confronting, arcadi2021muon,han2021leptonspecific,iwamoto2021winohiggsino,endo2021supersymmetric,crivellin2021consequences,hernandez2021fermion,Chakraborti:2021bmv}.  This demonstrates what a landmark result this is and the intense interest it is generating within the particle physics community.}}.

The deviation found already by the BNL measurement  also gave
rise to the question whether \update{it} could be due to
hypothetical, additional contributions to the hadronic vacuum
polarization.\footnote{%
  This question is further motivated by lattice QCD results on the
  hadronic vacuum polarization, see footnote \ref{footnoteLQCD}.
}
If such additional effects would exist, they could
indeed shift the SM
prediction for $\amu$ towards the \update{experimental value}, but would at the
same time worsen the fit for electroweak precision observables,
disfavouring such an explanation of the deviation  \cite{Passera:2008jk,Crivellin:2020zul,Keshavarzi:2020bfy,deRafael:2020uif,Malaescu:2020zuc}.

In spite of the vast number of works and the many varied
ideas, for most models the same general principles apply. Typically
the deviation requires new states with masses below the TeV scale or
not much above $1$ TeV to
explain the \update{experimental value} with perturbative couplings.
The models which allow large $\amu$ with particularly large masses
involve very large couplings and/or
introduce enhancements through new sources of muon chirality
flips (as we will describe in the next 
section).  
Therefore the absence of BSM signals at the LHC has led to tensions
with large $\amu$ in many models: either very large couplings and
heavy masses are needed or the stringent LHC limits have to be evaded
in other ways.

Not only LHC, but also dark matter searches can lead to tensions in many
models. The Planck experiment 
\cite{Aghanim:2018eyx,Tanabashi2018} observed the dark matter abundance 
of the universe to be:
\begin{equation} \label{eqn:DMRD}
	\Omega_{h^2}=0.1200\pm0.001.
\end{equation} 
Since BSM contributions to $\amu$ are often mediated by new weakly
interacting neutral particles, many interesting models also contain
dark matter candidate particles. Any dark matter candidate particle
with a relic density more than $0.12$ is over 
abundant and therefore strongly excluded.  
Further,  the negative results of
direct dark matter searches can lead to strong additional constraints on the
model parameter spaces.

In the present work we aim to provide a comprehensive picture
of the impact the new FNAL measurement has on BSM physics.
The models we investigate in detail represent a wide range of
possibilities. They cover models with new strongly or weakly
interacting particles, with extra Higgs or SUSY particles,
with or without a dark matter candidate, with or
without new chirality flips and with strong or weak constraints from the
LHC.
In all cases we provide a detailed description of the mechanisms
for the contributions to $\amu$; we then carry out detailed
investigations of the model parameter spaces, including 
applicable constraints from the LHC and dark
matter using state-of-the-art tools for evaluating constraints and LHC
recasting.
This allows us to answer which models and which model scenarios can
accommodate \update{the new FNAL measurement and the deviation
  $\damuNEW$} while satisfying all other constraints.

The rest of this paper is as follows.  In Sec.\ \ref{sec:BSMoverview}
we explain how the anomalous magnetic moment appears in quantum field
theories and emphasise the most important aspects which both make it
an excellent probe of BSM physics and make the observed anomaly very
difficult to explain simultaneously with current collider limits on
new physics. In Secs.\ \ref{sec:SingleField}, \ref{sec:TwoFields}, and
\ref{sec:ThreeFields} we present results for minimal 1-, 2- and
3-field extensions of the SM respectively that show the impact the new
FNAL result has on these models.  To provide a more global picture for
1- and 2-field extensions, in Sec.\ \ref{sec:one_field_overview} and
Sec.\ \ref{sec:two_field_overview} we also classify models of this
type systematically by quantum numbers and use known results to
summarise their status with respect to \update{explaining the BNL}
result, showing that this measurement severely restricts the set of
possible models.  This allows us to select models \update{with the best
  prospects} for detailed investigation, presenting results for the
two-Higgs doublet model (Sec.\ \ref{Sec:THDM}),
leptoquark models (Sec.\ \ref{Sec:Leptoquarks}) and two field extensions
with scalar singlet dark matter (Sec.\ \ref{sec:Z2symmetric}).  For three field models we perform a
detailed examination of models with mixed scalar singlet and doublet
dark matter (Sec.\ \ref{sec:2S1F}) and fermion singlet and doublet
dark matter (Sec.\ \ref{sec:2F1S}).  In Sec.\ \ref{sec:MSSM} we
discuss the impact of \update{the sharpened deviation} on the MSSM, which is
widely considered one of the best motivated extensions of the SM. This
section also contains a brief self-contained discussion of $\amu$ and the possible enhancement mechanisms in
the MSSM and explains in
detail our treatment of dark matter data and 
LHC recasting. All constraints are then applied
on the general MSSM, allowing all kinds of neutralino LSPs. 
Finally we present our conclusions in Section \ref{sec:Conclusions}.

\input{SectionOverview}

\input{SectionSingleFields}
\input{SectionTwoFields}
\input{three_fields}

\input{three_fields_2F1S}
\input{SectionSUSY}

	\section{Conclusions} \label{sec:Conclusions}

15 years after the BNL $\amu$ measurement showed a tantalizing
deviation from the SM theory value and following
tremendous theoretical work on improving and stabilizing the SM
prediction, the Fermilab E989 experiment has published its first 
measurement of $\amu$. The result is \update{a strong
  confirmation of the BNL result and the existence of a deviation to
  the SM. After including the FNAL result, the new world average results in
$\damuNEW=25.1\times10^{-10}$, a $4.2\sigma$ deviation. It strengthens
  the indications for the existence of BSM physics in the lepton
sector, possibly related to the muon mass generation mechanism.}

Which BSM scenarios can accommodate the new Fermilab $\amu$
measurement, and what are the required parameter values?  The present
paper provides a detailed survey of possible explanations to answer
this question. We focused on renormalizable models which were already
promising in view of the previous BNL result. We asked particularly in
what parameter space the models can accommodate the $\amu$ results,
taking into account up-to-date constraints from LHC and dark matter
searches, as well as other relevant constraints.  Our survey covered
simple extensions of the SM by one, two, or three new fields (required
to be full SM gauge multiplets), including e.g.\ leptoquark and
Two-Higgs doublet models, as well as a general version of the
MSSM. Secs.\ \ref{sec:SingleField}, \ref{sec:TwoFields} also contain
detailed overviews of the status of models, and 
summaries of our phenomenological results can 
be found at the end of each section.

A useful background information is that
the observed deviation $\damuNEW$ is \update{larger than } the electroweak contributions $\amu ^{\text{EW}}=15.36(0.10)\times10^{-10}$. BSM contributions to $\amu$ are typically suppressed as
$1/M_{\text{BSM}}^2$. Hence BSM models explaining
the deviation must have nontrivial properties, and many models are excluded as
explanations.
The common nontrivial feature of most viable explanations is enhancements in the
muon left--right chirality flip via new couplings and interactions.
  As explained around Eq.~(\ref{eqn:GeneralGM2Contribution}) the
  chirality flip enhancement is strongly related to the muon mass
  generation mechanism and causes related loop contributions to the
  muon mass.
As a side note, one obtains
a quite model-independent order-of-magnitude relationship: models in
which the muon mass correction does not exceed $100\%$ can 
explain $ \damuNEW$ only for
$M_{\text{BSM}}^{\text{FNAL}}\lesssim\update{2.1\text{\ TeV}}$ according to Eq.\ (\ref{FNALuppermassbound}).
But even in such models with chirality flip
enhancements an \update{explanation of $\damuNEW$} requires specific,
often ``non-traditional''
regions of parameter space.

Examples of excluded models include the red highlighted
models in Tables 
\ref{tab:one-field-summary}, \ref{tab:TwoFieldsDifferentSpin}, many
versions of leptoquark models or the
familiar type I, II versions of the 2HDM.
In the context of supersymmetry, familiar scenarios such as the
Constrained MSSM cannot explain the deviation.
Certain leptoquark and vector-like lepton models are examples which
generate very large chiral enhancements, but they need non-flavour universal
couplings, e.g.\ direct leptoquark couplings of muon to top-quark in the left-
and right-handed sector. 

An important outcome of the present study is that once the $\amu$ result
is combined with current data from LHC and dark matter
experiments, we obtain strong constraints on the detailed ways how BSM models
can be realized. The only viable models of the kind discussed in Sec.\ \ref{sec:TwoFields} that do not have chirality enhancements, are
particularly strongly constrained. \update{In these models, LHC data
  and $\damuNEW$ can be accommodated simultaneously in a small slice of
  parameter space, however it is impossible to also account for the full dark matter relic
  density, see Figs.\ \ref{fig:Min2FieldsLLSlice}--\ref{fig:Min2FieldsRRProfile}.} In contrast the simple 3-field models 2F1S and 2S1F
of Sec.\ \ref{sec:ThreeFields} are least constrained and can
accommodate $\amu$ and dark matter in a wide parameter region.  The required values of the new coupling constants, however, are large and
\update{it remains to be seen how such scenarios can arise in more complete theories.}

The 2HDM can accommodate \update{the observed} $\damuNEW$ 
while preserving minimal flavour violation, but only in the
lepton-specific type X version or the generalized flavour-aligned
2HDM. Even in these scenarios only a tiny parameter space remains,
where the new Yukawa couplings are close to 
their upper experimental limits and the new Higgs masses in a very
narrow range below $100$ GeV, see Fig.\ \ref{fig:THDMupdate}. Leptoquark masses are strongly
constrained by LHC to be significantly \update{above $1$ TeV}, pushing
the explanation of $\damuNEW$ close to the region violating the fine
tuning criterion on the muon mass, see Figs.\ \ref{fig:ScalarLeptoquarkSinglet}--\ref{fig:ScalarLeptoquarkProfiles}.

For SUSY scenarios, the general tension between LHC mass limits and
explanations of $\damuNEW$ is illustrated in
Fig.\ \ref{fig:briefsurveyplot}.
Still, the MSSM with Bino-like LSP and either
stau/slepton-coannihilation or chargino coannihilation can fully
explain $\damuNEW$ and the dark matter relic density, see
Fig.\ \ref{fig:Binosleptonscompressed}. Apart from the
constraints implied by the coannihilation mechanisms, \update{the parameter
space is wide open in the $M_1$--$\mu$ plane.} LHC and dark matter constraints however imply mass patterns, e.g.\ lower limits on the Higgsino
mass and two windows for the Wino mass. The case where both charginos are lighter than sleptons is very
strongly constrained and may be excluded by future data from $\amu$,
LHC and dark matter experiments (Fig.\ \ref{fig:chalightersleptonsa}). Further, in the largest part of viable parameter
space the simple GUT constraint on gaugino masses $M_1/M_2\approx1/2$
is strongly violated. The scenarios with a Higgsino-like LSP are strongly
constrained and can
accommodate the current $\amu$ \update{only in a limited range of masses}. Scenarios with Wino-like LSP emerge
as particularly interesting (Figs.\ \ref{fig:WHsleptonscompressed},
\ref{fig:chalightersleptonsb}). They can accommodate $\amu$ in a vast 
parameter space without significant constraints; however in these
scenarios the dark matter relic density can only be partially
explained without additional contributions to the relic density.

These results and discussions highlight the importance of $\amu$ not
only as a potential proof of BSM physics but also of a crucial constraint on
models.
Particularly in combination with current LHC and dark matter data, it
points to specific parameter regions of models and gives crucial clues
on how BSM physics can (or cannot) be realized.
Since many models involve muon chirality flip enhancements and/or
flavour non-universality, further experiments testing lepton flavour
violation, electric dipole moments, or lepton universality are
promising to uncover further properties of  BSM physics. Since the muon
chirality flip enhancements are related to the mass generation
mechanism for the muon, also
the measurement of the  Higgs--muon coupling at LHC or future lepton
colliders can provide \update{a test 
of various explanations of $\damuNEW$.} Further \update{ultimate} tests
may be performed at a multi-TeV muon collider
\cite{Capdevilla:2020qel,Buttazzo:2020eyl,Yin:2020afe,Capdevilla:2021rwo,Cheung:2021iev,Han:2021udl}.

\update{The new Fermilab $\amu$ measurement provides the best possible starting point for
  future $\amu$ determinations.}
Exciting further progress can be expected from the Run-2--4 results of
the FNAL $g-2$ experiment, the planned JPARC $g-2$ experiment
\cite{Mibe:2010zz,Abe:2019thb}, and from further progress on SM theory including the MUonE
initiative to provide alternative experimental input to the
determination of the hadronic contributions to $\amu$
\cite{Abbiendi:2016xup,Banerjee:2020tdt}.

 \section*{Acknowledgements}
 We thank the ColliderBit group from GAMBIT (and in particular Anders
 Kvellestad) for helpful discussions about available analyses and the
 statistical interpretation of their likelihoods and we also thank
 Felix Kahlhoefer for helpful comments on the statistical
 interpretation of both ColliderBit and DDCalc likelihoods. We also
 thank Adriano Cherchiglia for the consent to use internal results and data
 of Ref.\ \cite{Cherchiglia:2017uwv}. H.S-K. and D.S thank Hyesung Lee
 for the helpful discussion and comments on dark photon models. P.A.\ and D.J.\ thank Innes Bigaran for helpful discussions regarding leptoquark constraints applied in Ref.\ \cite{Bigaran:2020jil}.  D.J. thanks Susanne Westhoff for helpful clarifications regarding details about Ref.\ \cite{Freitas:2014pua}, and Ursula Laa for her helpful support with \smodels and \micromegas.  The work of P.A.\ is supported by the Australian Research Council Future Fellowship grant FT160100274.  P.A. also acknowledges the hospitality of Nanjing Normal University while working on this manuscript.  The work of P.A. and C.B. is also supported with the Australian Research Council Discovery Project grant DP180102209.  The work of C.B. was supported by the Australian Research Council through the ARC Centre of Excellence for Particle Physics at the Tera-scale CE110001104.
 This project was also undertaken with the assistance of resources and services from the National Computational Infrastructure, which is supported by the Australian Government.  We thank Astronomy Australia Limited for financial support of computing resources, and the Astronomy Supercomputer Time Allocation Committee for its generous grant of computing time. The work has further been supported by the high-performance computing cluster Taurus at ZIH, TU Dresden.
 The research placement of D.J. under which this work was done was
 supported by the Australian Government Research Training Program
 (RTP) Scholarship and the Deutscher Akademischer Austauschdienst (DAAD) One-Year Research Grant. The work
 of H.S.\ and D.S.\ is supported by DFG grant STO 876/7-1.

	\appendix

	\section{General $\amu$ Contributions} \label{app:MuonGm2Contributions}

\begin{figure}[tb]
		\centering
		\begin{subfigure}[t]{.25\textwidth}
			\includegraphics[width=!, height=0.15\textheight]{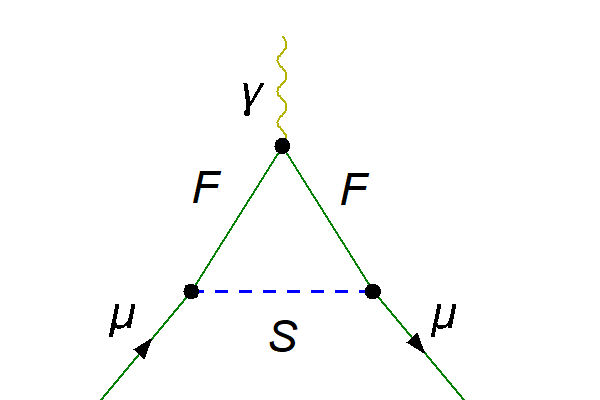}
			\caption{ \label{fig:FFSDiagram}}
		\end{subfigure}
		\hspace{0.05\linewidth}
		\begin{subfigure}[t]{.25\textwidth}
			\includegraphics[width=!, height=0.15\textheight]{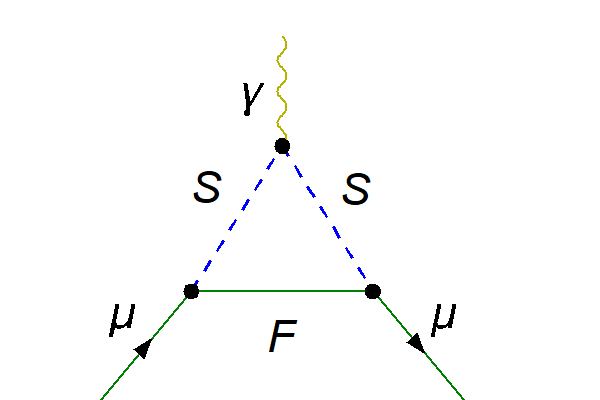}
			\caption{ \label{fig:SSFDiagram}}
		\end{subfigure}
		\hspace{0.05\linewidth}
		\begin{subfigure}[t]{.25\textwidth}
			\includegraphics[width=!, height=0.15\textheight]{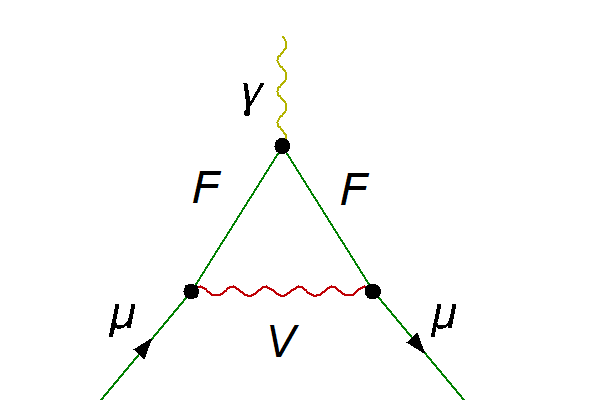}
			\caption{ \label{fig:FFVDiagram}}
		\end{subfigure}
		\begin{subfigure}[t]{.25\textwidth}
			\includegraphics[width=!, height=0.15\textheight]{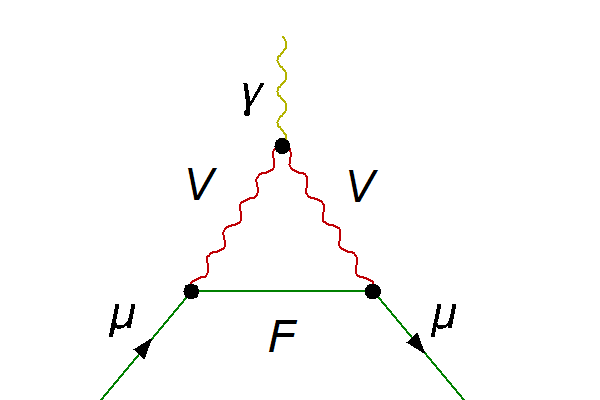}
			\caption{ \label{fig:VVFDiagram}}
		\end{subfigure}
		\hspace{0.05\linewidth}
		\begin{subfigure}[t]{.25\textwidth}
			\includegraphics[width=!, height=0.15\textheight]{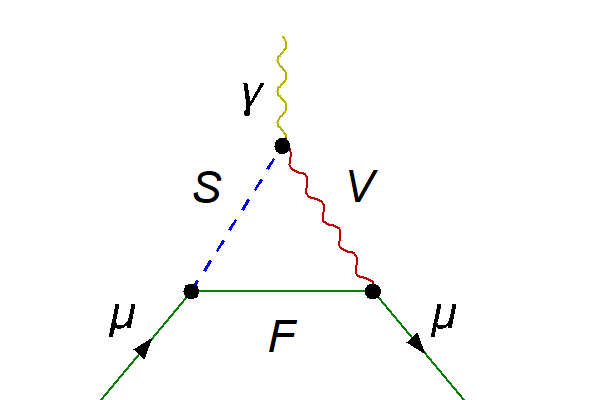}
			\caption{ \label{fig:VSFDiagram}}
		\end{subfigure}
		\hspace{0.05\linewidth}
		\begin{subfigure}[t]{.25\textwidth}
			\includegraphics[width=!, height=0.15\textheight]{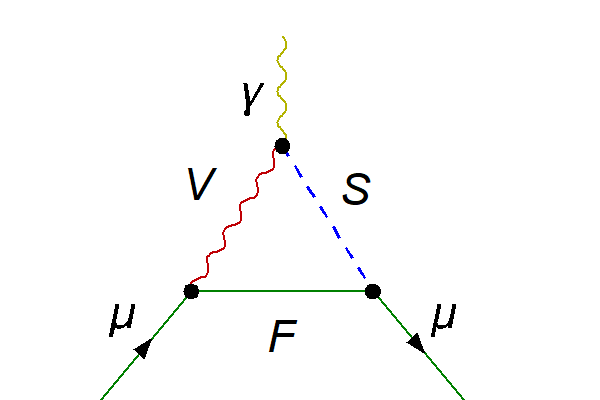}
			\caption{ \label{fig:SVFDiagram}}
		\end{subfigure}
		\caption{Diagrams producing general contributions to $\amu$. Reading left-to-right, top-to-bottom, the FFS diagram \ref{fig:FFSDiagram}, the SSF diagram \ref{fig:SSFDiagram}, the FFV diagram \ref{fig:FFVDiagram}, the VVF diagram \ref{fig:VVFDiagram}, the VSF diagram \ref{fig:VSFDiagram}, and the SVF diagram \ref{fig:SVFDiagram}. \label{fig:GeneralDiagrams}}
	\end{figure}

	Here we collect generic one-loop results for BSM contributions to $\amu$. We 
	consider a generic new fermion $F$, a new scalar $S$ and a new vector field $V$ 
	with masses $m_{F,S,V}$.  The muon can have the following general couplings to SM 
	and these BSM particles:
	\begin{equation} \label{eqn:GeneralCouplings}
		{\cal L}_\text{muon} = \lambda_L \overline{F_R}.S.L + \lambda_R \overline{F_L}.S.\mu + g_L \overline{F_L} \gamma^\nu V_\nu L + g_R \overline{F_R} \gamma^\nu V_\nu \mu + h.c.,
	\end{equation}
	where $L = (v_{\mu_L},\mu_L)^T$ corresponds to the $SU(2)_L$ left-handed lepton 
	doublet,
    and $\mu = \mu^\dagger_R$ corresponds to the conjugate of the right-handed muon.  From this Lagrangian, we can generate $6$ generic one-loop contributions, as shown 
    in Fig.\ \ref{fig:GeneralDiagrams}, where charge flows from left-to-right and the 
    fermions and scalars in the loop are assumed to have a negative charge.  Using 
    these couplings one can calculate (through Package-X in Ref.\ \cite{Patel2017})the general one-loop contributions in the $\xi=1$ (t'Hooft Feynman) gauge of the general R-$\xi$ class of gauges 
    (see Ref.\ \cite{Fujikawa1972}) read in the heavy BSM limit $M_{BSM}\gg m_\mu$:
%
    \begin{align} \label{eqn:GeneralFFSContribution}
		a^{\text{FFS}}_\mu &= \frac{Q_F m_\mu^2}{16 \pi^2 m_S^2} \bigg(
		- \frac{2 \lambda_L \lambda_R}{3} \frac{m_F}{m_\mu} \,F\!\begin{pmatrix}\frac{m_F^2}{m_S^2}\end{pmatrix}
		- \frac{(\lambda_L^2+\lambda_R^2)}{12} \,E\!\begin{pmatrix}\frac{m_F^2}{m_S^2}\end{pmatrix}
		\bigg),
\\ \label{eqn:GeneralSSFContribution}
		a^{\text{SSF}}_\mu &= \frac{Q_S m_\mu^2}{16 \pi^2 m_S^2} \bigg(
		  \frac{\lambda_L \lambda_R}{3} \frac{m_F}{m_\mu} \,C\!\begin{pmatrix}\frac{m_F^2}{m_S^2}\end{pmatrix}
		+ \frac{(\lambda_L^2+\lambda_R^2)}{12} \,B\!\begin{pmatrix}\frac{m_F^2}{m_S^2}\end{pmatrix}
		\bigg),
\\ \label{eqn:GeneralFFVContribution}
		a^{\text{FFV}}_\mu &= \frac{Q_F m_\mu^2}{16 \pi^2 m_V^2} \bigg(
		  \frac{4 (g_L g_R)}{3} \frac{m_F}{m_\mu} \,C\!\begin{pmatrix}\frac{m_F^2}{m_V^2}\end{pmatrix}
		- \frac{(g_L^2+g_R^2)}{3} \,M\!\begin{pmatrix}\frac{m_F^2}{m_V^2}\end{pmatrix}
		  \bigg),
\\ \label{eqn:GeneralVVFContribution}
		a^{\text{VVF}}_\mu &= \frac{Q_V m_\mu^2}{16 \pi^2 m_V^2} \bigg(
		- 3 (g_L g_R) \frac{m_F}{m_\mu} \,K\!\begin{pmatrix}\frac{m_F^2}{m_V^2}\end{pmatrix}
		- \frac{(g_L^2+g_R^2)}{6} \,J\!\begin{pmatrix}\frac{m_F^2}{m_V^2}\end{pmatrix}
		\bigg),
\\ \label{eqn:GeneralVSFContribution}
		a^{\text{VSF}}_\mu &= \frac{Q_V m_\mu^2}{16 \pi^2 m_V} \bigg(
		\frac{(\lambda_R g_L + \lambda_L g_R)}{2 m_\mu} \,N\!\begin{pmatrix}\frac{m_F^2}{m_V^2},\frac{m_F^2}{m_S^2}\end{pmatrix}
		\bigg),
\\ \label{eqn:GeneralSVFContribution}
		a^{\text{SVF}}_\mu &= \frac{Q_V m_\mu^2}{16 \pi^2 m_V} \bigg(
		\frac{(\lambda_L g_L + \lambda_R g_R)}{2 m_\mu} \,N\!\begin{pmatrix}\frac{m_F^2}{m_V^2},\frac{m_F^2}{m_S^2}\end{pmatrix}
		\bigg).
	\end{align}
    The one-loop functions are defined as:\footnote{%
      The loop functions are defined in agreement to loop functions familiar from the
      MSSM literature, see e.g.\ Ref.\ \cite{Fargnoli:2013zia} also
      for relationships to further loop functions. The SSF loop
      functions correspond to MSSM diagrams involving neutralinos and satisfy
      $B(x)=F_1^N(x)$, $C(x)=F_2^N(x)=6G_4(x)$. The FFS loop functions
      correspond to loops involving charginos and satisfy
      $E(x)=F_1^C(x)$ and $F(x)=F_2^C(x)=3G_3(x)$.} 
	\begin{align} \label{eqn:OneLoopFunctionB}
		B(x) &= \frac{2(1-6x+3x^2+2x^3-6x^2\log{x})}{(1-x)^4},
\\ \label{eqn:OneLoopFunctionC}
		C(x) &= \frac{3(1-x^2+2x\log{x})}{(1-x)^3},
\\ \label{eqn:OneLoopFunctionE}
		E(x) &= \frac{2(2+3x-6x^2+x^3+6x\log{x})}{(1-x)^4},
\\ \label{eqn:OneLoopFunctionF}
		F(x) &= \frac{3(-3+4x-x^2-2\log{x})}{2(1-x)^3},
\\ \label{eqn:OneLoopFunctionJ}
		J(x) &= \frac{7-33x+57x^2-31x^3+6x^2(3x-1)\log{x}}{(1-x)^4},
\\ \label{eqn:OneLoopFunctionK}
		K(x) &= \frac{(1-4x+3x^2-2x^2\log{x})}{(1-x)^3},
\\ \label{eqn:OneLoopFunctionM}
		M(x) &= \frac{(4-9x+5x^3+6(1-2x)x\log{x})}{(1-x)^4},
\\ \label{eqn:OneLoopFunctionN}
		N(x,y) &= \frac{y}{(y-1)} \frac{x (x-y)^2 \log{x}+(x-1) ((x-y)*(y-1)-(x-1) x \log{(x/y)})}{(1-x)^2(x-y)^2}.
	\end{align}
	Here $x = m_F^2 / m_S^2$ for FFS and SSF diagrams, $x = m_F^2
        / m_V^2$ for FFV, VVF, VSF, and SVF diagrams, and $y =
        m_F^2/m_S^2$ for VSF and SVF diagrams.  The above one-loop
        functions with one argument have the limits $\lim_{x\rightarrow 1}$OneLoop$(x) = 1$ (Except for $J(x)$, $K(x)$, and $M(x)$ which have the limits $7/5$, $2/3$, and $3/2$) and $\lim_{x\rightarrow \infty}$OneLoop$(x) = 0$, and have the following limits as $x \rightarrow 0$:

	\begin{align*}
		\underset{x\rightarrow 0}{\text{lim }} B(x) \rightarrow 2, &&
		\underset{x\rightarrow 0}{\text{lim }} C(x) \rightarrow 3, &&
		\underset{x\rightarrow 0}{\text{lim }} E(x) \rightarrow 4, &&
		\underset{x\rightarrow 0}{\text{lim }} F(x) \rightarrow \infty, \\
		\underset{x\rightarrow 0}{\text{lim }} J(x) \rightarrow 7, &&
		\underset{x\rightarrow 0}{\text{lim }} K(x) \rightarrow 1, &&
		\underset{x\rightarrow 0}{\text{lim }} M(x) \rightarrow 4.
	\end{align*}

        The formulas needed e.g.\ for SUSY diagrams in mass-insertion
        approximation are given by
        \begin{subequations}\label{FaDefinition}
          \begin{align}
  F_a(x,y) &= -\frac{F(x)-F(y)}{3(x-y)},  \\
  F_b(x,y) &= -\frac{C(x)-C(y)}{6(x-y)},\\
  \tilde{F}_a(x,y) &= -\frac{x F(x)- y F(y)}{3\left(\frac{1}{x}-\frac{1}{y}\right)}.
          \end{align}
          They have the limits $F_a(1,1)=1/4$ and
          $F_b(1,1)=\tilde{F}_a(1,1)=1/12$.
\end{subequations}

\input{SectionSUSYAppendix}

	\bibliographystyle{jhep}
	\bibliography{DouglasJacobMuonMoment,DouglasJacobMuonMomentElectroweak,DouglasJacobMuonMomentQED,DouglasJacobMuonMomentExperiment,DouglasJacobMuonMomentHadronic,DouglasJacobMuonMomentMSSM,DouglasJacobMuonMoment2HDM,DouglasJacobMuonMomentLeptoquarks,DouglasJacobPackages,DouglasJacobDarkMatter,DouglasJacobATLAS,DouglasJacobCMS,DouglasJacobLEP,CodeReferences,SectionSUSYbiblio,TheoryWPbiblio,Brookhaven_2001_BSM_solutions,Brookhaven_final_2006_BSM_solutions,Brookhaven_final_2006_BSM_solutions2,OnTheDay}

\end{document}

%% file: SectionOverview.tex

\section{Muon $g-2$ and physics beyond the SM}
\label{sec:BSMoverview}

In quantum field theory, the anomalous magnetic moment of the muon is
given by
\begin{align}
\amu &= -2m_\mu F_M(0)\,,
\label{amuFM}
\end{align}
where $m_\mu$ is the muon pole mass and $F_M(0)$ is the zero-momentum
limit of the magnetic moment form factor. The latter is defined via
the covariant decomposition of the 1-particle irreducible
muon--muon--photon vertex function $\Gamma^\mu(p,-p',q)$,
\begin{align}
  \label{covdecomp}
\bar u(p')\Gamma^{\mu}(p,-p',q) u(p) & =
-eQ\,
\bar u(p')\left[\gamma^\mu F_V(q^2) + (p+p')^\mu F_M(q^2) +
  \ldots\right] u(p)
\end{align}
with the on-shell renormalized electric charge $e$, $Q=-1$, on-shell momenta $p^2=p'{}^2=m_\mu^2$, on-shell spinors $u(p)$,
$u(p')$ and $q=p'-p$. The quantum field theory operator corresponding to $\amu$ connects
left- and right-handed muons, i.e.\ it involves a chirality flip. 

The observable $\amu$ is CP-conserving, flavour conserving, loop induced, and chirality flipping. 
These properties make it complementary to many other precision and collider observables.
In particular the need for a muon chirality flip has a pivotal influence on the BSM phenomenology of $\amu$. It requires
two ingredients.
\begin{itemize}
\item
  Breaking of chiral symmetry. There must be a theory parameter
  breaking the chiral symmetry under which the left- and right-handed
  muon fields transform with opposite phases. In the SM and the MSSM
  and many other models this chiral symmetry is broken only by the
  non-vanishing muon Yukawa coupling $y_\mu$.\footnote{%
    For the MSSM this statement is true if one follows the customary
    approach to parametrize the trilinear scalar soft SUSY-breaking
    parameters as $T_f\equiv A_f y_f$ by explicitly factoring out the
    respective Yukawa couplings.}
  In all these cases contributions to $F_M(0)$ are proportional to at
  least one power of the muon Yukawa coupling, where e.g.\ the MSSM Yukawa
  coupling is enhanced compared to the SM one.

  In some models, there are additional sources of breaking of the muon
  chiral symmetry. Examples are provided by the leptoquark model
  discussed below in Sec.\ \ref{Sec:Leptoquarks}, where the
  simultaneous presence of left- and right-handed couplings
  $\lambda_{L,R}$ and the charm- or top-Yukawa coupling breaks the
  muon chiral symmetry and leads to contributions governed by
  $\lambda_L\lambda_R m_{c,t}$. Similar mechanisms can also exist in
  the three-field models discussed below.
\item
  Spontaneous breaking of electroweak gauge invariance.
  Since the $\amu$ operator connects a left-handed lepton doublet and
  right-handed lepton singlet it is not invariant under electroweak
  (EW) gauge transformations. Hence any contribution to $F_M(0)$ also must
  be proportional to at least one power of some vacuum expectation
  value (VEV) breaking EW gauge invariance. In the SM, there is only
  a single VEV $v$, so together with the required chirality flip, each
  SM-contribution to $F_M(0)$ must be proportional to
  $y_\mu v$ and thus to the tree-level muon mass. However, e.g.\ in the
  MSSM, there are two VEVs $v_{u,d}$; hence there are contributions to
  $\amu$ governed by $y_\mu v_u$, while the tree-level muon mass is
  given via $y_\mu v_d$. This leads to the well-known enhancement by
  $\tan\beta=v_u/v_d$.
\end{itemize}
In addition, the gauge invariant operators contributing to $\amu$ are (at least) of dimension six; hence any BSM contribution to $\amu$ is suppressed by (at least) two powers of a typical BSM mass scale. 
  In conclusion, BSM contributions to $\amu$ can generically be
parametrized as
\begin{align} \label{eqn:GeneralGM2Contribution}
  \damu^{\text{BSM}} &=
  C_{\text{BSM}}\frac{m_\mu^2}{M_{\text{BSM}}^2} ,
\end{align}
where $M_{\text{BSM}}$ is the relevant mass scale and where
the coefficient $C_{\text{BSM}}$ depends on all model details
like origins of chirality flips and electroweak VEVs as well as further
BSM coupling strengths and loop factors.\footnote{We note that
  Eq.\ (\ref{eqn:GeneralGM2Contribution}) does not imply the naive
  scaling
  $\Delta a_e^{\text{BSM}}:\damu^{\text{BSM}}:\Delta a_\tau^{\text{BSM}}\approx
  m_e^2:m_\mu^2:m_\tau^2$ with the lepton generation since the
  coefficient $C_{\text{BSM}}$ does not have to be
  generation-independent. Still, the prefactor
  $m_l^2/M_{\text{BSM}}^2$ in $a_l$ implies that the muon magnetic
  moment is more sensitive to BSM physics than the electron magnetic
  moment and that typical models which explain e.g.\ the BNL deviation
  for $\amu$ give negligible contributions to $a_e$.
  For detailed discussions and examples for
  deviations from naive scaling in models with leptoquarks, two Higgs doublets or supersymmetry we refer to Refs.\ \cite{Giudice:2012ms,Crivellin:2018qmi}.} 

An interesting side comment is that
BSM particles will typically not only contribute to $\amu$ but also to the muon mass
in similar loops, and
those contributions depend on the same model details
and scale as $\Delta m_\mu^{\text{BSM}} /m_\mu \sim {\cal O}(C_{\text{BSM}})$.
The estimate $\damu^{\text{BSM}}\sim {\cal O}(\Delta m_\mu^{\text{BSM}}/m_\mu)\times
\frac{m_\mu^2}{M_{\text{BSM}}^2}$ is therefore valid in many models
\cite{Czarnecki:2001pv,Stockinger:1900zz}.
One may impose a criterion that these BSM corrections to the muon mass
do not introduce fine-tuning, i.e.\
do not exceed the actual muon mass.
In models where this criterion
is satisfied, $C_{\text{BSM}}$ can be at most of order unity and
a generic upper limit,
\begin{align}
  \label{eq:damulimit}
    \damu^{\text{BSM}} \lesssim {\cal O}(1)\frac{m_\mu^2}{M_{\text{BSM}}^2} ,
  \end{align}
is obtained \cite{Czarnecki:2001pv,Stockinger:1900zz}.
In this wide class of models, imposing this criterion then \update{implies an
order-of-magnitude upper limit} on the mass
scale for which the value $\damu$ can be
accommodated:\footnote{%
The case of vector-like leptons provides an interesting exception with
a slightly more complicated behaviour, see the discussions in
Refs.\ \cite{Kannike2012,Dermisek:2013gta,Dermisek:2020cod} and below
in Sec.\ \ref{sec:two_field_overview}. There, also tree-level BSM
contributions to the muon mass exist, and the ratio between $\damu$ and $\Delta m_\mu ^{\text{tree}}$ does not
scale as $1/M_{\text{BSM}}^2$ as above but as $1/(16\pi^2v^2)$. This
might seem to allow arbitrarily high masses, circumventing the bounds
(\ref{eq:damulimit},\ref{BNLuppermassbound},\ref{FNALuppermassbound}). However, even using only tree-level effects in
the muon mass, these references also find upper mass
limits from perturbativity and constraints on the Higgs--muon coupling.
}
\begin{align}\label{BNLuppermassbound}
  \text{BNL:}&&
  \damu^{\text{BSM}}&=\update{27.9 \times 10^{-10}} 
  &
  \Rightarrow
  \
  M_{\text{BSM}}&\lesssim
   {\cal O}(2)\text{TeV}\\
    \text{Including \update{FNAL}:}&&
  \damu^{\text{BSM}}&=\update{25.1 \times 10^{-10}} 
  &
  \Rightarrow
  \
  M_{\text{BSM}}&\lesssim
  \update{ {\cal O}(2.1)}\text{TeV}\label{FNALuppermassbound}
\end{align}

In Appendix \ref{app:MuonGm2Contributions} we collect the generic
one-loop Feynman diagrams which can contribute to $\amu$ in a general
renormalizable quantum field theory. The results are expressed in
terms of generic masses and couplings and reflect the above
discussion. Contributions containing the factor $m_\mu^2$ correspond
to chirality flips on the external muon line governed by the SM Yukawa
coupling and VEV; the other contributions correspond to chirality
flips via BSM couplings or fermion masses and require the simultaneous
presence of BSM couplings to left- and right-handed muons, which in
turn requires that some virtual states in the loop are not pure gauge eigenstates but mix
via electroweak VEVs.

%% file: SectionSingleFields.tex
\section{Single Field Extensions} \label{sec:SingleField}

In this section we discuss the impact of $\amu$ on simple single field
extensions of the SM. Such extensions can be interesting in their own
right or representative for more elaborate models with many new fields
and particles and illustrate the impact of the $\amu$ measurement. We
begin in Sec.\ \ref{sec:one_field_overview} with a general overview of
the status of one-field extensions, covering renormalizable models
with new spin $0$, spin $1/2$ or spin $1$ fields. In Secs.~\ref{Sec:THDM}-\ref{Sec:Leptoquarks} we
then show the impact of the latest data on the two most interesting
cases --- the two-Higgs doublet model and leptoquark models.

\subsection{Overview}\label{sec:one_field_overview}
\definecolor{viable}{rgb}{0., 1., 0.5}
\definecolor{ruledout}{rgb}{0.97, 0.51, 0.47}
\definecolor{UVCruledout}{rgb}{0.6, 0.4, 0.8}
\begin{table}
\begin{center}
\begin{tabular}{|c|c|c|c|} \hline
Model & Spin & $SU(3)_C \times SU(2)_L \times U(1)_Y$ & \update{Result for $\damu^\text{BNL}$, $\damuNEW$} \\ \hline
 1 &     0 &                 $({\bf 1},{\bf 1},1)$ & \cellcolor{ruledout} Excluded: $\Delta \amu < 0$ \\
 2 &     0 &                 $({\bf 1},{\bf 1},2)$ & \cellcolor{ruledout} Excluded: $\Delta \amu < 0$\\
 3 &     0 &          $({\bf 1},{\bf 2},-1/2)$ & \cellcolor{viable} Updated in Sec.\ \ref{Sec:THDM} \\
 4 &     0 &                $({\bf 1},{\bf 3},-1)$ & \cellcolor{ruledout} Excluded: $\Delta \amu < 0$ \\
 5 &     0 &   $({\bf \overline{3}},{\bf 1}, 1/3)$ & \cellcolor{viable} Updated Sec.\ \ref{Sec:Leptoquarks}. \\
 6 &     0 &   $({\bf \overline{3}},{\bf 1}, 4/3)$ & \cellcolor{ruledout} Excluded: LHC searches \\
 7 &     0 &   $({\bf \overline{3}},{\bf 3}, 1/3)$ & \cellcolor{ruledout} Excluded: LHC searches \\
 8 &     0 &              $({\bf 3},{\bf 2}, 7/6)$ & \cellcolor{viable} Updated Sec.\ \ref{Sec:Leptoquarks}. \\
 9 &     0 &              $({\bf 3},{\bf 2}, 1/6)$ & \cellcolor{ruledout} Excluded: LHC searches \\
10 & $1/2$ &                 $({\bf 1},{\bf 1},0)$ & \cellcolor{ruledout} Excluded: $\Delta \amu < 0$ \\
11 & $1/2$ &                $({\bf 1},{\bf 1},-1)$ & \cellcolor{ruledout} Excluded: $\Delta \amu$ too small \\
12 & $1/2$ &          $({\bf 1},{\bf 2},-1/2)$ & \cellcolor{ruledout} Excluded: LEP lepton mixing  \\
13 & $1/2$ &          $({\bf 1},{\bf 2},-3/2)$ & \cellcolor{ruledout} Excluded: $\Delta \amu < 0$ \\
14 & $1/2$ &                 $({\bf 1},{\bf 3},0)$ & \cellcolor{ruledout} Excluded: $\Delta \amu < 0$ \\
15 & $1/2$ &                $({\bf 1},{\bf 3},-1)$ & \cellcolor{ruledout} Excluded: $\Delta \amu < 0$ \\
16 &   $1$ &                 $({\bf 1},{\bf 1},0)$ & \cellcolor{viable} Special cases viable \\
17 &   $1$ &              $({\bf 1},{\bf 2},-3/2)$ & \cellcolor{UVCruledout} UV completion problems \\
18 &   $1$ &                $({\bf 1},{\bf 3}, 0)$ & \cellcolor{ruledout} Excluded: LHC searches \\
19 &   $1$ & $({\bf \overline{3}}, {\bf 1}, -2/3)$ & \cellcolor{UVCruledout} UV completion problems \\
20 &   $1$ & $({\bf \overline{3}}, {\bf 1}, -5/3)$ & \cellcolor{ruledout} Excluded: LHC searches \\
21 &   $1$ & $({\bf \overline{3}}, {\bf 2}, -5/6)$ & \cellcolor{UVCruledout} UV completion problems  \\
22 &   $1$ &  $({\bf \overline{3}}, {\bf 2}, 1/6)$ & \cellcolor{ruledout} Excluded: $\Delta \amu < 0$ \\
23 &   $1$ & $({\bf \overline{3}}, {\bf 3}, -2/3)$ & \cellcolor{ruledout} Excluded: proton decay\\
\hline
\end{tabular}
\caption{Summary of known results for gauge invariant single field extensions with one-loop contributions to the anomalous magnetic moment of the muon. These results are rather exhaustive due to systematic investigations and classifications in Ref.\ \cite{Freitas:2014pua,Queiroz:2014zfa,Biggio:2014ela,Biggio:2016wyy}.  Note however that while we present the results based on representations of SM gauge and Lorentz symmetries, the references make assumptions that can be important to the conclusions and are different in each paper. Thus the conclusions summarised in this table should be interpreted with care. For more information on models 1-2, 3-4, 5-9, 10-12, 13, 14-18 and 19-23 see references \cite{CoarasaPerez:1995wa, Biggio:2014ela, Gunion:1989in, Chiu:2014oma}, \cite{Freitas:2014pua, Biggio:2014ela}, \cite{Chakraverty:2001yg,Biggio:2014ela}, \cite{Biggio:2008in, Freitas:2014pua, Biggio:2014ela}, \cite{Biggio:2014ela}, \cite{Biggio:2008in, Freitas:2014pua, Biggio:2014ela, Biggio:2016wyy, Kelso:2014qka} and \cite{Biggio:2016wyy}, respectively.  We use color highlighting to give a visual indication  of the status of the model, namely green for viable explanations, red for excluded and purple for vector extensions excluded on the basis of their UV completions.   \label{tab:one-field-summary}  }
\end{center}
\end{table}

Before presenting our updated results for those cases in
Secs.\ \ref{Sec:THDM}-\ref{Sec:Leptoquarks}, we first classify the
single field extensions according to their spin and their SM
representations and charges, and discuss the known results to provide
a very important overview of what is possible and put our new results
in the appropriate context.  Single field models have been classified
or reviewed in a systematic manner in
Refs.\ \cite{Freitas:2014pua,Queiroz:2014zfa,Chiu:2014oma,Biggio:2014ela,Biggio:2016wyy},
with the results summarized in Table \ref{tab:one-field-summary}.

\update{The confirmation of a large positive deviation} from the SM
prediction in the anomalous magnetic moment of the muon rules out most
one-field extensions of the SM.  The reasons for this are
simple.  First to explain the \update{anomaly} these models must provide
a \update{positive contribution} to $\amu$, and this constraint alone rules
out a large number of the possible extensions. Secondly even if the
sign of the \update{contribution is positive, the models must have a chirality
flip in order for the contribution to be large enough with
perturbative couplings}. Without a chirality flipping enhancement,
contributions that explain $\amu$ \update{require the masses of the new
particles to be so light that they would already have been observed 
in collider experiments}.

Ref.\ \cite{Freitas:2014pua} considers scalars, fermions and vectors.
For fermions and scalars they considered gauge invariant extensions
with $SU(3)$ singlets, which may be $SU(2)$ singlets, doublets,
triplets ($Y=-1$) and adjoint triplets ($Y=0$) for fermions, and
doublets and triplets for scalars. They do not consider scalars
obtaining a VEV. They treated vector states as simplified models of
neutral and charged vector states without specifying any gauge
extension. They assume minimal flavour violating interactions with
leptons (see 2.2 of Ref.\ \cite{Freitas:2014pua} for details) for LEP
contact interaction limits and LHC searches, and perform the
calculation of $\damu$ at the one-loop level. They obtained a negative
contribution to $\amu$ from the scalar triplet, the neutral fermion
singlet, and fermion triplets with hypercharge $0$ or $-1$, and found
that while a charged fermion singlet can give a positive contribution
it is always too small to explain $\damu^{\textrm{BNL}}$. They found
scalar and fermion doublet scenarios that could accommodate
$\damu^{\textrm{BNL}}$ at the $1\sigma$ level were ruled out by LEP
searches for neutral scalars and LEP limits on mixing with SM leptons
respectively. For a single neutral vector boson, they find that the
region where $\damu^{\textrm{BNL}}$ can be explained within $1\sigma$
is entirely ruled out by LEP constraints from 4-fermion contact
interactions and resonance searches. They also consider a single
charged vector boson coupling to a right handed charged lepton and a
right handed neutrino,\footnote{Technically this is a two-field
  extension of the SM though they do not classify it as such.} and
find that in this case the region where $\damu^{\textrm{BNL}}$ can be
explained within $1 \sigma$ is ruled out by the combination of LEP
limits on contact interactions and LHC direct searches.  In summary
they find that all gauge invariant one-field extensions they
considered failed to explain the anomaly.  This paper's findings are
reflected in Table \ref{tab:one-field-summary}, except for the cases
of the scalar doublet (see Sec.\ref{Sec:THDM}) and the neutral vector
(see the discussion in Sec.\ \ref{sec:darkphoton} at the end of this
overview), where there is a lot of dedicated literature and it is
known that breaking the assumptions of Ref.\ \cite{Freitas:2014pua} can
change the result.

Refs.\ \cite{Chiu:2014oma, Biggio:2014ela, Queiroz:2014zfa} also take
a systematic approach.  Ref.\ \cite{Chiu:2014oma} considers scalar
bosons\footnote{They also consider vector states but assume additional
  fermions in that case.}.  Compared to results in the other papers
this adds the singly charged $SU(2)$ singlets to Table
\ref{tab:one-field-summary}. This result was also used in the
classification in Ref.\ \cite{Biggio:2014ela} (drawing also from
Ref.\ \cite{CoarasaPerez:1995wa}), along with doubly charged $SU(2)$
scalar singlets using results from Ref.\ \cite{Gunion:1989in} and
scalar leptoquarks (see Sec.\ \ref{Sec:Leptoquarks} for our update)
originally proposed in Ref.\ \cite{Chakraverty:2001yg}. They also add
a new result for a fermion $SU(2)$ singlet with hypercharge $-3/2$,
i.e.\ the fermion state $({\bf 1},{\bf 2},-3/2)$, showing that the
contribution is always negative above the LEP limit.  Otherwise their
classification overlaps with Ref.\ \cite{Freitas:2014pua} and the
conclusions are effectively consistent\footnote{They do however
  comment that they obtained minor differences to those from
  Ref.\ \cite{Freitas:2014pua} in the $\amu$ calculation for the
  $({\bf 1},{\bf 1},-1)$ and $({\bf 1},{\bf 2},-1/2)$ fermions, which
  alter the reason why they are excluded.  We checked these results
  and agree with the results of Ref.\ \cite{Freitas:2014pua}.}.
Ref.\ \cite{Queiroz:2014zfa} does not require $SU(2)_L$ invariance and
instead considers simplified models of Lorentz scalar, fermion and
vector states with results presented in terms of axial and vector
couplings, $g_a$ and $g_v$ and classify states according to
electromagnetic charges and $SU(3)$ representations.  The reference
presents plots of $\damu$ predictions against the mass of the new state for
specific cases of the couplings.  We checked that the results are
consistent with what we present in Table \ref{tab:one-field-summary},
but Ref.\ \cite{Queiroz:2014zfa} does not contain additional general
conclusions on the viability of each case.

While Refs.\ \cite{Freitas:2014pua} and \cite{Queiroz:2014zfa} used a
simplified models treatment of vector states,
Ref.\ \cite{Biggio:2016wyy} systematically classified vector
extensions according to SM gauge representations and considered the
implications of embedding these into a UV complete gauge extension of
the SM.  They found that only $({\bf 1},{\bf 1},0)$ may provide a
viable UV complete explanation of $\damu^{\textrm{BNL}}$, depending on
specific model dependent details (see Sec.\ \ref{sec:darkphoton} below
for more details on such explanations). Although the $({\bf 1},{\bf
  2},-3/2)$ vector state gives large contributions to $\amu$, they
rejected this since the UV completion into 331 models cannot provide a
$\damu^{\textrm{BNL}}$ explanation consistent with experimental limits
\cite{Kelso:2014qka}. A $({\bf 1},{\bf 3},0)$ vector state has no
chirality flip, so explanations are ruled out by LHC limits
\cite{Khachatryan:2014qwa}.  
$({\bf\overline{3}}, {\bf 1}, -2/3)$ and $({\bf \overline{3}}, {\bf
  2}, -5/6)$ have chirality flipping enhancements, but they reject
$({\bf \overline{3}}, {\bf 1}, -2/3)$ based on an $SU(4)_C\times
SU(2)_L \times U(1)_R$ UV completion and limits on the masses from
rare decays \cite{Kuznetsov:2012ai}, while the $({\bf \overline{3}},
{\bf 2}, -5/6)$ state is rejected based on an $SU(5)$ UV completion
and proton decay limits.  Models without chirality flip enhancements
($({\bf \overline{3}}, {\bf 1}, -5/3)$, $({\bf \overline{3}}, {\bf 2},
1/6)$ and $({\bf \overline{3}}, {\bf 3}, -2/3)$) \update{can all be
  ruled out by collider constraints or because they give the wrong
  sign}.  A summary of the constraints excluding each of the vector
leptoquarks are included in Table \ref{tab:one-field-summary}.

\subsubsection{Dark photon and dark $Z$ explanations}
\label{sec:darkphoton}
Before concluding this overview we now briefly discuss the particularly interesting case of an additional gauge field $Z_d$ with $({\bf 1}, {\bf 1}, {\bf 0})$ quantum numbers that arises from some additional $U(1)_d$ gauge symmetry.
The dark photon scenario assumes that the known quarks and leptons
have no $U(1)_d$ charge. The potential impact of dark photons on
$\amu$ has been extensively studied, after the first proposal in
Ref.\ \cite{Pospelov:2008zw}. 
Models with a general Higgs sector contain both kinetic mixing of the SM $B$-field and $Z_d$ and the mass mixing of the SM $Z$-field and $Z_d$.
As the mass mixing parameter is typically far smaller than the kinetic mixing one, the leading contribution to $\amu$ is proportional to the kinetic mixing parameter $\epsilon$.
The kinetic mixing term induces an interaction between the SM fermions and the dark photon,
and the region relevant for significant $\damu$  has first been found to be $10 ^{-6} < \epsilon ^2 < 10 ^{-4}$, with dark photon masses in the range between $1 \text{ MeV} \cdots 500 \text{ MeV}$~\cite{Pospelov:2008zw}.  However the electron anomalous magnetic moment result~\cite{Hanneke:2008tm} reduces the mass range to $20 \text{ MeV} \cdots 500 \text{ MeV}$~\cite{Davoudiasl:2012ig}, and the remaining range is  excluded by the following experimental results obtained from various dark photon production channels from A1 in Mainz~\cite{Merkel:2014avp} (radiative dark photon production in fixed-target electron scattering with decays into $e^+e^-$ pairs), BaBar~\cite{Lees:2014xha} (pair production in $e^+e^-$ collision with subsequent decay into $e ^+ e^-$ or $\mu ^+ \mu ^-$ pairs), NA48/2 at CERN~\cite{Batley:2015lha} ($\pi ^0$ decay modes via dark photon and subsequent decay into $e ^+ e^-$-pair) and from dark matter production via dark photon from NA46 at the CERN~\cite{Gninenko:2019qiv}.

As a result, pure dark photon models cannot  accommodate significant
contributions to $\amu$. Extensions, e.g.\ so-called ``dark $Z$'' models, open
up new possibilities but are also strongly constrained \cite{Davoudiasl:2012qa,Davoudiasl:2012ig,Davoudiasl:2014kua,Chen:2015vqy,Mohlabeng:2019vrz}.
Similarly, neutral $Z^\prime$ vector bosons  with direct gauge couplings to
leptons are also strongly constrained (as indicated in
Tab.~\ref{tab:one-field-summary}); for examples of remaining viable 
possibilities with significant contributions to $\amu$ we mention the
model with gauged $L_\mu - L_\tau$ quantum number and generalizations
thereof, see  Refs.\ \cite{Heeck:2011wj,Altmannshofer:2014pba,Altmannshofer:2016brv,Gninenko:2018tlp,Escudero:2019gzq,Amaral:2020tga,Huang:2021nkl}.  
Even in such viable models only rather small parts of the parameter
space are promising; in particular only specific windows for
the new vector boson masses can lead to viable explanations of
$\damuNEW$. In case of the $L_\mu -  
L_\tau$ model, the recent Ref.\ \cite{amaral2021distinguishing} has
shown that essentially only the mass range between $0.01\ldots0.1$GeV
remains, and that this parameter range can be further probed in the future by muon
fixed-target experiments and even by neutrino and dark matter search
experiments. For very heavy $Z'$ masses in the TeV region,
explanations of $\damuNEW$ are already disfavoured by LHC data, but
further constraints can ultimately be obtained at a muon collider
\cite{Huang:2021nkl}.
In case of ``dark $Z$'' models, the viable mass range is below the 1
GeV scale, and the promising parameter
space can be probed by measurements of the running weak mixing angle
at low energies at facilities such as JLab QWeak, JLab Moller, Mesa
P2, see e.g.\ Refs.\ \cite{Davoudiasl:2014kua,cadeddu2021muon}.

\subsection{Two-Higgs Doublet Model} \label{Sec:THDM}

As can be seen in Table \ref{tab:one-field-summary}, the two-Higgs
doublet model is one of the very few viable one-field explanations of
$\damu^{\textrm{BNL}}$. It is in fact the only possibility without
introducing new vector bosons or leptoquarks.  
The two-Higgs doublet model (2HDM) contains a charged Higgs $H^\pm$, a
CP-odd Higgs $A$, and two CP-even Higgs bosons $H,h$, where $h$ is
assumed to be SM-like (we assume here a CP-conserving Higgs potential, which
is sufficient to maximize contributions to $\amu$). To be specific we
list here the Yukawa Lagrangian for the neutral Higgs bosons in a form
appropriate for the 2HDM of type I, II, X, Y and the
flavour-aligned 2HDM, in the form of Ref.\ \cite{Cherchiglia:2016eui},
\begin{align}\label{yukawaalign}
{\cal{L}} _Y =& - \sum _{{\cal S}=h,H,A}\sum_{f}\,\frac{Y_{f}^{\cal S} m_f}{v}\,
{\cal S}{\bar{f}} P _{\text{R}} f + h.c.,
\\
  Y^{h}_{f} =& \sin(\beta-\alpha)+\cos(\beta-\alpha)\zeta_{f}, &
  Y^{A}_{d,l} =& -\zeta_{d,l}\,,\\
Y^{H}_{f} =& \cos(\beta-\alpha)-\sin(\beta-\alpha)\zeta_{f}, &
 Y^{A}_{u} =&+ \zeta_{u}\,.
\end{align}
where the Dirac fermions $f$ run over all quarks and leptons,
$(\beta-\alpha)$ is a mixing angle and $\sin(\beta-\alpha)=1$
corresponds to $h$ being SM-like. The
dimensionless Yukawa prefactors $\zeta_{f}$ depend on the 2HDM version
and will be specialized later.

The 2HDM has a rich
phenomenology with a plethora of 
new contributions to the Higgs potential and the Yukawa sector.
It differs from the
previously mentioned models in that two-loop contributions to
$\amu$ are known to be crucial. Typically the
dominant contributions arise via so-called Barr-Zee two-loop
diagrams. In these diagrams an inner fermion loop generates an effective
Higgs--$\gamma$--$\gamma$ interaction which then couples to the muon
via a second loop. If the new Higgs has a large Yukawa coupling to the
muon and if the couplings in the inner loop are large and the new
Higgs is light, the contributions to $\amu$ can be sizeable.
The Higgs mediated flavour changing neutral currents in the 2HDM can be avoided
by imposing either $\mathbb{Z}_2$ symmetry or flavour-alignment.

\begin{figure}
  \begin{subfigure}[]{0.5\textwidth}
    \begin{center}
      \includegraphics[scale=1.]{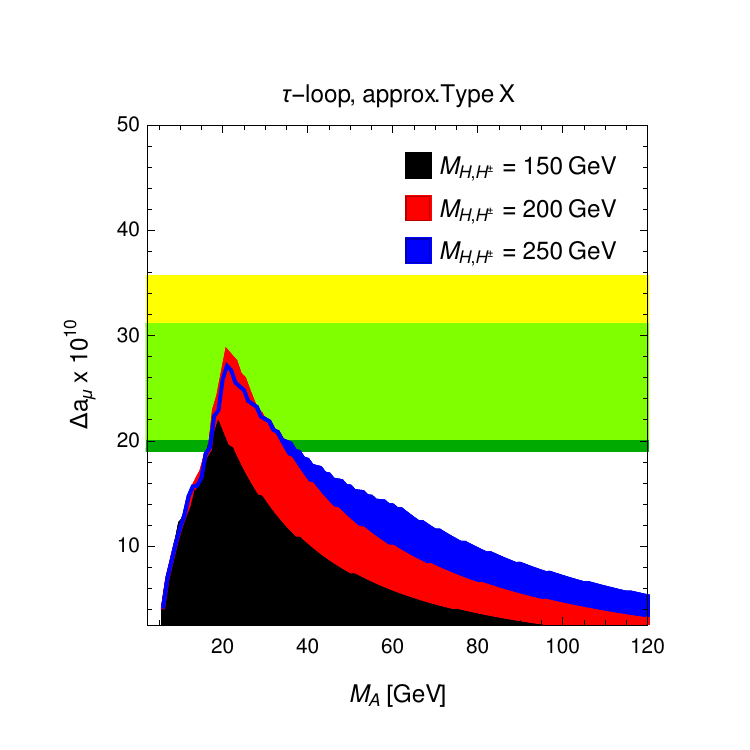}\caption{}\label{fig:THDMupdatea}
      \end{center}
  \end{subfigure}
  \begin{subfigure}[]{0.5\textwidth}
    \begin{center}
      \includegraphics[scale=1.]{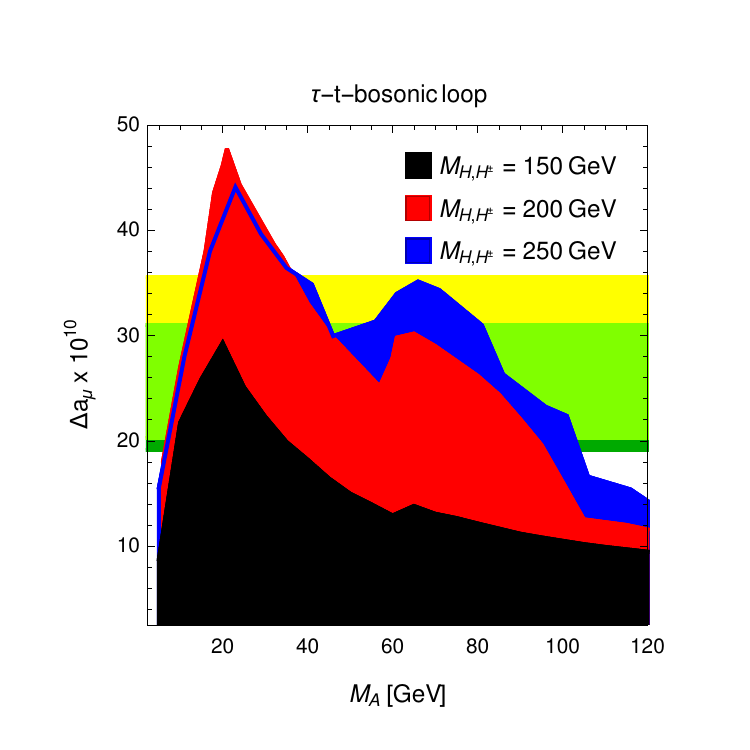}\caption{}\label{fig:THDMupdateb}
    \end{center}
  \end{subfigure}
  \caption{\label{fig:THDMupdate}
    The maximum results for $\damu$ in the two versions of the
    two-Higgs Doublet Model with minimal flavour violation, compared
    with the $1\sigma$ regions around $\damu^{\textrm{BNL}}$ (yellow) and
      new world average $\damuNEW$ (green); light green shows the overlap
      between the two regions. The maximum results are shown as functions of $M_A$, for
    three different values of $M_{H,H^\pm}$, as
    indicated: (a) lepton-specific/type X model (b) flavour-aligned two-Higgs Doublet Model. The results are based on
    Ref.\ \cite{Cherchiglia:2017uwv}. The left plot is technically
    obtained in the framework of the flavour-aligned model but taking
    only $\tau$-loop contributions, which coincides with the type X
    model.
}
\end{figure}

Fig.\ \ref{fig:THDMupdate} presents up-to-date results of the possible
contributions $\damu$ in both of these versions of the 2HDM. The
figure is based on results of Ref.\ \cite{Cherchiglia:2017uwv} and
compares them to the new world average   \update{$\damuNEW$ obtained from including}
the FNAL value. It
arises from scans of the model parameter space and shows the maximum
possible $\damu$ as a function of the most important parameters, the
two new Higgs masses $M_A$
and $M_H$, where the choice $M_H=M_{H^\pm}$ maximises $\damu$. The
reason why there are absolute upper limits on $\damu$ is a combination
of theoretical and experimental constraints, as discussed in the
following.

Fig.\ \ref{fig:THDMupdatea} shows the results for the 2HDM type
X, the so-called lepton-specific version of the 2HDM with
$\mathbb{Z}_2$ symmetry. 
A general analysis of all types of the 2HDM with discrete
$\mathbb{Z}_2$ symmetries and minimal flavour violation has been done in
Ref.\ \cite{Broggio:2014mna}, where only this lepton-specific
type X model survived as a possible source of significant
$\damu$. In this model, the parameters of Eq.\ (\ref{yukawaalign}) are
$\zeta_l=-\tan\beta$ for all charged leptons, while
$\zeta_{u,d}=\cot\beta$ for all quarks. The $\tan\beta$-suppression of
quark Yukawa couplings helps evading experimental constraints from
LEP, LHC
and flavour physics. In the type X 2HDM the main contributions arise from Barr-Zee
diagrams with an inner $\tau$-loop, which are
$(\tan\beta)^2$-enhanced. Hence important constraints arise 
from e.g.\ precision data on $Z\to\tau\tau$ and
$\tau$-decay \cite{Wang:2014sda,Abe:2015oca,Chun:2016hzs,Wang:2018hnw}
as well as from LEP data on the mass range $M_A\lesssim20$ GeV
\cite{Cherchiglia:2017uwv}. As 
Fig.\ \ref{fig:THDMupdatea} shows, 
\update{only a tiny parameter 
  space in the 2HDM type X remains a viable explanation of the
  observed $\damuNEW$.
  For a $1\sigma$ explanation,  $M_A$ must be in the small interval
  $20\ldots40$ GeV; the corresponding maximum values of the 
  $\tan\beta$ parameter, which governs the lepton Yukawa couplings in
  this model, are in the range $50\ldots100$.}
The masses of the new heavier Higgs bosons $M_{H,H^\pm}$ vary between
$150$ and $250$ GeV in the figure. Smaller values of these masses lead
to stronger constraints (since loop contributions to $\tau$-physics
are less suppressed), while larger values lead to a larger hierarchy
$M_A\ll M_{H,H^\pm}$ which leads to stronger constraints from
electroweak precision physics and theoretical constraints such as
perturbativity \cite{Chun:2016hzs,Cherchiglia:2017uwv}.
We mention also that the 2HDM type X parameter space with particularly large
contributions to $\amu$ can lead to peculiar $\tau$-rich final states
at LHC \cite{Chun:2015hsa,Iguro:2019sly} but can be tested
particularly well at a future lepton collider
\cite{Chun:2019sjo}  and is also compatible with CP violation and testable
contributions to the electron electric dipole moment \cite{Chun:2019oix}.

Fig.\ \ref{fig:THDMupdateb} shows results for the so-called
flavour-aligned two-Higgs doublet model, which is a more general but
still minimal flavour violating scenario. Here the parameters
$\zeta_l$, $\zeta_u$, $\zeta_d$ are independent, however assumed to be
generation-universal. The contributions to $\amu$
were first discussed in Ref.\ \cite{Ilisie:2015tra} and then
scrutinized in
Refs.\ \cite{Han:2015yys,Cherchiglia:2016eui,Cherchiglia:2017uwv}. Here
not only the $\tau$-lepton loop contributes in essentially the same
way as before, but also Barr-Zee diagrams with the top-quark in the
inner loop may contribute. To a smaller extent, also purely bosonic
two-loop diagrams can increase $\amu$. The plot takes into account all
contributions, based on
Refs.\ \cite{Cherchiglia:2016eui,Cherchiglia:2017uwv}.  In particular
the top-quark loop leads to a larger possible value of $\amu$. Its
contributions are bounded by constraints from LHC data and
$B$-physics. The LHC
constraints are weaker for $M_A>62$ GeV, where the decay $h\to AA$ is
kinematically impossible \cite{Cherchiglia:2017uwv}. This is reflected
in the behaviour of the maximum $\damu$ as a function of $M_A$ in the
plot. In case of the flavour-aligned 2HDM
\update{the world average deviation  
  $\damuNEW$ can be accommodated for $M_A$ up to $100$ GeV,
  if the
  heavy and charged Higgs masses are in the region $M_H = M_{H ^\pm} =200 \ldots 250$ GeV.} 
  In the parameter space which
  maximises $\damu$ in the flavour-aligned 2HDM the light $A$-boson has
  simultaneously significant Yukawa couplings to the top quark and to
  $\tau$-leptons, leading to a significant rate for the process $gg\to
  A\to \tau\tau$, which might be tested at future LHC runs.

Hence among the 2HDM versions without tree-level
  FCNC,  the well-known type I and type II versions are excluded as
  explanations of the deviation $\damuNEW$. In contrast, the
  lepton-specific type X model and the more general 
  flavour-aligned 2HDM can give significant contributions to
  $\damu$. In both cases, two-loop Barr-Zee diagrams with $A$-boson
  exchange and $\tau$-loop are important; in the flavour-aligned model
  also top-loops are important. The mass $M_A$ is severely constrained
  and the new Yukawa couplings must be at their upper experimental
  limits.
  Because of this, the 2HDM explanations of $\damuNEW$ are going to be
  further scrutinized at ongoing and future experiments: Any
  improvement of the LHC sensitivity to $gg\to A\to\tau\tau$ can
  either discover the $A$-boson or reduce the allowed parameter space
  visible in Fig.\ \ref{fig:THDMupdateb}.
  Likewise, improved measurements of $\tau$-decays can lead to
  reduced upper limits on the maximum $\tan\beta$ or $\zeta_l$,
  respectively, and reduce the viable parameter space in both
  plots. As mentioned above,
  Refs.\ \cite{Chun:2015hsa,Iguro:2019sly} have
  investigated how further future LHC measurements of processes such
  as $HA$ or $H^\pm A$ production with decay into multi-lepton final
  states can  impact the 2HDM
  explanations of $\damuNEW$. Lepton colliders even with modest
  c.o.m.\ energy offer additional coverage of the 2HDM parameter
  space via the Yukawa process $e^+e^-\to\gamma^*/Z^*\to\tau\tau A$
  \cite{Chun:2019sjo}, which is directly sensitive to the low-mass
  pseudoscalar Higgs boson relevant for $\amu$.

  Further, we mention that more exotic variants of the 2HDM which
involve neither 
$\mathbb{Z}_2$-symmetric nor general flavour-aligned Yukawa couplings
can open up additional possibilities.
E.g.\ large non-flavour aligned Yukawa couplings to $\tau$-leptons or
top quarks can allow large contributions to $\amu$ even for masses of
$M_A$ above $100$ GeV \cite{Iguro:2019sly,Li:2018aov}. In these cases, important
constraints arise from lepton flavour violating processes \cite{Iguro:2019sly} and
B-physics \cite{Li:2018aov}. Large, non-flavour aligned $\tau$-Yukawa
couplings also allow another window of significant contributions with
a very light CP-even Higgs with $M_H\lesssim1$ GeV \cite{Jana:2020pxx}. And a
muon-specific 2HDM can accommodate large $\damu$ with $\tan\beta$ of
order $1000$ \cite{Abe:2017jqo}.

\subsection{Scalar Leptoquarks} \label{Sec:Leptoquarks}
	In this subsection we update the results for the other single
        field models which could explain $\damu^{\textrm{BNL}}$,
        i.e.\ the scalar leptoquarks.  Scalar and vector leptoquarks
        that interact with SM leptons and quarks can appear as the
        only BSM particle in one-loop contributions to the anomalous
        magnetic moment of the muon. Scalar leptoquarks have been
        considered as a solution for the anomalous magnetic moment
        anomaly in Refs.\ \cite{Chakraverty:2001yg, Queiroz:2014zfa,
          Biggio:2014ela, Bauer:2015knc, Popov:2016fzr}, while vector
        leptoquarks have also been considered in
        Refs.\ \cite{Queiroz:2014zfa, Biggio:2016wyy}. Here we focus
        on studying scalar leptoquarks in detail, since one would
        expect vector leptoquarks to be associated with an extension
        of the gauge symmetries, which complicates the construction of
        these models, and taking the simplified model approach
        they
        may yield results which are rather misleading compared to what
        can be achieved in a realistic model.

	Requiring gauge invariant couplings to SM leptons and quarks restricts us to the five scalar leptoquarks \cite{Buchmuller:1986zs} shown in Table \ref{tab:one-field-summary} (Models 5--9).  Only two of these models\footnote{We follow the notation in Ref.\ \cite{Buchmuller:1986zs}.}, $S_{1}$ $({\bf \overline{3}},{\bf 1},1/3)$ and $R_{2}$ $({\bf 3},{\bf 2},7/6)$, have both left- and right-handed couplings to the SM fermions and can therefore have a chirality flip enhancing their one-loop contributions \cite{Chakraverty:2001yg,Queiroz:2014zfa,Biggio:2014ela}.  

Leptoquarks can in general have complicated flavour structure in their
couplings. Since our focus is on demonstrating the impact of the
anomalous magnetic moment experiment and demonstrating the various
ways to explain it, we prefer to simplify the flavour structure and
focus on the couplings that lead to an enhanced $\damu$ contribution.
We therefore restrict ourselves to muon-philic
leptoquarks that couple only to the second-generation of SM leptons,
evading constraints on flavour violating processes such as $\mu\rightarrow e\gamma$. Leptoquarks that induce flavour violation in the quark sector have
been widely considered in the literature as possible solutions to
flavour anomalies, and sometimes simultaneous explanations of $\amu$
and these anomalies (see e.g.\ Refs.\ \update{\cite{Bauer:2015knc, Popov:2016fzr}}).  
However we also do not consider these here for
the same reasons we choose to avoid lepton flavour violating couplings
and the same reasoning applies to simultaneous explanations of the
more recent $a_e$ anomaly \cite{Bigaran:2020jil}.

We found that it is possible to explain the $\damu^{\textrm{BNL}}$
\update{and $\damuNEW$ results with moderately sized}
perturbative couplings using leptoquarks that are both muon-philic
{\it and} charm-philic, i.e.\ leptoquarks that only couple to second
generation up-type quarks as well as only second generation charged
leptons. Specifically we found $\damu^{\textrm{BNL}}$ could be explained while satisfying LHC limits from direct searches as long as $\sqrt{|\lambda_L \lambda_R|}\gtrsim 0.4$, where $\lambda_L$ and $\lambda_R$ are the leptoquark couplings to the muons and the quarks.  However careful consideration of CKM mixing and
flavour changing neutral currents (FCNC) reveals stringent
constraints. While one may require that the new states couple only to
the charm and not the up-quark or top-quark, CKM effects will then
still generate couplings to the bottom and down-quark.  This effect is
very important and the impact of these for ``charm-philic'' leptoquark
explanations of $\damu^\textrm{BNL}$ has been considered in
Ref.\ \cite{Kowalska:2018ulj}.  There they find that constraints from
BR($K^+\rightarrow\pi^+\nu\overline{\nu}$) for the $S_1$ leptoquark,
or BR($K_L\rightarrow\mu^+\mu^-$) for the $R_2$ leptoquark, heavily
restrict one of the couplings that enter the $\amu$ calculation.  They
find this excludes fitting $\damu^{\textrm{BNL}}$ within $1\sigma$, but
in the case of the first model an explanation within $2\sigma$
remained possible, while for the second model explanations well beyond
$2\sigma$ were excluded.  They also consider the possibility that it
is the down-type couplings that are second generation only, and find
even more severe constraints in that case. Finally for a limited case,
they explore including a direct coupling to the top-quark and find
that quite large couplings to the top quark are needed to explain
$\damu^{\textrm{BNL}}$ within $1\sigma$.  Due to the strong flavour
constraints from coupling the leptoquark to the second generation of
SM quarks, we instead present results for top-philic leptoquarks,
i.e.\ using scalar leptoquarks which couple to the second generation
SM leptons, and the third generation of SM quarks.

Below is written the Lagrangian for both scalar leptoquarks, where here all fermions are written as 2-component left-handed Weyl spinors, for example $Q_3 = (t_{L},b_L)^T$ and $\mu_R^\dagger$, which follows the notation of Ref.\ \cite{Martin1997}. For simplicity we also define $\mu, t, b := \mu_R^\dagger, t_R^\dagger, b_R^\dagger$ below.
\begin{equation} \label{eqn:ScalarLeptoquarkSinglet}
    {\cal L}_{S_1} = -\begin{pmatrix}\lambda_{QL} Q_3\cdot L_2 S_{1} + \lambda_{t\mu} t \mu S_{1}^* + h.c.\end{pmatrix} - M_{S_1}^2 |S_{1}|^2 - g_{H S_1} |H|^2 |S_{1}|^2 - \frac{\lambda_{S_1}}{2} \begin{pmatrix}|S_{1}|^2\end{pmatrix}^2 ,
\end{equation}
\begin{equation} \label{eqn:ScalarLeptoquarkDoublet}
    {\cal L}_{R_2} = -\begin{pmatrix}\lambda_{Q\mu} R_{2}^\dagger Q_3 \mu + \lambda_{tL} L_2\cdot R_{2} t + h.c.\end{pmatrix} - M_{R_2}^2 |R_{2}|^2 - g_{H R_2} |H|^2 |R_{2}|^2 - \frac{\lambda_{R_2}}{2} \begin{pmatrix}|R_{2}|^2\end{pmatrix}^2.
\end{equation}       
where the dot product above denotes the $SU(2)_L$ product, so e.g.\ $Q_3 \cdot L_2 = t_L \mu_L - b_L \nu_{\mu L}$.  For the $S_{1}$ leptoquark one could also include $SU(3)_C \times SU(2)_L \times U(1)_Y$ gauge invariant renormalizable operators, $S_{1} Q_3 L_2$ and $S_{1} tb$ but unless these diquark couplings are severely suppressed or forbidden, they will give rise to rapid proton decay when combined with the leptoquark operators we consider here \cite{Arnold:2013cva,Queiroz:2014pra}. $R_{2}$ does not admit such renormalizable operators \cite{Arnold:2013cva} though there remain dangerous dimension 5 operators that would need to be forbidden or suppressed \cite{Queiroz:2014pra}. Since we are focused on $\amu$ we again simplify things by assuming all parameters are real, but note that if we were to consider complex phases then electric dipole moments would also be of interest, see e.g.\ Ref.\ \cite{Dekens:2018bci}.
      
Constraints on the masses of scalar leptoquarks with second and third
generation couplings to the SM leptons and quarks respectively can be
directly applied from $13$ TeV CMS
\cite{Sirunyan:2018ruf,Sirunyan:2018kzh} results, dependent on how
strong they couple to those fermions.  Given the above Lagrangians,
one can see that the scalar leptoquark singlet $S_1$ can decay to
either a top quark and muon or bottom quark and neutrino, while the
upper and lower components of the scalar leptoquark doublet decay as
$R^u_2$ to a top quark and muon and $R^d_2$ to either a top quark and
neutrino or a bottom quark and muon.  Thus for the leptoquark $S_1$
given in Eqs.\ (\ref{eqn:ScalarLeptoquarkSinglet}), the branching
fraction $\beta_{S_1} = Br(S_1 \rightarrow t\mu)$, is given by:
	\begin{align} \label{eqn:ScalarLeptoquarkSingletBR}
		\beta_{S_1} &= \frac{\lambda_{QL}^2 + \lambda_{t\mu}^2}{2\lambda_{QL}^2 + \lambda_{t\mu}^2}.
	\end{align}
For scalar leptoquark singlet $S_1$ the most stringent LHC limits when
coupling to third generation quarks and second generation leptons are
dependent on $\beta_{S_1}$ \cite{Sirunyan:2018ruf}.  Thus we can
calculate $\beta_{S_1}$ using selected values of the couplings between
$S_1$ and the fermions, and interpolate between them to find the
limits on the mass given in Ref.\ \cite{Sirunyan:2018ruf}.  Now for
$R_{2}$ in Eq.\ (\ref{eqn:ScalarLeptoquarkDoublet}), limits can be
placed on the upper component of the doublet, $R^u_2$, which decays
solely to $t\mu$.  In this case the mass limits from
Ref.\ \cite{Sirunyan:2018ruf} are applied where the branching ratio
for $R^u_2$ to decay to $t\mu$ is taken to be $\beta_{R ^u _2}=1$.
	
Further constraints can be placed on leptoquarks from the effective
coupling of a $Z$ boson to leptons.  The experimentally measured
effective couplings of the $Z$ boson to a pair of muons are given as
$g^{\mu\mu}_L = -0.2689\pm0.0011$, $g^{\mu\mu}_R = 0.2323\pm0.0013$
\cite{ALEPH:2005ab,Tanabashi2018} in the case of left- and
right-handed couplings.  The contribution from a scalar leptoquark
with couplings to any flavour of the SM fermions to the effective
couplings between $Z$ and muon, $\delta g^{\mu\mu}_{L,R}$, is given by
Eqs.\ (22,23) in Ref.\ \cite{Arnan:2019olv} for the leptoquarks $S_1$
and $R_2$ respectively.  Points with left-right effective couplings
more than $2\sigma$ away from the measured values are treated as
constrained.
	
Likewise, the effective coupling of the $Z$ boson to any two neutrinos
has been measured as the observed number of light neutrino species
$N_\nu = 2.9840\pm0.0082$ \cite{ALEPH:2005ab}.  The BSM contributions
from a scalar leptoquark to this are given by \cite{Arnan:2019olv}:
	\begin{equation} \label{eqn:ZnuEffectiveCouplingLQ}
		N_\nu = \sum_{i,j=e,\mu,\tau} \begin{pmatrix} |\delta^{ij} + \frac{\delta g^{ij}_{\nu L}}{g^{\textrm{SM}}_{\nu L}}|^2 + |\frac{\delta g^{ij}_{\nu R}}{g^{\textrm{SM}}_{\nu L}}|^2 \end{pmatrix}, 
	\end{equation} where $g^{\textrm{SM}}_{\nu L}$ are the SM couplings, and $\delta g^{ij}_{\nu L,R}$ are the BSM couplings between the $Z$ boson and the neutrinos given again in Eqs.\ (22,23) from Ref.\ \cite{Arnan:2019olv}.
	
Due to the large masses of the leptoquarks considered for this model,
it is reasonable to consider fine-tuning in the mass of the muon.
With large BSM masses and sizeable couplings to the SM, contributions
to the muon can be generated as detailed in
Sec.\ \ref{sec:BSMoverview}.  The specific constraint considered in
this paper for when the contribution to the muon mass is considered
not ``fine-tuned'' is
\begin{align} \label{eqn:FineTuningLimits}
\frac{1}{2} &< \frac{m_\mu^{\overline{\text{MS}}}}{m_\mu} <2,
	\end{align}   i.e.\ the relative difference between the $\overline{\text{MS}}$ and pole masses $m_\mu^{\overline{\text{MS}}}$ and $m_\mu $ should not exceed $100\%$. While not forbidden, one may consider explanations of large $\amu$ via ${\cal O}(>100\%)$ corrections to the pole mass $m_\mu$ as unattractive.

Since the chirality flip provides an enhancement to $\damu$
proportional to the squared mass of the heaviest quark that the
leptoquark couples to, in the case of a leptoquark coupling to the
SM's third quark generation the enhancement is $m_t/ m_\mu$. Note that
we actually found that the $m_c/m_\mu$ enhancement for charm-philic
leptoquarks was sufficient to allow explanations of
$\damu^{\textrm{BNL}}$ consistent with LHC limits (but not flavour
constraints) with perturbative couplings, therefore this much larger
enhancement should be more than sufficient, even with rather small
couplings or quite heavy masses.  In both
cases of the leptoquark models, the leptoquark contributions to $\amu$
are given by two kinds of diagrams of the FFS and SSF type,
\begin{align}
  \damu^{LQ}&=\damu^{\text{FFS}}+\damu^{\text{SSF}}\,.
\end{align}
For the scalar leptoquark singlet $S_{1}$, the contributions to $\amu$
specifically come from diagrams with top--leptoquark loops and are
given by:
	\begin{equation} \label{eqn:ScalarLeptuquarkSingletContribFFS}
	    \damu^{\text{FFS}} = \frac{3 m_\mu^2 Q_{t}}{32\pi^2 M_{S_1}^2}
		\begin{pmatrix}
		\frac{\lambda_{QL}^2 + \lambda_{t\mu}^2}{6} \,E\!\begin{pmatrix}\frac{m_t^2}{M_{S_1} ^2}\end{pmatrix}
		+ \frac{4\lambda_{QL} \lambda_{t\mu}}{3} \frac{m_t}{m_\mu} \,F\!\begin{pmatrix}\frac{m_t^2}{M_{S_1} ^2}\end{pmatrix}
		\end{pmatrix},
	\end{equation}
	\begin{equation} \label{eqn:ScalarLeptuquarkSingletContribSSF}
		\damu^{\text{SSF}} = \frac{3 m_\mu^2 Q_{S_1}}{32\pi^2 M_{S_1} ^2}
		\begin{pmatrix}
		- \frac{\lambda_{QL}^2 + \lambda_{t\mu}^2}{6} \,B\!\begin{pmatrix}\frac{m_t^2}{M_{S_1} ^2}\end{pmatrix}
		- \frac{2\lambda_{QL} \lambda_{t\mu}}{3} \frac{m_t}{m_\mu} \,C\!\begin{pmatrix}\frac{m_t^2}{M_{S_1} ^2}\end{pmatrix}
		\end{pmatrix},
	\end{equation} where the charges $Q_t=2/3$, $Q_{S_1}=1/3$ and the one-loop functions $B(x)$, $C(x)$, $E(x)$, and $F(x)$ are defined in Appendix \ref{app:MuonGm2Contributions}.  These contributions are described by the fermion-fermion-scalar (FFS) diagram \ref{fig:FFSDiagram} and the scalar-scalar-fermion (SSF) diagram \ref{fig:SSFDiagram} of Fig.\ \ref{fig:GeneralDiagrams} with the generic $F$ fermion lines replaced by a top quark and the scalar $S$ ones by $S_1$.  Similarly, the contributions from the scalar leptoquark doublet $R_{2}$ involving top/bottom and leptoquark loops are given by:
	\begin{align}
      \label{eqn:ScalarLeptuquarkDoubletContribFFS}
	  \damu^{\text{FFS}} &= \frac{3 m_\mu^2}{32\pi^2 M_{R_2} ^2}
	  \begin{pmatrix}
	  	Q_{t} \frac{4\lambda_{Q\mu} \lambda_{tL}}{3} \frac{m_t}{m_\mu} \,F\!\begin{pmatrix}\frac{m_t^2}{M_{R_2} ^2}\end{pmatrix}
	    - Q_{t} \frac{\lambda_{Q\mu}^2+\lambda_{tL}^2}{6} \,E\!\begin{pmatrix}\frac{m_t^2}{M_{R_2} ^2}\end{pmatrix}
	    - Q_{b} \frac{\lambda_{Q\mu}^2}{6} \,E\!\begin{pmatrix}\frac{m_b ^2}{M_{R_2} ^2}\end{pmatrix}
	  \end{pmatrix},\\
	  \label{eqn:ScalarLeptuquarkDoubletContribSSF}
	  \damu^{\text{SSF}} &= \frac{3 m_\mu ^2}{32\pi^2 M_{R_2} ^2}
	  \begin{pmatrix}
	  	Q_{R_2} ^u \frac{2\lambda_{Q\mu}\lambda_{tL}}{3} \frac{m_t}{m_\mu} \,C\!\begin{pmatrix}\frac{m_t ^2}{M_{R_2} ^2}\end{pmatrix}
	    - Q_{R_2} ^u \frac{\lambda_{Q\mu} ^2+\lambda_{tL} ^2}{6} \,B\!\begin{pmatrix}\frac{m_t ^2}{M_{R_2} ^2}\end{pmatrix}
	    - Q_{R_2}^d \frac{\lambda_{Q\mu}^2}{6} \,B\!\begin{pmatrix}\frac{m_b^2}{M_{R2}^2}\end{pmatrix}
	  \end{pmatrix},
	\end{align}
where $Q_{R_2}^d = 2/3$ is the charge of the lower component of the
leptoquark doublet, and $Q_{R_2}^u = 5/3$ is the charge of the upper
component, and $Q_b=-1/3$.  Note the colour factor $3$ in front of
each of the leptoquark contributions.  Each of the above contributions
is produced by a pair of diagrams which are of the FFS diagram type
\ref{fig:FFSDiagram} or the SSF diagram type \ref{fig:SSFDiagram} of
Fig.\ \ref{fig:GeneralDiagrams} with the generic $F$ fermion lines
replaced by a top or bottom quark and the scalar $S$ ones by $R_2^d$
and $R_2^u$, respectively.
	
    The important parameters for determining the above contributions to $\amu$ from scalar leptoquarks are the couplings $\lambda_L$, $\lambda_R$ between the leptoquarks and either the left- or right-handed top quarks, and the mass of the leptoquark $M_{LQ}$.  For either of these leptoquarks the dominant contribution to $\amu$, arises from the internal chirality flip enhancement and has the following approximate form,
	\begin{align} \label{eqn:ScalarLeptoquarkLOContrib}
		\damu^{LQ} &\approx \frac{\lambda_L \lambda_R m_\mu m_t}{8\pi^2 M_{LQ}^2} \bigg(Q_{t} \,F\!\begin{pmatrix}\frac{m_t^2}{M_{LQ}^2}\end{pmatrix} \pm \frac{Q_{LQ}}{2} \,C\!\begin{pmatrix}\frac{m_t^2}{M_{LQ}^2}\end{pmatrix}\bigg),
	\end{align}
	where the $-/+$ is for the singlet/doublet.  Comparing this to Eq.\ 
	(\ref{eqn:GeneralGM2Contribution}), we can see that $C_{BSM} = Q_t \lambda_L 
	\lambda_R m_t/ (8 \pi^2 m_\mu)$, with the ratio $m_t / m_\mu \approx 1600$, gives 
	the parametric enhancement to $\damu$ from the chirality flip.  We can also see how 
	allowing the leptoquark to couple to the top quark versus the charm quark leads to 
	a larger value of $\damu$.  While this can be used to qualitatively 
	understand our results, the numerical calculations were performed with \fs 2.5.0 
	\cite{Athron:2014yba,Athron:2017fvs} \footnote{where \sarah 4.14.1 
	\cite{Staub:2009bi,Staub:2010jh,Staub:2012pb,Staub:2013tta} was used to get 
	expressions for masses and vertices, and \fs also uses some numerical routines 
	originally from \softsusy \cite{Allanach:2001kg,Allanach:2013kza}.}. The \fs 
	calculation of $\damu$ includes the full one-loop contribution and the universal 
	leading logarithmic two-loop QED contribution \cite{Degrassi:1998es}, which tends 
	to reduce the result by $\approx 6\% - 10\%$.

\begin{figure}[ht!]
    	\centering
   		\includegraphics[width=0.32\textwidth]{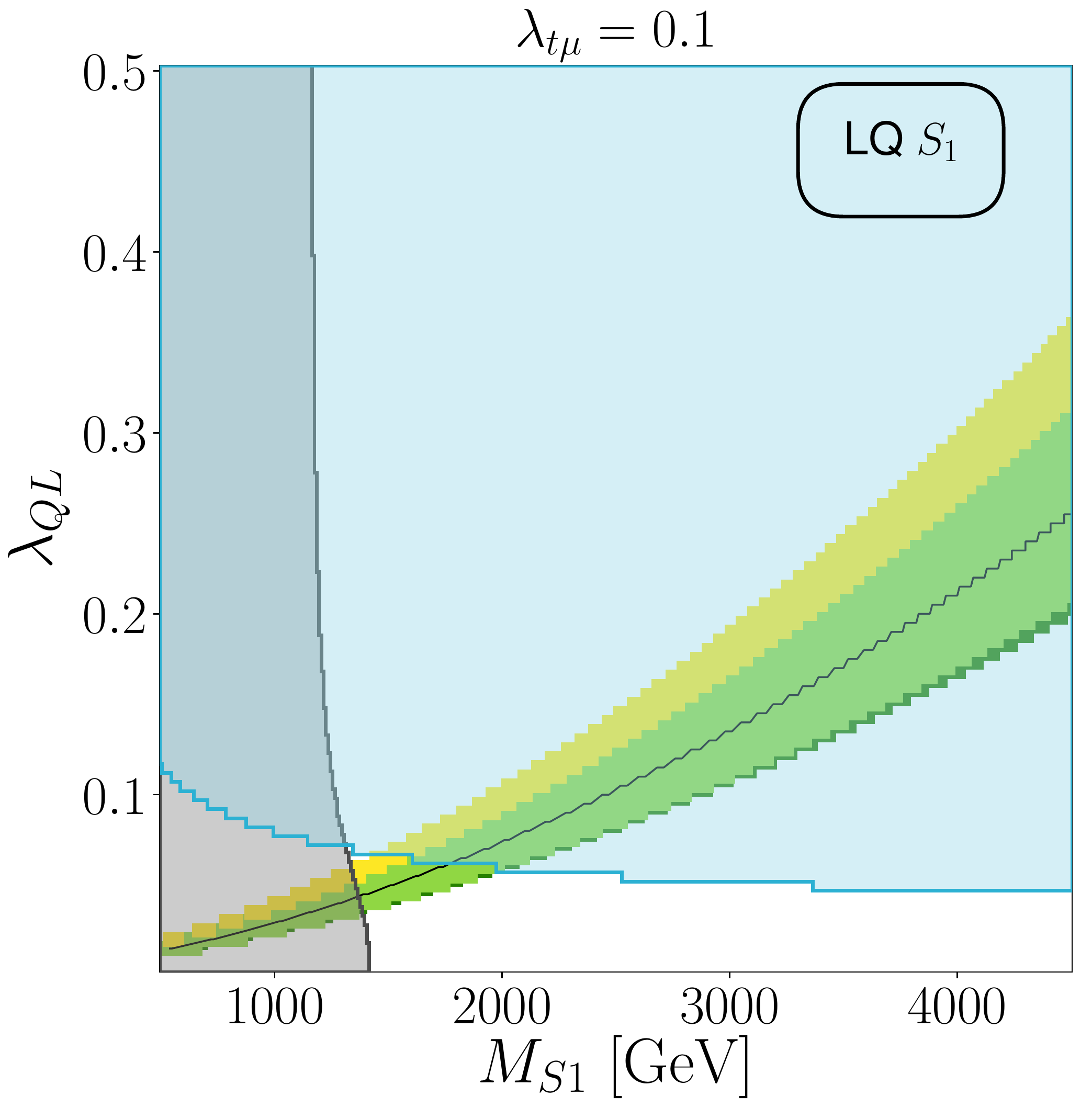}
   		\includegraphics[width=0.32\textwidth]{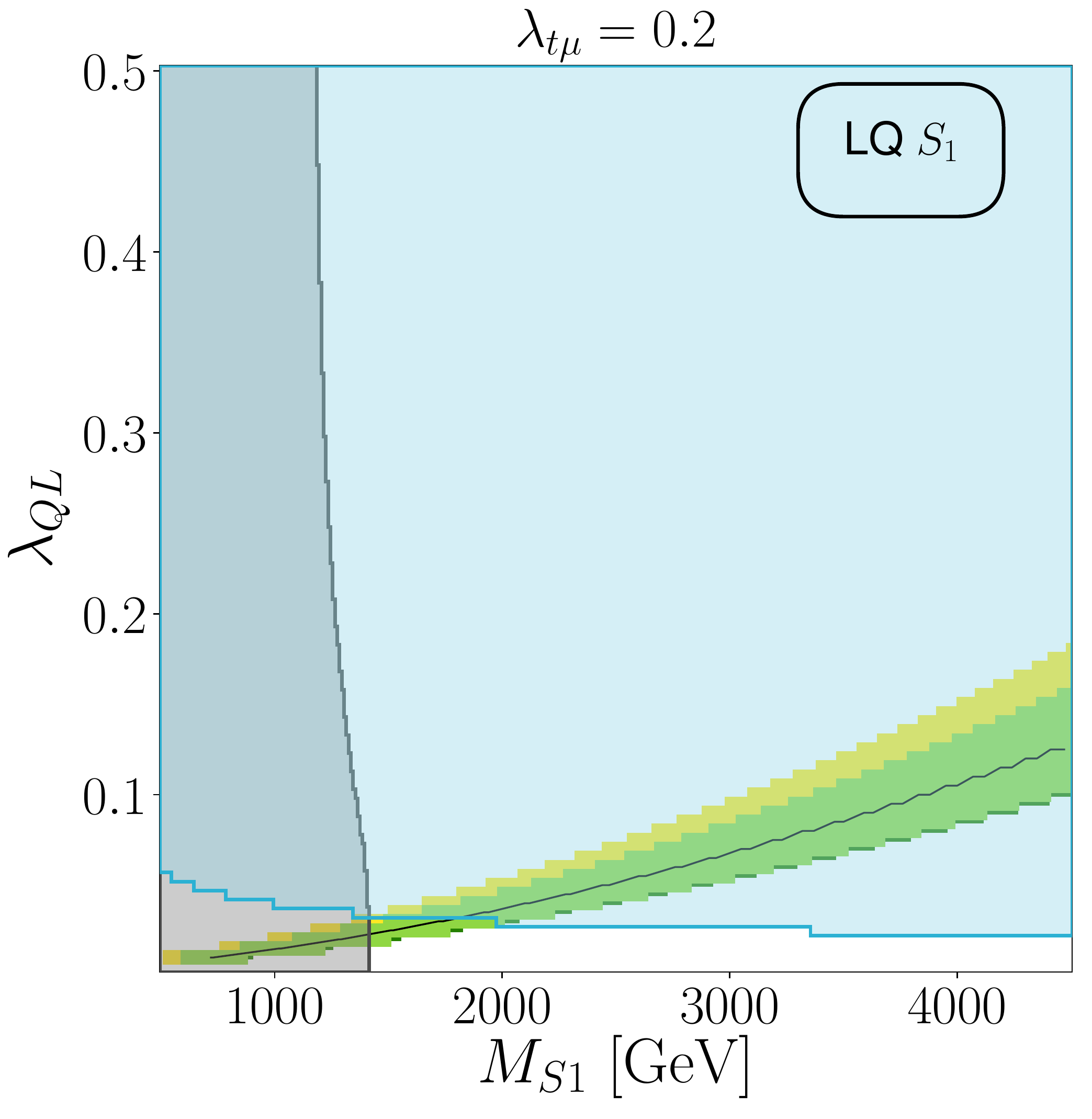}
   		\includegraphics[width=0.32\textwidth]{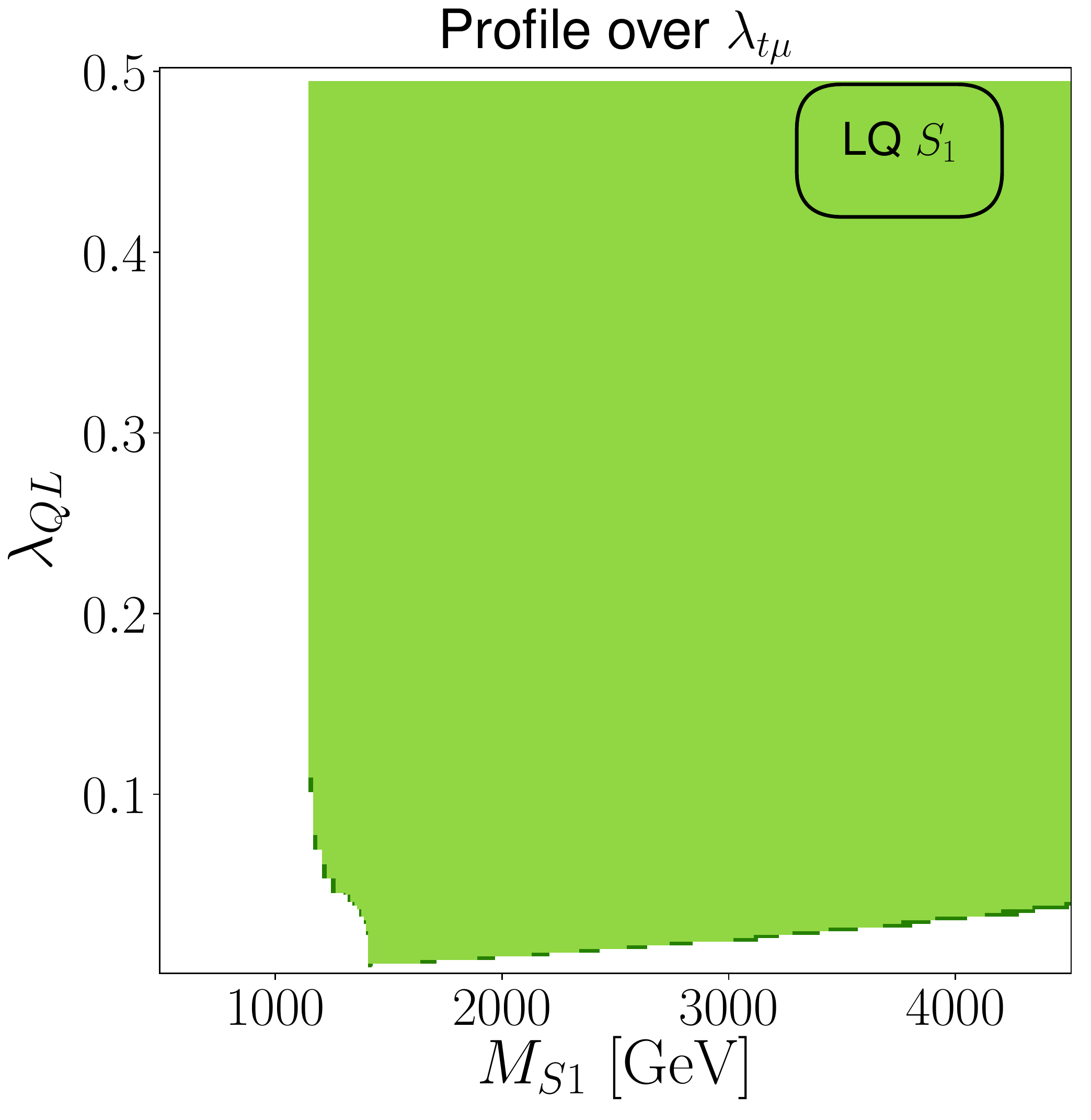}
    \caption{Scenarios which can explain $\damu^\text{BNL}$ and/or
      $\damuNEW$ in the model with an $SU(2)_L$ singlet
      scalar leptoquark $S_{1}$ defined in
      Eq.\ (\ref{eqn:ScalarLeptoquarkSinglet}). Results are shown in
      the $M_{S1}$-$\lambda_{QL}$ plane for fixed $\lambda_{t\mu}=0.1$
      (left panel), $\lambda_{t\mu}=0.2$ (middle panel) and by varying
      $\lambda_{t\mu}\in\update{[0,0.5]}$ (right panel). In
      the latter case we {\it profile} to find the regions in the
      $M_{S1}$-$\lambda_{QL}$ plane that for some value of
      $\lambda_{t\mu}$ in the scan, satisfy all experimental
      constraints and can explain the $\damu$ measurements within
      1$\sigma$.  Regions which can explain $\damu^\textrm{BNL}$ and
      $\damuNEW$ to within $1\sigma$ are yellow and green
      respectively, with the overlap between these two regions
      coloured lime.  The black line indicates points which produce a
      $\damu$ contribution matching \update{Eq.\ (\ref{eqn:avgDiscrepancy})}.  The
      grey regions are excluded by scalar leptoquark searches at the
      Large Hadron Collider \cite{Sirunyan:2018ruf}. Regions shaded
      cyan are disfavoured as they provide a relative contribution
      which shift the muon mass up by more than $100\%$ or down by
      more than $50\%$ as in Eq.\ (\ref{eqn:FineTuningLimits}).  In
      the right panel only solutions which have not been excluded by
      leptoquark searches or $Z\rightarrow\mu\mu$ or
      $Z\rightarrow\nu\nu$ constraints are displayed, i.e.\ any points
      ruled out by any of these experimental constraints are
      discarded. \label{fig:ScalarLeptoquarkSinglet} }
    \end{figure}

\begin{figure}[ht!]
		\centering
		\includegraphics[width=0.32\textwidth]{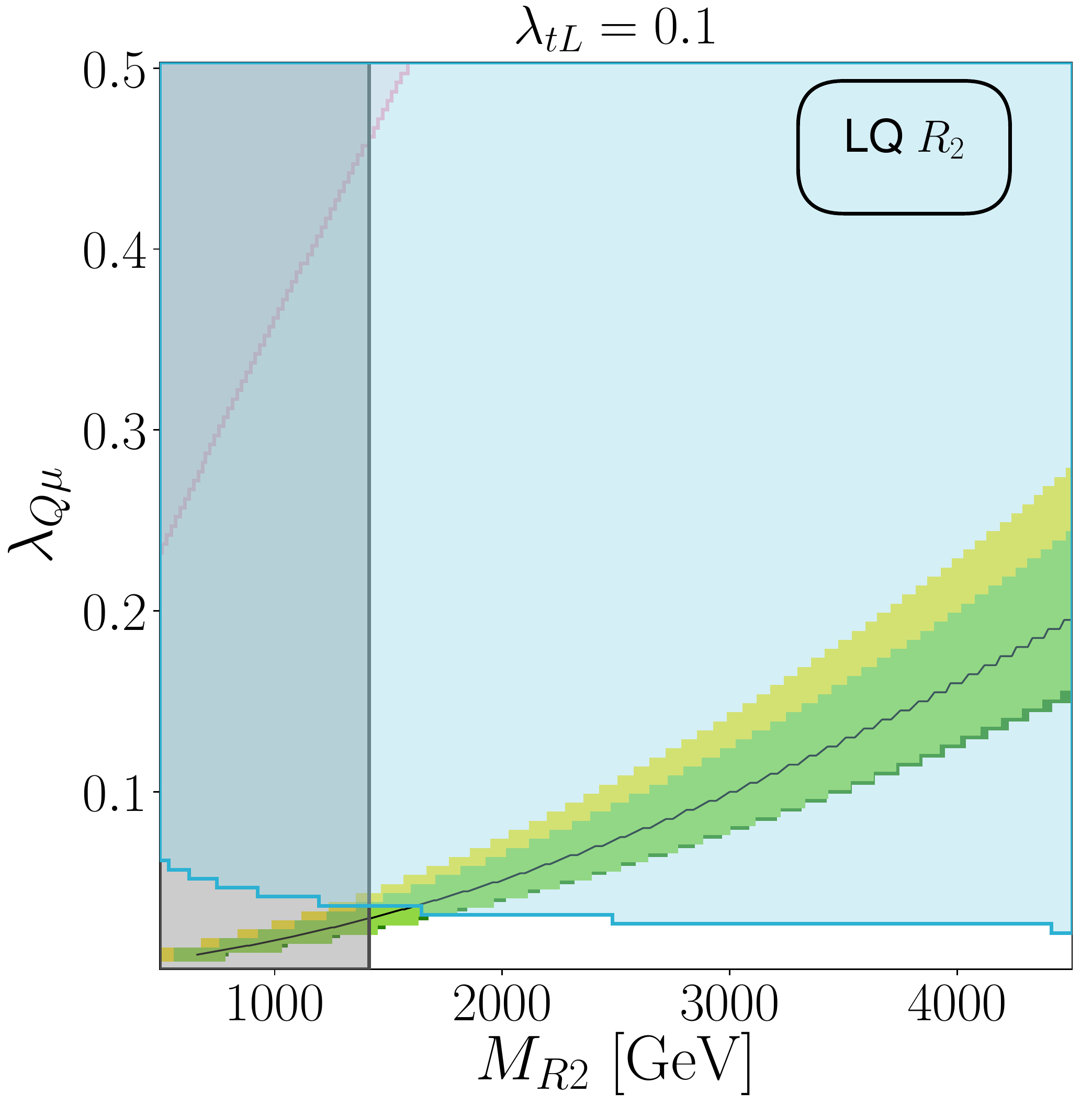}
		\includegraphics[width=0.32\textwidth]{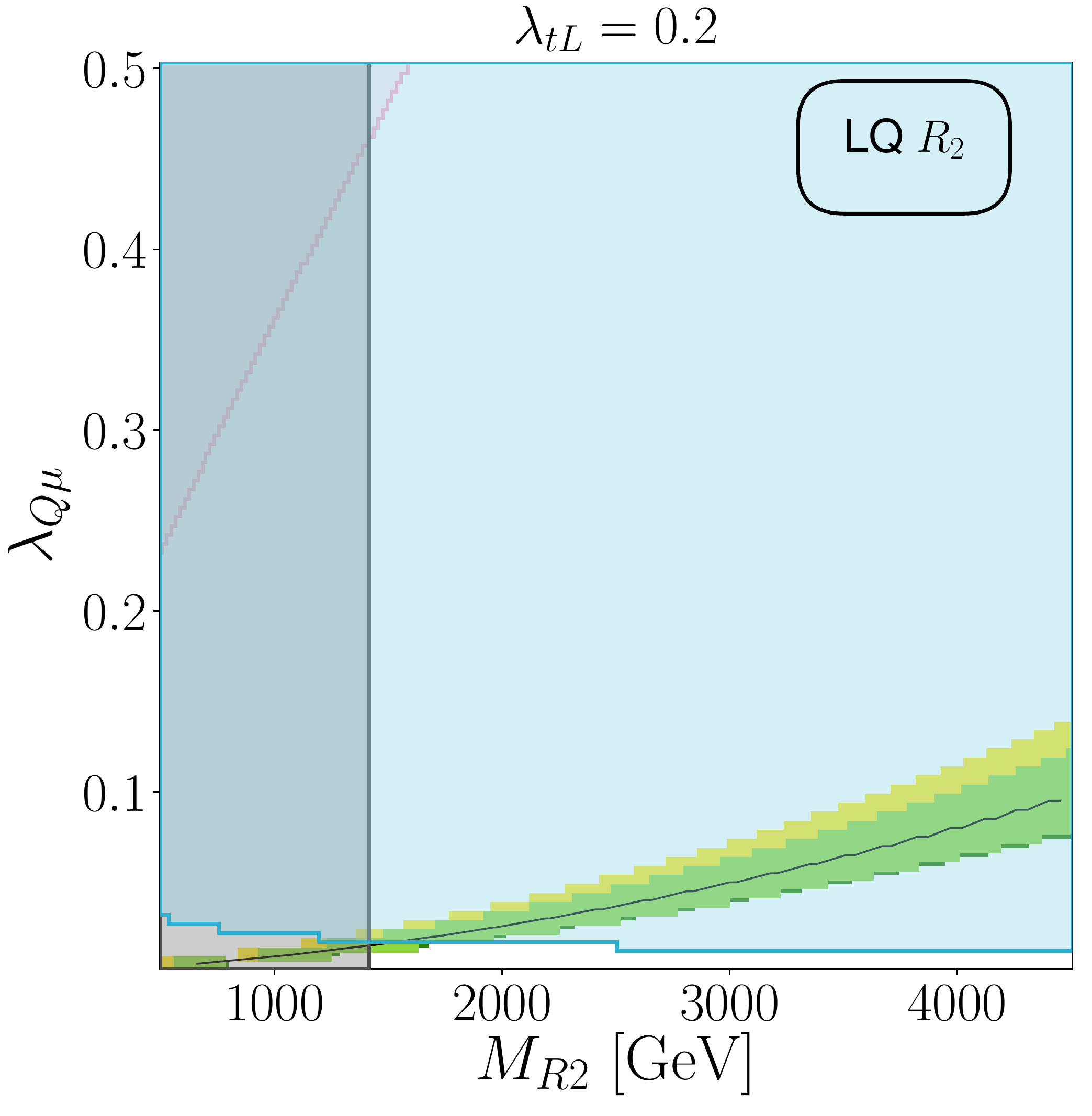}
		\includegraphics[width=0.32\textwidth]{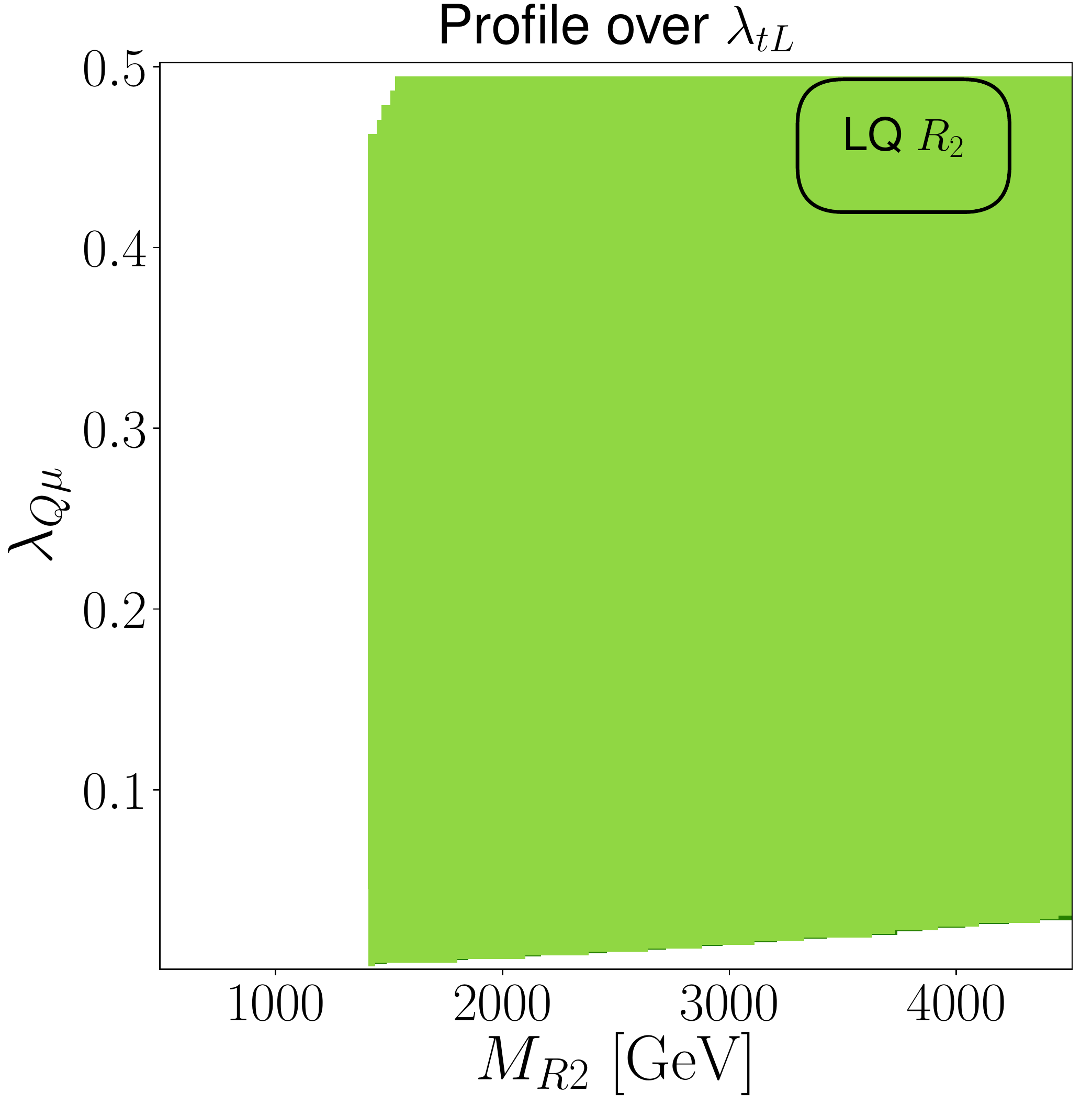}
		\caption{ Scenarios which can explain
                  $\damu^\text{BNL}$ and/or $\damuNEW$ in the
                  model with a scalar leptoquark $SU(2)_L$ doublet
                  $R_{2}$ defined in
                  Eq.\ (\ref{eqn:ScalarLeptoquarkDoublet}). Results
                  are shown in the $M_{R2}$-$\lambda_{QL}$ plane for
                  fixed $\lambda_{tL}=0.1$ (left panel),
                  $\lambda_{tL}=0.2$ (middle panel) and by varying
                  $\lambda_{tL}\in[0,0.5]$ (right
                  panel). In the latter case we {\it profile} to find
                  the regions in the $M_{R2}$-$\lambda_{QL}$ plane
                  where these measurements can be explained for some
                  value of $\lambda_{tL}$ in the scan and satisfy
                  experimental constraints. Colours are the same as in
                  Fig.\ \ref{fig:ScalarLeptoquarkSinglet}.  Additional
                  relevant exclusions in the parameter space from the
                  constraint $Z\rightarrow\nu\nu$ are shaded in pink
                  and in the right panel this is included in the list
                  of experimental constraints that points must
                  satisfy, in addition to those listed in
                  Fig.\ \ref{fig:ScalarLeptoquarkSinglet}. \label{fig:ScalarLeptoquarkDoublet}}
	\end{figure}

	The contributions to $\amu$ from the introduction of the scalar leptoquarks 
	$S_{1}$ and $R_{2}$ with Lagrangians given by Eqs.\ 
	(\ref{eqn:ScalarLeptoquarkSinglet},\ref{eqn:ScalarLeptoquarkDoublet}) are shown 
	in Figs.\ \ref{fig:ScalarLeptoquarkSinglet} and 
	\ref{fig:ScalarLeptoquarkDoublet}, respectively.  Results are shown in the 
	$M_{LQ}$-$\lambda_{QL}$ and $M_{LQ}$-$\lambda_{Q\mu}$ planes, scanning up to a 
	leptoquark mass of $M_{LQ}=4500$ GeV. We find that for both $S_{1}$ and $R_{2}$ 
	the observed $\amu$ discrepancy can be explained within $1\sigma$ using similar 
	ranges of couplings \update{and a leptoquark mass $M_{LQ} \gtrsim 1.1$--$1.5$ TeV}.

	The left and middle panels of
        Fig.\ \ref{fig:ScalarLeptoquarkSinglet} show where
        $\damu^\text{BNL}$ and $\damuNEW$ can be explained
        when the coupling to the right-handed top quark is fixed to
        $\lambda_{t\mu}=0.1,0.2$ respectively.  For $\lambda_{t\mu} =
        0.1$, the black line showing points which exactly explain the
        $\damuNEW$ discrepancy is a parabolic curve,
        following the quadratic relationship between leptoquark mass
        and coupling in Eq.\ (\ref{eqn:ScalarLeptoquarkLOContrib}).
        By increasing the coupling to $\lambda_{t\mu}=0.2$, a lower
        value of the coupling $\lambda_{QL}$ to the left-handed top
        quark is required to get the same contributions to $\damu$,
        and the region which \update{can explain $\damuNEW$}
        narrows and flattens as shown in the middle panel.  In both
        cases the new $\damuNEW$ value can be explained
        with a \update{marginally smaller} coupling than the
        previous BNL value due to the small \update{decrease} in the discrepancy.
        CMS searches for scalar leptoquarks \cite{Sirunyan:2018ruf},
        shown by grey shading, exclude regions with masses
        \update{$M_{S_1} \lesssim 1.1$--$1.5$ TeV}, dependent on the
        branching ratio $\beta_{S_1}$ in
        Eq.\ (\ref{eqn:ScalarLeptoquarkSingletBR}).  For lower
        $\lambda_{QL}$ couplings the strongest mass constraint of
        $M_{S_1} > 1420$ GeV arises, since $\beta_{S_1}\rightarrow1$
        as $\lambda_{QL}\rightarrow0$.  Additionally, the cyan region
        indicates points affected by the ``fine-tuning'' in the muon
        mass discussed in Sec.\ \ref{sec:BSMoverview} and defined in
        Eq.\ (\ref{eqn:FineTuningLimits}): here the loop contributions
        to muon mass calculated in \fs cause the muon mass to increase
        more than twice the pole mass or decrease less than half the
        pole mass.  Satisfying this fine-tuning criteria, places a
        rough upper limit of \update{$\lambda_{QL} \lesssim 0.06$} for
        $\lambda_{t\mu}=0.1$ and \update{$\lambda_{QL} \lesssim
          0.04$} for $\lambda_{t\mu}=0.2$.  Note that these fine-tuning
        conditions allow only a \update{small strip} of the parameter space to remain.
	
In the right panel of Fig.\ \ref{fig:ScalarLeptoquarkSinglet}, we
profile over $\lambda_{t\mu}$ to find the best fits to $\amu$, that is
we vary the coupling $\lambda_{t\mu}\in[0,0.5]$ and show the lowest
number of standard deviation within which $\amu$ can be explained in
the 2D plane, from all scenarios where
$\lambda_{t\mu}\in[0,0.5]$.  This panel has the same constraints as before, except for
the muon mass fine-tuning.  The LHC leptoquark searches impose a limit
of \update{$M_{S_1} \gtrsim 1.1$ TeV}, with slightly higher mass limits
implied by this for the lowest $\lambda_{QL}$ values since the
branching ratios depend on the leptoquark couplings. The exclusion in the bottom right of the plot is formed by our choice of restricting the $\lambda_{t\mu}$ to moderate values $\leq 0.5$.  If we allow $\lambda_{t\mu}$ to vary up to $\sqrt{4\pi}$ then no exclusion in the bottom right would be visible due to the large chirality flip enhancement.\footnote{In contrast if we plotted the charm-philic case the combination of LHC and large $\damu$ constraints would lead to a much larger exclusion in the bottom right of the plot than can be seen here, even when the equivalent coupling is varied up to the perturbativity limit.}  As the mass $M_{S_1}$ is increased, the minimum coupling $\lambda_{QL}$ which produces a contribution that can explain $\amu$ within $1\sigma$ increases, in line with
Eq.\ (\ref{eqn:ScalarLeptoquarkLOContrib}), with the \update{new world
  average} result able to be explained in \update{narrower region} of the
parameter space due to \update{the reduction in the uncertainty}.

 Similarly, for the scalar leptoquark doublet $R_{2}$, $\damu$ contours
for fixed values of the coupling to right-handed top quarks,
$\lambda_{tL}=0.1,0.2$, are shown in left and middle panels of
Fig.\ \ref{fig:ScalarLeptoquarkDoublet}, with a profile over
$\lambda_{tL}$ shown in the right panel.  Again, the $\damu$ contours
for fixed $\lambda_{tL}=0.1, 0.2$ follow the quadratic ratio given in
Eq.\ (\ref{eqn:ScalarLeptoquarkLOContrib}), with larger $\damu$ values
from a given leptoquark mass being obtained with the larger coupling
$\lambda_{tL}=0.2$.  This time, the mass constraints placed on the
leptoquark $R_{2}^u$ from the LHC are independent of the couplings, as
$R_2^u$ always decays to a top quark and muon.  Again imposing the
fine-tuning criteria has a huge impact, with the contributions from
the leptoquark reducing the muon mass $m_\mu$. 
\update{For $\lambda_{tL}=0.1$ only a small corner of
  parameter space can explain $\damuNEW$ without fine tuning, and for
  $\lambda_{tL}=0.2$ this shrinks to a tiny region.}  From
profiling over $\lambda_{tL}$, we find that again $\damuNEW$ can be
explained within $1\sigma$ for \update{masses $M_{R_2} \ge 1420$ GeV}.
For higher couplings $\lambda_{Q\mu} \gtrsim 0.47$, constraints from
$Z\rightarrow\nu\nu$ raise the minimum mass which can explain
$\damuNEW$. As with the previous model our choice to
  restrict ourselves to moderate $\lambda_{tL} \leq 0.5$ leads a lower
  bound on the $\lambda_{Q\mu}$ coupling that increases with mass as a
  smaller coupling $\lambda_{Q\mu}$ \update{cannot produce a large enough
  $\damuNEW$}.  As before there would be \update{no exclusion} in the bottom
  right region if we varied $\lambda_{tL}$ up to $\sqrt{4\pi}$.

	\begin{figure}[tb]
		\centering
        \includegraphics[width=0.32\textwidth]{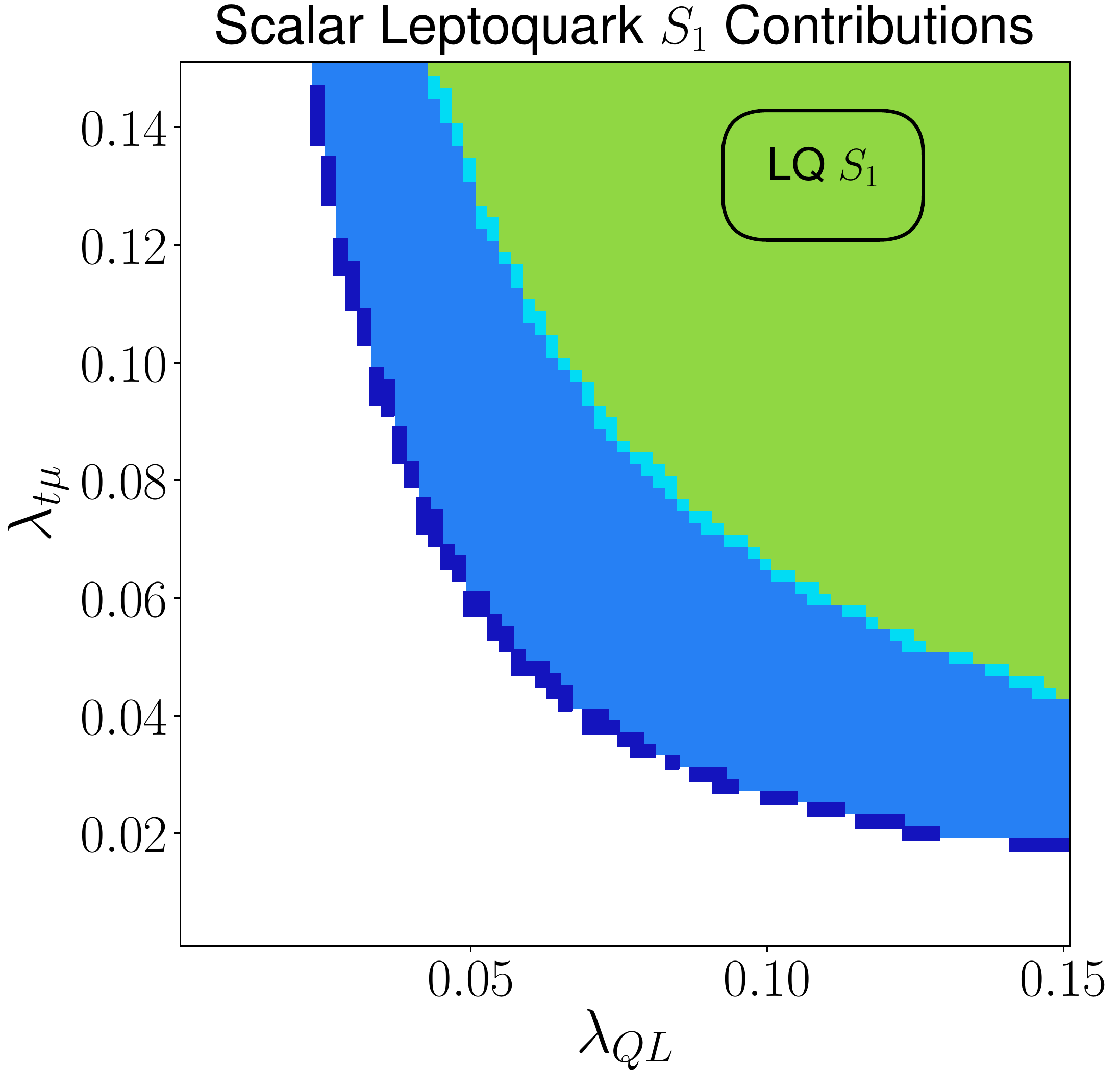}
	    \includegraphics[width=0.32\textwidth]{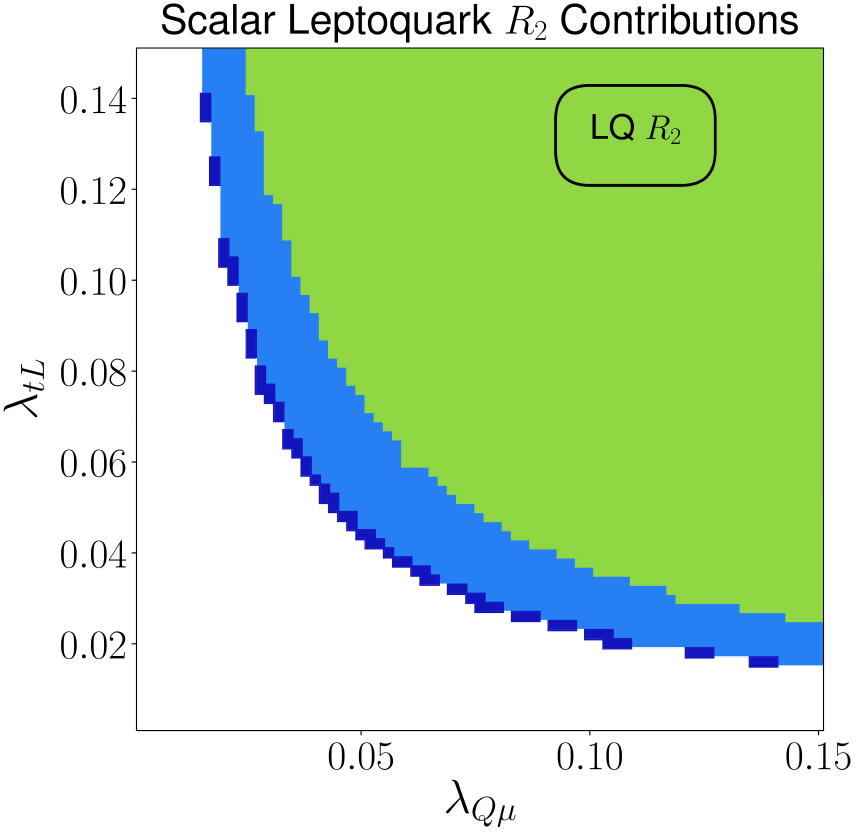}
	    \caption{ Results for $\damuNEW$ and $\damuBNL$ in the planes of the leptoquark couplings where we profile over all possible scalar leptoquark masses up to $4.5$ TeV.  More specifically in the left panel we show results for the $SU(2)_L$ singlet $S_{1}$ defined in Eq.\ (\ref{eqn:ScalarLeptoquarkSinglet}) in the $\lambda_{QL}$--$\lambda_{t\mu}$ plane and in the right panel we show the scalar leptoquark $SU(2)_L$ doublet $R_{2}$ defined in Eq.\ (\ref{eqn:ScalarLeptoquarkDoublet}) in the $\lambda_{Q\mu}$--$\lambda_{tL}$ plane.  We show couplings between the leptoquark and the SM fermions that can satisfy direct LHC searches and $Z\rightarrow\mu\mu$ and $Z\rightarrow\nu\nu$ constraints and can simulteneously explain $\damuNEW$ (green) or both $\damu^{\textrm{BNL}}$ {\it and} $\damuNEW$ (lime green) within $1\sigma$.  Additionally points which can also avoid the fine-tuning constraints in Eq.\ (\ref{eqn:FineTuningLimits}) on the muon mass are shaded in dark blue if they can explain $\damuNEW$, aqua if they can explain $\damu^{\textrm{BNL}}$ within $1\sigma$, and blue if they can explain {\it both}.  \label{fig:ScalarLeptoquarkProfiles}}
	\end{figure}

	In Fig.\ \ref{fig:ScalarLeptoquarkProfiles} the $\damu$
    predictions for both leptoquarks were profiled over the mass
    of the leptoquark with couplings ranging up to a cutoff of
    $0.5$, where we have again excluded points ruled out by the
    LHC searches and the decays $Z\rightarrow\mu\mu$ or
    $\nu\nu$. The regions which can explain \update{$\damuNEW$} follow the
    relationship given in
    Eq.\ (\ref{eqn:ScalarLeptoquarkLOContrib}), where the
    contribution has identical dependence on the couplings to the
    left- and right-handed top quarks.  As expected LHC limits on
    the leptoquark mass mean that there is a lower limit for the
    values of the couplings that can \update{explain the observed
    $\amu$ disagreement within $1\sigma$}.  However, this limit
    is extremely small so that only leptoquarks with couplings
    \update{$\lambda_{QL}\times\lambda_{t\mu} \lesssim 0.003$} are unable to produce large enough contributions to
    $\amu$ with a mass that has not be excluded by LHC searches in
    the $S_1$ model. As can be seen in the plot we obtain slightly lower limits on 
    the couplings in the $R_2$ model compared to the $S_1$ model.  This is due to 
    the higher limits from the LHC on the leptoquark singlet at low 
    couplings, as seen in Figs.\ \ref{fig:ScalarLeptoquarkSinglet}, 
    \ref{fig:ScalarLeptoquarkDoublet}.  Note that if we try to explain 
    $\damu^{\textrm{BNL}}$ or $\damuNEW$ while avoiding fine-tuning of the muon mass, 
    then we get an upper limit on the couplings of 
    \update{$\lambda_{QL}\times\lambda_{t\mu} \lesssim 0.006$} for the $S_1$ model and 
    \update{$\lambda_{tL}\times\lambda_{Q\mu} \lesssim 0.004$} for the $R_2$ model.  
    However, putting aside fine-tuning constraints the deviation between the SM 
    prediction and the observed value of the anomalous magnetic moment of the muon can 
    be explained within $1\sigma$, needing couplings of \update{reasonable size no 
    smaller than $\approx 0.05$ each}.

  There are many possibilities for detecting leptoquarks at collider
  experiments, including planned high-luminosity improvements at the
  LHC as well as several proposed colliders such as the HE-LHC
  \cite{Abada:2019ono}, ILC \cite{Behnke:2013xla}, CLIC
  \cite{Aicheler:2012bya}, CEPC \cite{CEPCStudyGroup:2018rmc}, and the
  FCC-ee \cite{Abada:2019zxq} and FCC-hh \cite{Benedikt:2018csr}.  The
  contributions of scalar leptoquarks to decays and flavour-changing
  processes such as $Z\rightarrow l_i l_j$, $Z\rightarrow \nu\nu$,
  $W\rightarrow l_i\nu$, $h\rightarrow l_i l_j$, $l_i\rightarrow l_j
  \gamma$, $l_i\rightarrow l_j \nu\nu$, and $l_i\rightarrow 3 l_j$
  have been examined by Ref.\ \cite{Crivellin:2020mjs}.  The authors
  found that the ILC, CLIC, CEPC, and FCC-ee may be sensitive to
  changes of the $Z\rightarrow l^+l^-$ couplings at the level of
  $10^{-5}-10^{-6}$ (relative to the standard coupling).  They also
  estimated the ILC and FCC-ee sensitivity to branching fractions of
  $Z\rightarrow ll'$ to be $10^{-8}-10^{-9}$.
  Ref.\ \cite{Crivellin:2020ukd} examined the Higgs decays
  $h\rightarrow\gamma\gamma$, $h\rightarrow gg$ and $h\rightarrow
  Z\gamma$, and showed that the scalar leptoquarks contributions to
  these can generally be measured with increased precision at the ILC,
  CLIC, CEPC, FCC-ee and FCC-hh.  They also showed that the value of
  the electroweak oblique observables, $S$ and $T$ (but not $U$), can
  change due to leading order contributions from scalar leptoquarks.
  They claimed that contributions to $T$ of size $0.01$ at the CEPC
  and $0.005$ at the FCC-ee could be measured.
  Ref.\ \cite{Dorsner:2018ynv} calculated the production
  cross-sections of scalar leptoquark pairs and a scalar leptoquark
  with a lepton in proton-proton collisions at the LHC, HL-LHC, and
  HE-LHC colliders.  The cross sections of leptoquark pair production
  at the HE-LHC and FCC-hh were also given by
  Ref.\ \cite{Bar-Shalom:2018ure}.  Ref.\ \cite{Bigaran:2020jil} was
  able to place an upper limit of $M_{LQ}<65$ TeV on the mass of the
  leptoquarks if one requires them to explain $\Delta a_e$
  simultaneously with $\damu$.  Additionally, other leptoquarks have
  also been shown to be detectable at collider experiments
  Ref.\ \cite{Hiller:2018wbv,Allanach:2019zfr,Altmannshofer:2020ywf,Asadi:2021gah,Huang:2021biu}.

  In summary leptoquarks are an exciting and well-motivated
  possibility for physics beyond the SM, which is not only motivated
  by $\amu$ but also by flavour anomalies. However, among all possible
  leptoquark quantum numbers, the $S_1$ and $R_2$ models are
  \update{the only two viable explanations of $\damu^{\text{BNL}}$ and
    $\damuNEW$}.  Furthermore, leptoquark couplings of the left- and
  the right-handed muon to top quarks are required. In this way,
  chirality flip enhancements by $m_t/m_\mu$ are possible.  Under
  these conditions, leptoquark masses above the LHC-limit of around
  \update{$1.4$~TeV} can accommodate $\damuNEW$ without violating
  flavour constraints. \update{The large $\damuNEW$ can even be
    explained with leptoquark couplings to the left- and right-handed
    muon that are as small as $\approx 0.05$, which is essentially
    unchanged from the BNL value, despite the small reduction in the
    central value.}  In principle, masses in the multi-TeV region can
  accommodate the measured $\amu$, if the couplings are sufficiently
  increased. However, if one takes the fine tuning criteria on the
  muon mass seriously then only very narrow regions of parameter space
  remain for natural leptoquark solutions of the observed $\amu$, just
  above the LHC limits.

%% file: SectionTwoFields.tex
\section{Two-Field Extensions} \label{sec:TwoFields}

In this section we consider simple extensions of the SM by two
fields. We focus on models where the new particles can appear together
in pure BSM loops that contribute to $\amu$, in which case the new
fields should have different spins.  It is not possible to get an enhancement through internal chirality flips in these models, restricting  explanations of $\amu$ to low mass regions, but compressed spectra in these regions may evade the LHC limits. In addition one of the new particles could be a
stable dark matter candidate, and then simultaneous explanation of
$\amu$ and dark matter is possible. For these reasons the case of two
fields with different spins behaves very differently from the one-field
extensions, and is representative for a range of more
elaborate models.

We begin in
Sec.\ \ref{sec:two_field_overview} with an overview 
of the status of two-field extensions. There we also comment on the
case with two fields of the same spin, including the interesting case
of vector-like leptons. From the overview we identify two specific
models that are promising in view of $\amu$,
dark matter and LHC constraints. These models will be analysed in detail in
Sec.\ \ref{sec:Z2symmetric} and compared against the latest $\amuNEW$ result and results from LHC and dark matter experiments.

\subsection{Overview}\label{sec:two_field_overview}

\definecolor{DM}{rgb}{0.6, 0.4, 0.8}
	
\begin{table}[t]
	\begin{center}
	\setlength\tabcolsep{1.5pt}
		\begin{tabular}{|rl|c|c|} \hline
			\multicolumn{2}{|c|}{{\scriptsize $(SU(3)_C \times SU(2)_L \times U(1)_Y)_\textrm{spin}$}} & $+\mathbb{Z}_2$ &  \update{Result for $\damu^\text{BNL}$, $\damuNEW$}\\ \hline
            \CombinedFieldsSNFC
            	& No  &\cellcolor{viable} Projected LHC 14 TeV exclusion, not confirmed \\
                && Yes &\cellcolor{viable} Updated Sec.\ \ref{sec:Z2symmetric} \\ \hline
			\CombinedFieldsSCFN & Both &\cellcolor{ruledout} Excluded: $\Delta \amu < 0$ \\ \hline
			\CombinedFieldsSDFN & Both &\cellcolor{ruledout} Excluded: $\Delta \amu < 0$ \\ \hline
			\CombinedFieldsSNFD
				& No  &\cellcolor{ruledout} Excluded: LHC searches\\
				&& Yes &\cellcolor{viable} Updated Sec.\ \ref{sec:Z2symmetric} \\ \hline
			 \CombinedFieldsSDFC 
				& No  &\cellcolor{ruledout} Excluded: LEP contact interactions \\
				&& Yes &\cellcolor{DM} Viable with under abundant DM \\ \hline
			\CombinedFieldsSCFD & Both &\cellcolor{ruledout} Excluded: $\Delta \amu < 0$ \\ \hline
			\CombinedFieldsSDFD & Both &\cellcolor{ruledout} Excluded:  LEP search \\ \hline
			\CombinedFieldsSDFA 
				& No  &\cellcolor{ruledout} Excluded: LHC searches \\
				&& Yes &\cellcolor{DM} Viable with under abundant DM \\ \hline
			\CombinedFieldsSDFT 
				& No  &\cellcolor{ruledout} Excluded: LHC searches + LEP contact interactions \\
				&& Yes &\cellcolor{DM} Viable with under abundant DM \\ \hline   
			\CombinedFieldsSAFD & Both &\cellcolor{ruledout} Excluded: $\Delta \amu < 0$ \\ \hline
			\CombinedFieldsSAFT 
				& No  &\cellcolor{ruledout} Excluded: LHC searches \\
				&& Yes &\cellcolor{DM} Viable with under abundant DM \\ \hline
			\CombinedFieldsSTFD & Both &\cellcolor{ruledout} Excluded: $\Delta \amu < 0$ \\ \hline
			\CombinedFieldsSTFA & Both &\cellcolor{ruledout} Excluded: $\Delta \amu < 0$ \\ \hline
			\CombinedFieldsFCVN & No   &\cellcolor{ruledout} Excluded: $\Delta \amu < 0$ \\ \hline
			\CombinedFieldsFDVN & No   &\cellcolor{ruledout} Excluded: $\Delta \amu < 0$ \\ \hline
			\CombinedFieldsFDVA & No   &\cellcolor{ruledout} Excluded: LHC searches + LEP contact interactions \\ \hline
			\CombinedFieldsFNVP & No   &\cellcolor{ruledout} Excluded: LHC searches + LEP contact interactions \\ \hline 
			\CombinedFieldsFDVC & No   &\cellcolor{ruledout} Excluded: LHC searches + LEP contact interactions   \\ \hline
			\CombinedFieldsFTVA & No   &\cellcolor{ruledout} Excluded: $\Delta \amu < 0$ \\ \hline
	\end{tabular}
	\caption{Summary of known results for gauge invariant
          extensions of two fields with different spin with one-loop
          contributions to the anomalous magnetic moment of the muon.
          These results are rather exhaustive due to systematic
          investigations and classifications in
          Refs.\ \cite{Freitas:2014pua,Kowalska:2017iqv,Calibbi:2018rzv}.
          Note that this summarises results in the literature where
          different assumptions have been made, see the text and the
          original references for details.  When there are multiple
          reasons for ruling a scenario out, we mention the most
          model- independent constraint.  We use color highlighting to
          give a visual indication of the status of the model, namely
          green for viable explanations, red for excluded and purple
          for extensions that are only viable with under abundant dark
          matter.
    	\label{tab:TwoFieldsDifferentSpin}}
		\end{center}
	\end{table}

With the possibilities for one-field extensions essentially exhausted,
it is natural to then consider including two new fields entering the
one-loop diagrams for $\amu$ together. The one-loop diagrams shown in
Fig.\ \ref{fig:GeneralDiagrams} of Appendix
\ref{app:MuonGm2Contributions}, have topologies FFS, SSF, VVF, FFV,
VSF and SVF (where F $=$ fermion, S $=$ scalar and V $=$ vector
boson). Clearly, having loops with only BSM particles requires that
the two new states have different spin. One of the new states must be
a fermion (allowed to be Dirac, Majorana or Weyl fermion), the other
either a scalar or a vector boson. This means the new states can't mix
and new fermions can't couple to both the left- and right-handed muon,
so no new chirality flips may be introduced. Therefore
$\damu^\textrm{BNL}$ explanations like this are heavily constrained
\update{and this has now been strongly confirmed with $\damuNEW$
  result}.  Nonetheless one important new feature is that these both
states couple to the muon together in trilinear vertices. If this is
the only way they couple to SM states, and the BSM states have similar
masses, then LHC limits could be evaded due to compressed spectra.
Furthermore if this is due to a $\mathbb{Z}_2$ symmetry where the BSM
states transform as odd, then it predicts a stable particle that could
be a dark matter candidate.

The status of $\damu^\textrm{BNL}$ from a pair of BSM fields with
different spins is summarised in Table
\ref{tab:TwoFieldsDifferentSpin}\update{, and these results have
  simply been confirmed by the new world average $\damuNEW$}.  This
table draws extensively from
Refs.\ \cite{Freitas:2014pua,Kowalska:2017iqv,Calibbi:2018rzv}.
Ref.~\cite{Freitas:2014pua} considers models without dark matter
candidates and requires minimal flavour violation, while
Refs.~\cite{Kowalska:2017iqv} and~\cite{Calibbi:2018rzv} consider
models with ${\mathbb{Z}}_2$ symmetry and dark matter candidates. Many
models can be immediately excluded because their contributions to
$\amu$ are always negative\footnote{The sign of $\damu$ from the
  one-loop diagrams can be understood analytically and
  Ref.\ \cite{Calibbi:2018rzv} also presents general conditions for a
  positive contribution, based on the hypercharge and the $SU(2)$
  representations.  
} with results in the literature in reasonable agreement regarding this.
Models with positive contributions to $\amu$ predict a sufficiently
large contribution only for very light states, and as a result
collider searches place severe constraints on these models.  However the collider search limits depend on the model
  assumptions and need to be understood in this context. The ${\mathbb{Z}}_2$ symmetry is particularly important as it regulates the interaction between the SM and BSM particles. Therefore, when necessary/required, the models in Table~\ref{tab:TwoFieldsDifferentSpin} are considered in two cases: with or without ${\mathbb{Z}}_2$ symmetry. Collider
  constraints effectively eliminate almost all the models without 
  $\mathbb{Z}_2$ symmetry, while requiring that the $\mathbb{Z}_2$
  symmetric models simultaneously explain dark matter and
  $\damu^\textrm{BNL}$ effectively restricts us to just the two models that we
  update in Sec.\ \ref{sec:Z2symmetric}. However first, in the
  following, we explain the assumptions used in our main sources to
  eliminate the remaining models that give a positive contribution to $\amu$ and the important
  caveats to these findings. 

The introduction of a vector-like fermion and a scalar or a vector
without any additional symmetries was dealt with by
Ref.\ \cite{Freitas:2014pua}, considering different $SU(2)_L$
representations, namely singlets, doublets, triplets or adjoint
triplets.  They quickly eliminate a scalar doublet and fermion doublet
combination, i.e.\ \TwoFieldsSDFD, without considering LHC constraints
because cancellations amongst the contributions mean $\damu$ is too
small for a $1\sigma$ explanation of $\damu^\textrm{BNL}$ while
satisfying basic assumptions like perturbativity and the $\approx 100$
GeV LEP limit \cite{Heister:2002mn,Abdallah:2003xe}.  For LHC searches
they look at Drell-Yan production, which depends only on the gauge
structure, but for decays they rely on $\mathbb{Z}_2$ violating
leptonic interactions where they assume a minimal flavour violating
structure. They apply also LEP constraints on contact interactions,
using the same assumptions for the lepton interactions. However one
should again note the caveats described at the beginning of
Sec.~\ref{sec:one_field_overview} on minimal flavour violation and
abscence of scalar VEV.  Adding only 8 TeV LHC searches effectively
eliminates three more $\damu^\textrm{BNL}$ explanations:
\TwoFieldsSNFD, \TwoFieldsSDFA and \TwoFieldsSAFT.\footnote{There is a
  very small $5$ GeV gap in the exclusion for \TwoFieldsSNFD and a
  small corner of parameter space of \TwoFieldsSAFT escaped 8 TeV
  searches, but given the LHC run-II projection in
  Ref.~\cite{Freitas:2014pua}, this should now be excluded unless there is a large excess in the data.  For this
  reason we describe these models as excluded in
  Table.~\ref{tab:TwoFieldsDifferentSpin}.}
  Constraints on $ee\ell\ell$ contact interactions derived from LEP
  observables alone further excludes \TwoFieldsSDFC, and in combination with LHC searches excludes \TwoFieldsSDFT and fermion-vector
  extensions \TwoFieldsFDVA, \TwoFieldsFNVC and \TwoFieldsFDVC. It is
  important to also note that the limits from contact interactions can
  also be avoided by cancellations from the contributions of heavy states that may appear as part of a more elaborate model, and are therefore quite model dependent.
  
Ref.\ \cite{Kowalska:2017iqv} considered scenarios with a new fermion
and scalar, where there is a $\mathbb{Z}_2$ symmetry under which the
BSM fields are odd, so that the models have a stable dark matter
candidate.  In all cases the dark matter candidate was the scalar.
The new couplings they introduce are renormalizable, perturbative, CP
conserving, gauge invariant and muon-philic (meaning that muons are
the only SM fermions the BSM states couple to). For the scalar-fermion
\TwoFieldsSNFC pair with charged fermion singlet, they find a region
where $\damu^{\textrm{BNL}}$ and the relic density can both be
explained within $2\sigma$.  LHC constraints exclude much of the
parameter space, but significant regions with compressed spectra
survive. The results for the pair \TwoFieldsSNFD are essentially the
same.  We update both models in Sec.~\ref{sec:Z2symmetric}. For an
inert scalar doublet, coupled with fermions that are $SU(2)_L$
singlets, doublets, triplets, i.e.\ \TwoFieldsSDFC, \TwoFieldsSDFD and
\TwoFieldsSDFT, they find a narrow region at, or just below, the Higgs
resonance region ($m_s\approx m_h/2$), that is consistent with both
the relic density and $\amu^{\textrm{BNL}}$ within
$2\sigma$\footnote{Explaining $\amu^\textrm{BNL}$ within $2\sigma$
  requires only adding a small part of the deviation.  As the authors
  comment in their text, for the doublet case the BSM $\damu$
  contribution is very small. A $1\sigma$ explanation should not be
  possible, so we regard this explanation excluded by LEP.}. LHC
searches almost exclude these scenarios, but in all cases a tiny
region where the fermion is about $100$ GeV survives. Since these
regions are so small, in Table \ref{tab:TwoFieldsDifferentSpin} we
only report that these models are viable with under abundant dark
matter. For an inert scalar doublet coupled with an adjoint triplet
fermion (\TwoFieldsSDFA) they find no region can be consistent
$\amu^{\textrm{BNL}}$ and the observed DM relic density.

 Ref.\ \cite{Calibbi:2018rzv} considered the same type
  of models, but substantially expanded the number by including a
  wider range of dark matter candidates and systematically classifying
  the representations in a manner similar to that shown in our Table
  \ref{tab:TwoFieldsDifferentSpin}.  For a given mass of the dark
  matter candidate that is sufficiently heavier than the $W$-boson mass,
  higher $SU(2)_L$ representations will have a comparatively large
  dark matter annihilation cross-section, they use this to place
  further analytically derived constraints on the models. Beyond the
  very fine tuned regions near the Higgs resonance\footnote{As an
    example of this they also present results with explanations near
    the Higgs resonance for the the scalar-fermion \TwoFieldsSDFC
    case. 
} the
  only models that could explain $\damu^{\textrm{BNL}}$ and the relic abundance of
  dark matter are scalar singlet dark matter with either a fermion singlet or fermion doublet, i.e.\ \TwoFieldsSNFC and
  \TwoFieldsSNFD. For these two surviving explanations of dark matter
  and $\damu^\textrm{BNL}$ they further apply constraints from the $8$
  TeV and $13$ TeV LHC searches and LEP limit on electroweak
  states. While in the latter model one can also get a neutral dark
  matter candidate if the fermion is lighter than the scalar, direct
  detection rules this out. They find that LHC searches heavily constrain the regions where relic density and $\damu^{\textrm{BNL}}$
  may be simultaneously explained, but cannot entirely exclude the possibility.
  
\subsubsection{Vector-like leptons}
Before ending this overview we briefly also discuss the case of models
with two fields of the same spin.  In this case mixing of fields and enhanced chirality flips are allowed, but a dark matter candidate is precluded.
Particularly interesting examples are extensions with vector-like leptons (VLLs), i.e.\ extension
by two new vector-like fermions with the same quantum numbers as the
left- and right-handed SM leptons. The muon-philic VLLs can couple
both to left- and right-handed muons and to the Higgs
boson; the contributions to $\amu$ behave similarly to the ones of
leptoquarks in Eq.\ (\ref{eqn:ScalarLeptoquarkLOContrib}), but the
chirality flip at the top quark is replaced by the new Yukawa coupling
of the VLL to the Higgs boson. Such models have been
discussed in detail in Refs.~\cite{Kannike2012,Dermisek:2013gta,Poh:2017tfo}.

A distinguishing property of such models is that significant new
contributions to the muon mass arise at tree level via the mixing with
the new VLL. The relation between loop contributions to
$\amu$ and to the tree-level muon mass 
is therefore more complicated than in the discussion of
Sec.\ \ref{sec:BSMoverview} leading to
Eq.\ (\ref{eq:damulimit}): for Higgs-loop contributions to $\amu$,
the ratio between $\damu$ and $\Delta m _\mu ^{\text{tree}}$ does not 
behave as $1/M_{\text{BSM}}^2$ but rather as $1/(16\pi^2 v^2)$ 
\cite{Kannike2012,Dermisek:2013gta}. Nevertheless, for 
couplings in line with electroweak precision constraints and
perturbativity, only masses up to the TeV scale are able to provide 
significant contributions to $\amu$, similar to the bound of
Eq.\ (\ref{BNLuppermassbound}).

Currently these models with a single VLL can accommodate
all existing limits, but there are two noteworthy constraints. On the
one hand, bounds from $\mu\to e\gamma$ require the VLL couplings to be
non-universal, i.e.\ very much weaker to the electron than to the muon
\update{(to allow large enough contributions for $\amuNEW$)}
\cite{Poh:2017tfo}. On the other hand,  the muon--Higgs coupling is
modified by the large tree-level contributions to the muon
mass. Already the current LHC constraints on the 
Higgs decay rate $h\to\mu^+\mu^-$ imply upper
limits on the possible contributions to $\amu$, and if future
experiments measure the $h\to\mu^+\mu^-$ rate to be in agreement with the SM prediction,
a deviation as large as $\damu^{\text{BNL}}$ cannot be explained
\cite{Poh:2017tfo}.  However extensions with more fields may relax the
mass limits and constraints from $h \to \mu ^+ \mu
^-$~\cite{Dermisek:2020cod,Dermisek:2021ajd}, given the lowest-order
calculations of these references.

Slightly generalized models, where vector-like fermions may also
carry quantum numbers different from ordinary leptons, have
been examined in Refs.\ \cite{Freitas:2014pua,Kannike2012}.  Ref.\ 
\cite{Freitas:2014pua} examined the introduction of a vector-like 
fermion $SU(2)_L$ doublet
{\OneFieldD$\!\!_{1/2}$} with either an $SU(2)_L$ 
singlet {(\OneFieldN$\!\!_{1/2}$ or \OneFieldC$\!\!_{1/2}$)} or a $SU(2)_L$ triplet
{(\OneFieldA$\!\!_{1/2}$ or \OneFieldT$\!\!_{1/2}$)}.  Each BSM fermion state is coupled to the Higgs boson 
and either the SM lepton $SU(2)_L$ doublet $L_i$, the SM lepton 
$SU(2)_L$ singlet $e_i$ with a flavour-universal coupling, or another 
BSM vector-like fermion.  BSM and SM fermions of the same charge then 
mix together, with the amount of mixing between the SM and fermion 
states constrained by LEP limits \cite{delAguila:2008pw}.  Thus they 
allow minimal flavour violation between the SM lepton generations, 
and some of the main constraints on the masses of the vector-like 
fermions comes from minimal flavour violation.  Ref.\ 
\cite{Kannike2012} additionally considered the introduction of a 
vector-like fermion $SU(2)_L$ doublet with hypercharge $Y=-3/2$ with 
either a charged fermion singlet or triplet (either \TwoFieldsFYFC or
\TwoFieldsFYFT).  Ultimately both consider that these mixing vector-like 
fermions can produce positive contributions to $\amu$ and cannot be 
ruled out even by $14$-TeV LHC projections.

\subsection{Scalar singlet plus fermion explanations} \label{sec:Z2symmetric}

Our overview has resulted in two kinds of two-field models of particular interest, 
shown in Table \ref{tab:TwoFieldsDifferentSpin}.  Both models are heavily constrained 
by LHC, but have not been excluded yet in the literature.  We denote the two models as 
Model L and Model R, which extend the SM by adding the following fields:  
\begin{align}
	\text{Model L:} \quad \phi, \ \psi _d = 
	\begin{pmatrix}
		\psi_d^+\\ \psi_d^0
	\end{pmatrix},\, & & 
	\text{Model R:} \quad \phi, \ \psi _s \equiv \psi_s^-,
\end{align}
where $\phi=\phi^0$ is an $SU(2)_L$ singlet scalar field with representation 
$({\mathbf{1}},{\mathbf{1}},0)_0$, $\psi_d$ a doublet fermion field with representation
$({\mathbf{1}},{\mathbf{2}},1/2)_{1/2}$, and $\psi_s$ a singlet fermion field with 
representation $({\mathbf{1}},{\mathbf{1}},-1)_{1/2}$.  These new fermions are 
vector-like, or Dirac fermions, thus not only are the Weyl spinors $\psi_d$ and 
$\psi_s$ introduced but also their Dirac partners $\psi^c_d = 
(\psi_d^{+c\dagger},\psi_d^{0c\dagger})$ and $\psi_s^c \equiv \psi_s^{-c\dagger}$.  
In Model L the new fields couple only to the left-handed muon, and in Model R to the 
right-handed muon.

The two models add the following terms to the SM Lagrangian, using
Weyl notation as in Eq.\ (\ref{eqn:ScalarLeptoquarkDoublet}):
\begin{align} \label{eqn:Min2FieldsLL}
  {\cal L}_{\text{L}} &= \begin{pmatrix}\lambda_L L\cdot \psi_d \phi - M_{\psi} \psi_d^c \psi_d + h.c.\end{pmatrix} - \frac{M_{\phi}^2}{2} |\phi|^2, \\
  \label{eqn:Min2FieldsRR}
	{\cal L}_{\text{R}} &= \begin{pmatrix}\lambda_R \mu \phi \psi_s - M_{\psi} \psi_s^c \psi_s + h.c.\end{pmatrix} - \frac{M_{\phi}^2}{2} |\phi|^2,
\end{align}
which can be written out into their $SU(2)_L$ components:
\begin{align} \label{eqn:Min2FieldsLLComponents}
  {\cal L}_{\text{L}} &= \begin{pmatrix} \lambda_L \nu_{\mu L} \psi_d^0 \phi^0 - \lambda_L \mu_L \psi_d^+ \phi^0 - M_{\psi} \psi_d^{0c\dagger}
    \psi_d^0 - M_{\psi} \psi_d^{+c\dagger} \psi_d^+ + h.c.\end{pmatrix} - \frac{M_{\phi}^2}{2} |\phi^0|^2, \\
  \label{eqn:Min2FieldsRRComponents}
	{\cal L}_{\text{R}} &= \begin{pmatrix}\lambda_R \mu_R^\dagger \psi_s^- \phi^0 -  M_{\psi} \psi_s^{-c\dagger} \psi_s^- + h.c.\end{pmatrix} -  \frac{M_{\phi}^2}{2} |\phi^0|^2.
\end{align}
We have not included additional renormalisable terms involving the
scalar singlet in the Higgs potential, which are not relevant for
$\amu$, but we do briefly comment on the impact they would have later
on, as they can affect the dark matter phenomenology.  This leaves
both models with just three parameters.

In the following we therefore present a detailed update of the
phenomenology of Model L and Model R, scanning over all three
parameters in each case to test whether or not dark matter and large
$\damu$ can be simultaneously explained.  We include the latest data
from Fermilab, and the most recent LHC collider searches included in
\smodels $1.2.3$ in
Ref.\ \cite{Kraml:2013mwa,Ambrogi:2017neo,Dutta:2018ioj,Ambrogi:2018ujg,Sjostrand:2006za,Sjostrand:2014zea,Buckley:2013jua}
and \checkmate $2.0.26$
\cite{Drees2015,Dercks2017,Read:2002hq,Cacciari:2005hq,Cacciari:2008gp,Cacciari:2011ma,deFavereau:2013fsa}.  It
should be noted that we deal only with BSM fields that couple to
second generation SM fermions.  Thus, flavour violation constraints on
our models can safely be ignored, including limits from contact
interactions.  This is in line with the methodology in
Ref.\ \cite{Calibbi:2018rzv}, but not \cite{Freitas:2014pua}.

    The BSM contributions from both of the two field dark matter models come from 
    the FFS diagram \ref{fig:FFSDiagram} of Fig.\ \ref{fig:GeneralDiagrams} with the 
    generic $F$ fermion lines replaced by $\psi^-$ and the scalar $S$ one by $\phi^0$.
    The contributions to $\amu$ from both of these models are
    given by identical expressions (where $\lambda_{L,R}$ are
    generally denoted by $\lambda$). The result is given by
    \begin{equation} \label{eqn:Min2FieldsLLContributions}
    	\damu^{\text{Model L}} =    	\damu^{\text{Model R}} = -Q_{\psi} \frac{\lambda^2}{32\pi^2} \frac{m_\mu^2}{6 M_{\phi}^2} \,E\!\begin{pmatrix}\frac{M_\psi^2}{M_\phi^2}\end{pmatrix},
    \end{equation}
    where $Q_\psi=-1$. It does not involve a chirality flip enhancement, which has a large impact on 
    the ability of these models to explain $\amu$ whilst avoiding collider constraints. 

	\begin{figure}[tb]
		\centering
		\includegraphics[width=0.32\textwidth]{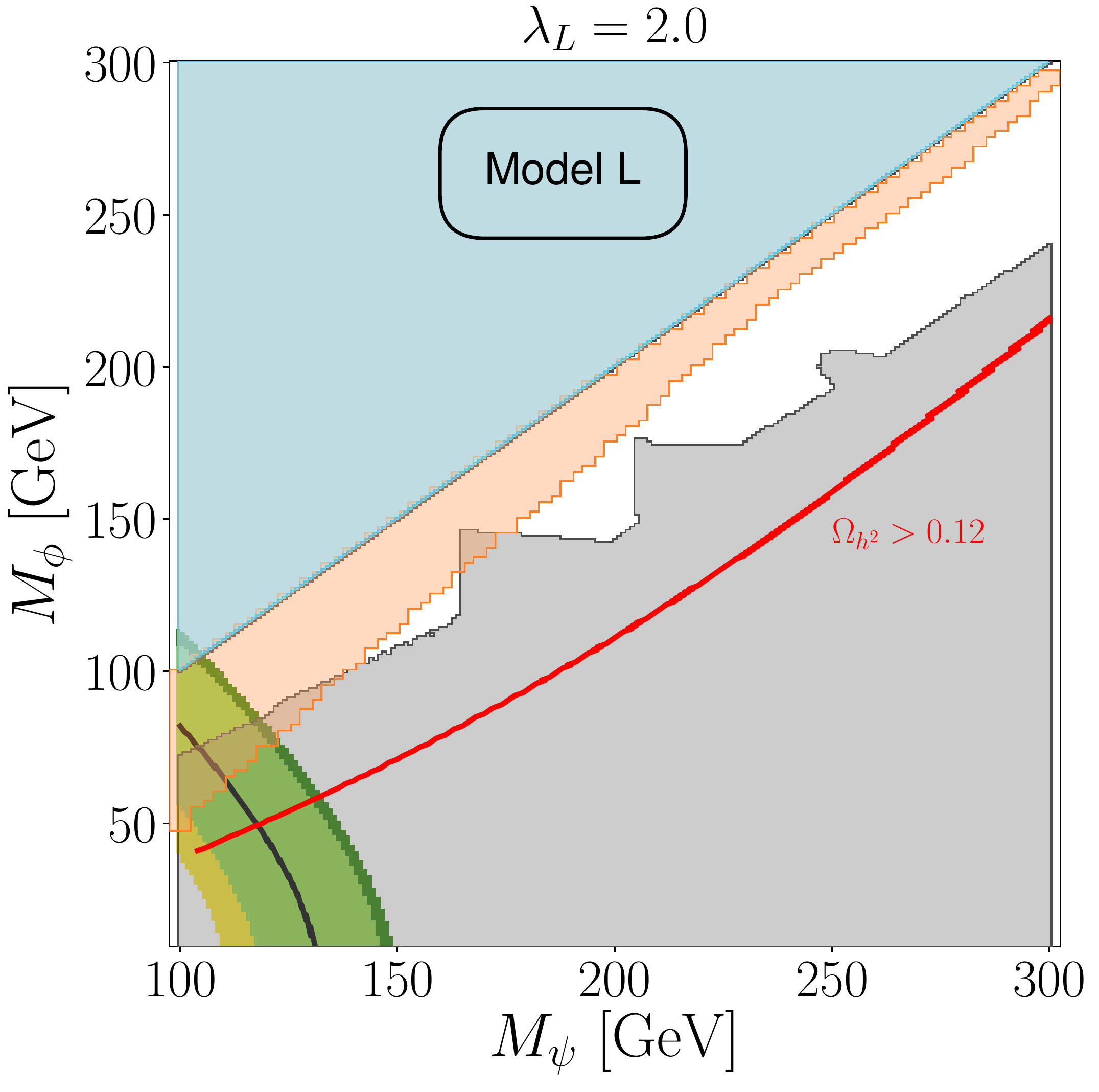}
		\includegraphics[width=0.32\textwidth]{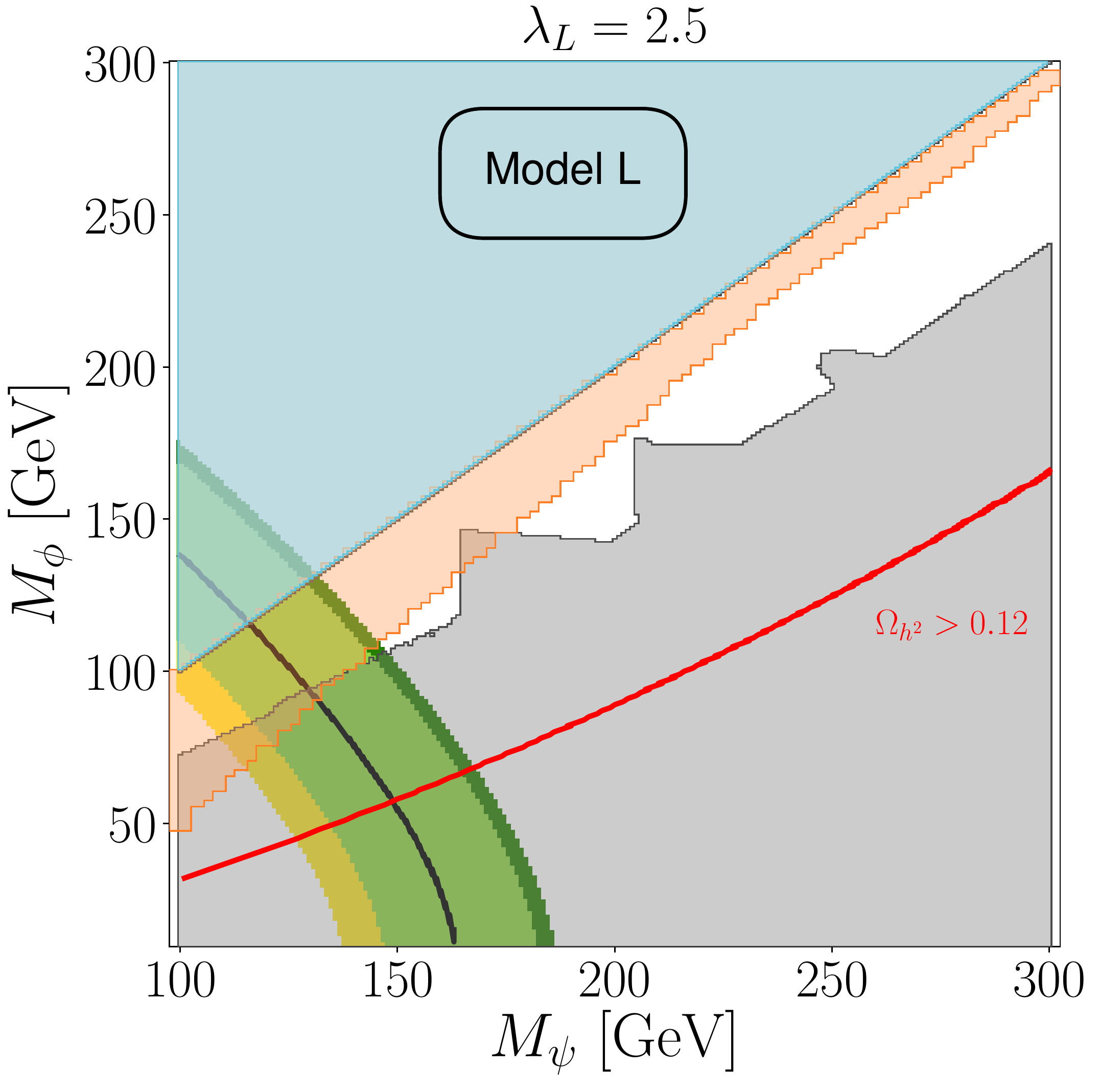}
		\includegraphics[width=0.32\textwidth]{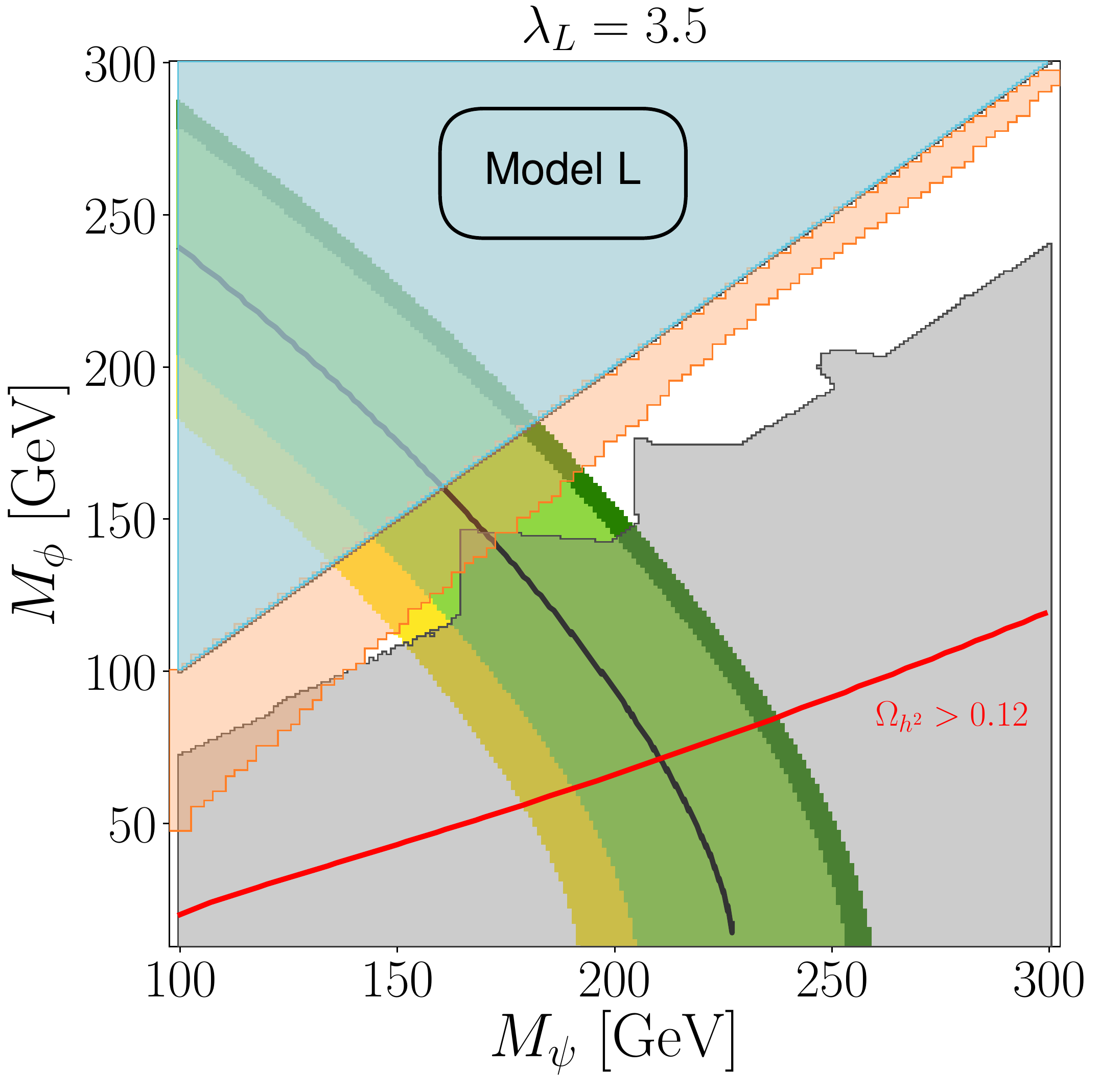}
		\caption{Results from Model L, scanning over the masses of the new fermion and scalar which couple to the left-handed muon.  Regions which can explain $\damu$ in \update{Eq.\ (\ref{eqn:avgDiscrepancy})} or Eq.\ (\ref{eqn:BNLDiscrepancy}) to within $1\sigma$ are coloured green and yellow respectively, with the overlap between the two regions is coloured lime.  The black line indicates masses which produce a $\damu$ contribution matching \update{Eq.\ (\ref{eqn:avgDiscrepancy})}.  The points where a dark matter candidate particle produces the observed relic abundance of $0.1200$, Eq.\ (\ref{eqn:DMRD}), are shown in red, with the region below being excluded due to having an over abundance of dark matter.  The region where $M_\psi < M_\phi$ is shaded cyan.  Regions excluded by $13$-TeV results at the LHC are shaded grey, with the exclusions from the soft leptons search \cite{Sirunyan:2018iwl} obtained using \checkmate shaded in orange.  Thus $\amu$ can only be explained in small slices of parameter space between the grey and orange regions.  As the coupling between the left-handed muon and the BSM particles is increased, higher masses are required to explain $\amu$.  \label{fig:Min2FieldsLLSlice} }
	\end{figure}

\begin{figure}[tb]
		\centering
		\includegraphics[width=0.32\textwidth]{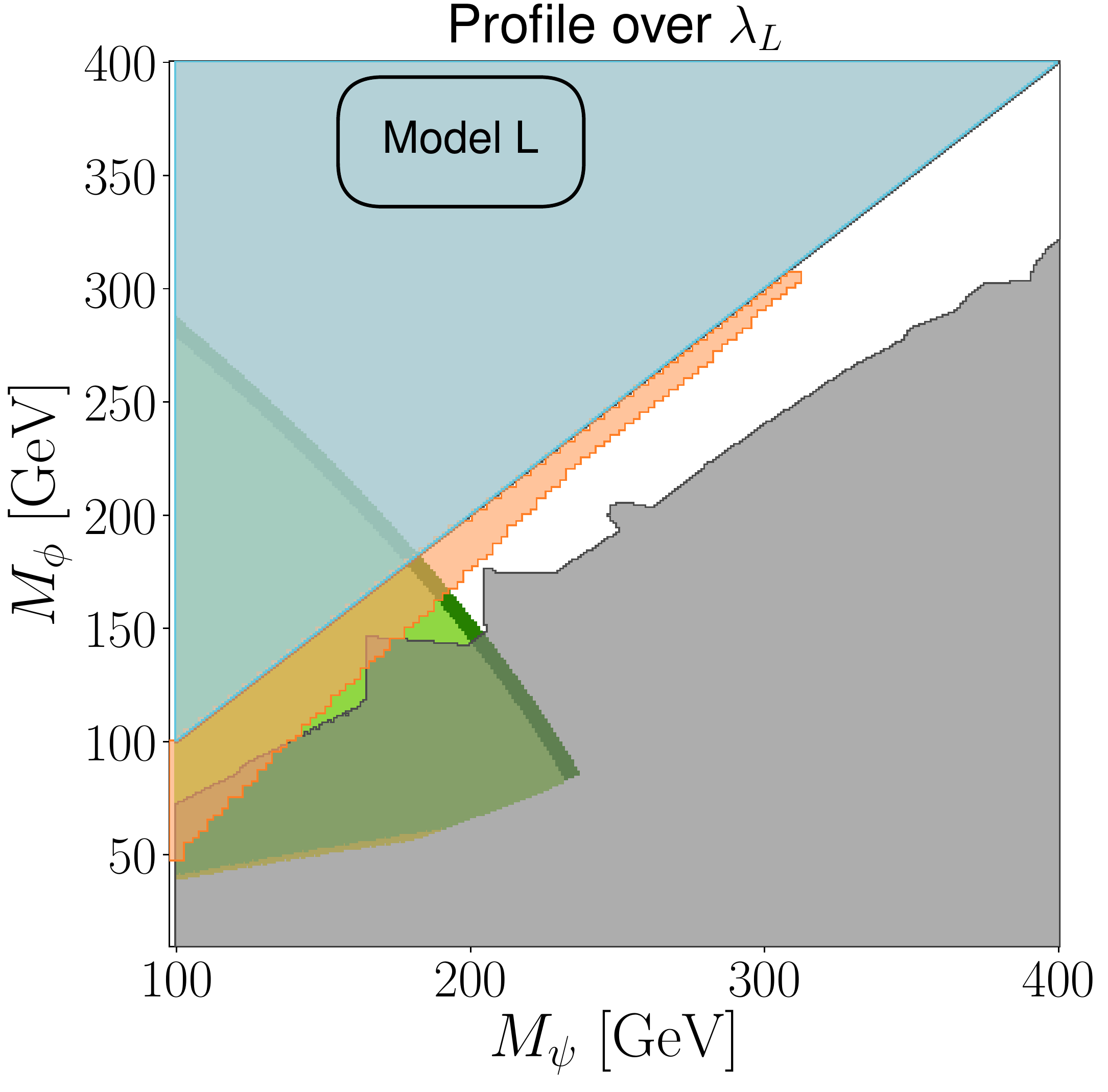}
		\includegraphics[width=0.32\textwidth]{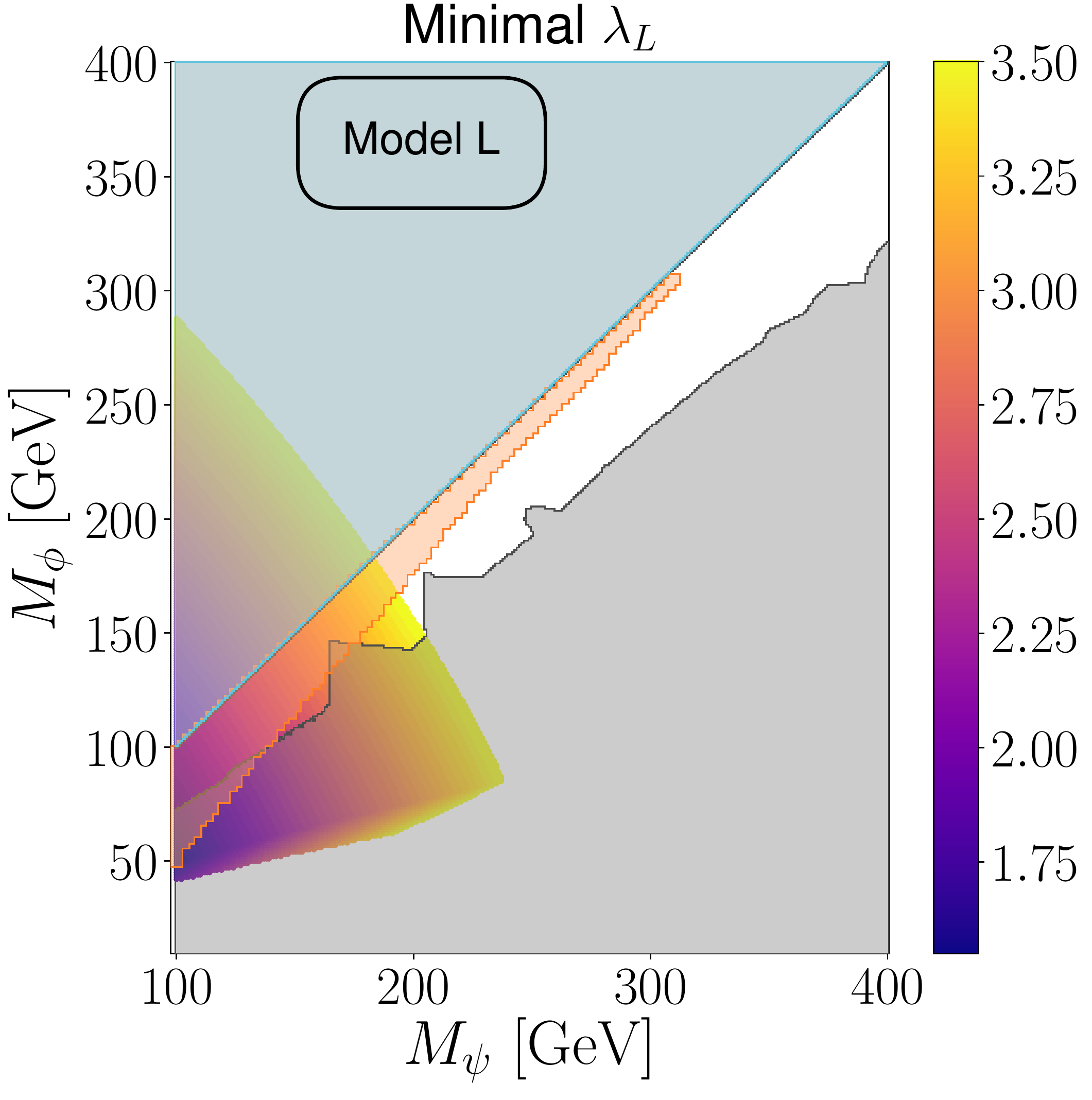}
		\includegraphics[width=0.32\textwidth]{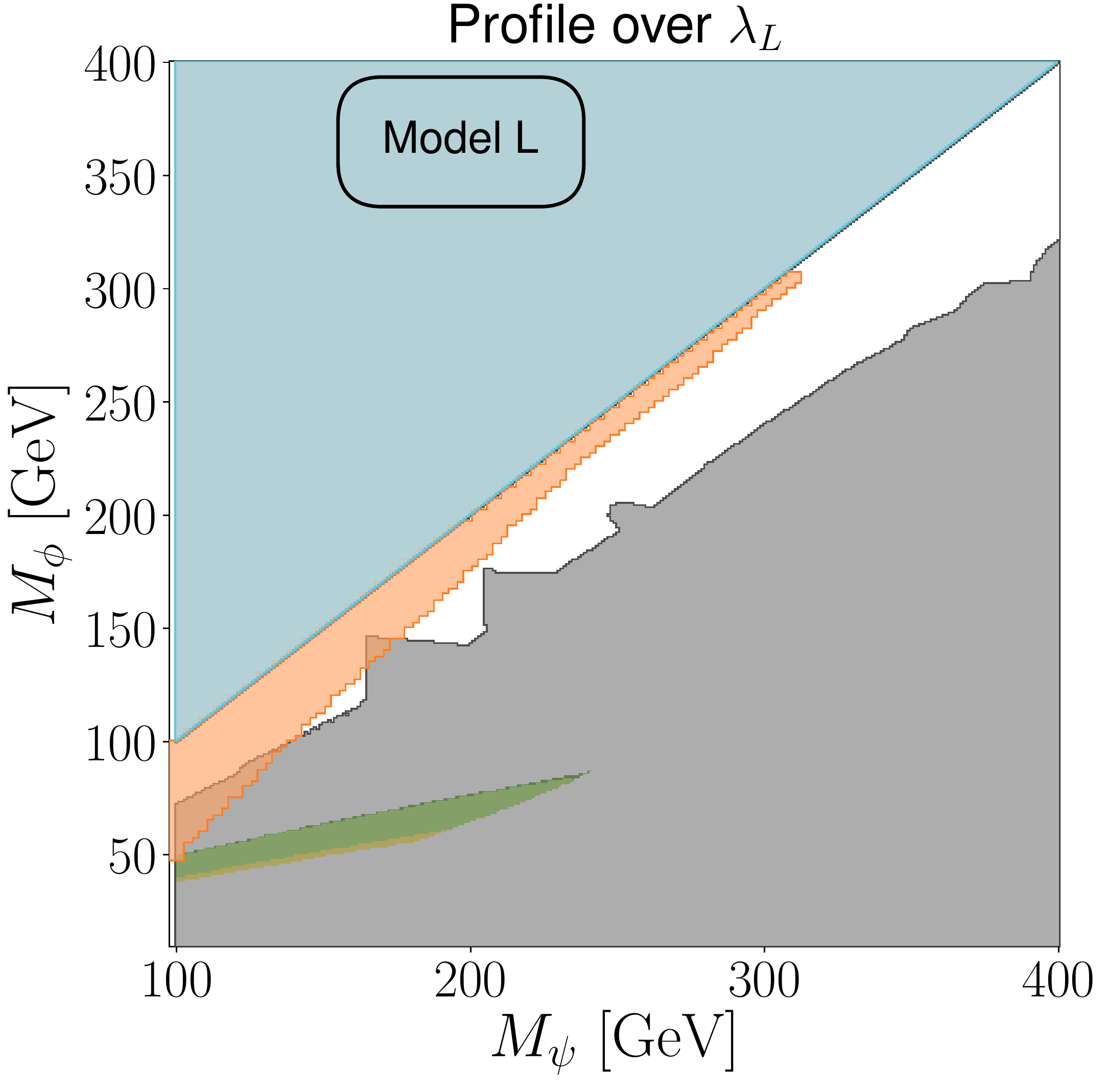}
		\caption{Several profiles of Model L.  Colours for the
                  first and third panels are the same as
                  Fig.\ \ref{fig:Min2FieldsLLSlice}, and the same LHC
                  limits (which do not depend on the coupling) are
                  overlaid.  The left panel shows the best
                  contributions to $\amu$ for each value of the BSM
                  masses, without having a dark matter candidate
                  particle with an over abundant relic density.
                  Likewise, the right panel only shows points which
                  have a dark matter relic density within $1\sigma$ of
                  the Planck observation \cite{Aghanim:2018eyx}. The
                  middle panel shows the smallest value of the
                  coupling $\lambda_L$ which can \update{explain $\amuNEW$
                  (Eq.\ (\ref{eqn:avgDiscrepancy}))} within
                  $1\sigma$ without producing over abundant dark matter.  \label{fig:Min2FieldsLLProfile} }
	\end{figure}

    The results for the first of the two field dark matter models,
    Model L, are shown in
    Figs.\ \ref{fig:Min2FieldsLLSlice}, \ref{fig:Min2FieldsLLProfile}.
    Scans were performed over a grid of scalar mass $M_\phi \in
    [0,400]$ GeV and fermion masses of $M_\psi \in [100,400]$ GeV,
    taking into consideration the LEP limit $M_\psi > 100$ GeV
    \cite{Heister:2002mn,Abdallah:2003xe} on charged fermion masses.
    Before showing results of the full scan, in
    Fig.\ \ref{fig:Min2FieldsLLSlice} we show $\damu$ in three slices
    of the parameter space in the $M_\psi$--$M_\phi$ ($M_\psi$ over
    $100...300$ GeV and $M_\phi$ over $0...300$ GeV) with fixed
    $\lambda_L = 2,2.5,3.5$.  Due to the lack of enhanced chirality
    flips a sufficiently large $\damu$ is obtained only when the
    coupling constant is large and the masses are relatively small.
    However due to the \update{new reduced} measurement of $\amu$, the
    discrepancy can be explained with \update{heavier} masses than
    before as indicated in the green curve.  For \update{$\lambda_L
      \le 2$}, the model cannot provide large enough $\damu$ to
    explain the anomaly within $1\sigma$ while avoiding the LEP
    $M_\psi > 100$ GeV limit and LHC limits (discussed below), while for very
    \update{large values of $\lambda_L = 2.5$ or $\lambda_L = 3.5$} it
    is possible to explain the anomaly but even when $\lambda_L$ is
    close to $\sqrt{4\pi}$, $\damuNEW$ can only be explained within
    $1\sigma$ \update{for masses below $260$ GeV}.  These results can
    be approximately reproduced using
    Eq.\ (\ref{eqn:Min2FieldsLLContributions}), though we have again
    performed the calculation with \fs 2.5.0
    \cite{Athron:2014yba,Athron:2017fvs}, which includes the full
    one-loop calculation and the universal leading logarithmic
    two-loop QED contribution \cite{Degrassi:1998es}.

    We do not consider scenarios with $M_\psi < M_\phi$ because such cases are like 
    Higgsino dark matter or the doublet case of minimal dark matter and as such will 
    be under abundant when the mass is below about $1$ TeV (see e.g.\ Ref.\ 
    \cite{ArkaniHamed:2006mb}).  Without a chirality flipping enhancement it will 
    not be possible to explain $\damu^{\text{BNL}}$ \update{or $\damuNEW$}
    with masses that are heavy enough to explain dark matter. Hence the dark matter candidate is given
    by the scalar singlet $\phi$.  Direct detection of 
    dark matter constraints then depend on the parameters of Higgs potential
    terms involving the singlet.  By including such terms, we found that direct detection 
    constraints rule out significant parts of the parameter space but
    can always be evaded by choosing the additional parameters to be small.
    Therefore, for simplicity, we neglect these  
    parameters in our final numerical results and do not show direct 
    detection constraints.

    The collider constraints are shown with overlayed shading.  The lower grey
    shaded region comes from searches for charginos, neutralinos,
    sleptons, and long-lived particles, using leptonic final
    states in the $8$-TeV searches \cite{Chatrchyan:2013oca,
      Aad:2014vma, Khachatryan:2014qwa, Khachatryan:2015lla} and the
    $13$-TeV searches \cite{Aaboud:2018jiw, Sirunyan:2018nwe,
      CMS:2016ybj, Aad:2019vnb}, included in \smodels $1.2.3$
    \cite{Kraml2014,Khosa2020} and excludes most of the light mass
    parameter space where $\amu$ can be explained.  Nonetheless there
    is still a considerable gap close to the $M_\phi = M_\psi$ line,
    which escapes these constraints, but may be closed by searches for
    compressed spectra.  We therefore also show in shaded orange the
    CMS search for the compressed spectra of soft leptons
    \cite{Sirunyan:2018iwl}, which was obtained using \checkmate
    $2.0.26$ \cite{Drees2015,Dercks2017} using \madanalysis
    \cite{Alwall:2014hca} and \pythia
    \cite{Sjostrand:2006za,Sjostrand:2014zea} to generate event
    cross-sections. As a result of these constraints there is little
    room for the model to evade collider constraints and explain the
    observed value of $\amu$, \update{though some gaps} do evade the
    collider constraints we apply.

    Finally we also consider the relic density of dark matter in this
    model using \micromegas $5.2.1$ \cite{Belanger2018}.  The red line
    in Fig.\ \ref{fig:Min2FieldsLLSlice} indicates where the model's
    dark matter candidate particle produces the Planck-observed relic
	density of $\Omega_{h^2}=0.1200$,
    Eq.\ (\ref{eqn:DMRD}). Along this red line the relic
      abundance of dark matter is depleted to the observed value
      through t-channel exchange of the BSM fermions. This mechanism is 
less effective below the line, where the mass splitting between the BSM scalar and fermion is larger,
 leading to over abundance. All points below
    the red line are strongly excluded by this over abundance, though this does not rule out any
    scenarios that are not already excluded by collider constraints.
Above the line the relic density is under abundant. Our
    results show \update{that it is not possible for any of these
      $\lambda_L$ values} to simultaneously explain \update{$\amuNEW$} and the
    relic density, while evading collider limits.  

	To give a more global picture of the status of the model in
        Fig.\ \ref{fig:Min2FieldsLLProfile} we also vary $\lambda _L
        \in [0,3.5]$.  In the left panel we show where on the
        $M_\phi$--$M_\psi$ plane it is possible to simultaneously fit
        the \update{BNL measurement or the new world average} for
        $\amu$ and avoid an over abundant relic density, for some
        value of $\lambda_L$ within this range.  As expected this
        shows that $\lambda_L$ may be adjusted so that for very light
        masses one may always explain $\amu$ within $1\sigma$, but as
        the masses are raised past \update{$200-300$ GeV}, this is no
        longer possible. As a result the combination of collider
        constraints shown in grey, cyan and orange \update{exclude all
          but two narrow regions of parameter space} in the compressed
        spectra region between the $M_\phi$ and $M_\psi$ masses. As
        shown in the middle panel of
        Fig.\ \ref{fig:Min2FieldsLLProfile}, where we plot the minimum
        value of $\lambda_L$ consistent with a $1\sigma$ explanation
       \update{of $\damuNEW$}, explaining the observed value requires
        \update{a very large coupling $\lambda_L \ge 2.5$}.  One may
        question the precision of our calculation for \update{such
          large values} of $\lambda_L$, however given the mass reach
        of the collider experiments which extends \update{well past}
        the $1\sigma$ region it is unlikely that including higher
        orders in the BSM contributions will change anything
        significantly.  Finally the right panel of
        Fig.\ \ref{fig:Min2FieldsLLProfile} shows $\amu$ results that
        are compatible with explaining the full observed dark matter
        relic density within $1\sigma$, having obtained this data from
        a targeted scan using \multinest 3.10
        \cite{Feroz:2007kg,Feroz:2008xx,Feroz:2013hea,Buchner2014}. Our
        results show that it is now impossible to simultaneously
        explain dark matter and $\damu^\text{BNL}$ and \update{the
          same is true with the updated $\damuNEW$ measurement}.
   
   	Compared to the results for this model shown in
        Ref.\ \cite{Calibbi:2018rzv}, we find that the most recent
        collider search(es) \cite{Aad:2014vma,Sirunyan:2018nwe}, are
        the most important for \update{narrowing the gaps} in the
        exclusion.  Currently, there is \update{very little room for
          the model to survive}.

	\begin{figure}[tb]
		\centering
		\includegraphics[width=0.32\textwidth]{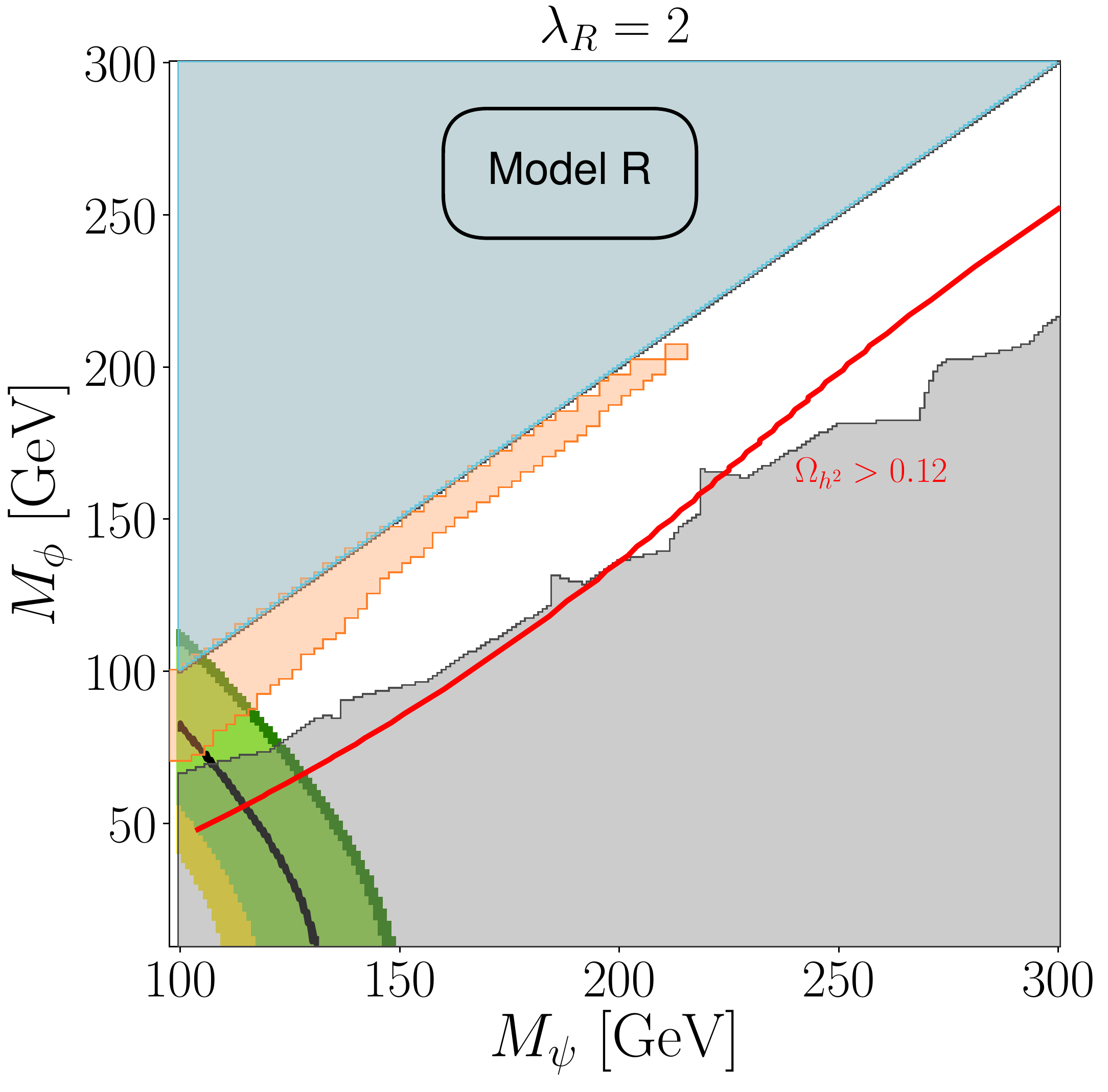}
		\includegraphics[width=0.32\textwidth]{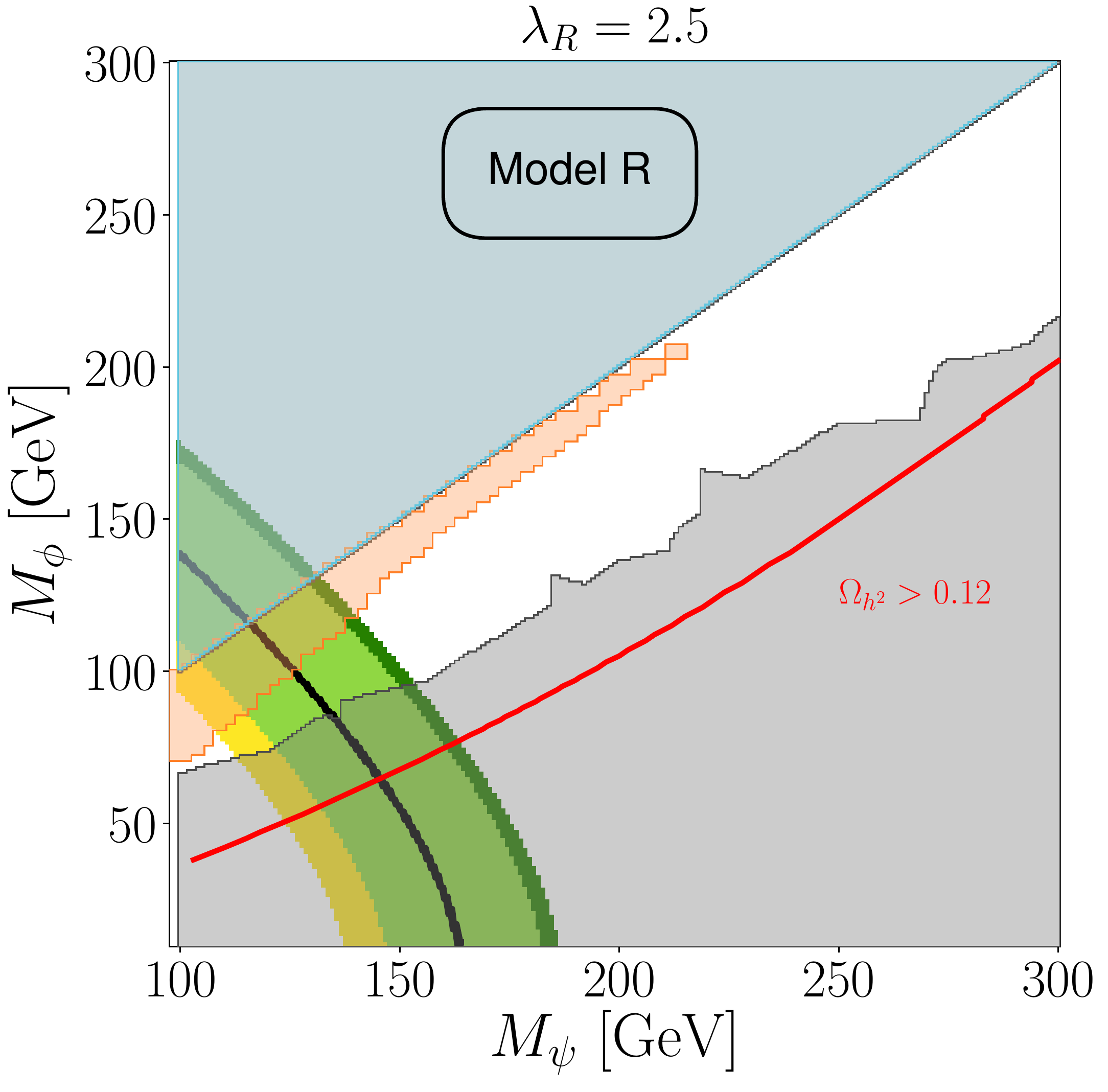}
		\includegraphics[width=0.32\textwidth]{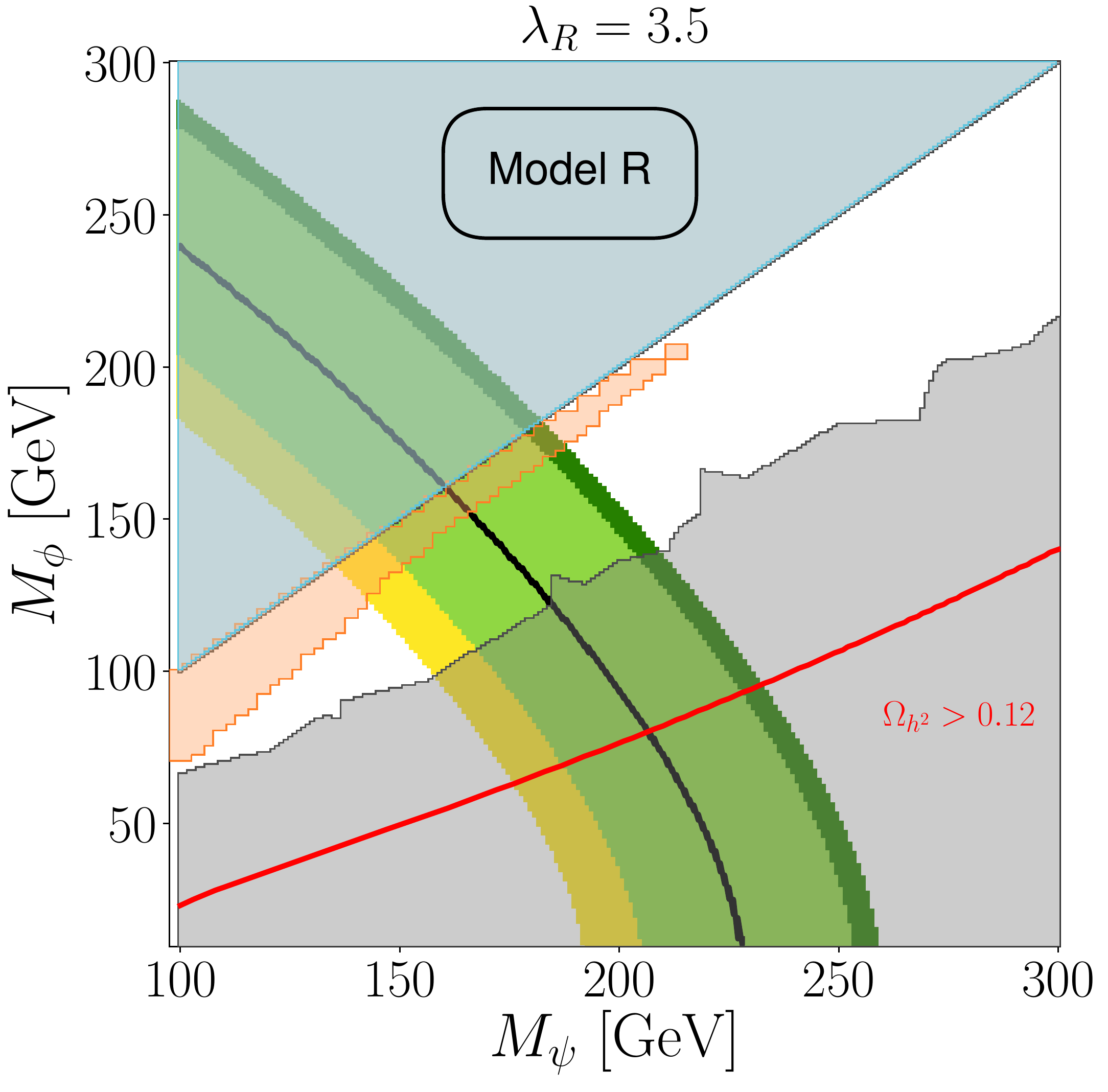}
		\caption{Results from Model R, scanning over the masses of the new fermion and scalar which couple to the right-handed muon.  Colours are the same as Fig.\ \ref{fig:Min2FieldsLLSlice}.  \label{fig:Min2FieldsRR}}
	\end{figure}

	\begin{figure}[tb]
		\centering
		\includegraphics[width=0.32\textwidth]{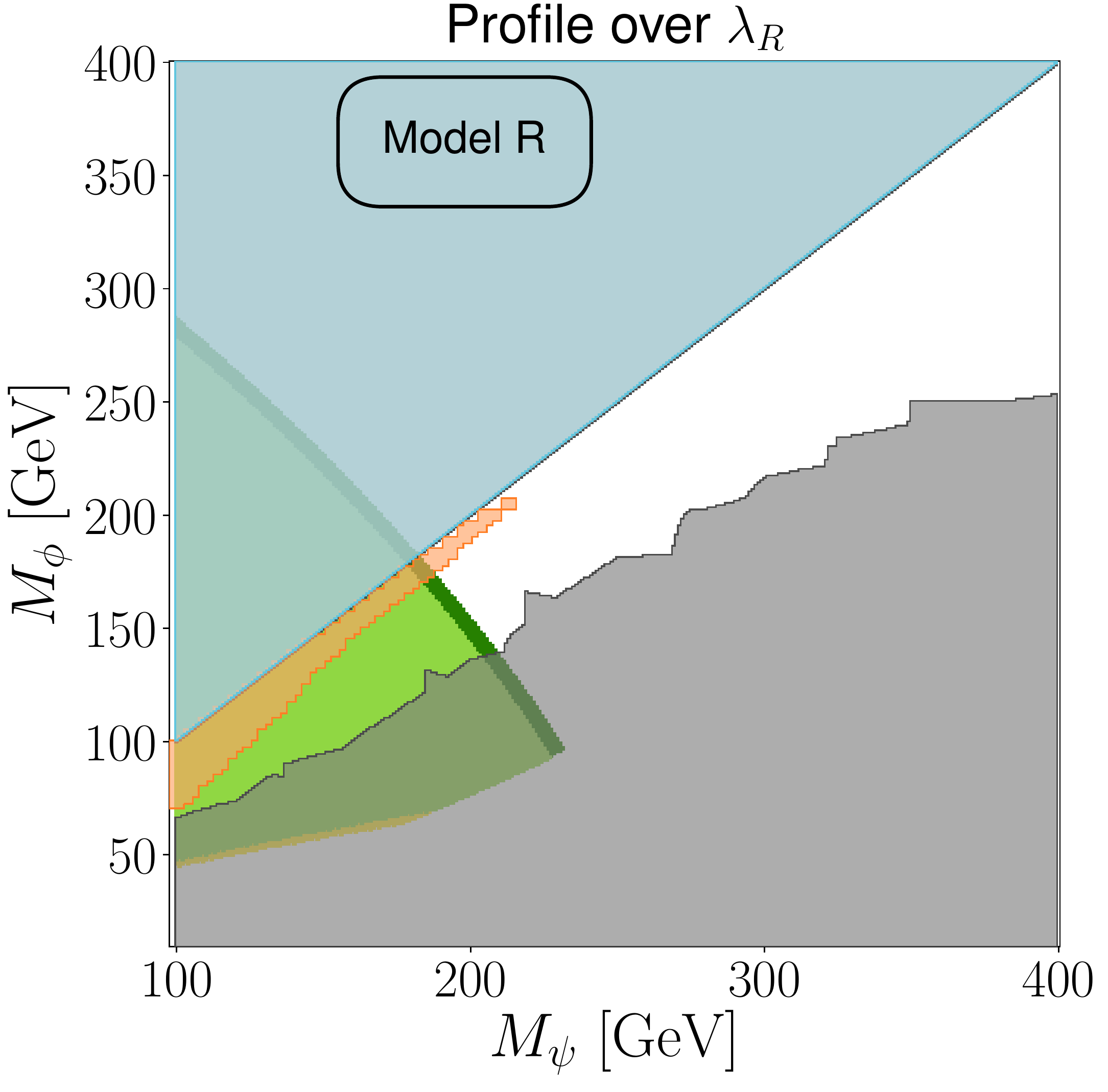}
		\includegraphics[width=0.32\textwidth]{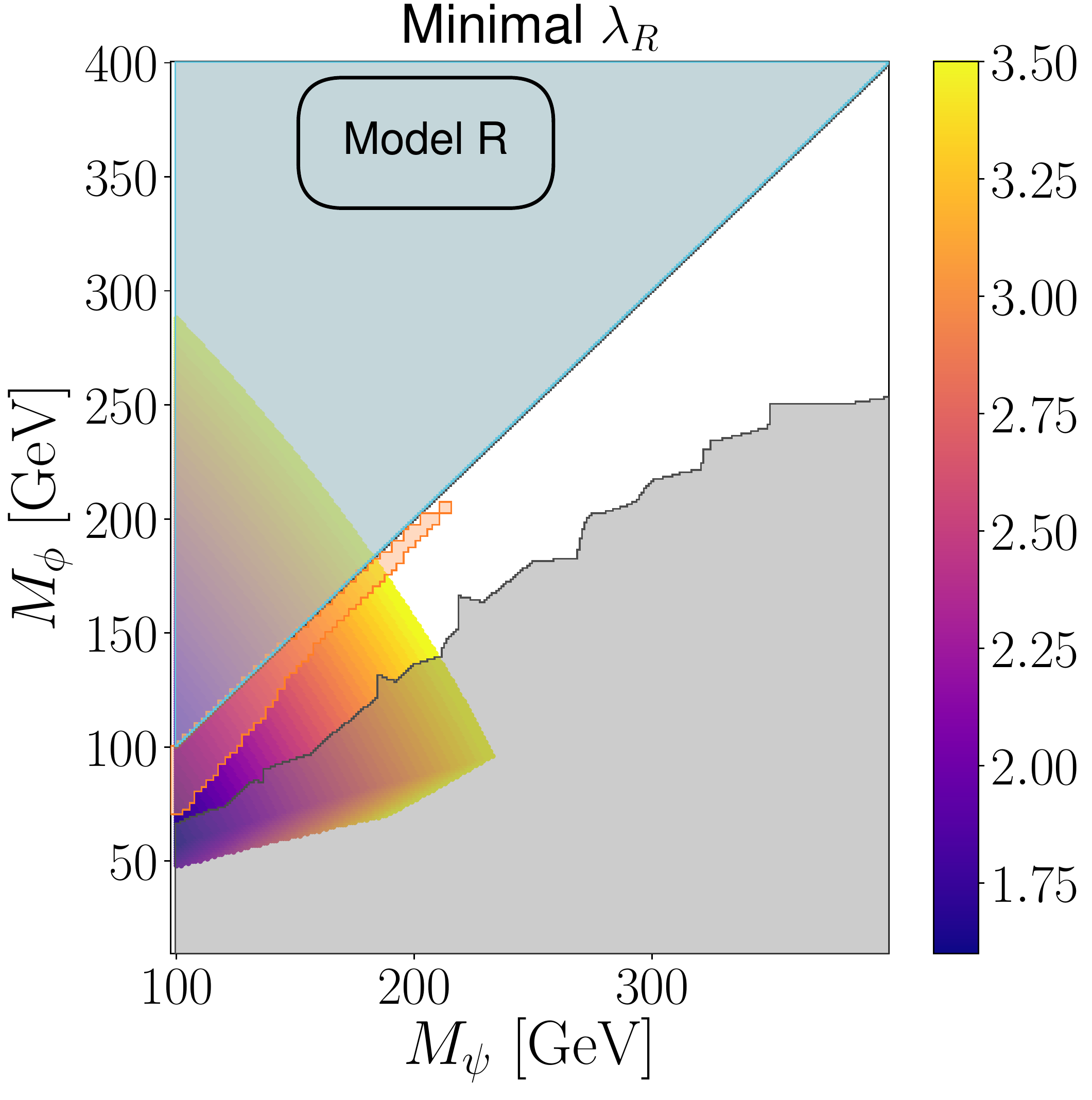}
		\includegraphics[width=0.32\textwidth]{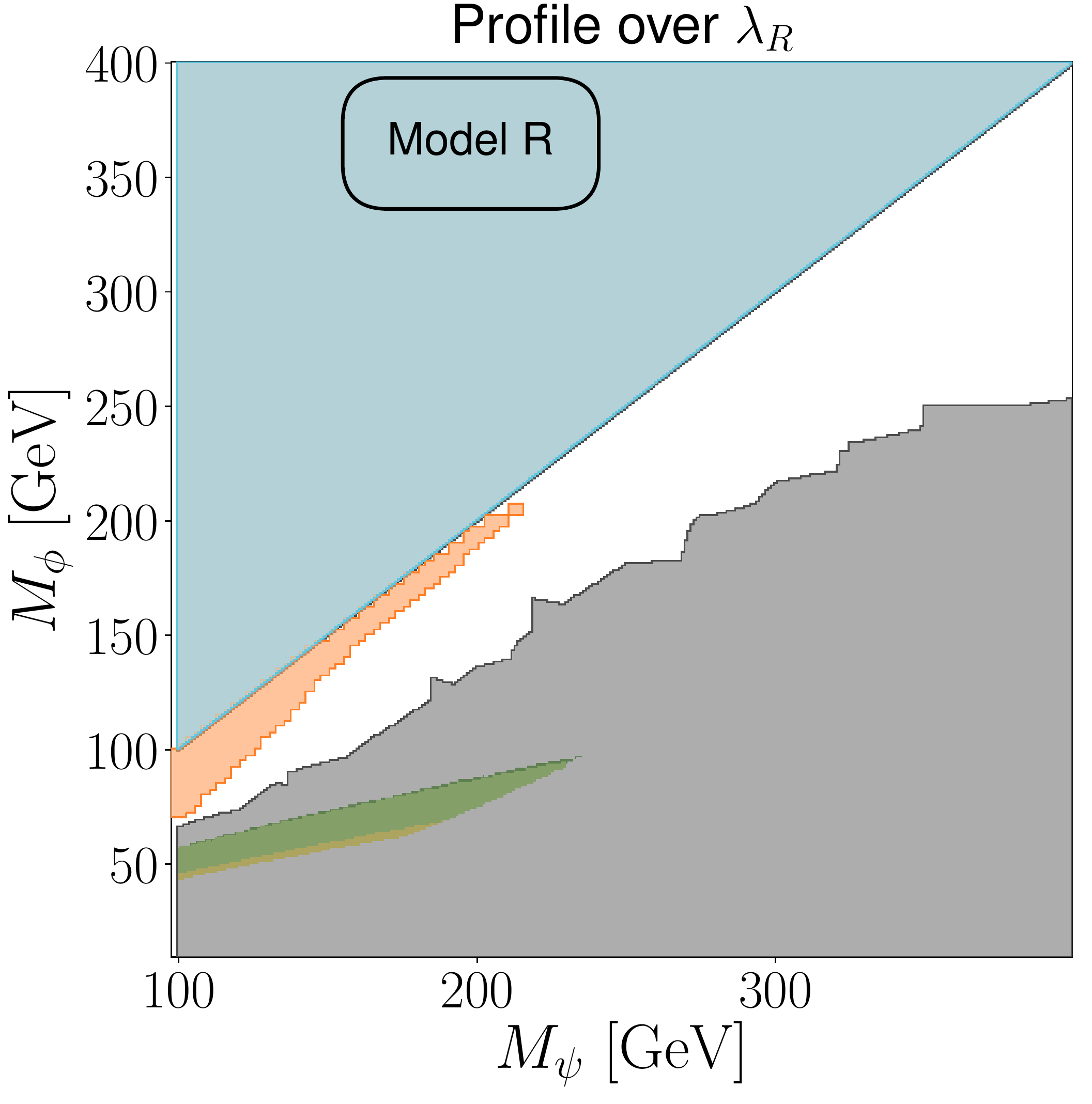}
		\caption{Several profiles of Model R.  Colours for the first and third panels are the same as Fig.\ \ref{fig:Min2FieldsLLSlice}.  The middle panel shows the lowest value of $\lambda_R$ which can \update{explain $\amuNEW$ from Eq.\ \update{(\ref{eqn:avgDiscrepancy})}} within $1\sigma$.  The panels are constructed in the same way as those in Fig.\ \ref{fig:Min2FieldsLLProfile}.  The same LHC limits as Fig.\ \ref{fig:Min2FieldsRR} (which do not depend on the coupling) are overlaid.  \label{fig:Min2FieldsRRProfile}}
	\end{figure}

	Results for Model R are shown in
        Figs.\ \ref{fig:Min2FieldsRR}, \ref{fig:Min2FieldsRRProfile}.
        Since the contributions to $\amu$ for this model are also
        governed by Eq.\ (\ref{eqn:Min2FieldsLLContributions}), the
        behaviour of $\damu$ is the same.  However as one can see in
        Figs.\ \ref{fig:Min2FieldsRR}, \ref{fig:Min2FieldsRRProfile}
        the collider constraints for Model R, which has no $SU(2)_L$
        interactions, are weaker than those for Model L, with the
        exclusions again coming from \smodels 1.2.3 through the
        $8$-TeV searches
        \cite{Aad:2014vma,CMS:hwa,CMS:2013dea,Khachatryan:2015lla} and
        the $13$-TeV searches
        \cite{Aaboud:2018jiw,Sirunyan:2018nwe,Aad:2019vnb,CMS:2016ybj,CMS:2017mkt}.
        The searches providing most of the constriction on the
        parameter space of Model R are the ATLAS searches
        \cite{Aad:2014vma,Aad:2019vnb} and the CMS searches
        \cite{CMS:2017mkt,Sirunyan:2018nwe}. As a result in the case
        of Model R there is still some room to explain the $\amu$
        measurement within currently applied collider constraints.
        However as can be seen in the left panel of
        Fig.\ \ref{fig:Min2FieldsRR} \update{even with $\lambda_R =
          2.0$, there is only a little room} to escape the combination
        of the standard electroweakino searches shaded grey and the
        soft lepton compressed spectra search shaded orange.  In the
        middle panel when $\lambda_R = 2.5$ one can see \update{a
          significantly larger, but still narrow} region where
        \update{$\amuNEW$} can be explained within $1\sigma$ while
        evading the collider limits.  In the right panel as
        $\lambda_R$ approaches the perturbative limit \update{both the
          value and area of the viable parameter space are larger.}
        In the left panel of Fig.\ \ref{fig:Min2FieldsRRProfile},
        where $\lambda_R$ is allowed to vary, one can see the full
        region of the $M_\phi-M_\psi$ plane where \update{$\amuNEW$ is
        explained} within $1\sigma$, while avoiding giving a relic
        density of dark matter which is too over abundant.  With the
        overlaid collider limits we can see the complete picture of
        parameter space where it remains possible to explain this
        observed $\amu$, while evading limits from colliders.
        \update{Avoiding collider constraints from the LHC requires
          compressed spectra, and comparing the green and lime green
          regions, one can see that the situation with the new world
          average has just marginally increased the masses that can be
          accommodated}.  However as shown by the middle panel of
        Fig.\ \ref{fig:Min2FieldsRRProfile}, \update{evading these
          limits requires a coupling of $\lambda_R \gtrsim 1.8$}.  One
        can also see that the relic density cannot be explained
        simultaneously with \update{$\amuNEW$} in any of the slices shown in
        Fig.\ \ref{fig:Min2FieldsRR}, and the right panel of
        Fig.\ \ref{fig:Min2FieldsRRProfile} does show that the regions
        which can explain dark matter relic density \update{and $\damuNEW$} to
        within $1\sigma$ are \update{fully ruled out} by collider
        constraints. 

  In summary, two-field models with different spin do not contain new
  sources of chirality flipping enhancements, and many models of this
  class are not able to generate significant positive contributions to
  $\amu$, see Table \ref{tab:TwoFieldsDifferentSpin}.  However the
  Model L and Model R investigated here do give positive
  contributions, and our results show they remain viable explanations
  of \update{the new world average
    deviation $\damuNEW$ after the FNAL measurement}, as can be seen from
  Figs.\ \ref{fig:Min2FieldsLLSlice}-\ref{fig:Min2FieldsRRProfile},
  with small parts of the parameters space escaping LHC
  constraints.  Model R is found to be
  slightly more viable due to reduced impact of LHC constraints, in
  particular those from compressed spectra searches.
In both models, however, it is \update{not possible} to simultaneously
  explain $\damuNEW$
  whilst producing the observed dark matter relic density
  \cite{Aghanim:2018eyx} and satisfying LHC collider constraints.  It
  is only possible to explain \update{$\damuNEW$} within $1\sigma$ with an under
  abundant dark matter candidate particle.  Further, the lack of
  chirality flip for either of these models constrain the masses to be
  \update{$M_\psi,M_\phi \le 210$ GeV} to both avoid collider constraints and explain 
  the value of $\damuNEW$, where compressed spectra searches are
  important for the viability of these models. 
Future data from \update{$\amu$ experiments and LHC} might entirely exclude these models.

%% file: three_fields.tex
\section{Three-Field Extensions } \label{sec:ThreeFields}

As we saw in Sec.\ \ref{sec:TwoFields} introducing two fields with different spins is the simplest way to explain dark matter and give new contributions to muon $g-2$, but collider constraints heavily constrain the region \update{where such models can explain the observed muon $g-2$}.  This is because in those cases it is not possible to induce a chirality flip in the muon $g-2$ diagram's loop, as both the new fermion and scalar only couple to either the left- or right-handed muon.

We therefore now consider the simplest models that can both have an
enhanced contribution to muon $g-2$ from a chirality flip and explain
dark matter. These are models with two fields of the same spin that
mix together and a third field with a different spin and a $\mathbb{Z}_2$ 
symmetry under which these new fields are odd, while the SM fields are
even. The leading contributions in supersymmetric 
models are in fact of this kind, as will be discussed in 
Sec.\ \ref{sec:MSSM}, but here we also consider three-field models 
that do not arise as part of the MSSM. As in previous sections we will restrict ourselves to
  renormalizable models with new scalars and new fermions.  Models
of this nature have previously been classified in
Ref.\ \cite{Calibbi:2018rzv}, and there it is demonstrated that due to
the chirality flip contribution there are many models which can
explain the $\amu$ anomaly in Eq.\ (\ref{eqn:BNLDiscrepancy}), with
models that have a new field with a much higher $SU(2)_L$
representation (up to $20$) being able to produce sizable
contributions to $\amu$.
Likewise, the larger contributions to $\amu$ from the chirality flip enables three field models to simultaneously produce a dark matter candidate with the observed dark matter relic abundance \cite{Planck2015} also with much higher $SU(2)_L$ representations compared to the two field models. 
This chirality flip enhancement also enables one to examine much higher 
mass scales compared to two field models, evading LHC collider constraints.  

In this section we present a detailed discussion of the phenomenology
of two kinds of three-field models, which we denote as 2S1F and 2F1S,
with two scalars and one fermion or vice versa. The quantum numbers
will be chosen such that the models represent a variety of possible
dark matter candidates. In model 2S1F, the dark matter is a scalar
singlet or doublet (or mixture) and may be compared to generic scalar
singlet or inert doublet models or to left- or right-handed
sneutrinos. In model 2F1S, the dark matter is a singlet or doublet
fermion (or a mixture) and may be compared to Bino or neutral
Higgsino. Model 2S1F has some similarities to the BLR
contribution in the MSSM, but the spin of the dark matter candidate is
different.
Model 2F1S contains the same states as the BHL contribution in the
MSSM. In both cases, the couplings are treated as arbitrary, not
respecting SUSY-like relations, which leads to a different
phenomenology, and we compare and contrast these two models with their
respective MSSM scenarios.  Both of these models were considered in
Ref.\ \cite{Calibbi:2018rzv}, showing particular slices of the
parameter space. Here we update these models with the latest
$\amuNEW$ result and direct detection constraints, and
perform a thorough sampling of the whole parameter space to show how
the data leads to meaningful constraints on the parameters of these
models.

\subsection{Three-field model with two scalars, one charged
  fermion and scalar dark matter}
\label{sec:2S1F}
We begin with the Model 2S1F which is the SM extended with three ${\mathbb{Z}}_2$-odd fields: one electrically charged fermion and two scalar fields,
	\begin{equation}
		\psi _s,\quad \phi_s ^0,\quad \phi_d= \begin{pmatrix} \frac{\phi^0_d + i A_\phi}{\sqrt{2}} \\ \phi_d^- \end{pmatrix}\,, 
	\end{equation}
        where  $\psi _s$ is an $SU(2)$ singlet charged Dirac fermion expressed in Weyl spinors with representation $({\bf{1}},{\bf{1}},1)_{1/2}$, $\phi _s ^0$ a scalar singlet with representation $({\bf{1}},{\bf{1}},0)_{0}$, and $\phi _d$ a scalar doublet with representation $({\bf{1}},{\bf{2}},-\frac{1}{2})_{0}$. 
	The relevant interactions are:
	\begin{equation} \label{eqn:Min3FieldsSLRF1}
		{\cal L}_{\text{2S1F}} = \begin{pmatrix}a_H H\cdot\phi_d \phi_s^0 + \lambda_L \phi_d \cdot L \psi_s + \lambda_R \phi_s^0 \mu \psi_s^c - M_{\psi} \psi_s^c \psi_s + h.c.\end{pmatrix} - M_{\phi_d}^2 |\phi_d|^2 - \frac{M_{\phi_s}^2}{2} |\phi_s^0|^2.
	\end{equation}

The new scalars couple to one of the left- and right-handed
muons each, along with the new fermion.  Then using a
trilinear scalar coupling between the two scalars and the
Higgs boson, the VEV of the Higgs bosons gives rise to mass
mixing terms between the two scalars, and induces a chirality
flip in the loop of muon $g-2$ processes.  The relevant
parameter combination for invoking this chirality flip
enhancement is $a_H\lambda_L\lambda_R$.
	
For Model 2S1F in Eq.\ (\ref{eqn:Min3FieldsSLRF1}), the
neutral scalar singlet $\phi_s^0$ and the CP-even part of the
neutral component of $\phi_d$ mix together into two flavours
of a neutral scalar $\phi^0_i$ for $i=1,2$:
\begin{align} \label{eqn:Min3FieldsSLRF1Mixing}
  {\cal L}_{\text{2S1F}} \ni -\frac{1}{2}\begin{pmatrix} \phi_s^0 & \phi_d^0 \end{pmatrix} \begin{pmatrix} M_{\phi_s}^2 & a_H v \\ a_H v & M_{\phi_d}^2 \end{pmatrix} \begin{pmatrix} \phi_s^0 \\ \phi_d^0 \end{pmatrix} 
	  &= -\frac{1}{2}\sum_{i=1,2} M^2_{\phi_i^0} |\phi_i^0|^2, \nonumber
\end{align}
where the mass eigenstates relate to the gauge eigenstates through the
mixing matrix $U_S$ as
\begin{align}
  \begin{pmatrix}\phi_1^0\\\phi_2^0
  \end{pmatrix}
   &= U_S^T\begin{pmatrix}\phi_s^0\\\phi_d^0
  \end{pmatrix}\,.
\end{align}
The dark matter candidate is then the lightest neutral scalar
$\phi^0_1$.  For this mixing to avoid producing a tachyon, the limit
$a_H v < M_{\phi_s} M_{\phi_d}$ must be respected. As in the previous
section on two-field extensions, we do not consider additional
renormalisable terms involving the new scalars that could appear in
the Higgs potential, but would not affect the $\amu$ prediction.
These terms can change the dark matter phenomenology though, adding
additional mechanisms to deplete the relic density, and new
interactions that can mediate direct detection.  However we do not
expect these to substantially modify our conclusions.

In this model the one-loop BSM contribution to the anomalous magnetic moment is 
	given by:
\begin{align} \label{eqn:amu_2S1F_model}
  \damu^{\text{2S1F}} = & \frac{m_\mu^2}{32 \pi^2} \bigg(
  \frac{\lambda_L^2}{12 M_{\phi_d}^2} \,E\!\begin{pmatrix} \frac{M_\psi^2}{M_{\phi_d}^2} \end{pmatrix}  
  + \sum_{i=1,2} \frac{\lambda_L^2|U_{S\,2i}|^2 + 2\lambda_R^2|U_{S\,1i}|^2}{12 M_{\phi_i^0}^2} \,E\!\begin{pmatrix} \frac{M_\psi^2}{M_{\phi_i^0}^2} \end{pmatrix}\nonumber \\
   &+ \frac{M_\psi}{m_\mu} \sum_{i=1,2} \frac{2\sqrt{2} \lambda_L \lambda_R U_{S\,1i} U_{S\,2i}}{3 M_{\phi_i^0}^2} \,F\!\begin{pmatrix} \frac{M^2_\psi}{ M^2_{\phi_i^0}} \end{pmatrix} \bigg).
\end{align}
These contributions come in the form of the FFS diagram
\ref{fig:FFSDiagram} of Fig.\ \ref{fig:GeneralDiagrams} with the
generic $F$ fermion lines replaced by $\psi^-$ the conjugate of
$\psi_s$, and the scalar $S$ one by $\phi^0_i$ and $A_\phi$,
respectively.

The second line of Eq.\ (\ref{eqn:amu_2S1F_model}) provides the
contribution with the chirality flip enhancement, and is therefore
typically the dominant contribution.  While a cursory glance at
Eq.\ (\ref{eqn:amu_2S1F_model}) might lead one to believe that this
model's chirality flip provides an apparent enhancement according to
the ratio $M_\psi/m_\mu$, the actual behaviour is more complicated.
The actual form of this enhancement can be seen by using the
simplification of the mixing angles
$U_{S\,11}U_{S\,21}=-U_{S\,12}U_{S\,22} \to a_H
v/(M_{\phi_1^0}^2-M_{\phi_2^0}^2)$.  Plugging this into
Eq.\ (\ref{eqn:amu_2S1F_model}) generates a difference between loop
functions which, together with the mass difference in the denominator,
can be written in terms of the loop function $\tilde{F}_a$ of the
appendix Eq.\ (\ref{FaDefinition}).  In this way, for $M_\psi \gg
m_\mu$, the contribution from this model simplifies to the chirality
flip contribution as:
\begin{align}
	\damu^{\text{2S1F}} \approx& \frac{m_\mu^2}{32 \pi^2}\frac{ M_\psi}{m_\mu} \sum_{i=1,2} \frac{2\sqrt{2} \lambda_L \lambda_R U_{S\,1i} U_{S\,2i}}{3 M_{\phi_i^0}^2} \,F\!\begin{pmatrix} \frac{M^2_\psi}{M^2_{\phi_i^0}} \end{pmatrix} \\
	\approx&
	\frac{m_\mu^2}{32 \pi^2}\frac{ M_\psi}{m_\mu}  \frac{2\sqrt{2} \lambda_L \lambda_R a_H v}{3( M_{\phi_1^0}^2-M_{\phi_2^0}^2)} \left(
	\frac{1}{M^2_{\phi_1^0}}  \,F\!\begin{pmatrix} \frac{M^2_\psi}{M^2_{\phi_1^0}} \end{pmatrix}
	-\frac{1}{M^2_{\phi_2^0}}\,F\!\begin{pmatrix} \frac{M^2_\psi}{M^2_{\phi_2^0}} \end{pmatrix}\right).\\
	\approx&
	-\frac{m_\mu^2}{32 \pi^2}\frac{ M_\psi}{m_\mu}  \frac{2\sqrt{2} \lambda_L \lambda_R a_H v}{M_{\psi}^4} \,\tilde{F}_a\!\begin{pmatrix} \frac{M^2_\psi}{M^2_{\phi_1^0}}, \frac{M^2_\psi}{M^2_{\phi_2^0}} \end{pmatrix}.
\end{align}
	Thus the final result for this chirality-enhanced contribution can be written in 
	the form
\begin{align}\label{eqn:2S1FLOContribution}
  \damu^{\text{2S1F}} \approx& -\frac{m_\mu^2}{32 \pi^2 M_{\psi}^2}\frac{}{}
  \frac{2\sqrt{2} \lambda_L \lambda_R a_H v}{m_\mu M_{\psi}} \,\tilde{F}_a\!\begin{pmatrix} \frac{M^2_\psi}{M^2_{\phi_1^0}}, \frac{M^2_\psi}{M^2_{\phi_2^0}} \end{pmatrix},
\end{align}
where e.g.\ $\tilde{F}_a(1,1)=1/12$.
	We see that the actual enhancement factor, beyond the typical loop and mass 
	suppression factor, is given by the ratio $\lambda_L \lambda_R a_H v/(m_\mu 
	M_{\psi})$.  As announced earlier it relies on the factor $\lambda_L \lambda_R 
	a_H$ and requires all three of these couplings to be non-zero.

We have presented this discussion of how the parametric behaviour of the chirality flip can be extracted from the exact result Eq.\ (\ref{eqn:amu_2S1F_model}) in quite some detail, and we remark that similar discussions can also be applied to other models, e.g.\ to the model 2F1S discussed later, or to MSSM contributions discussed in Sec.\ \ref{sec:MSSM}.\footnote{
    The enhancement may be  compared to the enhancement factor for the MSSM BLR 
    contribution shown in Eq.\ (\ref{eq:SUSYMIapprox}) in Sec.\ 
    \ref{sec:MSSM}, where $a_Hv$ corresponds to $m_\mu\mu\tan\beta$. In 
    the corresponding MSSM contribution, the couplings analogue to 
    $\lambda_{L,R}$ are gauge couplings, while these couplings are
    independent here.}

If either $\lambda_L$ or $\lambda_R$ are zero then clearly the above contribution vanishes, while it will also vanish if there is no mixing between the scalars, i.e.\ $a_H = 0$,  so that $U_{S\,1i}U_{S\,2i}$ vanishes.  If $\lambda_L = a_H = 0$ or $\lambda_R = a_H = 0$ then Eq.\ (\ref{eqn:amu_2S1F_model}) reduces to the same result for two fields with different spin, i.e.\ Eq.\ (\ref{eqn:Min2FieldsLLContributions}), while if $a_H$ is non-zero, then one gets a similar result, but with some dilution from the mixing.  Similarly if both $\lambda_L$ and $\lambda_R$ are non-zero, but $a_H=0$ then we simply get two copies of Eq.\ (\ref{eqn:Min2FieldsLLContributions}), with different couplings that are summed.  
Finally note that the mixing will also be heavily suppressed just by having a large enough mass splitting between the two scalar states, i.e.\ making the ratio $ (v a_H)^2 / (M_{\phi_s}^2 M_{\phi_d}^2)$ sufficiently small.

In the following we provide a detailed
phenomenological analysis of the Model 2S1F. The underlying question
is whether the model can accommodate the new $\amuNEW$
value simultaneously with 
current constraints from dark matter and LHC.
The model is complicated and depends on three masses and three couplings in a relevant way.  We focus on the
parameter space where $\lambda_L$, $\lambda_R$ and $a_H$ are all
non-zero, such that the model behaves differently from the models of
the previous section. Eq.\ (\ref{eqn:2S1FLOContribution}) then provides the dominant contribution and can be used to understand our results.  As in previous sections we do however use \fs 2.5.0 \cite{Athron:2014yba,Athron:2017fvs} to obtain the numerical results, where we augment the full one-loop BSM calculation with the universal leading logarithmic photonic $2$-loop contributions \cite{Degrassi1998}.    

Before entering details, we briefly explain relevant dark matter
processes. Without the $a_H$, $\lambda_L$ and $\lambda_R$ interactions
that are important for $\amu$, dark matter could behave either as pure
scalar singlet or as pure inert scalar doublet dark matter. In the
inert doublet case, already the standard SU(2)$_L$ electroweak gauge
interactions are sufficient to deplete the relic density to (below)
the observed value when it has a mass of (less than) about $600$ GeV
\cite{LopezHonorez:2006gr}. The preferred mass of scalar singlet dark
matter depends on the Higgs portal coupling, but outside of fine tuned
parameter regions and for couplings $\lambda_{L,R} < 1$, the best fit for the
data is found when the pure scalar singlet dark matter has a mass of
$\approx 1$--$3$ TeV \footnote{We do not consider the Higgs portal coupling needed for this, in our analysis, but the well studied
  phenomenology of scalar singlet dark matter may be useful in showing
  the kind of scenarios that we neglect here.}
\cite{Athron:2018ipf}.  Beyond these simple special cases, the model
here allows many additional depletion mechanisms: significant
$\lambda_{L,R}$ allow t-channel exchange of the new fermion, through
$\lambda_{L}$ or $\lambda_R$ if the scalar dark matter is doublet or
singlet dominated (or both if there is mixing).  If the new fermion
mass is sufficiently close, coannihilations through $\lambda_L$ and
$\lambda_R$ can also be active in depleting the relic density. The
coupling $a_H$ between singlet, doublet, and the SM Higgs can lead to
singlet--doublet coannihilation processes. This same coupling $a_H$ is
also relevant for direct detection constraints on dark matter, which
depend particularly on Higgs-mediated processes.  Importantly, the
mentioned additional depletion processes can compensate general
suppressions of cross sections for heavier masses. As a result even
heavier dark matter masses than for the inert doublet can become
possible.

As we will see from detailed scans, the  dark matter constraints
indeed only allow rather heavy masses, which are above LHC limits. 
For this reason we do not explicitly consider LHC
constraints in the following.

\begin{figure}[t]
	\centering
	\includegraphics[width=0.31\textwidth]{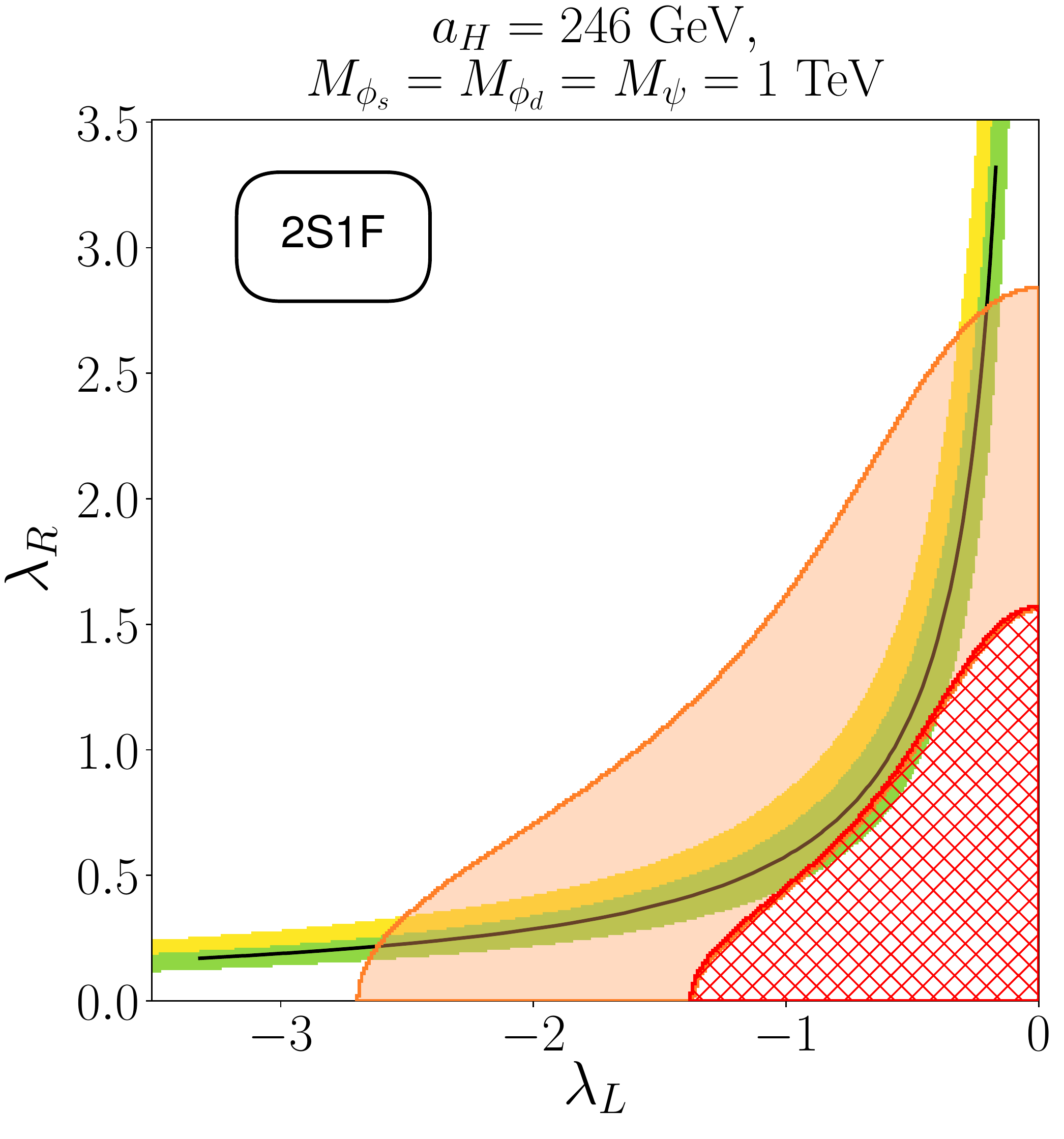}
`	\hspace{0.01\linewidth}
	\includegraphics[width=0.31\textwidth]{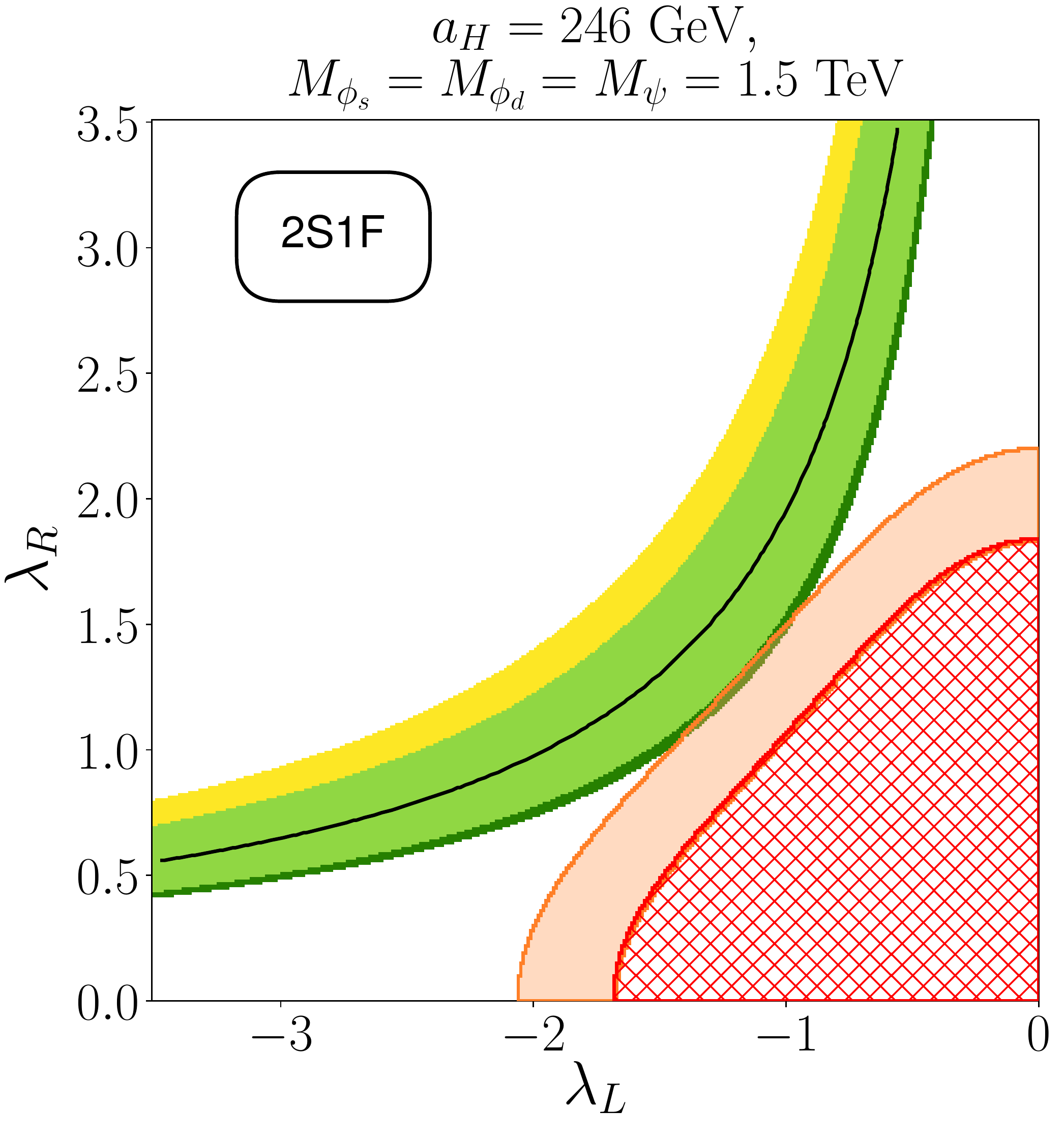}
	\hspace{0.01\linewidth}
	\includegraphics[width=0.31\textwidth]{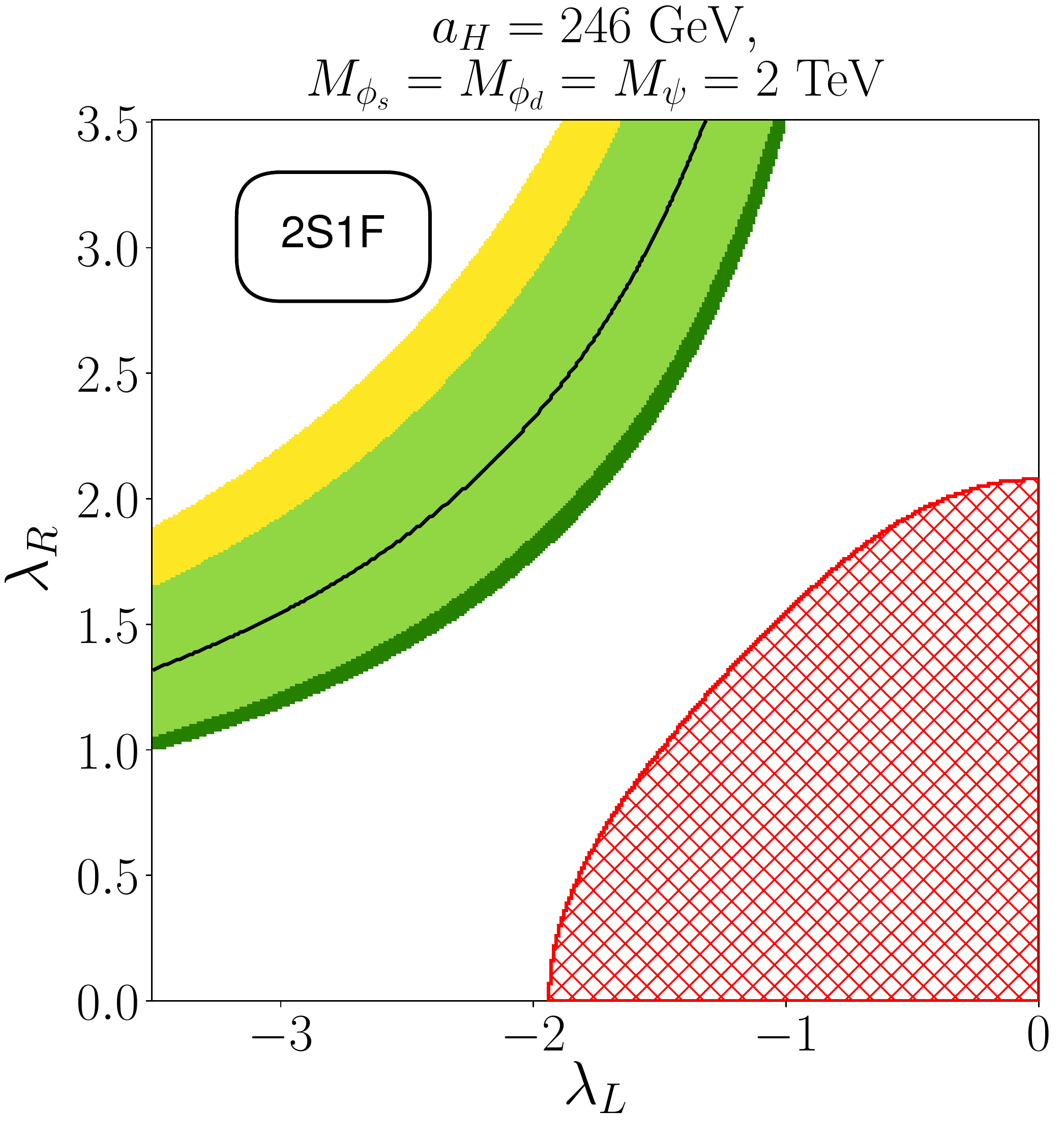}
	\caption{Results from Model 2S1F, scanning over the couplings of the new scalars and fermion to the left- and right-handed muon.  All BSM fields have identical masses as specified in the subfigures.  The trilinear coupling which parametrises the mixing of the neutral BSM scalars is fixed to $a_H = 246$ GeV.  Regions which can explain $\damu^\textrm{BNL}$ and $\damuNEW$ to within $1\sigma$ are coloured yellow and green respectively, while the overlap between these two regions is coloured lime.  The black line marks couplings which produce a contribution which exactly matches Eq.\ \update{(\ref{eqn:avgDiscrepancy})}.  The points where a dark matter candidate particle produces the observed relic abundance of $0.1200$, Eq.\ (\ref{eqn:DMRD}), are at the boundary of the hatched region shown in red, with the hatched region below and to the right being excluded due to having a relic abundance greater than $0.12$.  Regions excluded by searches for the direct detection of dark matter at the Xenon1T experiment\cite{Aprile:2017iyp,Aprile:2018dbl} are shaded in orange.   \label{fig:Min3FieldsSLRFRCouplingsEqualMasses} }
\end{figure}

In Fig.\ \ref{fig:Min3FieldsSLRFRCouplingsEqualMasses} we begin our
numerical analysis of this model.  The figure represents a baseline
behaviour.  In order to maximize contributions to $\amu$ via a large
chirality flip and mixing (see Eq.\ (\ref{eqn:2S1FLOContribution}) and
following discussion) we choose equal masses $M_{\phi_s}=M_{\phi_d}$
and fix $a_H=v=246$ GeV as a reference value, similar to scenarios
examined by Ref.\ \cite{Calibbi:2018rzv}.  We also choose $M_\psi$ to
be equal to the scalar masses. This is a rather special choice for
dark matter, as all mechanisms for depleting the relic density can be
active here, including scalar--fermion coannihilation. As a result,
the relic density will be quite suppressed and the Planck-observed
value can only be explained for small couplings.  On the other hand,
the dark matter direct detection will not be suppressed due to the
large mixing between the scalar and doublet, allowing interactions
with the nucleons through the moderately sized $a_H$ coupling.

	The behaviour of $\damu$ in the three panels of
        Fig.\ \ref{fig:Min3FieldsSLRFRCouplingsEqualMasses} clearly
        reflects the $\lambda_L\lambda_R a_H/M^2$ dependence of the
        chirality-flipping contributions shown in
        Eq.\ (\ref{eqn:2S1FLOContribution}).  In each panel the
        coupling dependence is approximately hyperbolic.
        Specifically, in the left panel for TeV-scale masses it is
        \update{possible} to \update{explain $\damuNEW$} with
        couplings \update{$|\lambda_L\lambda_R| \approx 0.8^2$}.  As the
        masses are increased to $1.5$ TeV and $2$ TeV in the middle
        and right panels respectively, the contribution to $\amu$ is
        suppressed.  Therefore at higher masses to explain
        $\damu^{\textrm{BNL}}$ \update{and $\damuNEW$} to within
        $1\sigma$ either larger couplings are required, as shown in
        the plot, or alternatively a larger  value of $a_H$ (which
        is fixed to 246 GeV on this plot) could be chosen.

	Each panel in Fig.\ \ref{fig:Min3FieldsSLRFRCouplingsEqualMasses} also shows a red 
	hatched region, which is excluded by over abundance of the
        dark matter relic density. At the boundary the observed value
	\cite{Aghanim:2018eyx} can be explained. In the non-hatched
        region the predicted relic density is under abundant. Further,
        the orange shaded regions are excluded by  direct 
detection searches using \DDCalc 2.2.0 \cite{Workgroup:2017lvb,Athron:2018hpc}.\footnote{%
In the calculation we have correctly rescaled the direct detection spin-independent 
(SI) cross-sections according to the abundance of dark matter
as 
\begin{align} \label{eqn:DDScaling}
	\sigma_{\textrm{SI,eff}} &= \sigma_{\textrm{SI}} \times   \frac{\Omega_{h^2}}{\Omega_{h^2,\textrm{Planck}}}, 
\end{align}
where
$\Omega_{h^2,\textrm{Planck}} = 0.1200$ is the dark matter relic abundance as observed 
by the Planck experiment, see Eq.\ (\ref{eqn:DMRD}), and $\Omega_{h^2}$ is the 
relic abundance produced by the dark matter candidate particle. 
}
As mentioned above, the equal mass scenario is constructed to maximize
the impact of relic density depletion mechanisms. The dark matter
candidate has significant singlet and doublet components, and the
depletion mechanisms are equally sensitive to both $\lambda_L$ and
$\lambda_R$, see  the Lagrangian Eq.\ (\ref{eqn:Min3FieldsSLRF1}). If
either coupling is sufficiently large the relic 
density becomes under abundant. As a result the red lines are
approximately parabolic in the $\lambda_L-\lambda_R$ plane, and for
	${\cal O}(1)$ values of the couplings the relic density can be
        explained. As the overall mass 
	scale is increased, the required couplings become slightly
        larger.    

The dark matter direct detection limits depend on spin-independent
cross sections between dark matter and nucleons in the detector. These
cross sections are mediated particularly by the Higgs 
boson, and the relevant coupling of dark matter to the Higgs boson
is particularly sizeable in the present equal mass case
with strong singlet--doublet mixing and a significant value of
$a_H$. It does not
depend on $\lambda_L$ and $\lambda_R$, but via the rescaling 
according to relic density, 
the resulting direct detection limits do vary across the $\lambda_L$--$\lambda_R$ plane.
These limits are generally strong, because of the unsuppressed coupling to
the Higgs boson.
They become however significantly weaker for increasing masses
because of the suppression of the cross sections, despite the slight
increase of the relic density.

The combination of all these constraints means that for our baseline
case, with equal masses, shown in the three panels of
Fig.\ \ref{fig:Min3FieldsSLRFRCouplingsEqualMasses} it is
\update{impossible to explain \update{$\damuNEW$} simultaneously} with the dark
matter relic density. \update{For masses of $1$ TeV, $\damuNEW$ could
  be explained simultaneously with the dark matter relic density,
  however direct detection constraints exclude a large part of the
  $\damu$ band and the entire region where the relic density is
  explained.} For masses of $1.5$ TeV, the direct detection limits
\update{constrain only a very little of the region which can explain
  $\damuNEW$, however the direct detection
  constraints still rule out an explanation of the relic density.}
For masses of $2$ TeV \update{the full relic density can be
  explained, however not simultaneously with $\damuNEW$.}  One may ask
whether just changing the value of the trilinear coupling $a_H$
between the Higgs boson and the two BSM scalar bosons could allow the
simultaneous explanation of large $\damu$ and dark matter. As
Eq.\ (\ref{eqn:2S1FLOContribution}) shows, increasing the value of
$a_H$ increases the size of the BSM contributions to $\amu$, while the
dark matter relic density is essentially independent of $a_H$ in these
scenarios.  Therefore the $\amu$-bands in
Fig.\ \ref{fig:Min3FieldsSLRFRCouplingsEqualMasses} would move down
and to the right, to smaller couplings, while the red hatched regions
would remain largely unchanged from adjusting $a_H$.  However the
direct detection constraints become stronger if $a_H$ is increased,
because it increases Higgs-nucleon interactions via the SM Higgs.  As
a result the strengthened direct detection limits would rule out
points with the observed dark matter relic density. Decreasing $a_H$ has the
opposite effect: the band of parameter space where
$\damu$ is explained moves up and to the left, further
away from the observed relic density, and weakening the direct
detection constraint will not help.  So in summary it is \update{not
  possible} in these equal mass scenarios to explain both the measured
value $\damu$ and the observed value of the relic density of dark
matter at the same time, while avoiding constraints from direct
detection. In the following we therefore consider further parameter
slices to obtain additional insight into the phenomenology of the
model.

	\begin{figure}[t]
		\centering
		\includegraphics[width=0.31\textwidth]{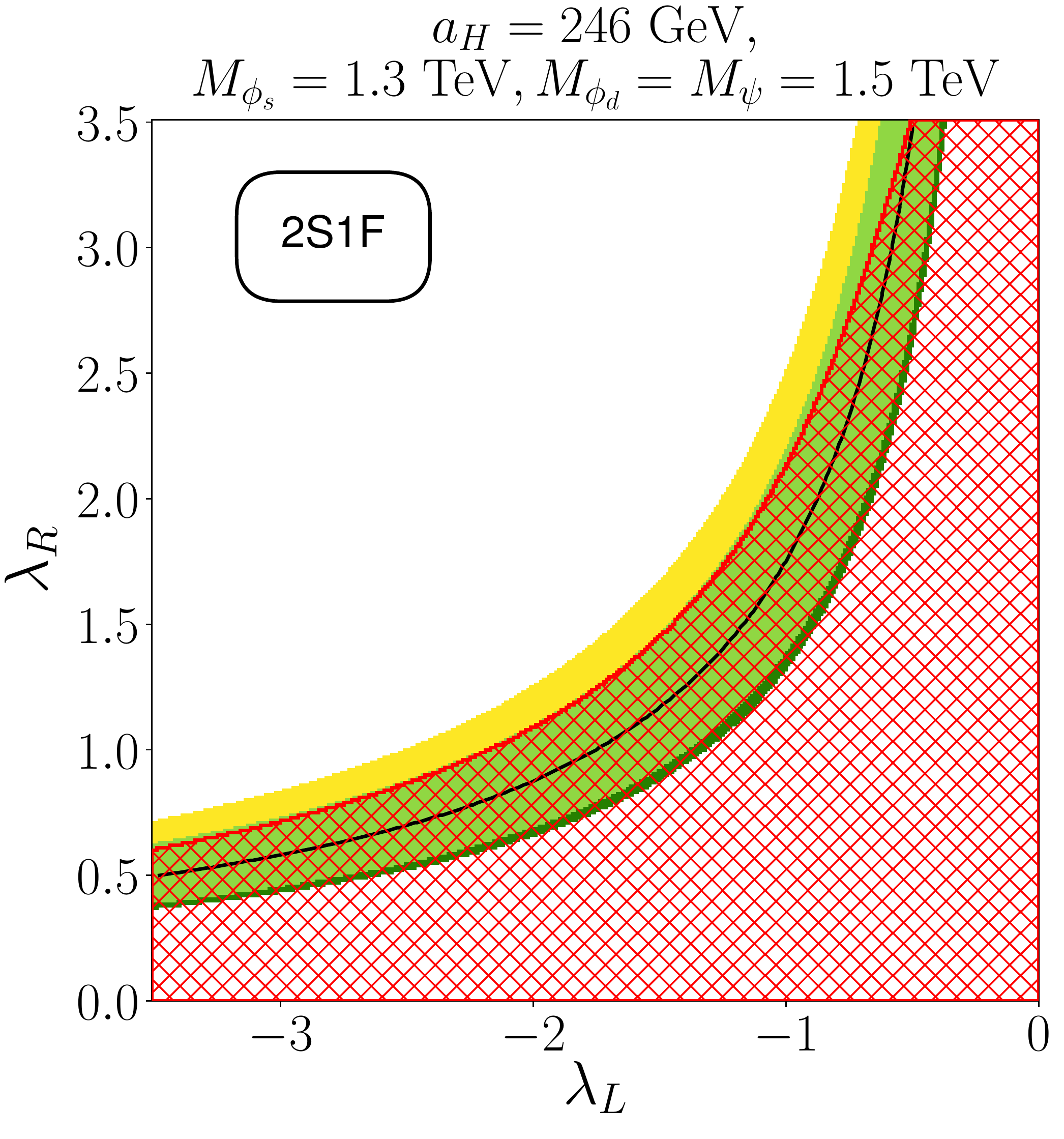}
		\hspace{0.01\linewidth}
		\includegraphics[width=0.31\textwidth]{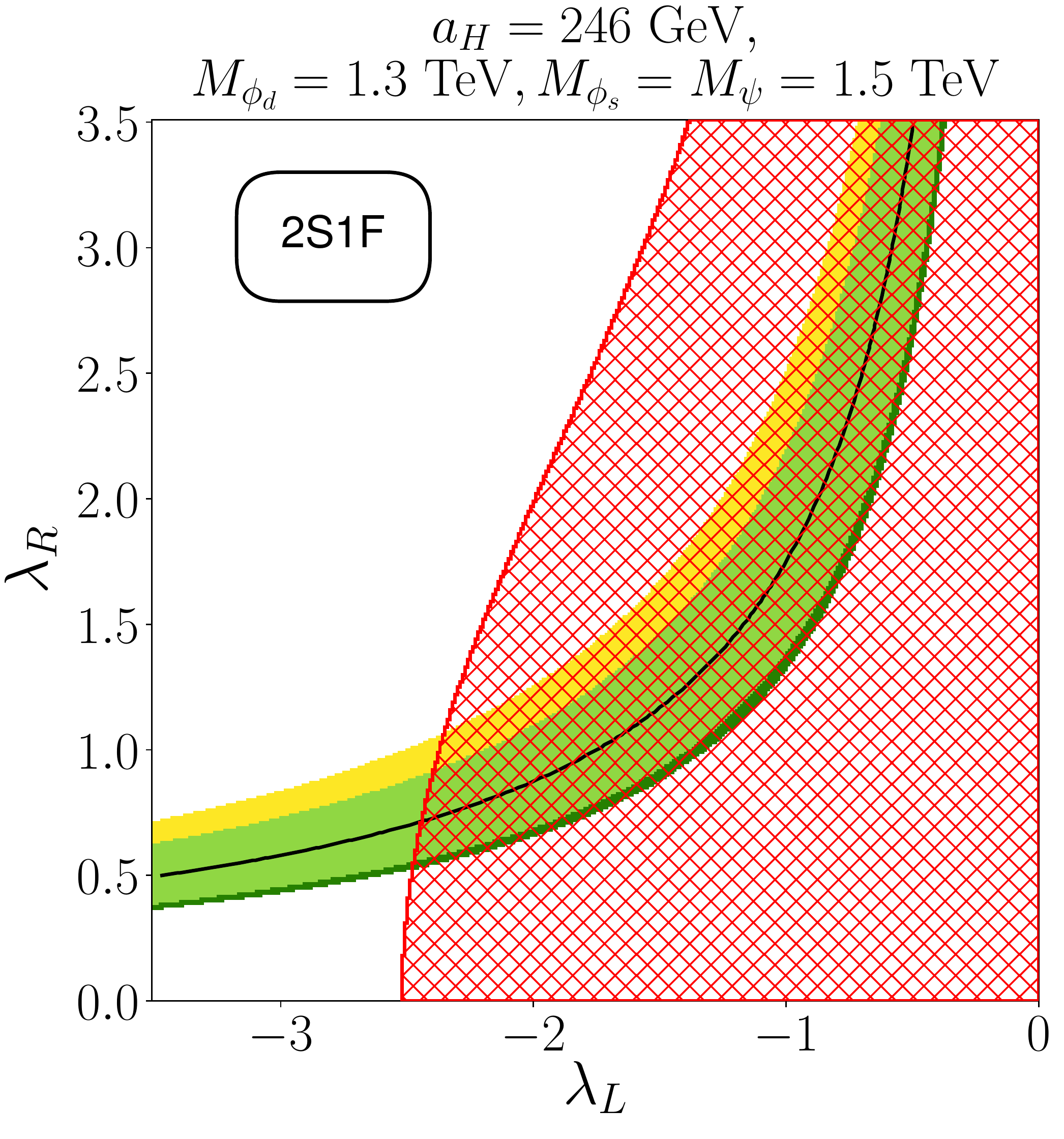}
		\hspace{0.01\linewidth}
		\includegraphics[width=0.31\textwidth]{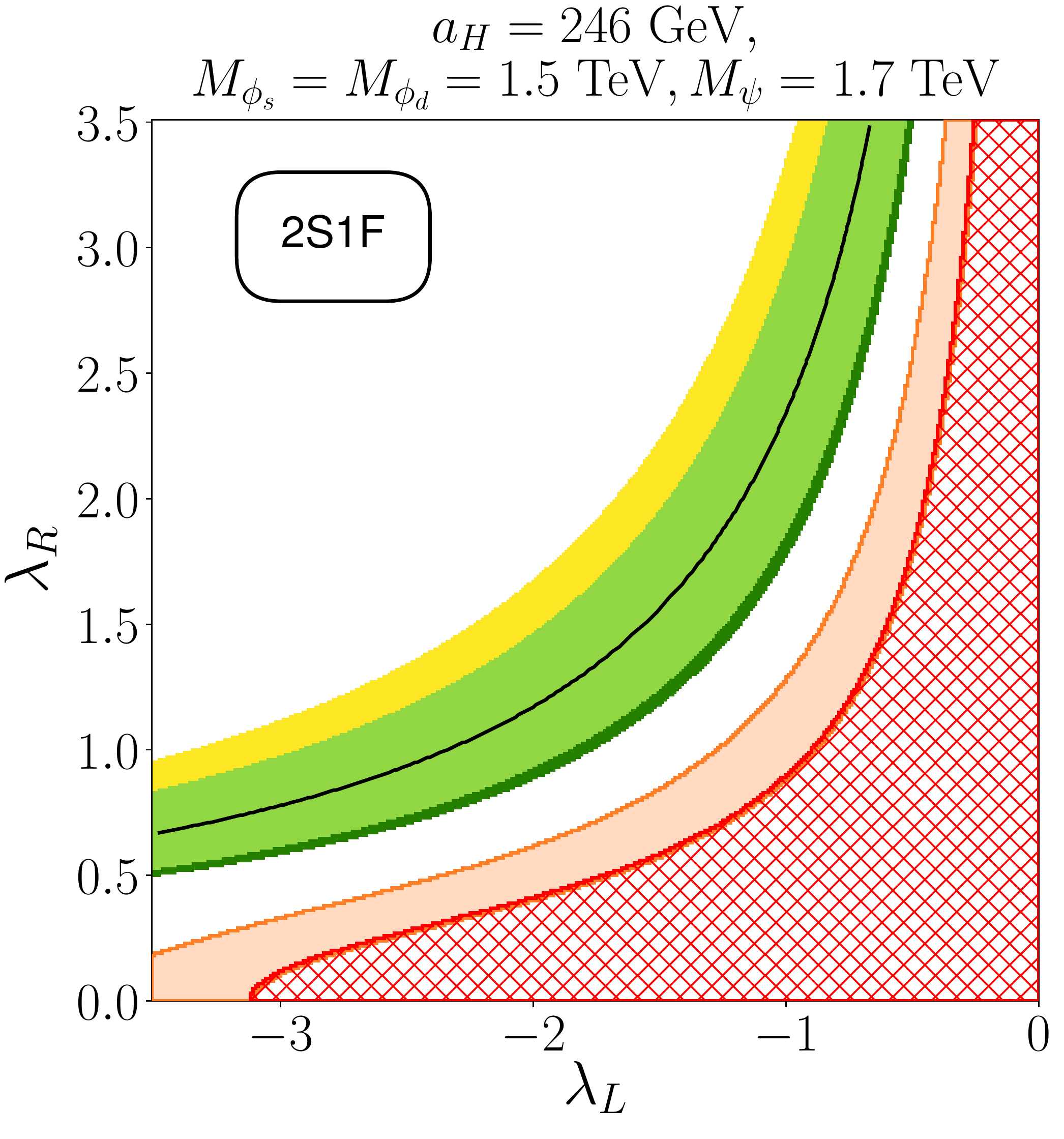}
		\caption{Results from Model 2S1F, scanning over the couplings of the new scalars and fermion to the left- and right-handed muon.  The trilinear coupling which parametrises the mixing of the neutral BSM scalars is fixed to $a_H = 246$ GeV.  Colours are the same as in Fig.\ \ref{fig:Min3FieldsSLRFRCouplingsEqualMasses}.  In these scenarios the mass of either the scalar singlet $\phi_s$, scalar doublet $\phi_d$, or fermion $\psi_s$ is displaced by $200$ GeV from the other BSM masses at $1500$ GeV. \label{fig:Min3FieldsSLRFRCouplingsSplitMasses} }
	\end{figure}

The situation can be changed substantially if we split the masses of 
the new states.  The relic density of each point will in general be
larger, since mass splittings will suppress several relic density
depletion mechanisms such as coannihilation channels. Fig.\ 
\ref{fig:Min3FieldsSLRFRCouplingsSplitMasses} investigates this for
three example cases. In each of the three panels the mass of the fermion singlet 
is $200$ GeV away from the dark matter scalar, which in turn
is either singlet- or doublet-dominated or a mixture.
All panels of  Fig.\ 
\ref{fig:Min3FieldsSLRFRCouplingsSplitMasses}  can be compared to the
middle  panel of Fig.\ 
\ref{fig:Min3FieldsSLRFRCouplingsEqualMasses}, where all masses were
equal to $1500$ GeV. The contribution to $\amu$ is similar in all
cases, with slight changes of the required couplings depending on the
increased/decreased masses.

The dark matter phenomenology changes
more dramatically.
Both the relic density and the direct detection constraints differ
strongly between the panels of Fig.\ 
\ref{fig:Min3FieldsSLRFRCouplingsSplitMasses} and compared to the
middle panel of Fig.\ \ref{fig:Min3FieldsSLRFRCouplingsEqualMasses}.
Specifically the direct detection constraint ruled out a full
explanation of the relic density in the equal mass case of
the middle panel of Fig.\ \ref{fig:Min3FieldsSLRFRCouplingsEqualMasses}. The same still happens
in the third panel 
of Fig.\ \ref{fig:Min3FieldsSLRFRCouplingsSplitMasses}, where the scalar
doublet and singlet masses are equal. The reason in all these cases is
the doublet--singlet mixing and the unsuppressed coupling of dark
matter to the Higgs boson via the
doublet--singlet--Higgs coupling $a_H$. In contrast, as soon as the
scalar doublet and singlet masses are different (left and middle
panels of  Fig.\ \ref{fig:Min3FieldsSLRFRCouplingsSplitMasses}), the
direct detection constraints are weakened, and as a result a full
explanation of the observed relic density is viable.

In detail the left panel of
Fig.\ \ref{fig:Min3FieldsSLRFRCouplingsSplitMasses} illustrates the
case where the dark matter candidate is dominantly a scalar singlet
with only about a $1\%$ admixture of doublet. The singlet mass is set
$200$ GeV lighter than the scalar doublet and the fermion masses.  The
relic density is driven up compared to the middle panel of
Fig.\ \ref{fig:Min3FieldsSLRFRCouplingsEqualMasses} since several dark
matter depletion mechanisms are suppressed.  Generally, the scalar
singlet participates only in few interactions and mostly annihilates
through t-channel exchange of the BSM fermion\footnote{Other processes also play a significant role including e.g.\ t-channel exchanged of scalar, which occurs through the small doublet admixture.}.  Although the singlet couples to
BSM fermions only through $\lambda_R$, in the cross-section for t-channel
exchange of the muon, contributions that depend only on $\lambda_L$
(or only on $\lambda_R$) are suppressed by either $m_\mu^2 / M_F^2$ or
by the relative velocity in comparison to the contribution
proportional to $\lambda_L^2 \lambda_R^2$.  This leads to the shape of the
red line that can be seen in the plot, where it follows a similar
trajectory to the $\amu$ contours, which is approximately a hyperbola.

\update{Strikingly in the left panel of
  Fig.\ \ref{fig:Min3FieldsSLRFRCouplingsSplitMasses} we find that the
  red curve, where the observed relic density can be explained, and
  the black curve, where $\damuNEW$ can be explained are very close to
  each other and follow the same trajectory.  However while a
  1$\sigma$ explanation of $\damuBNL$ could have been accommodated,
  the slightly higher couplings required to explain relic density
  measurements means that the over abundance of dark matter now rules
  out most of the parameter space where $\damuNEW$
  (Eq.\ (\ref{eqn:avgDiscrepancy})) can be explained, except for a
  small slice of parameter space with $\lambda_R > 2.5$. Nonetheless
  this could be changed with small adjustments to the mass splitting
  and in general it is possible to explain $\damuNEW$ and relic
  density at the same time.  Explanations of $\damuNEW$ are possible for
  masses above $1$ TeV, as long as the product $\lambda_L\lambda_R
  a_H$ is sufficient.}
	
	In the middle panel of
        Fig.\ \ref{fig:Min3FieldsSLRFRCouplingsSplitMasses} we instead
        focus on doublet-dominated dark matter, by setting the scalar
        doublet mass $200$ GeV lighter than both the scalar singlet
        and fermion (and again compared to the middle panel of
        Fig.\ \ref{fig:Min3FieldsSLRFRCouplingsEqualMasses}).  The
        impact on $\damu$ is \update{very similar} to what happened in
        the left plot as expected from
        Eq.\ (\ref{eqn:2S1FLOContribution}).  However, the change in
        the behaviour of dark matter is very different.  Now the dark
        matter candidate has generally many more interaction channels
        via the SU(2)$_L$ gauge interactions of the doublet
        component. Hence the relative importance of the
        $\lambda_{L,R}$ parameters is reduced compared to the
        singlet-dominated case. Although not visible in the plot, the
        overall variation of the relic density is much reduced
        compared to the left panel or the middle panel of
        Fig.\ \ref{fig:Min3FieldsSLRFRCouplingsEqualMasses}.  The
        residual dependence on $\lambda_L$ is still important to
        deplete the relic density sufficiently, but the residual
        dependence on $\lambda_R$ via the singlet-admixture is of
        minor importance. In this case the constraints from the relic
        density and from $\damu$ are essentially orthogonal in the
        parameter space and thus complementary.  Both constraints can
        be fulfilled, but agreement can only be achieved for very
        large values of $|\lambda_L|$ (\update{$\approx$ between $2.4$
          and $2.5$ when $\damuNEW$ is explained with $1\sigma$}).

In the right panel of Fig.\ \ref{fig:Min3FieldsSLRFRCouplingsSplitMasses} we raise 
the fermion mass by $200$ GeV, compared to the middle panel of Fig.\ 
\ref{fig:Min3FieldsSLRFRCouplingsEqualMasses}.  Here we can see that this slightly 
increases the size of the couplings required to explain the measured 
$\damu$ values.  As mentioned before, however, in this case with equal
scalar singlet and doublet masses the relic density must be under abundant
because of direct detection constraints.
Raising the fermion mass suppresses the coannihilation mechanisms involving the 
fermion, which are strongly dependent on the couplings $\lambda_L$ and 
$\lambda_R$, but t-channel fermion exchange can still be active as
suggested by the fact that the relic density curve is again a
hyperbola similar to the left panel. However since the singlet and doublet masses are equal in this case, the usual $SU(2)$ scalar doublet (co)annihilations have greater impact, reducing the required size of $\lambda_L\lambda_R$.  Due to the rescaling, the direct detection
constraints reflect this behaviour.
	With  \update{the chosen value of $a_H=246$ GeV} 
	\update{$\damuNEW$ can be explained} with an under abundant relic 
	density.

Just as for Fig.\ \ref{fig:Min3FieldsSLRFRCouplingsEqualMasses}, we can change $a_H$ 
to see if we can simultaneously explain $\amu$ and dark matter.  Currently in the 
\update{middle panel  of Fig.\ \ref{fig:Min3FieldsSLRFRCouplingsSplitMasses}, as well as 
the left panel with very large $\lambda_R$}, we can already explain $\amuNEW$ and 
dark matter simultaneously.  As we saw previously increasing $a_H$ increases the 
constraints from direct detection.  \update{For the left panel as nearly all} 
of the parameter space that can explain $\damuNEW$ to within $1\sigma$ 
is ruled out by over abundant dark matter, we can decrease $a_H$ to push the 
parameter space up into the under abundant regions.  However, by increasing $a_H$ 
for the middle panel, the observed dark matter relic density line would become more 
dependent on $\lambda_R$, as the two scalars mix more.  For the right panel, again 
increasing $a_H$ would cause the regions ruled out by direct detection to become 
more prevalent, while pulling the space which can explain $\damuNEW$ \update{closer to 
it}.  However, while decreasing $a_H$ would shrink the regions constrained by direct 
detection, \update{the red line does not shift close enough to the band which can 
explain $\damuNEW$}, which itself would move away up and to the left as $a_H$ is 
decreased.  So in the scenario with $M_\psi$ increased slightly from 
$M_{\phi_s}=M_{\phi_d}=1.5$ TeV, \update{it is never possible to explain $\damuNEW$ 	
and the dark matter relic density simultaneously, even if we adjust $a_H$.}  

\begin{figure}[t]
	\centering
	\includegraphics[width=0.32\textwidth]{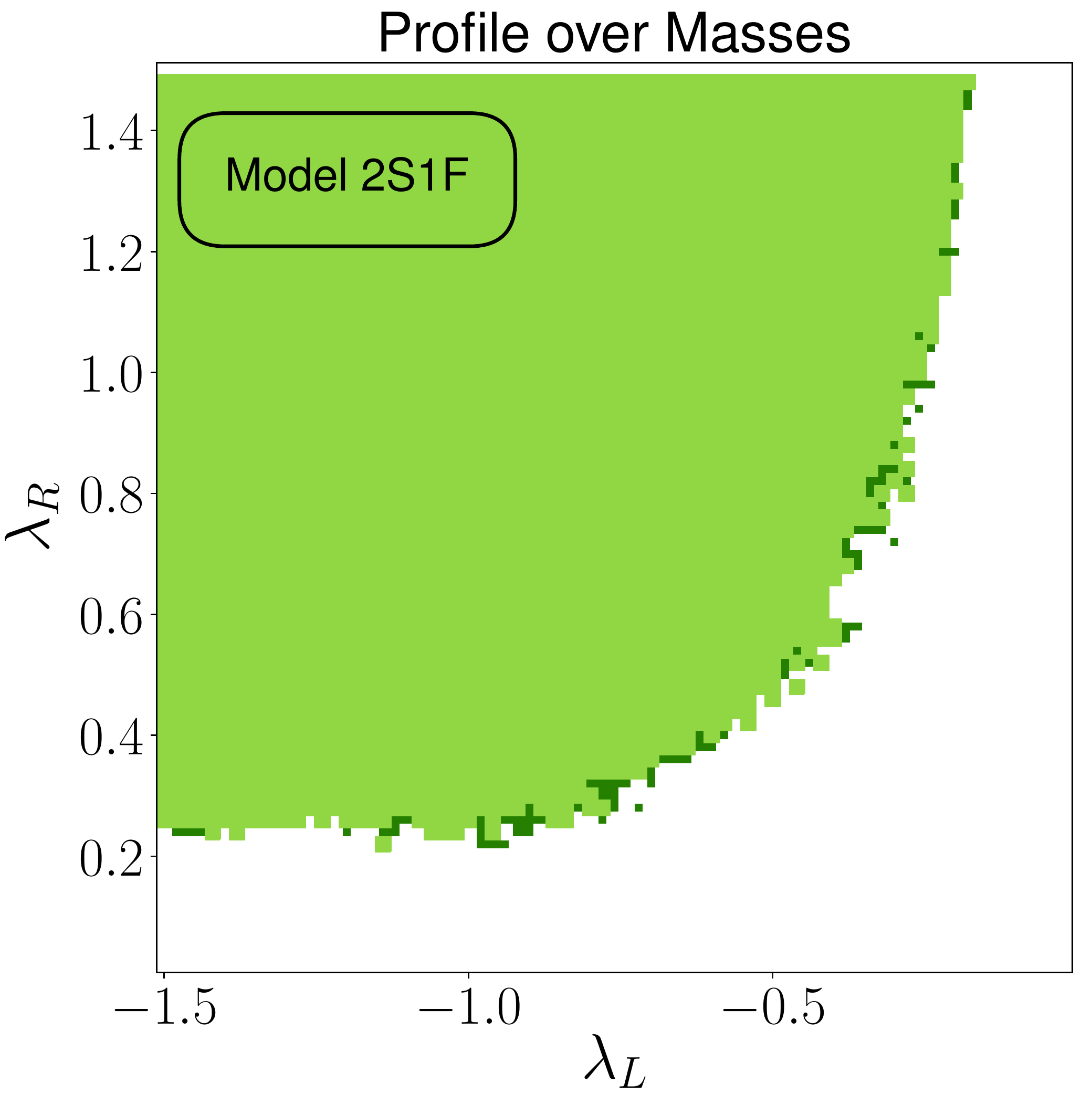}
	\includegraphics[width=0.32\textwidth]{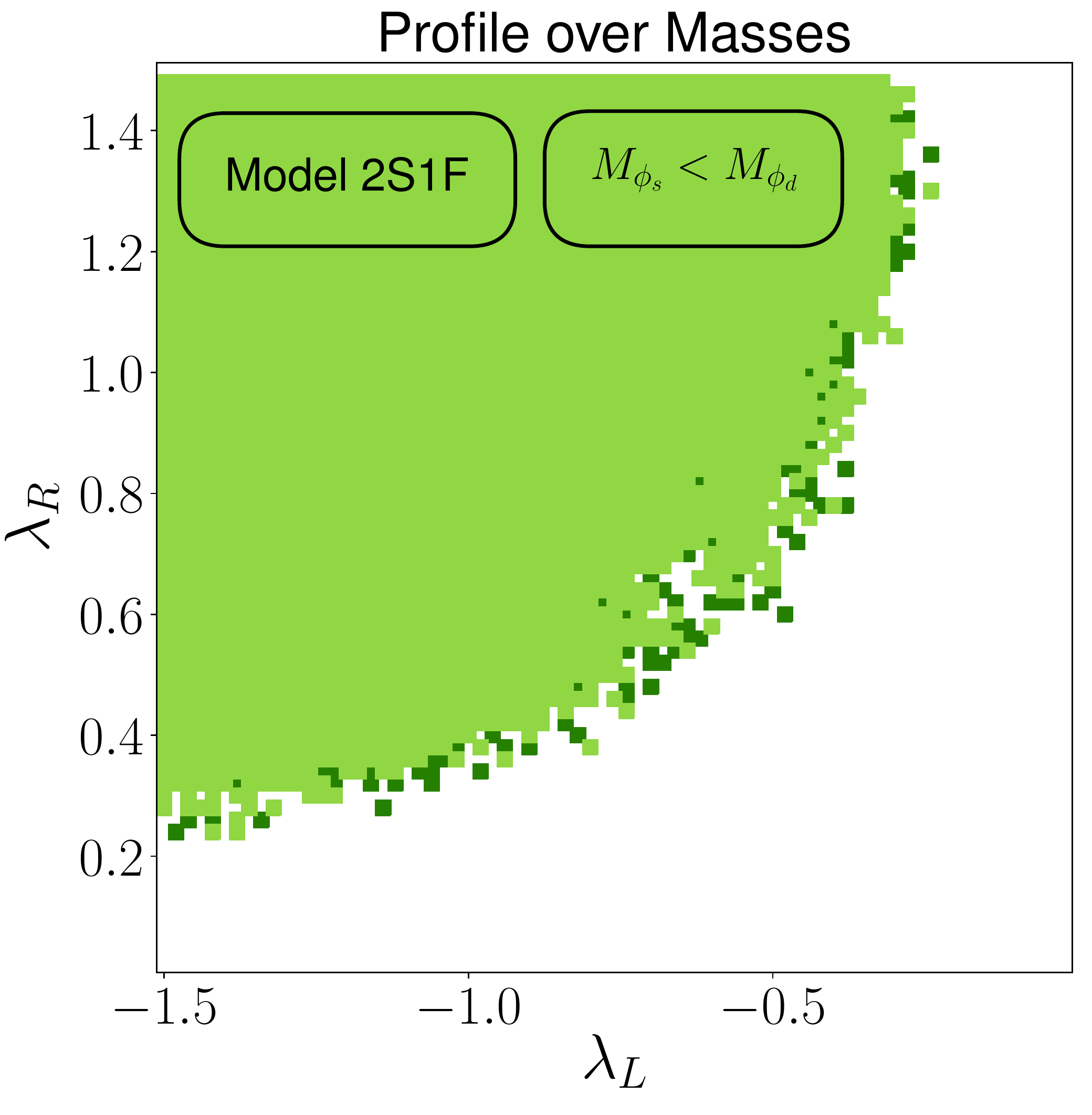}
	\includegraphics[width=0.32\textwidth]{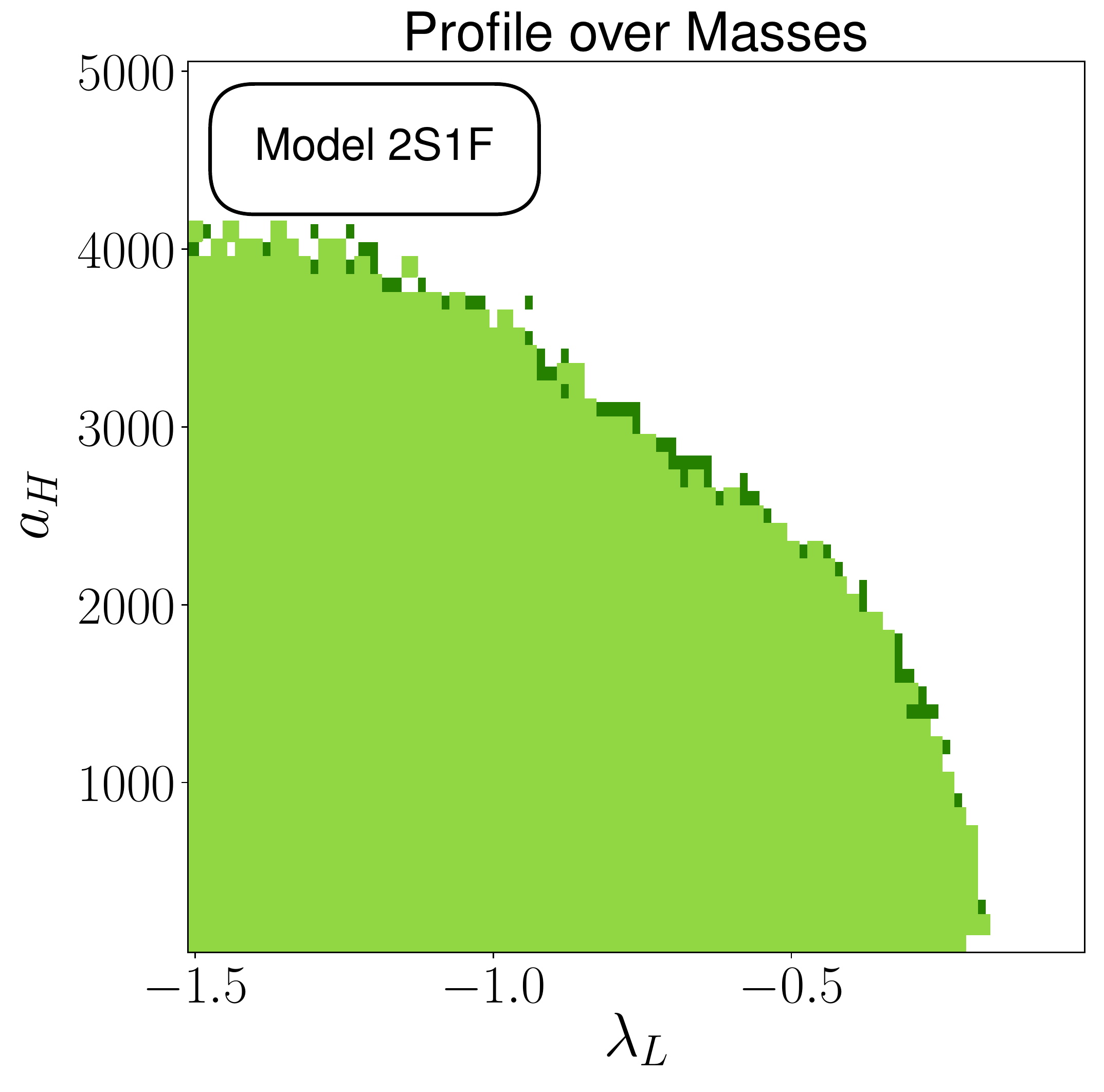}
	\caption{Profile over the $\lambda_L-\lambda_R$ plane or the
          $\lambda_L-a_H$ plane for Model 2S1F, scanning over the
          masses of the new scalars and fermion between $0$ and
          $5000$ GeV, as well as the mixing $a_H \le 5000$ GeV and
          coupling $\lambda_R \in [0,1.5]$.  Colours are the same as
          in Fig.\ \ref{fig:Min3FieldsSLRFRCouplingsEqualMasses}.  The
          scan targeted the observed value of the dark matter relic
          density $0.1200$, Eq.\ (\ref{eqn:DMRD}), the value of $\damu$ in Eq.\ (\ref{eqn:BNLDiscrepancy}), and the direct detection log likelihood provided by \DDCalc \cite{Workgroup:2017lvb,Athron:2018hpc}, where points which are excluded by direct detection or are more than $3\sigma$ away from the Planck observation are thrown away.  \label{fig:Min3FieldsSLRFRprofile} 
}
\end{figure}

In all the two-dimensional slices of parameter space we presented,
\update{explaining $\damuNEW$} and dark matter is only
possible with sizable interactions between the BSM states and muons,
$\lambda_L$ and $\lambda_R$. Indeed the relic density is over abundant
when both $\lambda_L$ and $\lambda_R$, are small because both the
t-channel exchange of the BSM fermion and coannihilations with the BSM
fermion are driven by these $\lambda_L$ and $\lambda_R$ couplings. At
lighter masses both these processes and (co)annihilation processes
involving $SU(2)$ interactions of the scalar doublet will become more
efficient as the mass suppression is reduced, with the $SU(2)$
interactions becoming more significant relative to $\lambda_L$ and
$\lambda_R$.  These effects imply that at much lighter masses it
should be possible to also deplete the relic density with much smaller
$\lambda_L$ and $\lambda_R$. At the same time, explaining
$\damu^{\text{BNL}}$ \update{or $\damuNEW$} with smaller
couplings is also possible when all the masses are small.  However
because it is possible for the observed relic density to be obtained
through $SU(2)$ interactions of a $600$ GeV scalar doublet, it is not
clear that with small couplings large $\damu$ and
dark matter can be explained simultaneously as the $\amu$ explanation may
then require larger values of $\lambda_L$ and $\lambda_R$ (or smaller
masses) than what is required for fitting the relic density.
Whether dark matter and the $\amu$
\update{anomalies} can be simultaneously explained for all
$\lambda_L$, $\lambda_R$ or if there is a lower bound on these
couplings just from these requirements remains an open question.

Therefore we now turn to a complete exploration of the parameter space
of this model, varying $M_{\phi_s}$, $M_{\phi_d}$, $M_\psi$, $a_H$,
$\lambda_L$, and $\lambda_R$ to see the impact of the new
$\damuNEW$ measurements and address the question posed
above.  We vary $M_i \in [0,5000]$, $a_H \in [0, 5000]$,
$\lambda_{L,R} \in [0, 1.5]$ in a number of \multinest
\cite{Feroz:2007kg,Feroz:2008xx,Feroz:2013hea,Buchner2014} scans with a likelihood constructed to target
$\Omega_{h^2}=0.1200$, different possible\footnote{Due to the time
  required for such large dimensional scans, these were performed
  before the release of the FNAL Muon g-2 results \update{\cite{PhysRevLett.126.141801}} and without
  knowledge of the result. Therefore for we performed scans for a
  range of results that we considered plausible given the previous
  $\amu^{\textrm{BNL}}$ measurement.} values of $\amu$ and solutions
which evade direct detection limits. In total we collected about \update{15
million} samples targeting several different values in the range $\damu \underset{\sim}{\in} [0,42\times10^{-10}]$.

In the left panel of Fig.\ \ref{fig:Min3FieldsSLRFRprofile} we present
results in the $\lambda_L$--$\lambda_R$ plane.  This shows that it is
only possible to explain the $\amu^{\text{BNL}}$ measurement and dark
matter simultaneously when \update{$|\lambda_L \lambda_R| \ge 0.22$}.
With the new \update{$\amu^{\text{FNAL}}$ measurement} this limit
update {changes very little}.  Couplings smaller than this cannot
simultaneously explain \update{$\amuNEW$} and dark matter. For the case
of scalar doublet dominated dark matter the $SU(2)$ interactions will
deplete the relic density to below the observed value whenever
$m_{\phi_d} \lesssim 600$ GeV and this limit on the lightest mass
makes explaining large $\damu$ impossible for such small
couplings. For pure scalar singlet dark matter, note that if the
scalar doublet is so heavy it completely decouples, then the model
effectively reduces to that of Model R from the two field section.
There we already found that a simultaneous explanation of dark matter
and the $\amu$ measurements is only possible for very large
$\lambda_R$ (higher than the values shown in
Fig.\ \ref{fig:Min3FieldsSLRFRprofile}) and not possible at all when
collider limits are also taken into account.

If instead the dark matter has a significant admixture of both the
scalar singlet and the scalar doublet, then the situation is a bit
more complicated. With both singlet and doublet mixing, the relic
density can deplete through processes involving both $\lambda_L$ and
$\lambda_R$ couplings and $a_H$ in addition to the $SU(2)$
interactions of the doublet component. One can compensate increased
splitting in the masses by raising $a_H$ to keep $\damu$ fixed to the
measured value, or similarly one can reduce the mass splitting to
compensate for reducing $a_H$ but in either case the relic density is
still depleted too efficiently since both raising $a_H$ and reducing
the mass splitting can enhance annihilation cross-sections.
Increasing $a_H$ and reducing the mass splitting also increases direct
detection cross-sections, increasing the tension with large $\damu$
further. Therefore when there is significant mixing between the scalar
singlet and doublet, requiring that the measured relic density is
obtained implies masses that are too large (for any given $a_H$ value)
for large $\damu$ to be explained with small couplings. We also looked
at scenarios where the dark matter is more singlet or more doublet in
nature separately.  We found quite similar limits on
$\lambda_L\lambda_R$ emerge in both cases, though the more singlet DM
case had a \update{marginally higher limit}, as can be seen in the
\update{middle} panel of Fig.\ \ref{fig:Min3FieldsSLRFRprofile}.

\update{The tension with $a_H$ is also shown in the plot on the right
  panel of Fig.\ \ref{fig:Min3FieldsSLRFRprofile}.  There one can see
  that there is an upper limit on $a_H$, coming from the dark matter
  constraints discussed above. The value of $a_H$ can also be
  restricted by the need to avoid tachyonic scalars that would appear
  when $a_H v < M_{\phi_s} M_{\phi_d}$.  However we find it is the
  limits from dark matter that leads to the constraint shown
  in the right panel of
  Fig.\ \ref{fig:Min3FieldsSLRFRprofile}.}

 As a result we find a lower limit on $|\lambda_L\lambda_R|$ that
 increases with the value for $\damu$ used as a constraint.  In this
 way we directly see the impact of the new $\damuNEW$ result on this
 model.  We do not expect collider limits to affect our results here,
 since we do not find any solutions explaining both dark matter and
 $\damu^{\text{BNL}}$ \update{or $\damuNEW$} where the lightest of the
 two scalar masses $M_{\phi^0_1}$ is below around $500$ GeV for our
 choice of coupling range ($\lambda_{L,R} \le 1.5$).  Our scan results
 also indicate interesting structure in the mass planes. We reserve
 the detailed discussion and presentation of such mass plane results
 for a future dedicated global fit of the model.  Nonetheless it is
 clear from our results that the combination of parameters and masses
 that are allowed are significantly impacted by the interplay between
 dark matter and $\amu$ constraints.  The $\amu^{\textrm{BNL}}$ and
 $\amu^{\textrm{FNAL}}$ measurements are therefore very important for
 the phenomenology of models like this and should be included in all
 such studies, and in any future global fits of any models like this.

%% file: three_fields_2F1S.tex
\subsection{Three-field model with two fermions, one charged scalar
  and fermionic dark matter}
\label{sec:2F1S}
Having looked at scalar dark matter in the previous section we now
consider a model with a fermionic dark matter candidate.  For this we
choose a minimal three-field model with two new fermions and one new
scalar field, which we call Model 2F1S.   Specifically this model extends the SM by adding three $\mathbb{Z}_2$-odd fields
\begin{align}
  \phi_d = \begin{pmatrix} \phi^+_d \\ \phi_d^0 \end{pmatrix}, ~ & \psi_d = \begin{pmatrix} \psi^0_d \\ \psi_d^- \end{pmatrix}, ~ \psi _s ^0\,, 
\end{align}
where $\phi _d$ is a scalar doublet with representation
$({\bf{1}},{\bf{2}},\frac{1}{2}) _0$, $\psi _d$ a Dirac fermion doublet with
representation $({\bf{1}},{\bf{2}},-\frac{1}{2})_{1/2}$ expressed in
Weyl spinors $\psi_d$ and its  Dirac partner $\psi _d
^c =(\psi_d^{0c\dagger},\psi_d^{-c\dagger})$,
and $\psi _s ^0$ a
neutral singlet Weyl fermion with representation
$({\bf{1}},{\bf{1}},0)_{1/2}$. 

In principle the model allows  scenarios with
scalar dark matter, however we do not consider such scenarios here. We
use the model as an illustration of fermionic dark matter, in which
case the dark matter candidate is predicted to be a mixture of the
fermion singlet and the neutral doublet
component. Ref.\ \cite{Calibbi:2018rzv} also studied the model in
certain parameter slices. Here we intend to determine the full status
of the model, studying the detailed parameter dependence and
performing general scans of the parameter space.

The relevant interactions are:
	\begin{align} \label{eqn:Min3FieldsFLRS1}
		{\cal L}_{\text{2F1S}} = \bigg(& \lambda_{1} H \cdot
                \psi_d \psi_s^0 - \lambda_{2} \psi_d^c \psi_s^0 H + \lambda_L L_2\cdot \phi_d \psi_s^0 + \lambda_R \psi_{d} \mu \phi_d^\dagger \nonumber \\ 
		&- M_{\psi_d} \psi_d^c \psi_d - \frac{M_{\psi_s}}{2} \psi_s^0 \psi_s^0 + h.c.\bigg) - M_{\phi}^2 |\phi_d|^2,
	\end{align}
where the sign in front of the $\lambda_2$ term reflects our
definition of $\psi_d^c$ as an $SU(2)$ anti-doublet.

The model has four new coupling parameters. $\lambda_{L,R}$ are the
couplings to the left- and right-handed muon, similarly to the case of
the previous Model 2S1F.  The couplings $\lambda_{1,2}$ govern the
mixing of the BSM fermions via the Higgs VEV, similarly to the $a_H$
parameter in the previous case. The products
$\lambda_i\lambda_L\lambda_R$ are responsible for the new source of
muon chirality flips (for $i=1,2$).  To reiterate, this model can also
be compared to the Bino-Higgsino-left smuon (BHL) system for the MSSM,
see Eq.\ (\ref{eq:SUSYMIapprox}) below. There $\psi_d$ corresponds to
the Higgsino, $\psi_s$ to the Bino, and $\phi_d$ to the left smuon
doublet, and the couplings $\lambda_{1},\lambda_{2}$ and $\lambda_L$
would all correspond to the Bino coupling $g_1$, while $\lambda_R$
would correspond to the muon yukawa coupling.  A difference is that
the MSSM counterparts of the couplings $\lambda_{1,2}$ would couple
the Higgsinos to the up-type and down-type Higgs doublets, and the
resulting contributions to the mass mixings would involve the
$\tan\beta$ parameter, while here there is just one single Higgs
doublet and the mass mixing is simpler.

The mixing of the BSM fermions affects the singlet $\psi^0_s$ and the
neutral components of the fermion doublet $\psi_d^0$ and its Dirac
partner $\psi_d^{0c\dagger}$.  They mix into three different flavours
of neutral fermions $\psi^0_i$ with Majorana mass terms via the mixing
matrix $V_F$:
\begin{equation} \label{eqn:Min3FieldsFLRS1Mixing}
   \begin{split}
	{\cal L}_{\text{2F1S}} & \ni  -\frac{1}{2}\begin{pmatrix} \psi_s^{0} & \psi_d^{0} & \psi_d^{0c\dagger} \end{pmatrix} \begin{pmatrix} M_{\psi_s} & \frac{\lambda_{1} v}{\sqrt{2}} & \frac{\lambda_{2} v}{\sqrt{2}} \\ \frac{\lambda_{1} v}{\sqrt{2}} & 0 & M_{\psi_d} \\ \frac{\lambda_{2} v}{\sqrt{2}} & M_{\psi_d} & 0 \end{pmatrix}  \begin{pmatrix} \psi_s^0 \\ \psi_d^{0} \\ \psi_d^{0c\dagger} \end{pmatrix} 
  = -\frac{1}{2}\sum_{i=1,2,3} M_{\psi_i^0} \psi_i^0\psi_i^0 \,,
   \end{split}
\end{equation}
	where
\begin{equation}
   \begin{pmatrix} \psi_1^0 \\ \psi_2^{0} \\ \psi_3^{0} \end{pmatrix} =             V_F^T\begin{pmatrix} \psi_s^0 \\ \psi_d^{0} \\ \psi_d^{0c\dagger} \end{pmatrix}\,.
\end{equation}

The contribution to $\amu$ from this model comes from two diagrams,
the FFS diagram of Fig.\ \ref{fig:FFSDiagram} and the SSF diagram of
Fig.\ \ref{fig:SSFDiagram}, with the generic $F$ fermion lines
replaced by $\psi^-$ and $\psi_i^0$ and the scalar $S$ ones by
$\phi^0_d $ and $\phi^{+*}_d$, respectively.  Among these diagrams,
only the SSF diagram with neutral fermion exchange can lead to an
enhanced chirality flip. The full contributions are given by:
\begin{align}\label{eqn:amu_2F1S_model}
  \damu^{\text{2F1S}} =&
  \frac{m_\mu^2}{32 \pi^2 M_\phi^2} \bigg(\frac{\lambda_R^2}{6} \,E\!\begin{pmatrix}\frac{M^2_{\psi_d}}{M^2_\phi} \end{pmatrix}
        - \sum_{i=1}^3 
	\frac{\lambda_R^2 |V_{F\,2i}|^2 + \lambda_L^2 |V_{F\,1i}|^2}{6}\,B\!\begin{pmatrix} \frac{M^2_{\psi_i^0}}{M^2_\phi}\end{pmatrix} \nonumber\\ 
        & +\sum_{i=1}^3  \frac{M_{\psi_i^0}}{m_\mu}\frac{2 \lambda_L \lambda_R V_{F\,1i} V_{F\,2i}}{3}  \,C\!\begin{pmatrix}\frac{M^2_{\psi_i^0}}{M^2_\phi}\end{pmatrix}\bigg).
\end{align}
The term in the second line corresponds to the additional source of
chirality flip in the model. It can be analyzed similarly to the term
in Eq.\ (\ref{eqn:2S1FLOContribution}).  The actual enhancement
factor, beyond the loop and mass suppression factor, is given by the
coupling combination $\lambda_{1,2}\lambda_L\lambda_R/y_\mu$ where the
$\lambda_{1,2}$ are contained in the product of the mixing matrix
elements $V_{F\,1j} V_{F\,2j}$ of the neutral fermions.  If this
coupling combination vanishes, the new chirality flip contribution and
the third term disappear, and we end up with the contribution of Model
L from Sec.\ \ref{sec:TwoFields},
Eq.\ (\ref{eqn:Min2FieldsLLContributions}).
	
\begin{figure}[t]
  \centering
  \includegraphics[width=0.31\textwidth]{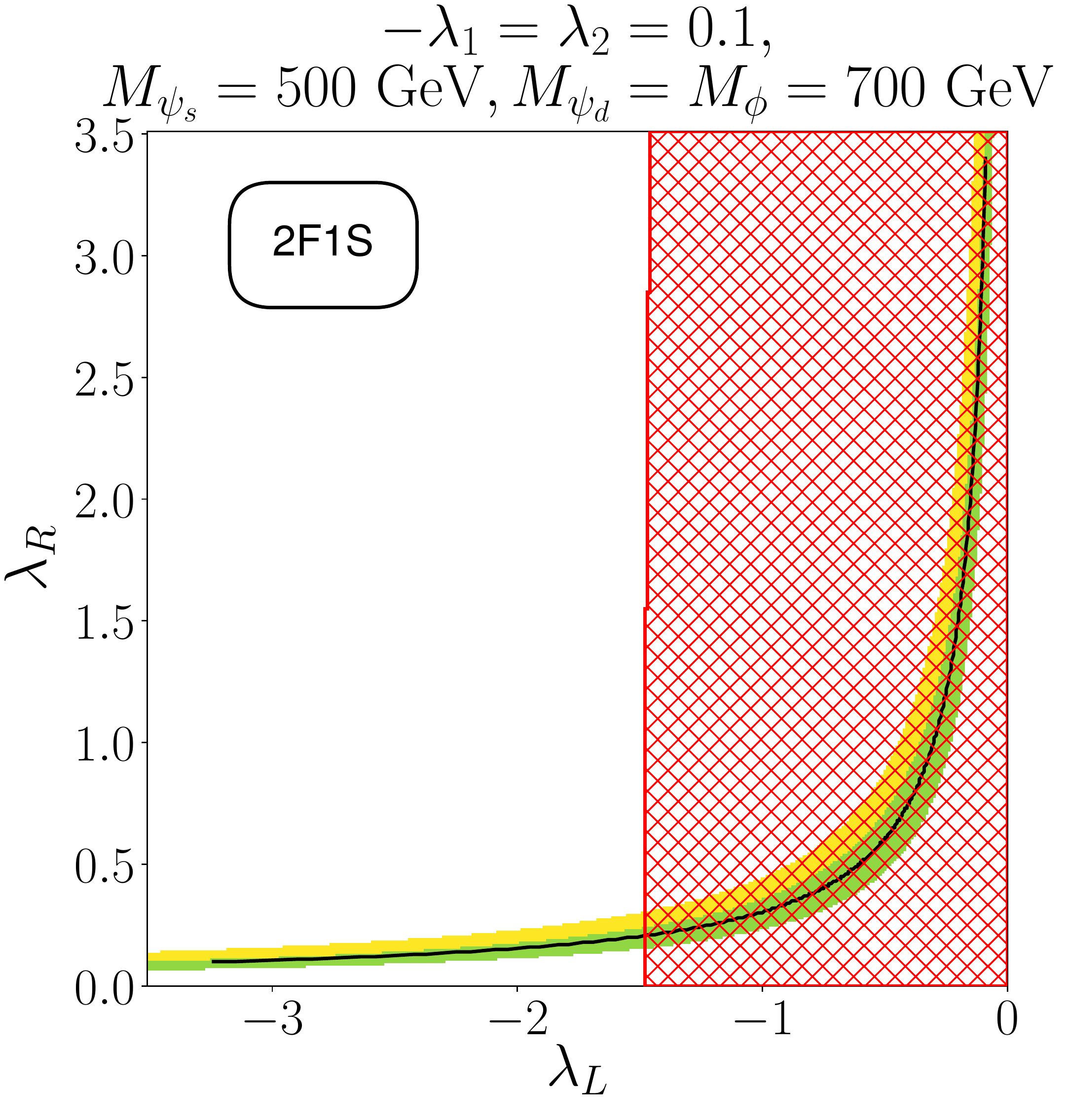}
  \hspace{0.01\linewidth}
   \includegraphics[width=0.31\textwidth]{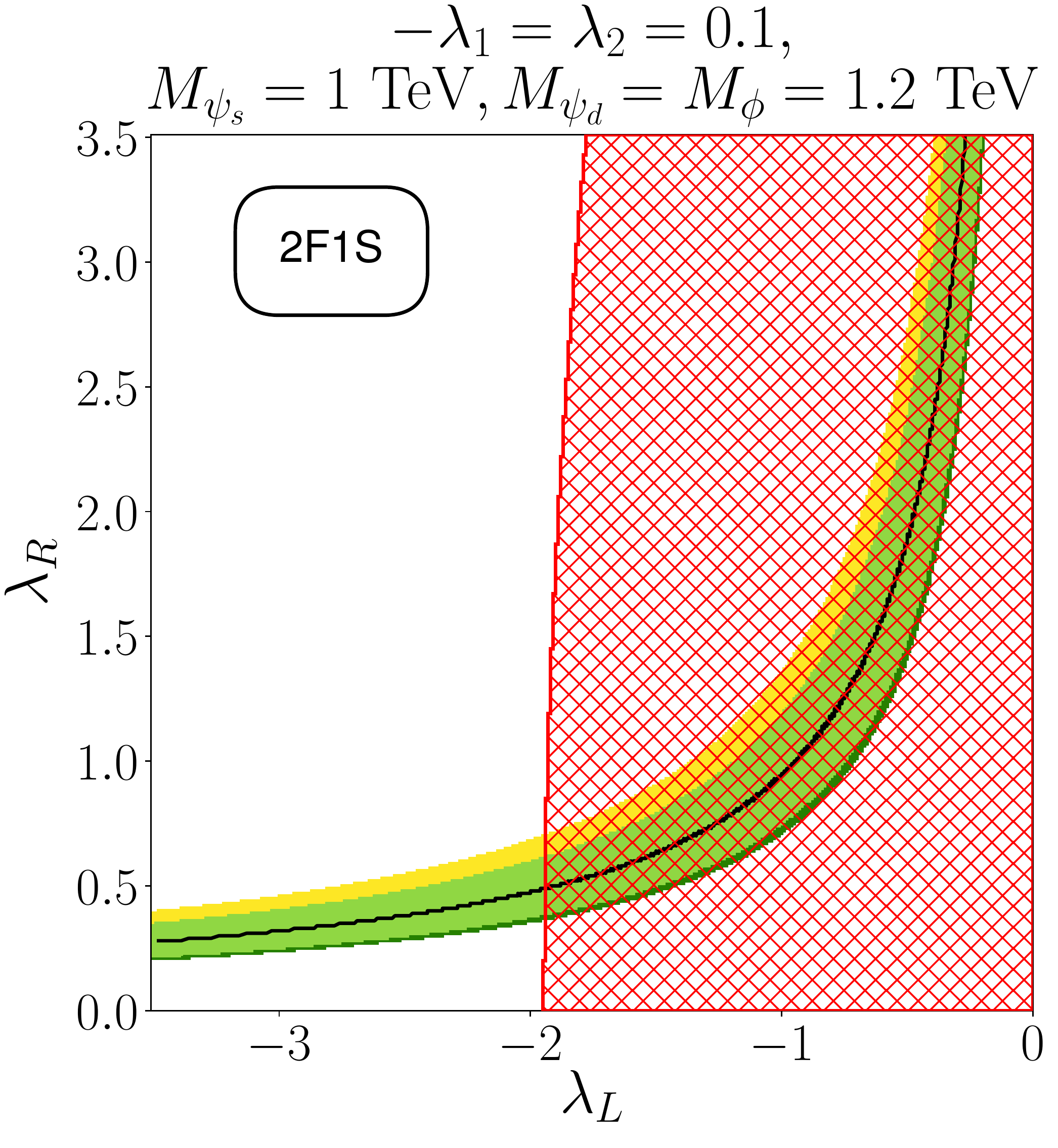}
  \hspace{0.01\linewidth} 		
\includegraphics[width=0.31\textwidth]{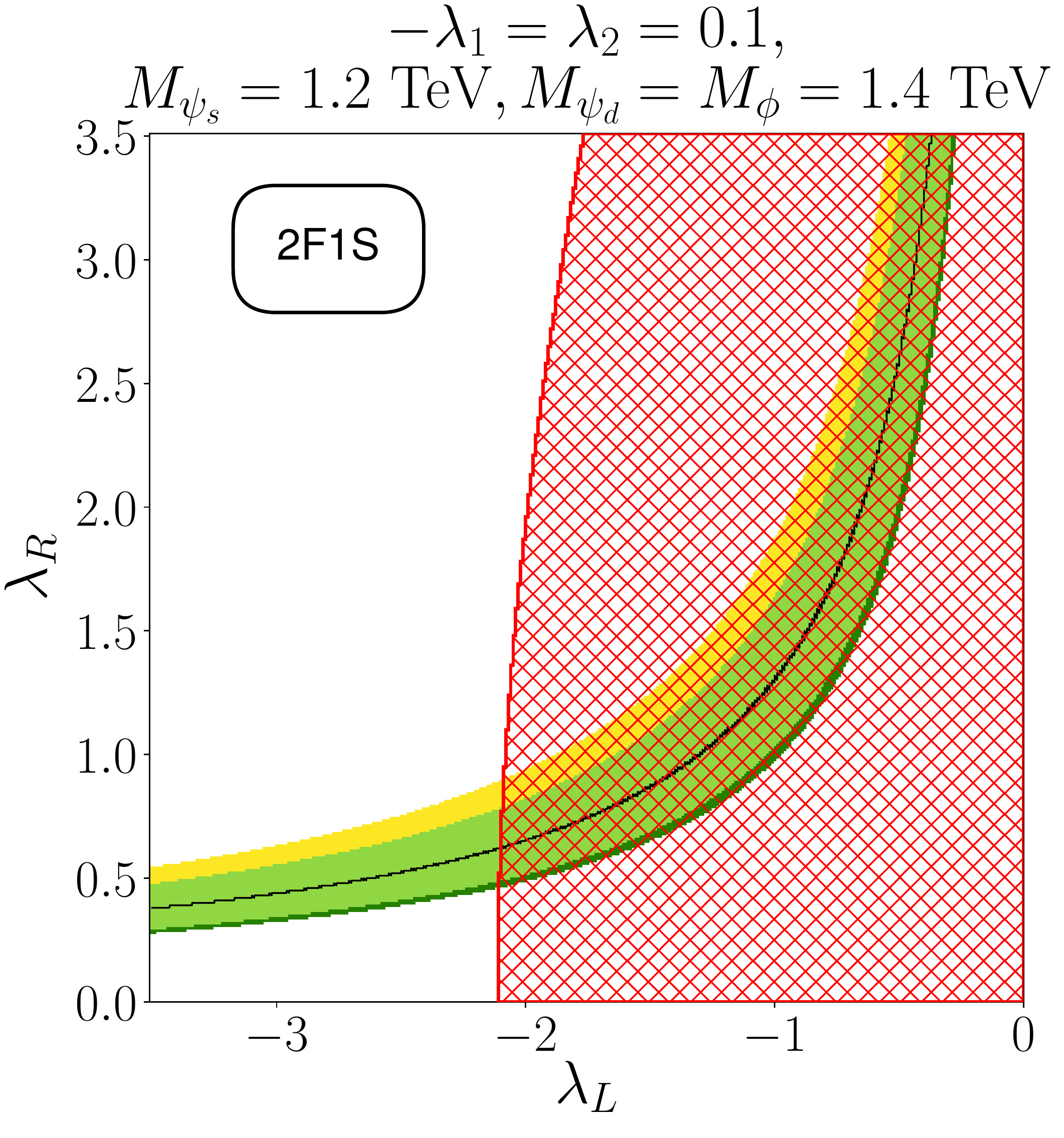}
\caption{Results from Model 2F1S, scanning over the couplings of the
  new fermions and scalar to the left- and right-handed muon.  Colours
  are the same as in
  Fig.\ \ref{fig:Min3FieldsSLRFRCouplingsEqualMasses}.  The mixing
  couplings are fixed to $-\lambda_{1}=\lambda_{2}=0.1$.  In this
  scenario the masses of the scalar doublet $\phi_d$ and fermion
  doublet $\psi_d$ are raised by $200$ GeV above the mass of the
  fermion singlet $\psi_S$.  Points to the right of the red line
  produce an over abundance of dark matter and are strongly
  excluded. \label{fig:Min3FieldsFLRSRCouplingsSplitMasses}}
\end{figure}

The case of a pure fermion $SU(2)$ doublet dark matter candidate is
very well known as this corresponds to Higgsino dark matter in the
MSSM, or the lowest representation of minimal dark matter
\cite{Cirelli:2005uq}.  In that case it is possible to deplete the
relic density to the measured value, using only $SU(2)$ gauge
interactions, when the fermion doublet has a mass of $1$ TeV
\cite{ArkaniHamed:2006mb} while above (below) it will be over (under)
abundant.  This can be adjusted by mixing with singlet, to obtain
well-tempered dark matter\cite{ArkaniHamed:2006mb} where the rapid
depletion of dark matter from $SU(2)$ interactions are diluted through
mixing with the singlet, though well-tempered dark matter is heavily
constrained by direct detection (see e.g. Fig.\ 8 of
Ref.\ \cite{Athron:2016gor}).  As in the previous example explaining
$\damu^{\text{BNL}}$ \update{and $\damuNEW$ after including the FNAL result} will 
require significant values of the $\lambda_L$, $\lambda_R$ and $\lambda_i$
couplings, leading to additional depletion mechanisms.  Therefore
scenarios where all the input masses are set to be equal will only explain dark
matter at very heavy masses.

Instead we first focus on scenarios where the dark matter is dominantly
singlet in nature in
Fig.\ \ref{fig:Min3FieldsFLRSRCouplingsSplitMasses}, showing two
dimensional parameter slices in the $\lambda_L$--$\lambda_R$ plane,
with the overall mass scale increasing between the panels from left
to right.  Specifically each of the three panels shows scenarios where
the fermion doublet $\psi_d$ and scalar doublet $\phi_d$ are always
$200$ GeV higher than the mass of the fermion singlet
$\psi_s^0$.  To simplify things further, we choose fairly small values
for the mixing parameters $-\lambda_{1}=\lambda_{2}=0.1$, since large
singlet--doublet mixing increases dark matter direct detection
constraints and depletes the relic density.  The other couplings are
varied in the range $-\lambda_L,\lambda_R \in [0,3.5]$.

As a function of $\lambda_L,\lambda_R $, we get a \update{similar}
curve for \update{$\amuNEW$} to the ones shown in 
Figs.\ \ref{fig:Min3FieldsSLRFRCouplingsEqualMasses}, \ref{fig:Min3FieldsSLRFRCouplingsSplitMasses},
due to the dominant chirality flip contribution of
Eq.\ (\ref{eqn:amu_2F1S_model}) having the same dependence on
$\lambda_L \lambda_R$ as Eq.\ (\ref{eqn:2S1FLOContribution}). The
nearly vertical red lines indicate points in agreement with the Planck
observed \cite{Aghanim:2018eyx} dark matter relic abundance; points to
the right are over abundant and excluded.  As can be seen in all
panels of Fig.\ \ref{fig:Min3FieldsFLRSRCouplingsSplitMasses}, all
points allowed by the dark matter relic density are also in agreement
with direct detection constraints due to the small values of the
$\lambda_i$ coupling \footnote{This can be compared to MSSM
  scenarios where such a small mass splitting between the Bino and Higgsino would face more severe constraints from direct detection since the gauge interactions that control the Bino-Higgsino-Higgs vertices are larger than $\lambda_i$ couplings chosen for this example.}.

Going from left to right in the three panels of
Fig.\ \ref{fig:Min3FieldsFLRSRCouplingsSplitMasses}, the singlet mass
is increased from $500$ GeV to $1000$ GeV and $1200$ GeV, while the
mass splitting remains $200$ GeV.  In the left panel the dark matter
relic abundance depends only on $\lambda_L$, the coupling of the
singlet to the muon and the BSM scalar, due to the large mixing
suppression of any $\lambda_R$ contribution through the fermion
doublet component of the dark matter.\footnote{Note that in contrast
  to the scalar dark matter case discussed in the previous section, in
  the annihilation cross-section for the t-channel of the BSM scalar,
  terms that depend only on $\lambda_L$ are not suppressed compared to
  terms that depend on both $\lambda_L$ and $\lambda_R$.} In the
middle and right panels the relative doublet content of the dark
matter candidate rises, opening up additional annihilation mechanisms
through the $SU(2)$ interactions and the $\lambda_R$ coupling.  Thus when
$\lambda_R$ is large t-channel exchange of the BSM scalars may play a role
through $\lambda_R$, leading to the curvature of
the red line, which is most pronounced in the right panel.  Therefore,
while in the left panel avoiding an over abundance of dark matter gives
a simple bound of \update{$\lambda_L < -1.5$}, in the middle and right
panel larger values of $|\lambda_L|$ are required to deplete the relic
density to the observed value or below, and the precise limit also
depends on $\lambda_R$. In all panels \update{$\damuNEW$} 
can be explained simultaneously with the relic density
if $\lambda_R$ has the rather small values \update{$\lambda_R\in[0.1,0.7]$}.

\begin{figure}[t]
  \centering
  \includegraphics[width=0.31\textwidth]{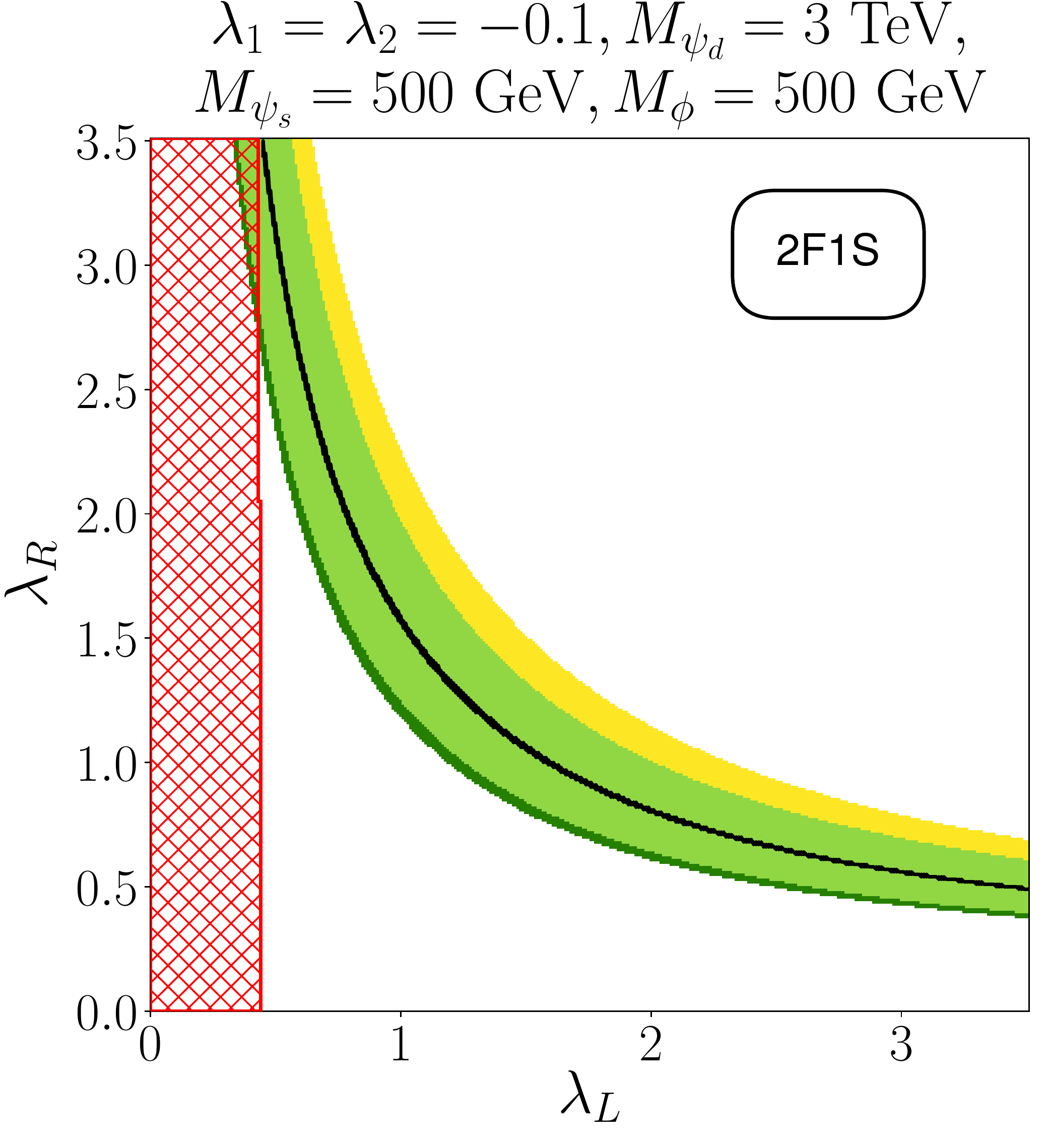}
  \hspace{0.01\linewidth}
  \includegraphics[width=0.31\textwidth]{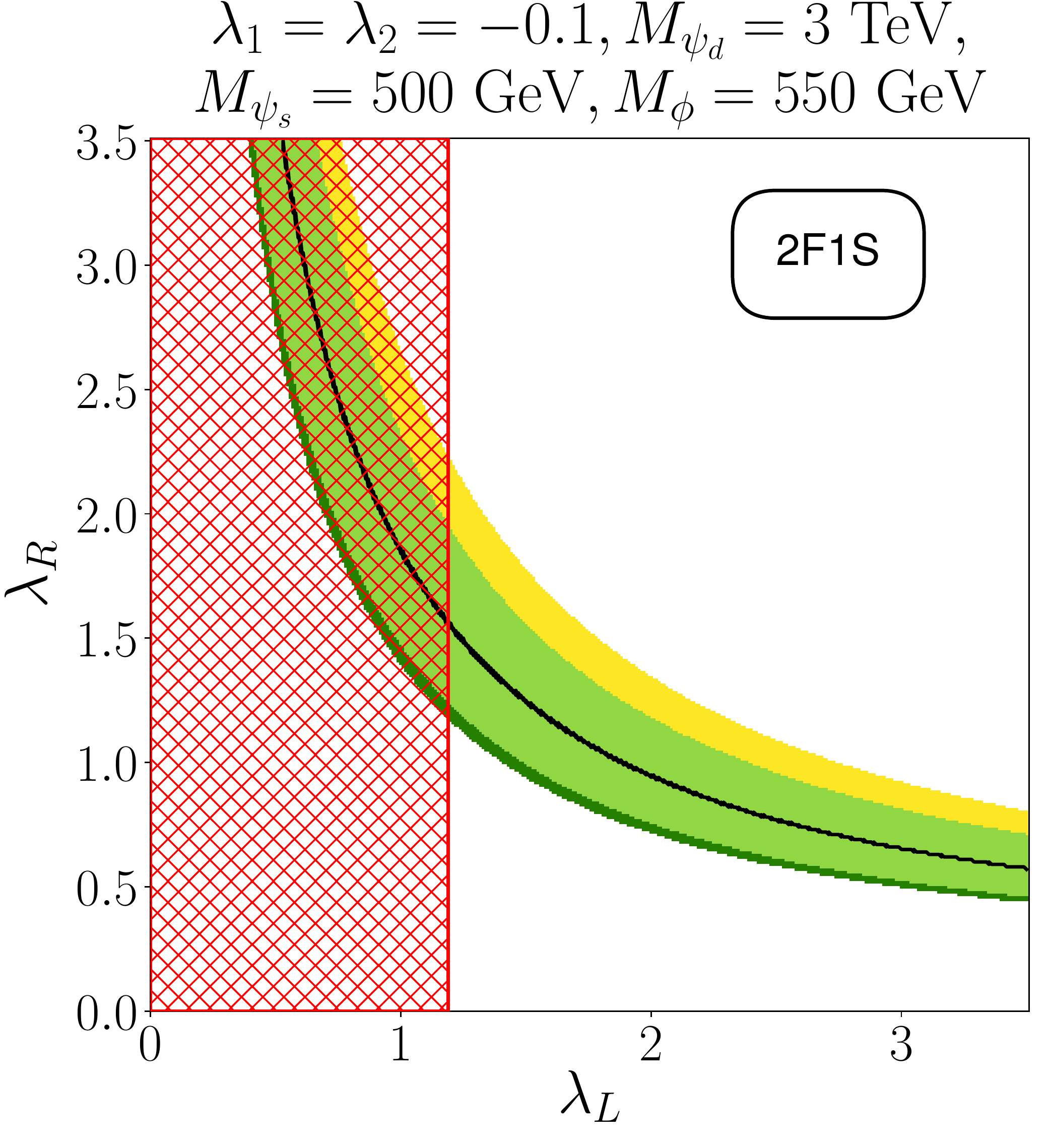}
  \hspace{0.01\linewidth}
  \includegraphics[width=0.31\textwidth]{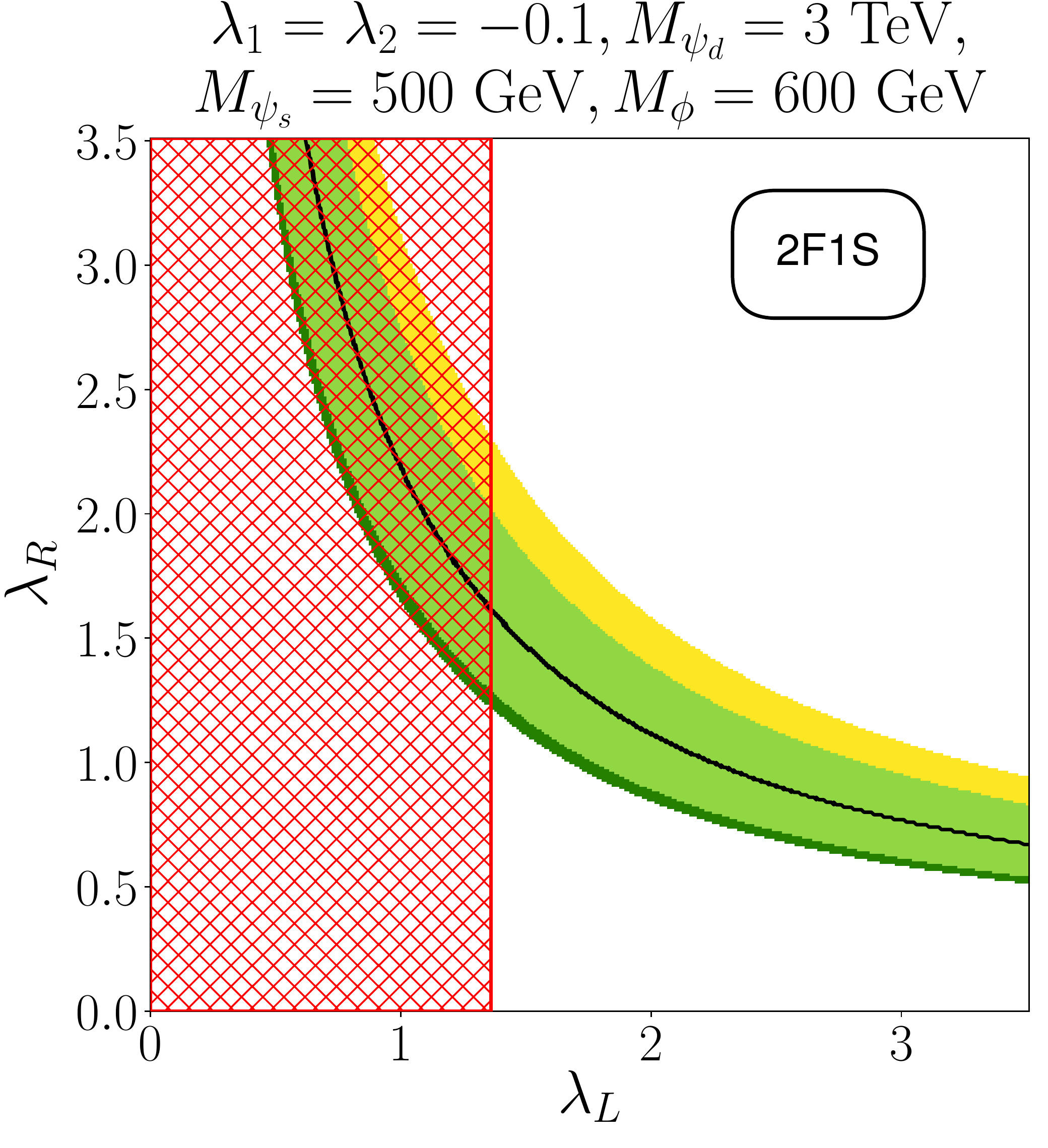}
\caption{ Results from Model 2F1S, scanning over the couplings of the
  new scalars and fermion to the left- and right-handed muon.  Colours
  are the same as in
  Fig.\ \ref{fig:Min3FieldsSLRFRCouplingsEqualMasses}.  The mixing
  couplings are fixed to $\lambda_{1}=\lambda_{2}=-0.1$.  In this
  scenario the fermion doublet has its mass fixed at $M_{\psi_d}=3000$
  GeV, the fermion singlet is fixed at $M_{\psi_s}=500$ GeV, and the
  scalar doublet has its mass slowly increased from $M_{\psi_s}$.
  Points below or to the left of the red line produce an over abundance
  of dark matter and are strongly
  excluded. \label{fig:Min3FieldsFLRSRCouplingsLargerDoubletSinglet}}
	\end{figure}
        
The mass splitting between the singlet fermion $\psi_s$ and the scalar
doublet $\phi$ has a large impact on the relic density.  In
Fig.\ \ref{fig:Min3FieldsFLRSRCouplingsLargerDoubletSinglet} we show
this impact by varying this mass splitting over three
$\lambda_L$--$\lambda_R$ planes, specifically we fix $M_{\psi_s} =
500$ GeV and set $M_\phi = 500$ GeV, $550$ GeV and $600$ GeV in the
left, middle and right panels respectively. At the same time we keep
the fermion doublet mass fixed at a heavy scale $M_{\psi_d} = 3000$
GeV, well above that of the dark matter candidate, so that in this
case it will not play any role in the depletion of dark matter relic
density\footnote{The exact choice for the mass is not important and,
  for this reason, the plots of
  Fig.\ \ref{fig:Min3FieldsFLRSRCouplingsLargerDoubletSinglet}
  actually allow understanding of the behaviour of the model in a
  wider range of doublet masses.}.  This choice also suppresses the $\damu$
prediction and means that very large couplings are required to explain
$\damu^\text{BNL}$ \update{and $\damuNEW$}. We also
choose $\lambda_{1,2}=-0.1$ and  $\lambda_L>0$. The relative
sign change between $\lambda_1$ and $\lambda_2$ leads to destructive
interference between different terms and thus a further slight suppression of
the overall contributions to $\damu$.
  
In all three panels of
Fig.\ \ref{fig:Min3FieldsFLRSRCouplingsLargerDoubletSinglet} the relic
density depends only on $\lambda_L$ due to the decoupling of the
doublet state, while $\damu$ depends on both $\lambda_L$ and
$\lambda_R$ since the chirality flip enhancement is proportional to
$\lambda_L\lambda_R$.  Just as before, \update{in all cases we can explain
the discrepancy in $\amu$ and provide a dark matter candidate
particle simultaneously}.  Between the three panels the $\damu$ result only
changes a little, as expected due to the small mass increases. However
the dark matter relic density depends strongly on the value of the
mass splitting. In the left panel where $M_\phi = M_{\psi_s}$, most of
the parameter space has an under abundance of dark matter. The very
small mass splitting between the dark matter fermion $\psi_1^0$ and
the BSM scalar opens up coannihilation channels which were suppressed
in Fig.\ \ref{fig:Min3FieldsFLRSRCouplingsSplitMasses}.  Due to these
highly efficient coannihilations the relic density is depleted for
much smaller values of $\lambda_L$ than in the previous figures. When
$M_\phi$ is increased to $550$ GeV in the middle panel, the
(co)annihilations through $\lambda_L$ become less efficient. As a
result a much larger $\lambda_L\approx 1.1$ is required to deplete the
relic density to the observed value via t-channel processes as in the
case of Fig.\ \ref{fig:Min3FieldsFLRSRCouplingsSplitMasses}.
Increasing the scalar mass further to $600$ GeV (right panel) an even
larger $\lambda_L\approx 1.4$ is required to produce a relic abundance
that matches the Planck observation.  However while the bound from an
over abundant relic density changed a lot between $500$ GeV and $550$
GeV, as we move away from the optimal window for coannihilations,
subsequently increasing $M_\phi$ above $600$ GeV has a much weaker
impact.  If we increase the scalar mass further the relic density limit gradually
increases until somewhere between $2.3$--$2.4$ TeV it is no longer
possible to deplete the relic density to the observed value or below
with perturbative couplings.
	
\begin{figure}[t]
  \centering
  \includegraphics[width=0.32\textwidth]{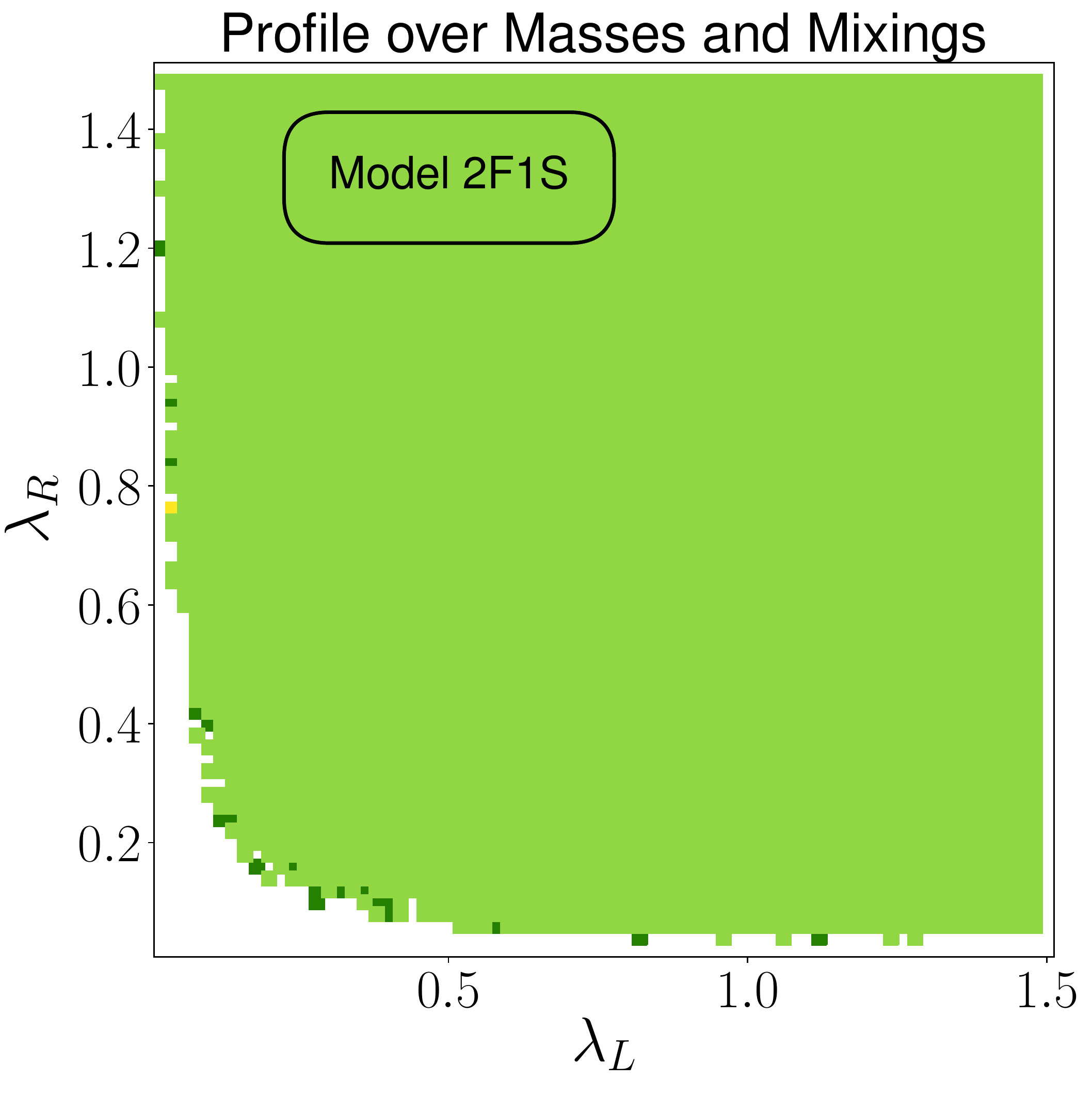}
  \includegraphics[width=0.32\textwidth]{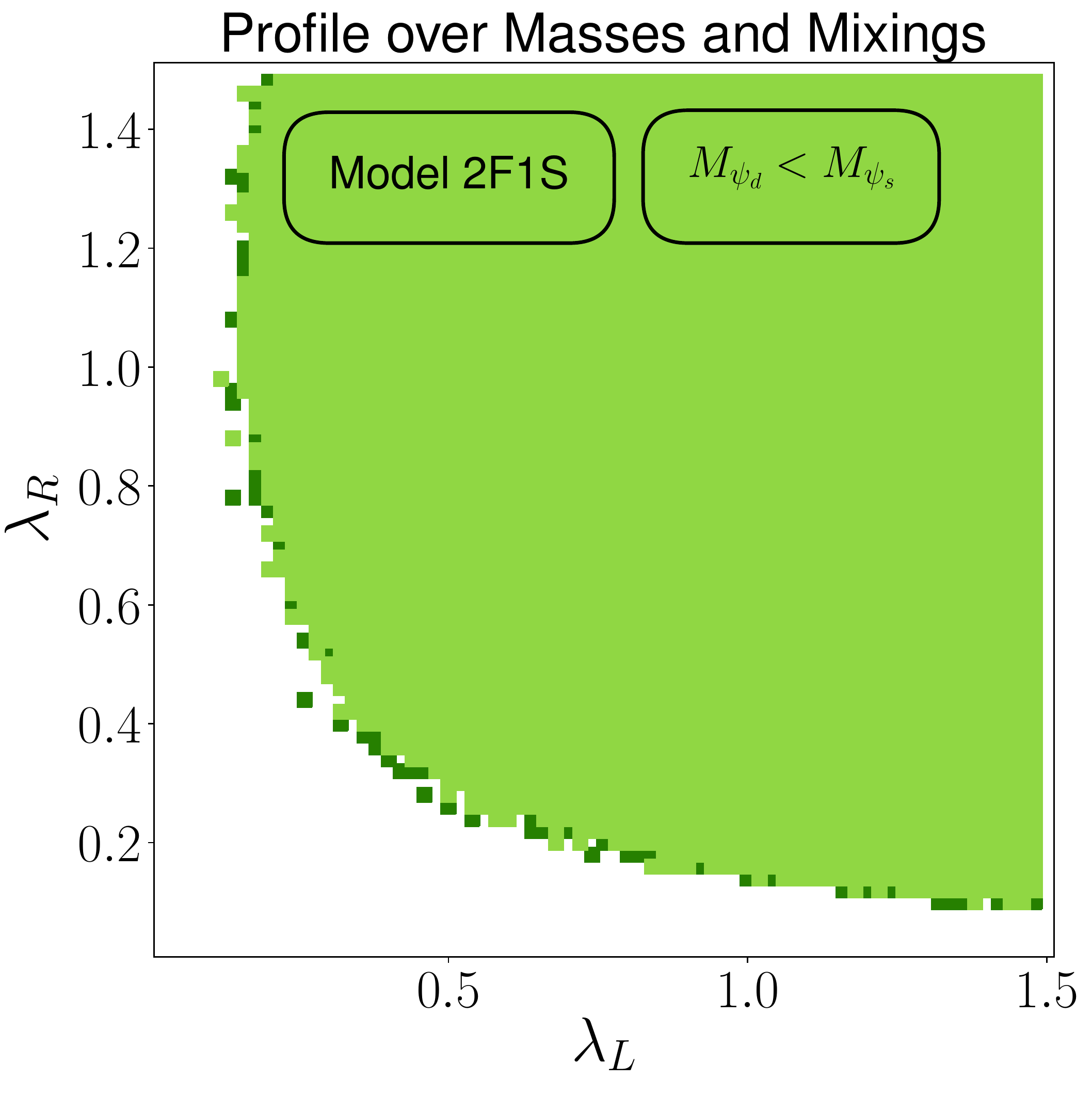}	
  \caption{Viable points in the $\lambda_L-\lambda_R$ plane obtained from scanning over the masses of the new scalar and fermions between $0$ and $5000$ GeV, as well as the Yukawa couplings $\lambda_{L,R} \in [0,1.5]$ and $\lambda_{1,2}\in [0,3.5]$ for Model 2F1S.  Colours are the same as in Fig.\ \ref{fig:Min3FieldsSLRFRCouplingsEqualMasses}.  Scan targeted the observed value of the dark matter relic density $0.1200$, Eq.\ (\ref{eqn:DMRD}), the $\amu$ value \update{in Eq. (\ref{eqn:BNLDiscrepancy})}, and the direct detection log likelihood provided by DDCalc \cite{Workgroup:2017lvb}, where points which are excluded by direct detection limits or are more than $3\sigma$ away from the Planck observation are thrown away.  
  \label{fig:Min3FieldsFLRSRprofile}}
\end{figure}

In Sec.\ \ref{sec:2S1F} the fact that a $600$ GeV scalar dark matter
candidate naturally depletes the relic density to the observed value
played a critical role in leading to a lower bound on
$|\lambda_L\lambda_R|$ from the combination of $\damu$ and dark matter
constraints.  Since a pure fermion doublet naturally depletes the relic
density to the observed value when it has a mass of about $1.1$ TeV, we
may anticipate a similar result in this model. In addition we have also
seen that when the doublet fermion is very heavy, large
couplings are required, while mixed singlet-doublet fermion dark
matter scenarios should be strongly constrained by direct detection.
Therefore we now perform a scan over the full parameter space of the
Model 2F1S to see if this also leads to a lower bound on $|\lambda_L \lambda_R|$.

We sample all free parameters in this model ($\lambda_{1},
\lambda_{2}, \lambda_{L}$, $\lambda_{R}$, $M_\phi$, $M_{\psi_d}$, and
$M_{\psi_s}$) using \multinest \cite{Feroz:2007kg,Feroz:2008xx,Feroz:2013hea,Buchner2014}, with mass range $M_i
\in [0,5000]$ GeV, and the BSM Yukawa couplings having the values
$\lambda_{L,R} \in [0,1.5]$ and $\lambda_{1,2} \in [0,3.5]$, with the
results shown in Fig.\ \ref{fig:Min3FieldsFLRSRprofile}.  Again, several values of 
$\damu$ where targeted in the range $[0,42]\times10^{-10}$, with a total of \update{62 
million} points scanned.  In the left panel of the figure, we can see that there is a 
lower bound on the couplings \update{$|\lambda_L \lambda_R| \gtrsim 0.036$} below which 
we cannot simultaneously explain the measured value of $\amu$ whilst
producing a dark matter candidate particle with the observed relic
density that escapes direct detection limits. As we anticipated in the
discussion above, pure singlet fermion dark matter requires larger
couplings to explain large $\damu$, while fermion doublet dark
matter annihilates too effectively for light masses.

Significant singlet-doublet fermion mixing can dilute the $SU(2)$
gauge interactions of the doublet, but does not avoid this problem as
it means that t-channel exchange of the BSM scalars will be active
through both $\lambda_L$ and $\lambda_R$, and increasing the mixing
also increases direct detection.  Reducing $\lambda_L \lambda_R$ while
increasing $\lambda_i$ to keep the $\damu$ prediction fixed makes
direct detection limits more relevant. As a result the combination of
these constraints leads to the lower limit $|\lambda_L\lambda_R|$
shown in the plot. Interestingly we also find a more \update{severe}
limit on \update{$|\lambda_L \lambda_R|\approx0.14$} for dark matter
that is mostly doublet in nature, as shown in the right panel of
Fig.\ \ref{fig:Min3FieldsFLRSRprofile}.

As in the previous case this shows how the $\amu$ measurements have
have a significant impact on the model, implying a siginifcant
constraint on the couplings.  \update{However since the new world
  average is quite close to the BNL value there is little difference
  between the limits implied by the $\damuBNL$ and $\damuNEW$
  results}.  The masses in
our samples are mostly heavy enough to easily evade collider
limits\footnote{The lightest mass scenarios in our samples {\it are}
  relevant for the lower bound on the couplings we obtained and our
  samples include some points where the lightest BSM fermion is
  between $300$--$500$ GeV.  In principle therefore LHC data could
  imply further constraints on our surviving samples.  However we
  tested all such points from the scan targetting the
  $\amu^{\text{BNL}}$ with \smodels 1.2.3 and did not find any further
  exclusion. Therefore we expect our limit from just dark matter and
  Muon g-2 to be robust. }.  However collider limits should be
included in a full global fit of the model, but we leave that and a
presentation of the full impact of all constraints on all parameters
and masses to future work.  Again we stress that our results
demonstrate that the \update{FNAL measurement plays a
  critical role} in the phenomenology of these models and should be
included in all phenomenological studies of models of this type and
future global fits.

As discussed earlier this model can be compared to the BHL scenario of
the MSSM.  There since $\lambda_L\lambda_R\lambda_{i}$ corresponds to
$g_1^2 y_\mu$, $\amu$ can only be explained with masses less than
$200$ GeV, due to the weaker enhancement \cite{Endo:2017zrj}.  Similarly the
$|\lambda_L\lambda_R|$ bound corresponds to $g_1 y_\mu$, which should
be well below the bound for this model and would suggest that the BHL
scenario alone cannot explain both the measured $\amu$ and dark
matter.  However it is important to note that the comparison is not
perfect due to the fact that the MSSM contains two Higgs doublets,
which gives an additional enhancement from the ratio of the two Higgs
VEVs, $\tan \beta$.  As a result in the BHL scenario, a dominantly
bino dark matter candidate can simultaneously explain dark matter and
account for $\damu^{\text{BNL}}$ and $\damuNEW$ despite having 
couplings below the lower bound for this in the Model 2F1S.  On the
other hand, the Model 2F1S has much greater freedom to explain dark
matter and large $\damu$ at much higher masses than is possible in the
BHL scenario. The fact that the $\lambda_R\neq y_\mu$ in this model is of great
significance, but on top of that the freedom to push all couplings up
to values much larger than that of $g_1$ and to vary the interactions
independently, gives rise to a rich and distinct phenomenology.  

Therefore our results show that this class of simple models provide an
interesting way to explain both dark matter and large $\damu$.  They
are complementary to the explanations that come from the MSSM and
appear to be significantly less constrained.  However these models
have not been constructed from any principle other than to explain
the phenomenology, and as a result the deeper motivation is unclear.  It
would be very interesting to consider whether models that are
motivated from more fundamental principles can have such scenarios as a
low energy effective field theory. 

%% file: SectionSUSY.tex
\newcommand{\amuWHL}{\Delta a_{\mu}^{1{\rm L}}(\text{WHL})} 
\newcommand{\amuWHnu}{\Delta a_{\mu}^{1{\rm L}}(\text{WHL(cha)})} 
\newcommand{\amuWHmuL}{\Delta a_{\mu}^{1{\rm L}}(\text{WHL(neu)})} 
\newcommand{\amuBHmuL}{\Delta a_{\mu}^{1{\rm L}}(\text{BHL})}
\newcommand{\amuBHmuR}{\Delta a_{\mu}^{1{\rm L}}(\text{BHR})}
\newcommand{\amuBmuLmuR}{\Delta a_{\mu}^{1{\rm L}}(\text{BLR})} 
\newcommand{\txl}{\text{L}}
\newcommand{\txr}{\text{R}}
\newcommand{\txw}{\text{W}}

\section{Supersymmetry and the Minimal Supersymmetric Standard Model} 
\label{sec:MSSM}
In the present section we consider the Minimal Supersymmetric Standard
Model (MSSM). It is one of
the most promising extensions of the SM and offers potential
explanations of EW naturalness and of dark matter as the lightest
supersymmetric particle (LSP), it is compatible
with unification of gauge couplings --- and it could easily explain
the deviation $\damu$ \update{based on the initial BNL
  measurement. However a tension has developed because of LHC
and dark matter searches, which severely constrain the MSSM parameter
space. } \update{The tension remains in place
  in view of the new FNAL result (\ref{Eq:FNALresult}) if the MSSM is
  required to accommodate the deviation $\damuNEW$.}

Here we present an update and investigate the parameter space of the MSSM which can
accommodate \update{this deviation, given the new FNAL measurement
Eq.\  (\ref{Eq:FNALresult}) and the SM theory update 
(\ref{amuSMWP}).} We focus particularly on the  
combined constraints from LHC 
and dark matter direct detection (DMDD) searches.

The MSSM contributions have been extensively studied in the
past. For analyses and reviews of the basic behaviour we refer to
Refs.\ \cite{Moroi:1995yh,Martin:2001st,Stockinger:2006zn,Cho:2011rk};
higher-precision calculations were done in Refs.\
\cite{Arhrib:2001xx,Chen:2001kn,Heinemeyer:2003dq,Heinemeyer:2004yq,Feng:2006ei,Marchetti:2008hw,vonWeitershausen:2010zr,Fargnoli:2013zda,Fargnoli:2013zia}.
For recent studies
in the light of LHC run-II data we refer to
Refs.\ \cite{Hagiwara:2017lse,Choudhury:2017acn,Endo:2020mqz,Chakraborti:2020vjp,Horigome:2021qof}
(which involve a detailed recasting of LHC data) and
\cite{Yin:2016shg,Zhu:2016ncq,Hussain:2017fbp,Altin:2017sxx,Choudhury:2017fuu,Endo:2017zrj,Yanagida:2017dao,Frank:2017ohg,Bagnaschi:2017tru,Wang:2018vxp,Chakraborti:2017dpu,Ning:2017dng,Li:2017fbg,Pozzo:2018anw,Tran:2018kxv,Yang:2018guw,Wang:2018vrr,Cox:2018qyi,Cox:2018vsv,Bhattacharyya:2018inr,Ibe:2019jbx,Abdughani:2019wai,Dong:2019iaf,Kotlarski:2019muo,Yanagida:2020jzy,Yang:2020bmh,Han:2020exx,Cao:2019evo,Yamaguchi:2016oqz,Yanagida:2018eho,Shimizu:2015ara,Yin:2016pkz}
(which involve model building, the construction of specific scenarios
and/or constraints from dark matter or electroweak precision observables). Earlier
studies of the SUSY phenomenology of $\amu$ of the LHC run 1 era
can be found in
Refs.\ \cite{Cho:2011rk,Endo:2011gy,Endo:2011xq,Endo:2011mc,Kim:2012fc,Ibe:2012qu,Zhang:2013hva,Bhattacharyya:2013xma,Akula:2013ioa,Evans:2013uza,Ibe:2013oha,Endo:2013lva,Endo:2013bba,Iwamoto:2014ywa,Kersten:2014xaa,Badziak:2014kea,Gogoladze:2014cha,Bach:2015doa,Ajaib:2015yma,Kowalska:2015zja,Khalil:2015wua,Gogoladze:2015jua,Baez:2015sqj,Harigaya:2015jba,Harigaya:2015kfa,Chakrabortty:2015ika,Padley:2015uma,Chowdhury:2015rja,Li:2016ucz,Okada:2016wlm,Belyaev:2016oxy,Kobakhidze:2016mdx,Belanger:2017vpq,Fukuyama:2016mqb}.
Refs.\ \cite{Badziak:2019gaf,Endo:2019bcj} study simultaneous SUSY
explanations of $(g-2)$ of both the muon and the electron.\footnote{%
\label{aefootnote}
  There used to be a slight disagreement between the experimental number for
  the electron $g-2$ \cite{Hanneke:2008tm} and the SM theory evaluation
  \cite{Aoyama:2012wj,Aoyama:2017uqe} based on the
   measurement of the fine-structure constant of
   Ref.\ \cite{Parker:2018vye}.
   However, a more
   recent measurement of the 
   fine-structure constant of Ref.\ \cite{Morel:2020dww} leads to good agreement
   between theory and experiment for the electron $g-2$. Here we do
   not consider the electron $g-2$ further.} We will provide more detailed comparisons to the
literature in the following subsections.

The MSSM
contributions to $\amu$ are mainly generated
by left- and right-handed smuons ${\tilde{\mu}}_{\txl,\txr}$, 
sneutrino ${\tilde{\nu}}_{\mu\text{L}}$, 
Bino ${\tilde{B}}$, Winos ${\tilde{W}}_{1,2,3}$, and 
Higgsinos ${\tilde{H}}_{u,d}$ 
which are the SUSY partners of the muon, muon-neutrino, and the
electroweak SM gauge and Higgs bosons. 
${\tilde{B}}$, ${\tilde{W}}_{1,2,3}$ and ${\tilde{H}}_{u,d}$ 
form neutralinos $\chi ^0 _{1,2,3,4}$ and charginos $\chi ^\pm _{1,2}$ mass eigenstates. 
They appear in the one-loop
diagrams of  Fig.\ \ref{fig:FFSDiagram} (with charginos and
sneutrinos) and of Fig.\ \ref{fig:SSFDiagram} (with neutralinos and
smuons). 

\begin{figure}[t]
  \includegraphics[scale=1.]{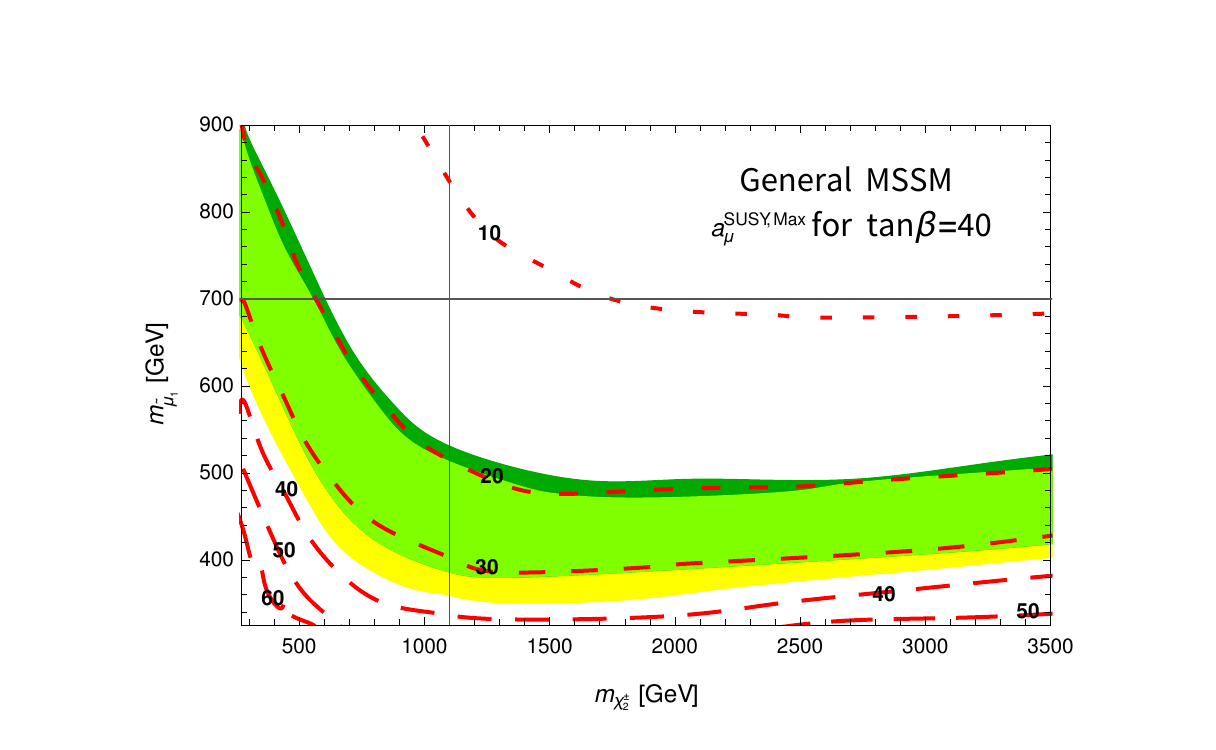}
  \includegraphics[scale=1.]{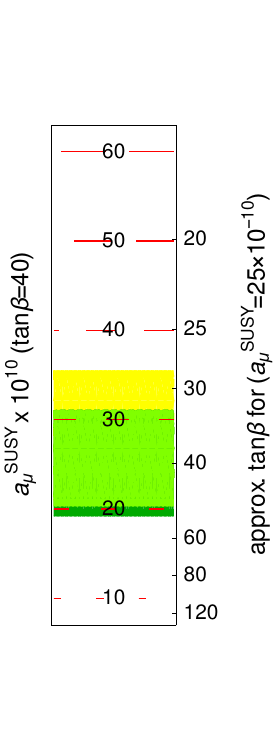}
  \caption{\label{fig:briefsurveyplot}
    The theoretical maximum MSSM contribution $\amu^{\text{SUSY,Max}}$ for
    $\tan\beta=40$ in
    the plane of the heaviest chargino and the lightest smuon
    mass. (For each point in the plane, the actual value of the MSSM
    contribution can take any value between $0$ and $\pm
    \amu^{\text{SUSY,Max}}$, depending on the signs of parameters and
    details such as other masses and mixings.) The yellow/green
    coloured regions show where
    $\amu^{\text{SUSY,Max}}$ (for $\tan\beta=40$) is within the $1\sigma$ bands corresponding
    to the BNL and new deviations $\damu^{\textrm{BNL}}$ and
    $\damuNEW$, see Eqs.\
    (\ref{eqn:BNLDiscrepancy},\ref{eqn:avgDiscrepancy}), and their
  overlap.
    The red dashed contour lines can be interpreted in two
    ways. Firstly, they directly correspond  to certain values of
    $\amu^{\text{SUSY,Max}}$ for $\tan\beta=40$, as indicated in the left
    axis of the legend plot. Secondly, thanks to the approximate
  linearity in $\tan\beta$, each contour can be used to estimate the
  required $\tan\beta$ value for which $\amu^{\text{SUSY,Max}}$ agrees
  with the deviation  $\damuNEW$ (keeping other input
  parameters fixed). These $\tan\beta$ values can be
  read off from the right axis of the legend plot (the values are
  approximate since the linearity is not exact).
      As an example of the reinterpretation we take the point  $m_{\chi^\pm_2} = 1750$ GeV and
    $m_{\tilde{\mu}_1}= 700$ GeV.  For $\tan\beta =40$ we get
      $\amu^{\text{SUSY,Max}} =10 \times 10^{-10}$. The required
      $\tan\beta$ value to get $\damuNEW$ \update{would be around
        100}, as read off from the right axis. 
    The
      results for $\amu^{\text{SUSY,Max}}$ were obtained from a scan using
      GM2Calc \cite{Athron:2015rva} in which all relevant SUSY masses
      are varied independently between $100$\,GeV and 4\,TeV.
      The black lines
    indicate the maximum LHC reach for charginos and sleptons of
    $1100$ and $700$ GeV reported in
    Refs.\ \cite{Aaboud:2018jiw,Aad:2019vnb}, respectively.
  }
\end{figure}

To provide a brief overview and set the stage for the following
discussion we present Fig.~\ref{fig:briefsurveyplot} and the following
approximation of 
the leading MSSM contributions to $\amu$,
\begin{subequations}
\label{eq:SUSY1Lnumerical}
\begin{align}
\amuWHL
&\approx21\times10^{-10}\text{sign}(\mu M_2) \left(\frac{500\text{ GeV}}{M_{\text{SUSY}}}\right)^2\frac{\tan\beta}{40},
\label{eq:SUSYWHL}\\
\amuBHmuL &\approx1.2\times10^{-10}\text{sign}(\mu M_1)\left(\frac{500\text{ GeV}}{M_{\text{SUSY}}}\right)^2\frac{\tan\beta}{40},\\
\amuBHmuR &\approx-2.4\times10^{-10}\text{sign}(\mu M_1)\left(\frac{500\text{ GeV}}{M_{\text{SUSY}}}\right)^2\frac{\tan\beta}{40},\\
\amuBmuLmuR
&\approx2.4\times10^{-10}\text{sign}(\mu M_1)\left(\frac{500\text{ GeV}}{M_{\text{SUSY}}}\right)^2\frac{\tan\beta}{40}\frac{\mu}{500\text{ GeV}}. \label{eq:SUSYBLR}
\end{align} 
\end{subequations}
The letters B,W,H,L, and R are the abbreviations for Bino, Winos, Higgsinos, and Left-and Right-handed 
smuons respectively. 
As indicated by the letters in brackets, each contribution depends on three of these states and their 
respective masses.
In each formula the three appropriate masses have been set to a common
scale $M_{\text{SUSY}}$ and $M_{\text{SUSY}}\gg M_Z$ is assumed. The approximations
highlight the linearity in $\tan\beta=v_u/v_d$, the ratio of the two
MSSM Higgs VEVs.

The physics of the individual contributions is as follows. In the
first three lines the muon chirality is flipped at the Yukawa coupling
to a Higgsino; the Higgsino is then converted to a gaugino via a Higgs
vacuum expectation value. The appearance of the enhanced VEV is always
accompanied by a factor of the Higgsino mass $\mu$. This explains the
appearance of one Higgsino, one gaugino (Bino or Wino) and of the factors $\tan\beta$
and $\mu$ in the numerators. 

The BLR
contribution in the fourth line is special.
Here the muon chirality is flipped at the smuon line via the insertion
of a smuon-left-right flip, which is governed by a product of the
Higgsino mass $\mu$ and the enhanced VEV in the smuon mixing
matrix. This explains the appearance of left- and right-handed smuons
and of the factors $\tan\beta$ and $\mu$ in the numerator. 
In contrast to the other contributions, the BLR contribution is
approximately linearly enhanced by the Higgsino mass $\mu$ since this
parameter does not appear as the mass of a virtual Higgsino. The signs
of all contributions are 
determined by the signs of $\mu$ and the gaugino masses
$M_{1,2}$. In the following we will only consider positive signs of
all these parameters, leading to positive MSSM contributions to $\amu$.

Figure
\ref{fig:briefsurveyplot} shows the theoretical maximum
of the SUSY   contributions (i.e.\ MSSM minus SM contributions)  $\amu^{\text{SUSY,Max}}$
in the plane of $m _{\chi ^\pm _2}$ and $m_{{\tilde{\mu}} _1}$, where $\chi ^\pm _2$ is the heavier 
chargino and $\tilde{\mu} _1$ the lighter smuon (for similar plots in
different planes see
Refs.\ \cite{Byrne:2002cw,Stockinger:2006zn,Badziak:2014kea}).
It fixes $\tan\beta=40$ and allows all SUSY masses to vary
independently between $100$\,GeV and 
4\,TeV, for further details see the caption.\footnote{%
  Widening the range of masses would not change the plot since the
  maximum $\amu^{\text{SUSY}}$ is obtained in the bulk of the mass range, not at
  its boundary. This is a reflection of the fact that the leading
  contributions approximated in Eq.\ (\ref{eq:SUSY1Lnumerical}) are
  suppressed if Bino, Wino or Higgsino masses are too small.}
The red dashed lines and the yellow/green coloured regions correspond
to the indicated values of $\amu^{\text{SUSY,Max}}$; specifically the
yellow and green colours correspond to the $1\sigma$ regions for the
BNL deviation and the new  deviation including FNAL in
Eqs.\ (\ref{eqn:BNLDiscrepancy},\ref{eqn:avgDiscrepancy}),
respectively;  the bright green colour corresponds to the overlap
region. As indicated on the right of the legend plot, an alternative
interpretation of the contour lines and regions is possible. Thanks to
the approximate linearity in $\tan\beta$, each value for $\amu^{\text{SUSY}}$ with
$\tan\beta=40$ can be translated into approximate $\tan\beta$-values
for which the other values of $\amu^{\text{SUSY}}$ can be obtained. In the legend
plot this translation is done for $\damuNEW=\update{25.1\times10^{-10}}$,
the new experimental average deviation from the SM. E.g.\ on the contour
labeled \update{``20''} we have $\amu^{\text{SUSY,Max}}=20\times10^{-10}$
for $\tan\beta=40$, and equivalently we would get
$\amu^{\text{SUSY,Max}}=\damuNEW$
for $\tan\beta\approx\update{50}$. On the contour
labeled \update{``50''}, we have $\amu^{\text{SUSY,Max}}=\update{50}\times10^{-10}$
for $\tan\beta=40$, and the world average deviation can be
explained for $\tan\beta\approx\update{20}$.

The figure and the formulas show that large contributions in the
ballpark of the BNL deviation are of
course possible but require upper limits on the SUSY masses. E.g.\
$\amu^{\text{SUSY}}>20\times10^{-10}$  \update{(which is just at the
  lower $1\sigma$ level of the deviation)} can 
only be explained (for $\tan\beta=40$ and $\mu\le4$\,TeV) if 
\begin{itemize}
\item either both chargino masses are lighter than around 1.1\,TeV 
(vertical black line in Fig.~\ref{fig:briefsurveyplot})
\item
  or one smuon is lighter than around 700\,GeV (horizontal black line  in Fig.~\ref{fig:briefsurveyplot}).
\end{itemize}
This \update{already illustrates the potential
  tension} between $\amu$ and LHC data since the
LHC-searches are sensitive to chargino and smuon masses above these
values.  \update{In addition to LHC, constraints from dark matter searches enforce
lower limits on combinations of the chargino and neutralino masses.}
In more detail, the WHL contributions (\ref{eq:SUSYWHL}) are by far
most important in the largest part of parameter space.
However for large $\mu$, the BLR contribution can become sizable as
well. In Fig.\ \ref{fig:briefsurveyplot} the slight rise of
$\amu^{\text{SUSY,Max}}$ with increasing $m_{\chi^\pm_2}$ is due to the
BLR contribution. In contrast, the BHR and
particularly the BHL 
contributions are not important for the maximum value
$\amu^{\text{SUSY,Max}}$, and in general they are practically always strongly suppressed, unless there
are extremely large mass splittings between left- and right-handed
smuons. We do not consider such cases in the present paper; hence in
the following only the WHL and BLR contributions are important.\footnote{%
  Specifically the BHR contribution can be decisive if
  $\tan\beta\gg50$ and if $m_\txl\gg m_\txr$ \cite{Bach:2015doa}. For further
  dedicated investigations focusing on parameter situations in which
  the BHL or BHR contributions dominate we refer to
  Ref.\ \cite{Endo:2017zrj}.\label{footnoteBHR}}

In the following subsections we will first introduce the SUSY
contributions to $\amu$ in more detail (Sec.~\ref{sec:SUSYcontributions}), survey relevant
LHC and dark matter constraints (Sec.~\ref{sec:SUSYconstraints}) and present a detailed
phenomenological analysis and brief summary
(Secs.~\ref{sec:SUSYphenosetup},~\ref{sec:SUSYpheno} and Sec.~\ref{sec:SUSYSummary}).

\subsection{SUSY parameters and contributions to $\amu$}
\label{sec:SUSYcontributions}

In the following we define our notation and provide details on the
MSSM contributions to $\amu$. Our notation is essentially the same as
e.g.\ in Refs.\ \cite{Martin:2001st,Stockinger:2006zn,Cho:2011rk}. To
fix it we provide the mass matrices of  the charginos and
neutralinos and the smuons:
\begin{align}
X &= 
\left(\begin{array}{cc}
M_2 & M_W\sqrt2 \sin\beta\\
 M_W \sqrt2 \cos\beta & \mu
      \end{array}
\right)
,
\\
Y &=
\left(\begin{array}{cccc}
M_1 & 0   & -M_Z s_\txw \cos\beta & M_Z s_\txw \sin\beta \\
0   & M_2 & M_Z c_\txw \cos\beta & -M_Z c_\txw \sin\beta \\
-M_Z s_\txw \cos\beta &M_Z c_\txw \cos\beta & 0 & -\mu \\
M_Z s_\txw \sin\beta & -M_Z c_\txw \sin\beta& -\mu & 0
      \end{array}
\right),
\\
M_{\tilde{\mu}}^2 &= \left(\begin{array}{lr}
m_\mu^2 + m_\txl ^2 + M_Z ^2 \cos2\beta\left(-\frac12+s_\txw ^2\right)
&
m_\mu (-\mu\tan\beta+A_\mu^*)
\\
m_\mu (-\mu^*\tan\beta+A_\mu)
&
m_\mu^2 + m_{\txr}^2 + M_Z ^2\cos2\beta\ (-s_\txw ^2)
			\end{array}
\right),
\label{smuonmassmatrix}
\end{align}
where the fundamental SUSY parameters appearing in these expressions are $\tan\beta$ and
the two gaugino (Bino and Wino) mass parameters $M_{1,2}$, the
Higgsino mass $\mu$ and the
left-/right-handed smuon masses $m_{\txl, \txr}$. The smuon mass matrix also involves the trilinear
soft SUSY-breaking 
parameter $A_\mu$, which however will not play any role in the present
paper. The other appearing parameters are
the SM parameters $m_\mu,M_{W,Z}$ and $s_\txw =\sqrt{1-c_\txw ^2}$.
In our numerical treatment the fundamental SUSY parameters are defined
as running $\overline{\text{DR}}$-parameters at the scale 1 TeV, and
the
\code{MSSMEFTHiggs_mAmu} spectrum generator, created with\footnote{%
  \fs is a generic spectrum generator generator, and the
  FlexibleEFTHiggs extension \cite{Athron:2016fuq,Athron:2017fvs,Kwasnitza:2020wli} improves the Higgs mass
  calculation by resummation of large logarithms. The version used
  within \gambit\ is the one of Ref.\ \cite{Athron:2017fvs}. \fs also uses some numerical routines originally from
  \cite{Allanach:2001kg,Allanach:2013kza} and uses \sarah 4.14.1
  \cite{Staub:2009bi,Staub:2010jh,Staub:2012pb,Staub:2013tta}.}
{\fs} \cite{Athron:2014yba,Athron:2017fvs} and incorporated in \gambit-1.3,
 is used for
the precise evaluation of the spectrum of mass eigenvalues including
higher-order corrections.

The 1-loop contributions of the MSSM to $\amu$ have been
systematically and comprehensively studied in Ref.\ \cite{Moroi:1995yh}, for
reviews see
Refs.\ \cite{Martin:2001st,Stockinger:2006zn,Cho:2011rk}. A wide range
of higher-precision
calculations of 2-loop contributions is
available. Including higher-order corrections, the full known
SUSY contributions (i.e.\ the difference between MSSM
and SM contributions) to $\amu$ can 
be written as
\newcommand{\amuSUOL}{a_\mu^{\rm 1L\, SUSY}}
\newcommand{\amuFSf}{a_\mu^{{\rm 2L,} f\tilde{f}}}
\newcommand{\amuTLa}{a_\mu^{\rm 2L(a)}}
\newcommand{\amuphotonic}{a_\mu^{\rm 2L,\ photonic}}
\begin{align}
\amu^{\text{SUSY}} &= \left[\amuSUOL  +\amuTLa
+
\amuphotonic 
 +\amuFSf\right]_{t_\beta\text{-resummed}}
\ ,
\label{amuSUSYdecomposition}
\end{align}
Refs.\ \cite{Arhrib:2001xx,Chen:2001kn,Heinemeyer:2003dq,Heinemeyer:2004yq,Cheung:2009fc}
evaluated all 2-loop diagrams $\amuTLa$ in which a SUSY loop is inserted into a
SM-like loop, including so-called Barr-Zee
diagrams. Given current experimental constraints these diagrams are
very small. Refs.\ \cite{Degrassi:1998es,vonWeitershausen:2010zr}
computed leading QED-logarithms and the full 2-loop QED corrections $\amuphotonic$;
Refs.\ \cite{Marchetti:2008hw,Bach:2015doa} showed how to take into
account $n$-loop higher-order terms enhanced by $(\tan\beta)^n$, i.e.\
carry out a $\tan\beta$-resummation.
Finally Refs.\ \cite{Fargnoli:2013zda,Fargnoli:2013zia} computed
genuine SUSY 2-loop corrections $\amuFSf$ to the SUSY 1-loop diagrams which include
non-decoupling effects from e.g.\ heavy squarks. Each of
these three kinds of corrections can shift the 1-loop contributions by
around 10\%. All mentioned 1-loop and 2-loop contributions in Eq.\ (\ref{amuSUSYdecomposition})
are
implemented in the code GM2Calc \cite{Athron:2015rva}, which is used
in
our later phenomenological evaluations.\footnote{%
  For higher-order calculations in extensions of the MSSM see
  Refs.\ \cite{Su:2020lrv,Liu:2020nsm,Dong:2019iaf,Zhao:2014dxa
  }.} 

The exact one-loop expression for $\amu^{\text{1L SUSY}}$ can be
found in most mentioned references; a full overview of all
contributions including higher orders is given in
Ref.\ \cite{Athron:2015rva}. 
Here we provide the 1-loop
contributions in mass-insertion 
approximation, which allows to directly read off the main parameter
dependences. Following the form given e.g.\ in
Refs.\ \cite{Cho:2011rk,Fargnoli:2013zia,Bach:2015doa} they 
read
\begin{align}
\amu^{\text{1L SUSY}} &\approx \amuWHL + \amuBHmuL \nonumber + \amuBHmuR +
\amuBmuLmuR, 
\end{align}
with
\begin{subequations}
\label{eq:SUSYMIapprox}
\begin{align}
\amuWHL &= \frac{g_{2}^{2}}{8 \pi ^{2}} \frac{m_{\mu} ^{2} M_{2}}{m_\txl ^{4}}\,\mu \tan\beta\, 
F_{a}\left(\frac{M_{2}^{2}}{m_\txl ^{2}},\frac{\mu^{2}}{m_\txl ^{2}}\right) \nonumber\\* 
&- \frac{g_{2}^{2}}{16 \pi ^{2}} \frac{m_{\mu}^{2} M_{2}}{m_\txl ^{4}}\,\mu \tan\beta\,
F_{b}\left(\frac{M_{2}^{2}}{m_\txl ^{2}},\frac{\mu^{2}}{m_\txl ^{2}}\right),\\*
\amuBHmuL &= \frac{g_{1}^{2}}{16 \pi ^{2}}\frac{m_{\mu}^{2} M_{1}}{m_\txl ^{4}}\,\mu \tan\beta\,
F_{b}\left(\frac{M_{1}^{2}}{m_\txl ^{2}},\frac{\mu^{2}}{m_\txl ^{2}}\right), \\* 
\amuBHmuR &= - \frac{g_{1}^{2}}{8 \pi ^{2}} \frac{m_{\mu}^{2} M_{1}}{m_\txr ^{4}}\,\mu \tan\beta\, 
F_{b}\left(\frac{M_{1}^{2}}{m_\txr ^{2}},\frac{\mu^{2}}{m_\txr ^{2}}\right), \\* 
\amuBmuLmuR &= \frac{g_{1}^{2}}{8 \pi ^{2}} \frac{m_{\mu}^{2}}{M_{1}^{3}}\,\mu \tan\beta\, 
F_{b}\left(\frac{m_\txl ^{2}}{M_{1}^{2}},\frac{m_\txr ^{2}}{M_{1}^{2}}\right).
\end{align} 
\end{subequations}
For SUSY masses significantly above $M_Z$ this is a very good
approximation (of the 1-loop contributions). Physics explanations and
numerical examples have 
already been given around Eq.\ (\ref{eq:SUSY1Lnumerical}).  Next to
$\tan\beta$ and the $SU(2)\times U(1)$ gauge couplings $g_{1,2}$,
these contributions depend on the five independent SUSY mass
parameters $M_{1,2}$, $\mu$ and $m_{\txl, \txr}$ introduced above.
The appearing loop functions are normalized as $F_a(1,1)=1/4$,
$F_b(1,1)=1/12$ and can be found in the mentioned references as well
as in Appendix \ref{app:MuonGm2Contributions}.

The linear enhancement in $\tan\beta$ already explained in
Sec.\ \ref{sec:BSMoverview} is apparent.\footnote{%
  The leading higher-order effects from QED-logarithms and from $n$-loop
  $(\tan\beta)^n$-effects can be approximately taken into account by
  multiplying the formulas by
  \cite{Degrassi:1998es,vonWeitershausen:2010zr,Marchetti:2008hw,Bach:2015doa}
  $\left(1-\frac{4\alpha}{\pi}\log\frac{M_{\text{SUSY}}}{m_\mu}\right)/(1+\Delta_\mu)$,
  where $\Delta_\mu$ is a correction to the muon Yukawa coupling and
  $M_{\text{SUSY}}$ the appropriate SUSY mass scale.}  The $\tan\beta$
enhancement is accompanied by explicit factors of the Majorana gaugino
masses $M_{1,2}$ and the MSSM Higgsino mass parameter $\mu$.\footnote{As a side
remark, in SUSY models with continuous R-symmetry such as the MRSSM,
such Majorana gaugino masses and the $\mu$-parameter are zero. Hence $\amu^{\text{SUSY}}$ is not $\tan\beta$-enhanced,
leading to distinctly different $\amu$ phenomenology
\cite{Kotlarski:2019muo}.
\label{footnoteMRSSM}}

As indicated, all contributions in Eq.~(\ref{eq:SUSYMIapprox}) involve
three different SUSY masses; the generic behaviour is
$\propto1/M_{\text{SUSY}}^2$.  The BLR-contribution in
Eq.\ (\ref{eq:SUSYMIapprox}) is 
special because it is linearly enhanced by large $\mu$;
this enhancement arises via the smuon mixing off-diagonal element in
Eq.\ (\ref{smuonmassmatrix}).

Specific constraints on this BLR-contribution have
been very thoroughly investigated in Ref.\ \cite{Endo:2013lva}. 
Most importantly, vacuum stability requires that staus, the
superpartners of $\tau$-leptons, do not receive
a charge-breaking vacuum expectation value, and this provides a
constraint on the relation between the off-diagonal and diagonal
elements of the stau mass matrix similar to
Eq.\ (\ref{smuonmassmatrix}). As a quantitative example,
Ref.\ \cite{Endo:2013lva} finds that in case of universal left- and
right-handed stau masses, the Higgsino mass has an upper limit,
specifically 
\begin{align}\label{VacStabEndo}
m_{\tilde{\tau}_\txl}=m_{\tilde{\tau}_\txr}&=300\text{
  GeV (}600\text{ GeV)}
&\Rightarrow&&
\text{ }\mu&\lesssim1\text{
  TeV (}2\text{ TeV)}\,.
\end{align}
In our later plots we will show the appropriate
constraint in the approximate form of Eq.\ (14) of
Ref.\ \cite{Endo:2013lva}.

\subsection{LHC and dark matter constraints on explanations of $\damuNEW$}
\label{sec:SUSYconstraints}

Our aim is to analyze MSSM contributions to $\amu$ in the context of
constraints from LHC and dark matter searches. Here we list the
relevant constraints.

The relevant LHC constraints can be grouped into ``standard'' searches
for electroweak particles (charginos/neutralinos and sleptons) and
searches optimized for compressed spectra. The relevant searches are
the following:
\begin{itemize}
\item
  Chargino/neutralino searches with decay into sleptons:
The strongest chargino/neutralino mass limits are obtained from the
pair production channel $pp\to{\chi} ^\pm _1 {\chi} ^0
_2$ with subsequent decay via on-shell sleptons into three charged
leptons and two LSPs. In simplified-model
interpretations, in which 100\%\ decay branching ratios are assumed
and the slepton mass is halfway between the LSP- and the chargino
mass, the limits extend up to \cite{Aaboud:2018jiw,Sirunyan:2017lae}
\begin{subequations}\label{Charginosleptonlimits}
  \begin{align}
    \text{Chargino/slepton channel:}\quad
\parbox{0cm}{\mbox{$        \chi_1^\pm\chi_2^0\to\tilde{l}_\txl \nu\tilde{l}_\txl l(\tilde{\nu}\nu),l\tilde{\nu}\tilde{l}_\txl 
    l(\tilde{\nu}\nu) \to l\nu\chi_1^0,
    ll(\nu\nu)\chi_1^0$}}
    &&
    &&
\\\label{maximumChamasslimit}
    &&  m_{\chi^\pm_1}&\approx1100\text{ GeV}
  &&(\text{for
    $m_{\text{LSP}}\approx0\ldots500\text{ GeV}$}),\\
&&  m_{\text{LSP}} &\approx700\text{ GeV}
&& (\text{for
    $m_{\chi^\pm_1}\approx900\ldots1000 \text{ GeV}$}).
  \end{align}
\end{subequations}
Further important, slightly weaker limits are
obtained from the pair production channel $pp \rightarrow {\chi}
^\pm _1 {\chi} ^\mp _1$ with subsequent decay via on-shell sleptons into
two charged leptons and two LSPs. The limits in simplified-model
interpretations reach up to $m_{\chi^\pm_1}\approx 1000$ GeV
\cite{Aad:2019vnb}.
All limits of this kind depend on a significant mass splitting
$\gtrsim100$ GeV between
the chargino and LSP masses; for smaller mass splittings the limits
become much weaker.
\item
  Chargino/neutralino searches with decay into other particles:
The above limits are absent if the charginos cannot decay into
sleptons e.g.\ because sleptons are too heavy.
In such cases further limits
are applicable.
One such limit is obtained from the channel 
$pp\to{\chi}^\pm _1 {\chi} ^0 _2$ assuming subsequent decays into on-shell $W$ and Higgs
bosons plus LSP. This  limit extends up to \cite{Aad:2019vvf}
\begin{subequations}\label{eq:ChaWhlimits}
\begin{align}
\text{Chargino/$Wh$-channel:}\quad
\parbox{0cm}{\mbox{$        \chi_1^\pm\chi_2^0\to Wh\chi_1^0\chi_1^0,
    W\to l\nu, h\to b\bar b$}}
    &&
    &&
\\&&  m_{\chi^\pm_1}&\approx 750\text{ GeV}
&& (\text{for
    $m_{\text{LSP}}\approx0\ldots100\text{ GeV}$}),\\
&&  m_{\text{LSP}} &\approx250\text{ GeV}
&&(\text{for
    $m_{\chi^\pm_1}\approx600 \text{ GeV}$}),
\end{align}
\end{subequations}
where Wino-like charginos and Bino-like LSP are assumed. If this
assumption is not met, the production cross section is lower and/or
the decay branching ratios are reduced \cite{Canepa:2020ntc}.
Similar but
slightly weaker limits are obtained in Ref.\ \cite{Sirunyan:2018ubx}.
A complementary limit is obtained in Ref.\ \cite{Aaboud:2017nhr} which
searched for 
$pp\to{\chi}^\pm _1{\chi}^\mp _1,{\chi}^\pm _1 {\chi} ^0 _2$ with
subsequent decays via $\tilde{\tau}$-sleptons (staus) into $\tau$-leptons. The limit
in a simplified-model interpretation assuming Wino-like chargino
reaches up to
\begin{subequations}\label{eq:Chastaulimits}
\begin{align}
  \text{Chargino/stau-channel:}\quad
\parbox{0cm}{\mbox{$\chi_1^\pm\chi_2^0\to\tilde{\tau}\nu\tilde{\tau}\tau(\tilde{\nu}\nu),
  \tau\tilde{\nu}\tilde{\tau}\tau(\tilde{\nu}\nu)
  \to\tau\nu\chi_1^0,  \tau\tau(\nu\nu)\chi_1^0,$}}    &&
    &&
\\
\parbox{0cm}{\mbox{$
  \chi_1^\pm\chi_1^\mp\to2\times \tilde{\tau}\nu(\tilde{\nu}\tau)\to 2\times\tau\nu\chi_1^0$}}
    &&
    &&
\\
&&  m_{\chi^\pm_1}&\approx 760\text{ GeV}
&& (\text{for
    $m_{\text{LSP}}\approx0\ldots200\text{ GeV}$}),\\
&&  m_{\text{LSP}} &\approx300\text{ GeV}
&&(\text{for
    $m_{\chi^\pm_1}\approx600\ldots700 \text{ GeV}$}).
\end{align}
\end{subequations}
Figure 8 of Ref.\ \cite{Aaboud:2017nhr} and the recasting study of
Ref.\ \cite{Hagiwara:2017lse} show that the limit is rather robust against
changes of the stau masses and mixings and against the Higgsino content
of the chargino.
There exist further chargino searches with decays into $W$ and $Z$
bosons
\cite{Aaboud:2018jiw,Sirunyan:2017qaj,Aad:2019vnb,Aaboud:2018sua,Aad:2019vvi},
however the 
resulting limits are weaker and do not lead to excluded regions of the
parameter spaces we will consider.
\item
  Slepton searches: Searches for the direct production of
  slepton pairs
  $\tilde{l}\tilde{l},(\tilde{l}=\tilde{e},\,\tilde{\mu})$ with
  subsequent decay into leptons plus LSP have been analyzed in
  Ref.\ \cite{Aad:2019vnb} (based on $139\text{\,fb}^{-1}$ data) and
  \cite{Aaboud:2018jiw,Sirunyan:2018nwe} (based on
  $36\text{\,fb}^{-1}$ data). The limits extend up to
  \begin{subequations}\label{eq:sleptonsearch}
    \begin{align}
    \label{eq:sleptonsearch2019}
\text{Slepton \cite{Aad:2019vnb}}:\quad
\parbox{0cm}{\mbox{$\tilde{l}_{\txl, \txr}\tilde{l}_{\txl, \txr}\to l^+ l^-\chi_1^0\chi_1^0,$}}    &&
    &&
\\&&  m_{\tilde{l}}&\approx 700\text{ GeV}
&& (\text{for
    $m_{\text{LSP}}\approx0\ldots300\text{ GeV}$}),\\
&&  m_{\text{LSP}} &\approx400\text{ GeV}
&&(\text{for
    $m_{\chi^\pm_1}\approx550\ldots650 \text{ GeV}$}),\\
  \text{Slepton \cite{Aaboud:2018jiw,Sirunyan:2018nwe}}:&& m_{\tilde{l}}&\approx 500\text{ GeV}
&& (\text{for
    $m_{\text{LSP}}\approx0\ldots300\text{ GeV}$}),\\
  &&  m_{\text{LSP}} &\approx300\text{ GeV}
  &&(\text{for
    $m_{\chi^\pm_1}\approx500 \text{ GeV}$}).
    \end{align}
  \end{subequations}
\item
  Searches for SUSY particles with compressed-mass spectra:
  The compressed mass spectrum scenarios are investigated through the
  chargino-neutralino pair production modes ${\chi} ^\pm _1
  {\chi} ^0 _2/{\chi}^\pm_1{\chi}^\mp_1$ with decays via virtual
  $W$/$Z$ bosons and slepton pair production
  $\tilde{l}\tilde{l}$ with decays into leptons. In simplified-model
  analyses the limits depend on the nature of the
  charginos/neutralinos (Higgsino- or Wino-like) and reach up to
  masses of around 250 GeV and mass splittings to the LSP between
  $1\ldots50$ GeV~\cite{Aad:2019qnd,Sirunyan:2019zfq,Sirunyan:2018iwl}. 
\end{itemize}
Our technical setup for checking against these constraints is as
follows. The LHC searches with the highest mass reach, i.e.\ the
chargino/neutralino  searches
of Refs.\ \cite{Aaboud:2018jiw,Sirunyan:2017lae,Sirunyan:2017qaj}, and the slepton
searches of Ref.\ \cite{Aaboud:2018jiw}, and the compressed-mass
searches of Ref.\ \cite{Sirunyan:2018iwl}
are checked using \colliderbit \cite{Balazs:2017moi}, a recasting tool
  within the \gambit-1.3 
  software framework
  \cite{Athron:2017ard,Workgroup:2017bkh,Balazs:2017moi,Workgroup:2017lvb,Workgroup:2017htr,Workgroup:2017myk}. 
  This framework was extended and applied to the chargino/neutralino sector already in
  Ref.\ \cite{Athron:2018vxy}, where also a full description
  of all included analyses can be found. For
  each signal region (SR) of each analysis, \gambit/\colliderbit
  evaluates the theory prediction of the signal yield
  $S_{\text{SR}}$. It then constructs the log-likelihood differences
  $\ln{\cal L}_{\text{SR}}\equiv\ln{\cal
    L}(n|s=S_{\text{SR}},b)-\ln{\cal L}(n|s=0,b) $  for
  the computed signal yield  and the observed
  event number $n$ and background expectation $b$ reported by the
  experiments. It also determines for each analysis which signal
  region SR$^{\text{max}}$ has the highest expected sensitivity. For each given analysis
  \gambit/\colliderbit outputs a single ``effective'' log-likelihood
  difference
    $\ln{\cal L}_{\text{eff}}\equiv\ln{\cal L}_{\text{SR$^{\text{max}}$}}$.
    We refer to
    Refs.\   \cite{Athron:2017ard,Workgroup:2017bkh,Balazs:2017moi,Workgroup:2017lvb,Workgroup:2017htr,Workgroup:2017myk}
    for a detailed description of the procedure and cross-checks with
    original LHC analyses.
    To obtain a conservative LHC exclusion contour in the following
    plots of this section we proceed as follows. For each parameter
    point we take the largest effective $(-2\ln{\cal L}_{\text{eff}})$ of
    any implemented analysis and employ $(-2\ln{\cal L}_{\text{eff}})^{\text{Max}}\ge6$ as the criterion for
    exclusion by the LHC recasting analysis.

    In App.\ \ref{app:SUSYLHC}
    we provide further extensive details on the behaviour of the recasting
    for the parameter regions and  the LHC analyses most relevant
    for our discussion. We discuss both cases with high sensitivity and
    low sensitivity.
    In particular we validate the recasting procedure
    by reproducing the exclusion contour of
    the ATLAS chargino/neutralino search of Ref.\ \cite{Aaboud:2018jiw},
    Fig.\ 8c.  We further verify that the exclusion contour obtained as
    described above via $(-2\ln{\cal L}_{\text{eff}})^{\text{Max}}$ is also fully consistent with the contour where
    the predicted signal yield for at least one signal region is equal
    to the respective ATLAS 95\% C.L.\ upper limit. We checked that
    the same would be true for all 
    following plots of the paper.\footnote{%
      In exceptional cases the comparison is not possible. In
      particular, the CMS analysis of Ref.\ \cite{Sirunyan:2018iwl} for
      compressed spectra gives the maximum contribution $(-2\ln{\cal
    L}_{\text{eff}})^{\text{Max}}$ in some parameter regions; for this analysis no individual
      95\% C.L.\ upper signal limits are available.
    }

As we will see the remaining mentioned LHC searches affect only a
minor portion of the relevant SUSY parameter space. Hence we do not
implement them in
\gambit/\colliderbit but take
them into account by directly 
using the 
simplified-model interpretations of the original ATLAS/CMS
references. This is a very conservative approach which likely slightly
overestimates the true LHC constraints on the MSSM, 
but it is motivated by the desire to find parameter regions which are
definitely viable. 
Specifically, we apply the following constraints.
Figs.\ 14, 16 of Ref.\ \cite{Aad:2019qnd} are applied for 
Wino- or Higgsino-like charginos or sleptons as appropriate.
Fig.\ 6 of  Ref.\ \cite{Aad:2019vvf} based on the $Wh$-channel
chargino/neutralino search is applied to the lightest chargino in case
of the hierarchy $M_1<M_2<\mu$ and if the chargino is lighter than
sleptons and staus. Fig.\ 7b of Ref.\ \cite{Aaboud:2017nhr} based
on the stau-channel search is applied to the lightest chargino if its
decay into stau is dominant and to the heavier chargino in case
$\mu>M_2$ if the decay into stau is kinematically possible.
As a further check we also directly apply the strong constraints of Figs.\ 7b, 7c of
Ref.\ \cite{Aad:2019vnb} based on  direct slepton searches (see
Eq.\ (\ref{eq:sleptonsearch2019})) and chargino searches with decays
to sleptons; however as we will see they have no impact on our
parameter spaces.

The following constraints from dark matter physics are relevant:
\begin{itemize}
\item
  Dark Matter Relic Density (DMRD): We assume the LSP to be the
  lightest neutralino, and unless noted otherwise we assume the LSP to
  be stable. In this case the LSP contributes to the dark matter relic
  density, and we require that the LSP relic density is in agreement with
  or smaller than the observed dark matter relic density. We have to
  distinguish the cases of dominantly Bino- or Wino- or Higgsino-like
  LSP. In the case of Wino- or Higgsino-like LSP and LSP-masses below 1
  TeV, the relic density is always smaller than the observed value,
  thus leading to no constraints for our analysis (but to the
  requirement of additional, non-MSSM components of dark matter, see
  e.g.\ Ref.\ \cite{Bramante:2015una,Roszkowski:2017nbc} for  recent
  reviews). Ref.\ \cite{Chakraborti:2017dpu} has shown that
  coannihilation effects can increase the Higgsino- or Wino-like relic density
  if the mass splitting between sleptons and the LSP is significantly
  below $10\%$. However the extent is not sufficient to be of interest
  in the parameter regions of interest for $\amu$, where we consider
  LSP masses below around 500 GeV.

  In the case of a Bino-like LSP in the considered mass range, the
  relic density is typically too large unless a specific mechanism
  acts to enhance the dark matter annihilation and to suppress the
  relic density. In the mass range of Bino masses of around
  200\ldots600 GeV there are three possibilities: stau-coannihilation,
  other slepton-coannihilation, and Wino-coannihilation
  \cite{Roszkowski:2017nbc,Ellis:1998kh,Ellis:1999mm,Nihei:2002sc,Harigaya:2014dwa}.  
  As we will see, in each of our scenarios with Bino-like LSP, one of
  these possibilities can be realized without further impact on
  LHC-exclusion of parameters or on values of $\amu$. Hence in
  summary we do not need to explicitly apply DMRD-constraints on our analysis of $\amu$ in
  the MSSM.
\item
  Dark Matter Direct Detection (DMDD): If the LSP is stable there are
  constraints from the non-observation of dark matter in direct
  detection experiments. The most
  stringent constraints are obtained from the XENON1T experiment
  \cite{Aprile:2017iyp}; similar but weaker limits are obtained from
  XENON100, PandaX-II and
  LUX \cite{1207.5988,Tan:2016zwf,1608.07648}. We evaluate these
  constraints using \darkbit and \DDCalc
  \cite{Workgroup:2017lvb,Athron:2018hpc}\footnote{%
    The actual calculations can be done using internal code as well as
    interfaces to the public codes DarkSUSY
    \cite{Bringmann:2018lay} and  \micromegas \cite{Belanger2018}. For
    the calculations presented here we choose internal code and the interface to
    DarkSUSY.}
  within the \gambit
  software framework \cite{Athron:2017ard,Workgroup:2017bkh,Balazs:2017moi,Workgroup:2017lvb,Workgroup:2017htr,Workgroup:2017myk}, using the provided log-likelihood
  functions as described in Ref.\ \cite{Workgroup:2017lvb}. Since the
  XENON1T limits are the strongest, we will only use those in our
  phenomenological analysis and consider a parameter point excluded at
  the $90\%$ confidence level if $2\ln{\cal L}(\sigma=0)-2\ln{\cal L}(\sigma,m_{\text{LSP}})>1.64$ for the XENON1T analysis.
  The required
  calculations depend on the dark matter relic density. In the case of a
  Bino-like LSP, which allows an explanation of the observed value, we
  set the relic density to the observed value. In the case of Higgsino- or
  Wino-like LSP, in which case the relic density is smaller than the
  observed one, we use the relic density computed by \darkbit.

  It is well known that the phenomenological impact of these
  constraints is that strong gaugino--Higgsino mixing of the LSP is
  not viable, except in ``blind spots'' which are characterized by
  particular ratios $\mu/m_{\text{LSP}}$, require negative $\mu$ and depend on $\tan\beta$ and
  the CP-odd Higgs boson mass $M_A$\ \cite{Huang:2014xua}. As we will see, in case of a
  Bino-like LSP, lower limits on the Higgsino mass, and in case of a Higgsino-like LSP, lower limits on the
  Wino mass $M_2$ are implied. In contrast, the limits obtained in
  case of a Wino-like LSP turn out to be weaker.\footnote{%
Technically, we determined these limits from dark matter direct
detection in a separate computation using the \gambit/\darkbit
framework. The limits may be of interest in their own right.
In the parameter space relevant for our later plots with
$\tan\beta=40$
we
obtained the following simple functional forms of the limits, correct
to within $2\%$:
\begin{align*}
  \text{Bino-like LSP:}&&\mu&> 467\text{ GeV}+0.157M_1(1+M_1/159\text{ GeV})\\
  \text{Wino-like LSP:}&&\mu&> 34.2\text{ GeV} +1.46 M_2\\
  \text{Higgsino-like LSP:}&&M_2&>207\text{ GeV} +1.83\mu
\end{align*}
\label{footnotedarkmatterapprox}  }
\end{itemize}

\subsection{Setup of the phenomenological analysis}
\label{sec:SUSYphenosetup}

Our phenomenological analysis focuses on a wide parameter space of the
MSSM (without flavour mixing in the sfermion sector). Our setup is as
follows (all input parameters are defined as 
$\overline{\text{DR}}$-parameters at the scale $1$\,TeV):
\begin{itemize}
\item
  The SUSY 1-loop contributions to $\amu$ depend on five mass parameters
  $M_1$, $M_2$, $\mu$, $m_\txl$, $m_\txr$. We treat them as independent,
  except for setting the two slepton masses equal, $m_\txl=m_\txr$. Allowing
  different smuon masses would neither lead to enhanced $\amu^{\text{SUSY}}$ nor
  significantly alter LHC and dark matter limits unless the mass
  splittings are extreme (see footnote \ref{footnoteBHR}).
  Since the influence of the trilinear scalar $A$-parameters is very
  small, we set $A_f=0$ for all sfermion flavours.
\item
  In all plots we fix
  \begin{align}
    \tan\beta&=40
  \end{align}
  as a reference value. Since $\amu^{\text{SUSY}}$ is essentially linear in
  $\tan\beta$ it is easily possible to re-interpret plots for
  other values of $\tan\beta$, as already explained in the caption of
  Fig.\ \ref{fig:briefsurveyplot}.
\item
  We assume R-parity conservation, such that the lightest SUSY
  particle (LSP) is stable and forms (a component of) dark matter. We
  assume the LSP to be the lightest
  neutralino. 
\item
  The selectron masses are set equal to the smuon masses, and
  generally we assume absence of flavour mixing in the slepton
  sector.\footnote{%
    Possible slepton flavour mixing between staus and smuons can lead
    to additional contributions to $\amu$ via enhanced chirality
    flips of the kind
    $\tilde{\mu}_\txl\to\tilde{\tau}_\txl\to\tilde{\tau}_\txr\to\tilde{\mu}_\txr$. Such
    effects have been studied in
    Refs.\ \cite{Moroi:1995yh,Baez:2015sqj}; they enhance in
    particular the BLR-contributions and thus allow explanations of
    large $\amu$ with higher values of $M_1$, without conflict to
    bounds from the non-observation of $\tau\to\mu\gamma$.
    In contrast, slepton flavour mixing between selectrons and smuons
    does not enhance $\amu$, but it leads to specific
    correlations to flavour-violating decays like $\mu\to e\gamma$ and
    $\mu\to e$ conversion, which in turn show interesting differences
    between the MSSM
    \cite{Chacko:2001xd,Kersten:2014xaa,Calibbi:2015kja} and other
    models such as the MRSSM which does not allow a $\tan\beta$
    enhanced dipole contributions
    \cite{Kotlarski:2019muo}. \label{footnoteLFV}
  }
  This is
  taken into account in evaluating
  the LHC search limits.
  The precise values of stau masses and mixings
  are not important for any of the
  considered observables except for the dark matter relic density in
  case of a Bino-like LSP. In the case of a
    Bino-like LSP, the LSP-relic density is generally too high, and we
    assume coannihilation with either staus, other sleptons or Winos
    to suppress the relic density to an 
    acceptable value. Stau-coannihilation is generally possible as
    long as the Bino-mass is 
    below around 600\,GeV \cite{Ellis:1998kh,Ellis:1999mm}. It
    requires that one stau is sufficiently
    light but it does not fix the stau-masses and 
    mixings uniquely. Since no other considered observables depend on
    their precise values, we fix the stau mass parameters
    to
    $2M_1$ in cases where stau-coannihilation is relevant. Due to stau
    mixing one mass eigenvalue is then close to the LSP
    mass.
    In cases where stau-coannihilation is not relevant we fix the stau
    mass parameters to $2$ TeV, representing heavy staus. In the first
    case
    the chargino/neutralino
    limits based on Ref.\ \cite{Aaboud:2017nhr}, see
    Eq.\ (\ref{eq:Chastaulimits}), are relevant and taken into
    account.
\item
  Squark and gluino masses are set to $6$ TeV and the CP-odd
  Higgs mass is set to $M_A=2$ TeV. In this way, all
  respective LHC limits are evaded, and the mass of the SM-like Higgs
  boson is in the right ball-park. Such heavy squarks imply small,
  positive 2-loop corrections to $\amu^{\text{SUSY}}$ \cite{Fargnoli:2013zda,Fargnoli:2013zia}; we do not
  finetune the squark masses and mixings to fit the SM-like Higgs mass
  to agree with the measured values exactly, since there is no unique
  way to do it and since the impact on $\amu^{\text{SUSY}}$ and all other
  observables considered here is negligible.
\end{itemize}

Thus the essential parameters for our discussion are the four mass
parameters
\begin{align}
  M_1, M_2,\mu,m_{\txl, \txr},
\end{align}
where $M_1$ will be similar to the LSP-mass $m_{\text{LSP}}$ in case
the LSP is Bino-like; the two chargino masses are essentially given by
$M_2$ and $\mu$, respectively.

Despite having only four parameters, the parameter space is complex
and there are many possible organizing principles. We may distinguish
24 different mass orderings and compressed or non-compressed
spectra. We may classify according to the nature of the LSP and how
dark matter is generated and according to which contributions to
$\amu^{\text{SUSY}}$ (WHL and/or BLR in
Eqs.\ (\ref{eq:SUSY1Lnumerical},\ref{eq:SUSYMIapprox})) are dominant. 
We find that a very effective way to organize the discussion is to
divide the parameter space in three distinct kinds of scenarios,
denoting $m_{\tilde{l}}\equiv m_{\txl, \txr}$ as the generic 1st and 2nd
generation slepton mass, see also Tab.\ \ref{tab:SUSYscenarios}:
\begin{itemize}
  \item \underline{Scenario with heavy chargino and smuons:}
    \begin{align}
      m_{\tilde{l}}&>700 \text{\, GeV},&m_{\chi^\pm_{1,2}}>1100\text{\,GeV}\,.
    \end{align}
    These mass ranges correspond to the maximum reach of the LHC searches for sleptons and charginos found in Refs.\ \cite{Aaboud:2018jiw,Aad:2019vnb} and are indicated by black lines in Fig.\ \ref{fig:briefsurveyplot}.
      The scenario thus evades all LHC constraints in the simplest
      possible way and is denoted as ``LHC-unconstrained'' in
      Tab.\ \ref{tab:SUSYscenarios}. The following scenarios are
      defined to allow lighter SUSY particles. 
  \item
    \underline{Scenarios with lighter sleptons:} First we consider
    scenarios with slepton masses significantly below $700$ GeV, which
    is possible if the sleptons are close in mass with the LSP, i.e.\
    \begin{align}
      m_{\text{LSP}}\lessapprox m_{\tilde{l}}<700\text{\,GeV}\,,
    \end{align}
    where the $\lessapprox$ symbol here denotes ``lighter but not much
    lighter'' to evade LHC limits, i.e.\ within around $100$ GeV.\footnote{%
      More explicitly, the notation $      m_{\text{LSP}}\lessapprox
      m_{\tilde{l}}$ may be written as $      m_{\text{LSP}}<
      m_{\tilde{l}}< m_{\text{LSP}}+\Delta m_{\text{LHC}}$, where
      $\Delta m_{\text{LHC}}$ is the mass gap allowed by LHC
      searches. The value of this allowed mass gap depends on the
      details of the considered spectrum but is typically of order $100$ GeV.}
    There are
    three sub-scenarios depending on the nature of the LSP, which we denote as
    \begin{align}
      (\tilde{B}\tilde{l}),\ (\tilde{W}\tilde{l}),\ (\tilde{H}\tilde{l})
    \end{align}
    scenarios, respectively. In all these scenarios the chargino
    masses are obviously either heavier than the slepton masses or
    lighter but not by a significant amount.
  \item
    \underline{Scenarios with charginos lighter than sleptons:}
    Next we consider scenarios where both charginos are lighter than
    the sleptons:
    \begin{align}
            m_{\text{LSP}}<m_{\chi^\pm_{1,2}}<m_{\tilde{l}}\,.
    \end{align}
    Here weaker LHC limits on charginos apply, such that chargino masses below
    $700$ GeV are viable. We assume slepton masses to be
    non-compressed with the LSP, such that the LHC
    constraints on sleptons of
    Refs.\ \cite{Aad:2019vnb,Aaboud:2018jiw,Sirunyan:2018nwe} are
    relevant (although the 
    simplified-model interpretations illustrated in
    Eqs.\ (\ref{eq:sleptonsearch}) do not necessarily apply). Again we
    can distinguish several 
    sub-scenarios, depending on the nature of the LSP. We denote them
    as
    \begin{align}
      (\tilde{B}\tilde{W}\tilde{H}),\ (\tilde{H}\tilde{W}/\tilde{W}\tilde{H})
    \end{align}
    scenarios. In the $(\tilde{B}\tilde{W}\tilde{H})$ scenario, no
    particular mass-ordering between the two \update{chargino mass parameters} $M_2$ and
    $\mu$ is implied; the $(\tilde{H}\tilde{W}/\tilde{W}\tilde{H})$
    scenarios will be discussed together; the LSP is either Wino- or
    Higgsino-like and no particular value 
    of $M_1$ is implied except that the Bino is not the
    LSP.
\end{itemize}

\begin{table}
  \begin{center}
    \begin{tabular}{|l|l|l|l|l|}
      \hline
      Scenario & Hierarchy & LSP/DM & Dominant
      $\amu$ 
      \\\hline
      ``LHC-unconstrained'' & $m_{\tilde{l}}>700\text{\,GeV,
      }m_{\chi^\pm_{1,2}}>1100\text{\,GeV}$&Bino, Wino, Higgsino&WHL+BLR\\\hline
      $(\tilde{B}\tilde{l})$ & $M_1\lessapprox m_{\txl, \txr}\lesssim\mu,M_2$ &
      Bino/$\tilde{\tau}$- or $\tilde{l}$- or $\chi^\pm$-coann. &
      WHL+BLR\\\hline
      $(\tilde{W}\tilde{l})$ & $M_2\lessapprox m_{\txl, \txr}\lesssim\mu,M_1$ &
      Wino & WHL+BLR\\\hline
      $(\tilde{H}\tilde{l})$ & $\mu\lessapprox m_{\txl, \txr}\lesssim M_1,M_2$ &
      Higgsino & WHL\\\hline
      $(\tilde{B}\tilde{W}\tilde{H})$ & $M_1<M_2,\mu<m_{\txl, \txr}$
      & Bino/$\tilde{\tau}$- or $\chi^\pm$-coann.
      & WHL \\\hline
      $(\tilde{W}\tilde{H})$ & $M_2<\mu,M_1<m_{\txl, \txr}$ &
      Wino& WHL\\\hline
      $(\tilde{H}\tilde{W})$ & $\mu<M_2,M_1<m_{\txl, \txr}$ &
      Higgsino& WHL\\\hline
    \end{tabular}
  \end{center}
  \caption{\label{tab:SUSYscenarios}
Overview of the scenarios defined in section \ref{sec:SUSYphenosetup}
and analysed in Sec.\ \ref{sec:SUSYpheno}.  Here  the $\lessapprox$
symbol denotes ``lighter but not much 
    lighter'', i.e.\ sufficiently close to evade LHC
    limits, see main text; the $\lesssim$ symbol denotes ``lighter or slightly
    heavier''.
  In each case, the nature of the dark matter candidate is
  indicated. In cases of Bino-like LSP we assume the dark matter
  density to agree with observation, which in turn requires one of the
  indicated coannihilation mechanisms to be present. In the other
  cases we only assume that the predicted  dark matter density does not surpass
  the observed one.}
\end{table}

\subsection{Phenomenological analysis}
\label{sec:SUSYpheno}

We will now discuss each scenario of Table
\ref{tab:SUSYscenarios}  in turn. In each case we will
evaluate the possible values for $\amu^{\text{SUSY}}$ and the constraints from LHC and 
dark matter, and we will determine resulting viable parameter regions.
Before explaining our results we briefly discuss the status of related
studies in the
literature. The general, phenomenological MSSM was analyzed in view of the BNL result
$\damu^{\text{BNL}}$ versus LHC run-II in
Refs.\ \cite{Hagiwara:2017lse,Cox:2018qyi,Abdughani:2019wai,Endo:2020mqz,Chakraborti:2020vjp},
in Refs.\ \cite{Cox:2018qyi,Abdughani:2019wai,Chakraborti:2020vjp}
including dark matter.
All
these references require a  Bino-like LSP and consider parameter regions similar to our
$(\tilde{B}\tilde{l})$
scenario.\footnote{%
  While finalizing this paper,
  Ref.\ \cite{Chakraborti:2021kkr} appeared, which also contains
  results on the general MSSM, comparing $\amu$ to results from dark
  matter and LHC experiments in cases with Wino-like and Higgsino-like
  LSP. It uses the same approach as
  Ref.\ \cite{Chakraborti:2020vjp} and shows similar complementarities
  to our study.
}
Refs.\ \cite{Hagiwara:2017lse,Endo:2020mqz} also consider 
the general $(\tilde{B}\tilde{W}\tilde{H})$ scenario but do not consider dark matter. The LHC run-II data is
treated with an increasing level of detail, and slightly different
restrictions on the allowed masses are employed.
Ref.\ \cite{Chakraborti:2020vjp} uses LHC recasting
of a similar set of constraints as discussed in
Sec.\ \ref{sec:SUSYconstraints}, but with different recasting
tools. It assumes stau-masses equal to the other 
slepton masses, i.e.\ generation universality, but allows differences
between left- and right-handed sfermions $m_\txl\ne m_\txr$. This leads to slightly weaker
LHC limits on $M_2$. Ref.\ \cite{Hagiwara:2017lse} carries out a
recasting of chargino search channels via staus, as mentioned in
Sec.\ \ref{sec:SUSYconstraints}. Refs.\ \cite{Cox:2018qyi,Abdughani:2019wai,Endo:2020mqz} treat LHC-data
in different simplified ways without recasting. Ref.\ \cite{Abdughani:2019wai}
focuses on scans of two parameter regions (both within our
$(\tilde{B}\tilde{l})$ scenario), in which different states are decoupled.
Ref.\ \cite{Endo:2020mqz} focuses on the two cases $\mu= M_2$ and $\mu=2M_2$
but allows for arbitrary slepton masses.
The results of these references are that Bino-like LSP with
either chargino-coannihilation or slepton/stau-coannihilation can provide 
viable explanations of dark matter and $\damu^{\text{BNL}}$, and
Refs.\ \cite{Abdughani:2019wai,Chakraborti:2020vjp} specify upper 
limits on the LSP mass.

Our study aims to provide an up-to-date and coherent analysis of the
general MSSM in view of the 
Fermilab result for $\amu$, dark matter data and our LHC recasting. We will 
treat all scenarios of Table
\ref{tab:SUSYscenarios} including Bino-, Wino- or Higgsino-like LSP 
 and provide details on allowed and preferred patterns of SUSY masses.

\subsubsection{Scenario with heavy charginos and smuons, above all LHC
limits}

As a first basic
scenario we consider sleptons and charginos heavier than $700$ GeV and
$1100$ GeV, respectively, 
denoted as  ``LHC-unconstrained'' in Tab.\ \ref{tab:SUSYscenarios}.
In Fig.~\ref{fig:briefsurveyplot} this region corresponds to the upper
right quadrant of the plot, delineated by the black lines.
The choice of this region is motivated by
the maximum LHC reach for charginos and sleptons
found in Refs.\ \cite{Aaboud:2018jiw,Aad:2019vnb}. In other words, in the considered region the LHC limits are
trivially fulfilled. Of course, the maximum LHC constraints were obtained for the simple special case of a massless
Bino-like LSP $\chi ^0 _1$, intermediate sleptons, and heavier
charginos. The scenarios discussed later on will involve smaller masses and evade the LHC limits through
choices of specific mass patterns, hierarchies and mass splittings.

Here we will first discuss the behaviour of $\amu^{\text{SUSY}}$ in the upper right quadrant of 
Fig.~\ref{fig:briefsurveyplot}, i.e.\ for $m_{\tilde{\mu}_1}\ge700$ GeV,
$m_{\chi^\pm_{2}}\ge1100$ GeV. The figure already shows that $\amu^{\text{SUSY}}$ is severely
limited for such high masses. Overall we obtain
$\amu^{\text{SUSY}}\le13\times10^{-10}$ (for  $\tan\beta=40$ and
$\mu\le4$ TeV), and the maximum $\amu^{\text{SUSY}}$ quickly drops 
for even heavier slepton masses.
\update{Hence  the  deviation observed at BNL
  (\ref{eqn:BNLDiscrepancy}) could at most be 
explained at the $2\sigma$-level.
for these values of  $\tan\beta$ and $\mu$.
 This remains the case also with the
slightly smaller value and uncertainty of $\damuNEW$.}\footnote{%
  Allowing for $\tan\beta\to\infty$ \cite{Bach:2015doa} or ultra-high values of
  $\mu$ \cite{Endo:2013lva} changes the picture. In both cases the linearity in
  $\tan\beta$ and $\mu$ visible in Eqs.\ (\ref{eq:SUSYMIapprox}) is replaced by a
  saturation resulting from resummed higher-order effects. In such
  extreme parameter regions it is possible to obtain
  $\amu^{\text{SUSY}}\approx20\times10^{-10}$ for LSP masses above 1 TeV with
  $\mu=100$ TeV and $\tan\beta=40$ \cite{Endo:2013lva} or even
  $\amu^{\text{SUSY}}>30\times10^{-10}$ for LSP masses above 1 TeV with
  $\tan\beta\to\infty$ \cite{Bach:2015doa}. Similarly, the scenario
  of Ref.\ \cite{Crivellin:2010ty} realizes radiative muon mass
  generation in the MSSM with non-holomorphic soft SUSY breaking
  parameters and allows LSP masses above 1 TeV while explaining
  $\damu^{\text{BNL}}$.
  \label{footnoteTBMUlarge}}

Despite the small contributions to $\amu$, scenarios with heavy
sparticles can be well motivated. E.g.\ the original focus point
scenario \cite{Feng:1999zg,Feng:1999mn} naturally involve sleptons
above the TeV scale, and in models with universality boundary
conditions below the GUT scale, squarks and sleptons are rather close
in mass \cite{Ellis:2007ac} and LHC-constraints on squarks imply
sleptons above the TeV scale \cite{Costa:2017gup}.  
Another attractive scenario where all sparticle masses are above the
TeV scale is given by a Higgsino-like LSP with Higgsino mass around $1$
TeV. This case leads to an explanation of the observed dark matter
relic density without further tuning of masses or mass splittings and
can be realized in the constrained MSSM or in more general variants of the MSSM (see
e.g.\ Ref.\ \cite{Roszkowski:2017nbc}). However all such scenarios
restrict $\amu^{\text{SUSY}}$ to values below around $10\times10^{-10}$ for
$\tan\beta\lesssim50$.

In the following we will focus on the alternative, ``non-standard''
scenarios which allow lighter sparticle masses.

\subsubsection{$(\tilde{B}\tilde{l})$-scenario with light sleptons and
  Bino }
%

\begin{figure}[t]
  \begin{subfigure}[t]{0.405\textwidth}
    \centering \includegraphics[width=\textwidth]{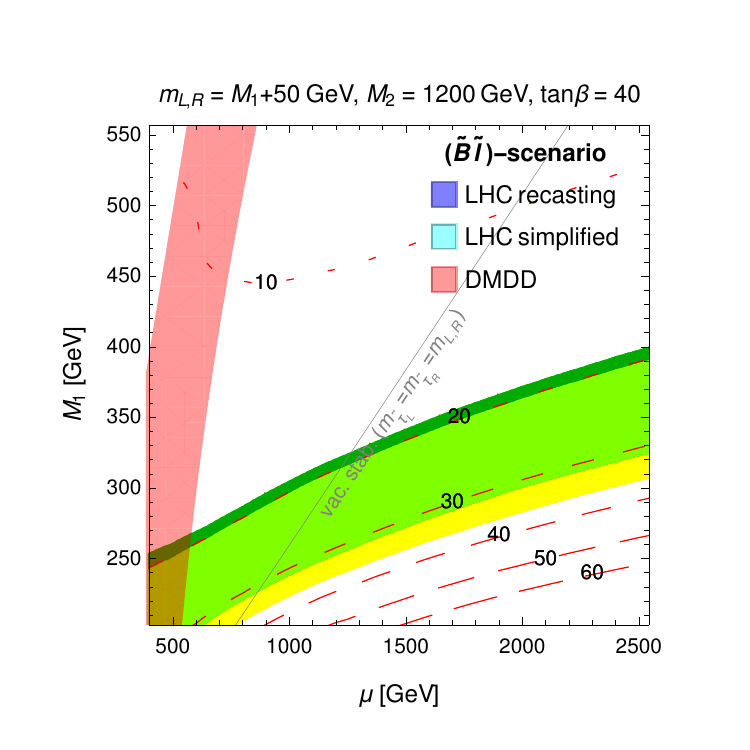}
    \caption{}
    \label{fig:Binosleptonscompresseda}
  \end{subfigure}
  \begin{subfigure}[t]{0.405\textwidth}
    \centering\includegraphics[width=\textwidth]{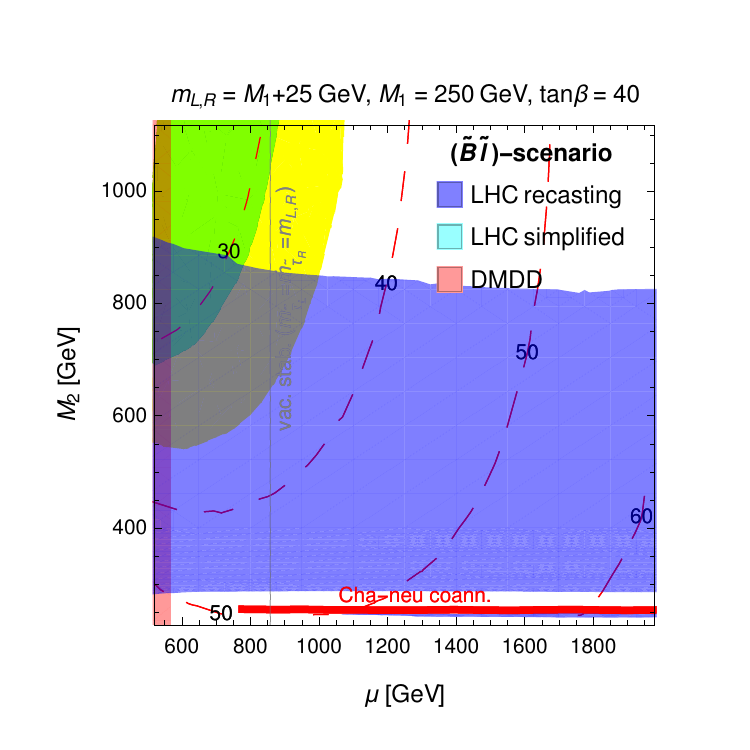}
    \caption{}
    \label{fig:Binosleptonscompressedb}
  \end{subfigure}
  \begin{subfigure}[t]{0.18\textwidth}
    \centering\includegraphics[scale=1.]{SUSYplots/PlotAmuLegend.pdf}
  \end{subfigure}
  \caption{\label{fig:Binosleptonscompressed} 
    $(\tilde{B}\tilde{l})$-scenario with either $M_2=1200$ GeV fixed
    or $M_1=250$ GeV fixed. For the remaining parameter values see the
    plots.
    The red dashed contours correspond to values of
    $\amu$ as indicated in the legend on the right; the yellow/green coloured
    regions correspond to the $1\sigma$    bands corresponding
    to the BNL deviation (\ref{eqn:BNLDiscrepancy}) and the new
    deviation including FNAL (\ref{eqn:avgDiscrepancy}), and their overlap. For the
    $\tan\beta$-reinterpretation see caption of Fig.\ \ref{fig:briefsurveyplot}.
    The (light) red shaded regions are excluded by dark matter direct detection
    if the LSP is assumed stable; the blue shaded region corresponds
    to the limits from the LHC recasting, see
    Fig.\ \ref{fig:ReproduceATLASplot} and text for details. These
    plots do not contain regions excluded by
    the additional LHC limits implemented in a
    simplified way.
    The red thick
    solid line in the right plot corresponds to the parameter strip
    where chargino-neutralino coannihilation is possible; directly
    below this strip a tiny region is excluded by the LHC-constraint
    from compressed masses, Ref.\ \cite{Sirunyan:2018iwl}, but we verified that this
    does not exclude the chargino-neutralino coannihilation region.
    The gray thin line corresponds to the vacuum
    stability constraint of Ref.\ \cite{Endo:2013lva}; it applies in case the
    left- and right-handed stau-masses are set equal to the
    smuon/selectron masses and excludes the points to the right,
    i.e.\ with larger $\mu$. 
}
\end{figure}

The $(\tilde{B}\tilde{l})$-scenario is characterized by a Bino-like LSP and
sleptons significantly lighter than $700$ GeV. Given current LHC
constraints it is viable if the mass splitting between sleptons and
the LSP is sufficiently small, $m_{\txl, \txr}-M_1\lesssim100$\,GeV (for
masses below around 200 GeV and very small splittings additional
constraints from compressed-mass searches become relevant).
The dark matter relic density can be correctly achieved by either ${\tilde{l}}$/${\tilde{\tau}}$ or
$\chi ^{\pm}$-coannihilation by appropriate fine-tuning of parameters as discussed below. 
The scenario is illustrated in
Fig.\ \ref{fig:Binosleptonscompressed}. The left plot
Fig.\ \ref{fig:Binosleptonscompresseda} shows results in the
$\mu$--$M_1$-plane. The Wino mass is fixed to the
rather high value $M_2=1200$ GeV, safely but not too far above the
chargino mass limit (\ref{maximumChamasslimit}), and  the
mass splitting $m_{\txl, \txr}-M_1=50$ GeV. The right plot 
Fig.\ \ref{fig:Binosleptonscompressedb} shows results in the
$\mu$--$M_2$-plane, while the Bino and slepton masses are fixed to the
rather light values $M_1=250$ GeV and
$m_{\txl, \txr}=275$ GeV. In both plots the shown quantities are not very
sensitive to the choice of the mass splitting, so the plots are
representative for a wider range of values for $m_{\txl, \txr}-M_1$.

The red dashed lines and yellow/green coloured regions show the contours of
$\amu$ and $1\sigma$ regions corresponding to the  measurements from
BNL only and the average including \update{FNAL}. The behaviour of $\amu$ in this
$(\tilde{B}\tilde{l})$-scenario is dominated by the WHL and BLR
contributions of
Eqs.\ (\ref{eq:SUSY1Lnumerical},\ref{eq:SUSYMIapprox}) and can be well
understood via these approximations. The
WHL contributions dominate in the left plot at large $M_1$ and
very small $\mu$ and in the right plot at $\mu\lesssim1 $ TeV; in these
regions $\amu^{\text{SUSY}}$ decreases with increasing $\mu$. The BLR contributions
are linearly enhanced by $\mu$ and dominate at large $\mu$ in both
plots. As the plots show very large $\amu^{\text{SUSY}}$ can be obtained both for large $\mu$,
where the BLR-contribution dominates, and for small $\mu$ with 
WHL-dominance.

LHC-constraints obtained by the recasting described in Sec.\ \ref{sec:SUSYconstraints} are
displayed by the blue shaded region in the plots. 
The parameter space of the left plot
Fig.\ \ref{fig:Binosleptonscompresseda} is entirely allowed (more
details on the recasting are exhibited in the Appendix), even where
the Higgsino-like chargino is light. The right plot
Fig.\ \ref{fig:Binosleptonscompressedb} shows the expected large excluded
region approximately for $300\text{ GeV}<M_2<900\text{ GeV}$. It is
excluded by the chargino/slepton channel search of
Refs.\ \cite{Aaboud:2018jiw,Sirunyan:2017lae}, see
Eq.\ (\ref{Charginosleptonlimits}). The recasting shows again
that the Wino-like charginos are significantly more constrained than
Higgsino-like charginos and that the exclusion region is smaller than
in the simplified-model interpretation of
Eq.\ (\ref{maximumChamasslimit}). An additional strip of parameter
space at around $M_2\approx M_1-5$ GeV (in which case the mass
eigenvalues satisfy $m_{\chi^\pm_1}-m_{\chi^0_1}\approx15$ GeV) is
excluded by the recasting of the CMS compressed-mass search of
Ref.\ \cite{Sirunyan:2018iwl}.

Regarding dark matter it is well known that in case of a Bino-like LSP
in the considered mass range the relic density is too high unless some
coannihilation mechanism is active. In our case there are three
options: \mbox{chargino-,} stau- or slepton-coannihilation (see also the review
\cite{Roszkowski:2017nbc}).
The possibility of chargino-coannihilation takes place in the
parameter space where 
$m_{\chi^\pm_1}-m_{\chi^0_1}\approx25$ GeV in the right plot, shown as
the thick red line around $M_2=255$ GeV. Here the relic density
takes the measured value (\ref{eqn:DMRD}) without further tuning of the slepton
masses. Everywhere else in the two plots the relic density can be
correctly explained via slepton- or stau-coannihilation
by slightly finetuning the slepton and/or stau
masses.
Since there is no unique way to achieve the required coannihilation
we do not carry out this finetuning but fix the parameters as
described above for the evaluation of all other observables.
In this way the plot is representative for all these cases.\footnote{%
  The other observables 
  have been evaluated by setting the stau masses to $2M_1$ in Fig.~\ref{fig:Binosleptonscompresseda} 
  and to $2000$ GeV in Fig.~\ref{fig:Binosleptonscompressedb}. 
  In order to achieve
  stau-coannihilation at least one stau mass has to be small and close
  to the LSP-mass. None of the plotted observables would change
  significantly, except that the LHC-constraints in the right plot
  would become slightly weaker since a larger variety of decay modes
  would exist for the charginos \cite{Chakraborti:2020vjp}. In this
  sense both plots are representative for a variety of cases and
  conservative in case of stau-coannihilation.}

Assuming now that the relic density is correctly explained,
constraints from dark matter direct detection become relevant.
The constraints from direct detection experiments are shown as the
(light) red shaded bands; they exclude a large portion of the parameter space with
small $\mu$,
implying $\mu\gtrsim600\ldots 800$ GeV in the plot. This reflects the
well-known need for small gaugino--Higgsino mixing and a significant mass gap between the LSP and the
Higgsino mass $\mu$, see also Sec.\ \ref{sec:SUSYconstraints} and
footnote \ref{footnotedarkmatterapprox}.

The thin solid gray line in the plots corresponds to the vacuum
stability constraint of Ref.\ \cite{Endo:2013lva} on stau-mixing already explained
around Eq.\ (\ref{VacStabEndo}). It excludes the large-$\mu$ region to its right
under the condition that both left- and
right-handed stau masses are as light as the
smuon/selectron masses. This upper limit on $\mu$ thus applies in particular if
stau-coannihilation and $m_{\tilde{\tau}_\txl}\approx m_{\tilde{\tau}_\txr}$
is assumed. Then the limit on $\mu$
significantly reduces the region in which the BLR-contributions to
$\amu^{\text{SUSY}}$ dominate. The vacuum stability constraint can be evaded and larger
$\mu$ and large $\amu^{\text{SUSY}}$ for heavier $M_1$ remain possible under the
assumption that   one
stau or both staus are heavier. This can be compatible with a dark matter explanation
either in the case of selectron/smuon--coannihilation, or in the case
of stau--coannihilation with strongly non-universal
left-/right-handed staus.

In summary, the $(\tilde{B}\tilde{l})$-scenario allows the following three
parameter regions with large $\amu^{\text{SUSY}}$.
\update{
The first region is in the lower left of Fig.\ \ref{fig:Binosleptonscompresseda} and the
upper left  of Fig.\ \ref{fig:Binosleptonscompressedb} between the dark matter and
vacuum stability constraints on $\mu$. It involves $\mu$ around the 1
TeV scale and is allowed by all constraints
even if we assume completely universal sleptons
$m_{\txl, \txr}\approx m_{\tilde{\tau}_{\txl, \txr}}$.
Here dark
matter constraints can be explained by stau and/or slepton
coannihilation and the $\amu$ result can be accommodated easily via
the large BLR contributions. The new world average result
   for $\damu$ including FNAL can
  be explained well for $M_1\lesssim300$ GeV and $\tan\beta=40$.
As shown by Fig.\ \ref{fig:Binosleptonscompressedb},
$M_2$ can be as low as around $900$ GeV for our choice of $M_1=250$ GeV, while for
larger $M_1$ the LHC-limit on $M_2$ would relax slightly, and
WHL-contributions could further increase $\amu^{\text{SUSY}}$. }
\update{
The second region is to the right of the vacuum stability lines where
$\mu$ is in the multi-TeV region and   $\amu^{\text{SUSY}}$ is further
increased by the BLR contributions.
The region is viable if at least one stau is sufficiently heavier. The
dark matter relic density 
can then be generated via slepton
or $\tilde{\tau}_1$ coannihilation.
Here an explanation of $\damu$ is widely
  possible. The LSP mass can be heavier than $300$ GeV, and the large white regions in
  Fig.\ \ref{fig:Binosleptonscompressedb} and lower right corner of Fig.\ \ref{fig:Binosleptonscompresseda} mean that $\damuNEW$ can
  be explained for $\tan\beta<40$ according to the right axis
   of the legend plot.
   The third region is the
 patch of parameter space close to the lower border of
Fig.\ \ref{fig:Binosleptonscompressedb}. Here the
Bino-like LSP, the Wino and the sleptons are all light and 
close in mass. This patch of parameter space allows in particular to
generate dark matter via chargino-coannihilation, and it leads to very
large $\amu^{\text{SUSY}}$ for any value of $\mu$ above the dark matter
limit. Here again the updated deviation $\damuNEW$ can be
  explained for $\tan\beta<40$.}

The recent model-building literature has put forward a variety of
constructions leading to
our second parameter region with light sleptons and very heavy $\mu$,
where $\amu^{\text{SUSY}}$ is dominated by the
BLR-contributions. These constructions
are  particularly motivated in view of 
the LHC constraints on the coloured SUSY particles and the Higgs mass.
Clearly, a straightforward conclusion from such constraints is that
gluino and top-squark masses are in the (multi-)TeV region. Via 
renormalization effects these masses can enter the electroweak symmetry
breaking relations, and in many models very large $\mu$ in the  multi-TeV
region is then necessary in order to cancel such effects and allow a
Higgs-VEV compatible with observations
\cite{Yanagida:2020jzy,Ibe:2019jbx,Yanagida:2017dao}.

Many concrete constructions are inspired by universality ideas but
involve some degree of non-universality to accommodate all existing
constraints. One class of such models with multi-TeV-scale $\mu$
involves non-universal sfermion masses.
General models with non-universal
sfermion masses (but universal gaugino masses) have been constructed and investigated in
Refs.\ \cite{Okada:2016wlm,Tran:2018kxv}, where 
$(\tilde{B}\tilde{l})$-like scenarios with $\mu\sim10$ TeV were
identified as promising. A specific kind of sfermion non-universality
was considered in Refs.\ 
\cite{Ibe:2019jbx,Ibe:2013oha,Hussain:2017fbp}, where
the third generation of sfermions is assumed heavier than the
first two, but universality between squarks and sleptons at some high
scale is retained. Up-to-date LHC
constraints on gluino and Wino masses then imply that non-universal
gaugino masses are almost unavoidable unless one allows an unstable
charged slepton LSP \cite{Ibe:2019jbx}.

Another class of models retains universality of all scalar
soft SUSY-breaking masses but allows non-universal gaugino masses. In
this case again, the scenario of Fig.\ \ref{fig:Binosleptonscompressed}
with very large $\mu$ is the only option to obtain significant $\amu^{\text{SUSY}}$
\cite{Akula:2013ioa,Gogoladze:2014cha,Chakrabortty:2015ika,Wang:2018vrr}. 

We mention that the scenario with $\mu$ in the multi-TeV region is
also important to obtain large $\amu^{\text{SUSY}}$ in the context of various
specific model constructions, such as models based on Pati-Salam
symmetry \cite{Belyaev:2016oxy} (here, at the same time light Winos
are preferred), models with usual GUT constraints but extra vectorlike
matter fields \cite{Choudhury:2017acn,Choudhury:2017fuu},
and in a hybrid gauge-gravity mediation model with only four free
parameters \cite{Zhu:2016ncq}.
\subsubsection{$(\tilde{W}\tilde{l})$-scenario with light sleptons
  and Wino}
\label{sec:WinoSlepscenario}

%
\begin{figure}[t]
  \begin{subfigure}[t]{0.405\textwidth}
    \centering\includegraphics[width=\textwidth]{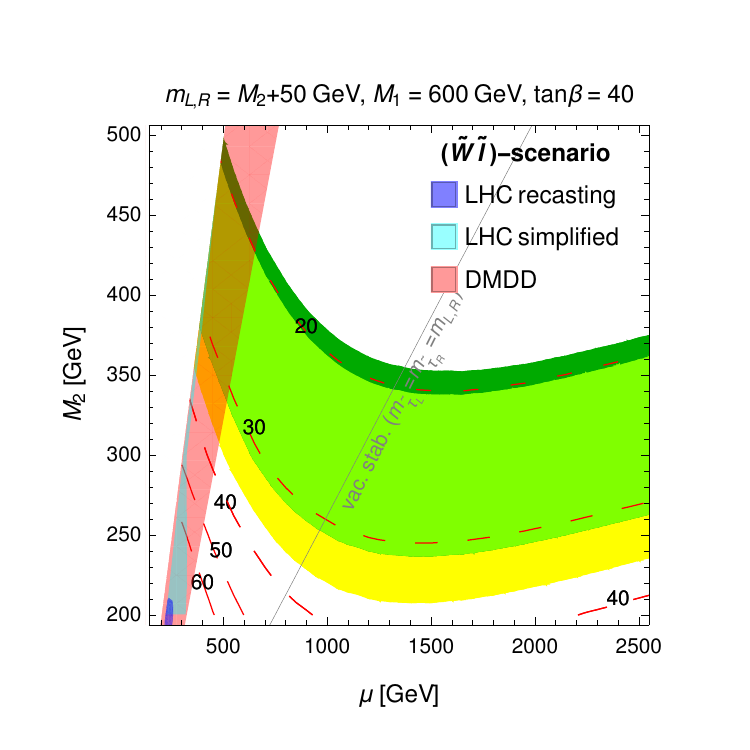}
    \caption{}
    \label{fig:WHsleptonscompresseda}
  \end{subfigure}
  \begin{subfigure}[t]{0.405\textwidth}
    \centering\includegraphics[width=\textwidth]{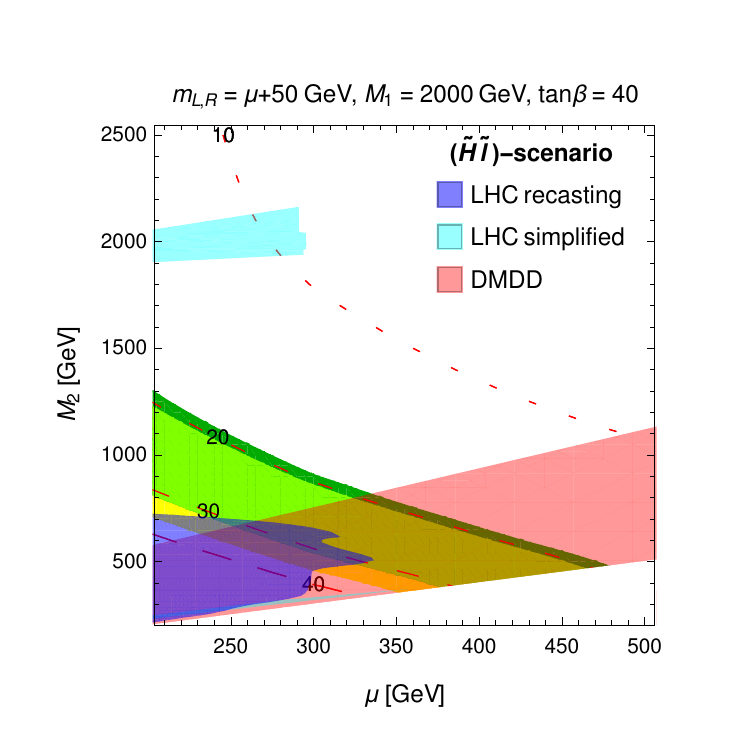}
    \caption{}
    \label{fig:WHsleptonscompressedb}
  \end{subfigure}
  \begin{subfigure}[t]{0.18\textwidth}
    \centering\includegraphics[scale=1.]{SUSYplots/PlotAmuLegend.pdf}
  \end{subfigure}
  \caption{\label{fig:WHsleptonscompressed} (a)
    $(\tilde{W}\tilde{l})$-scenario. (b)
    $(\tilde{H}\tilde{l})$-scenario. For parameter values see the
    plots and the text.
    The red dashed contours correspond to values of
    $\amu^{\text{SUSY}}$ as indicated in the legend on the right; the yellow/green coloured
    regions correspond to the $1\sigma$    bands corresponding
    to the BNL deviation (\ref{eqn:BNLDiscrepancy}) and the new
    deviation including FNAL (\ref{eqn:avgDiscrepancy}), and their overlap. For the    
    $\tan\beta$-reinterpretation see caption of Fig.\ \ref{fig:briefsurveyplot}.
    The red shaded region is excluded by dark matter direct detection
    if the LSP is assumed stable; the blue shaded regions correspond
    to the limits from the LHC recasting, see
    Fig.\ \ref{fig:ReproduceATLASplot} for details. The cyan shaded
    region corresponds to the additional LHC limits implemented in a
    simplified way; in both plots the slepton search
    (\ref{eq:sleptonsearch}), Ref.\ \cite{Aad:2019vnb} excludes a
    narrow strip at small $\mu$ and $M_2$, where the slepton--LSP mass splitting
    is largest. In the the right plot the compressed-mass searches of
    Ref.\ \cite{Aad:2019qnd} exclude another small region at large
    $M_2$, which enters the LSP mass via mixing. The thin solid gray line corresponds to the vacuum
    stability constraint of Ref.\ \cite{Endo:2013lva}; it applies in case the
    left- and right-handed stau-masses are set equal to the
    smuon/selectron masses and excludes the points to its right,
    i.e.\ with larger $\mu$.
}
\end{figure}

The $(\tilde{W}\tilde{l})$-scenario involves a Wino-like LSP and
sleptons significantly lighter than 700 GeV. Current LHC-constraints
on sleptons allow this scenario provided the  slepton--LSP mass
splitting is sufficiently small. The scenario is illustrated in
Fig.~\ref{fig:WHsleptonscompresseda} 
in the $\mu$--$M_2$-plane.
The mass splitting is chosen as $m_{\txl,\txr}=M_2+50$ GeV, but the
plotted quantities are not very sensitive to this choice.
By definition of the scenario, the Bino mass is
assumed to be heavier than the Wino mass. Since the Bino mass is also
not strongly constrained by LHC data we fix it to $M_1=600$ GeV, an
intermediate value which
is always heavier than $M_2$ in the plot 
but still allows significant BLR-contributions to $\amu^{\text{SUSY}}$.

The behaviour of $\amu^{\text{SUSY}}$ (red dashed lines and yellow/green coloured
  regions) in this scenario with a Wino-like LSP is
similar to the previous one in the case of a Bino-like LSP. The
$\amu^{\text{SUSY}}$-contours in
Figs.~\ref{fig:Binosleptonscompresseda},~\ref{fig:Binosleptonscompressedb} 
and~\ref{fig:WHsleptonscompresseda} 
have a similar shape. For small $\mu$
the WHL-contributions of 
Eqs.\ (\ref{eq:SUSY1Lnumerical},\ref{eq:SUSYMIapprox}) dominate, and
for $\mu\gtrsim1500$\,GeV the
BLR-contributions to $\amu^{\text{SUSY}}$ dominate. An important difference is that the Wino-like
LSP scenario allows very low   Wino masses without the need for
finetuning $M_1\approx M_2$ as e.g.\ in
Fig.~\ref{fig:Binosleptonscompressedb}. 
Hence the WHL-dominance region is wider in the 
$(\tilde{W}\tilde{l})$-scenario. In all of this WHL-dominance region
the actual choice of $M_1=600$ GeV is inconsequential.
This choice is important for $\mu\gtrsim1500$\,GeV, where
the BLR-contributions dominate and $\amu^{\text{SUSY}}$ rises with $\mu$. Higher
choices of $M_1$ would reduce $\amu^{\text{SUSY}}$ in this region.

The recasting of the ATLAS chargino search 
\cite{Aaboud:2018jiw} excludes only a tiny blue shaded region in the
plot at $M_2,\mu\lesssim220$ GeV, where both charginos and the
sleptons are similar in mass.
In addition the cyan shaded narrow strip at small $\mu$ corresponds to
the additional LHC limits implemented in a simplified way as discussed
in Sec.\ \ref{sec:SUSYconstraints}. The specific analysis relevant
here is
the slepton search (\ref{eq:sleptonsearch}),
Ref.\ \cite{Aad:2019vnb}. It excludes this cyan parameter strip,
 where the  splitting between the slepton and LSP mass eigenvalues  is
largest.

The plot shows the dark matter direct detection limit as a red shaded band.
It is well-known that a Wino-like LSP cannot
produce the observed relic density unless the Wino mass is in the
multi-TeV region, for a recent account see
Ref.\ \cite{Beneke:2020vff}. In the mass region of interest for us, we
obtained a
Wino-LSP relic density which is typically a factor $10\ldots100$ smaller than
the observed relic density.
Nevertheless,
the LSP--nucleon cross sections depend on the Wino--Higgsino mixing
and are rather high. Hence the dark matter direct searches imply
significant lower limits on the Higgsino mass $\mu$ of around
$300\ldots800$ GeV, 
see also footnote
\ref{footnotedarkmatterapprox}. 
This limit could only be circumvented by dropping the assumption of a
stable LSP, e.g.\ by assuming R-parity violation of LSP-decays into
light gravitinos.

The  thin solid gray line in the plot corresponds to the vacuum
stability constraint of Ref.\ \cite{Endo:2013lva}. It applies if
the left- and right-handed stau-masses are both set equal to the
smuon/selectron masses.  In such a case of slepton universality
an upper limit on $\mu$ exists which
essentially eliminates the region in which the BLR-contribution to
$\amu^{\text{SUSY}}$ dominates.

In summary, the $(\tilde{W}\tilde{l})$-scenario can easily accommodate
$\damu$ \update{as large as the deviation $\damuNEW$ or even larger}.
\update{Specifically e.g.\ the new world
  average (\ref{eqn:avgDiscrepancy}) can be explained for $\tan\beta=40$ with universal slepton
  masses and an LSP mass around $M_2=350$ GeV and $\mu=800$ GeV.
Higher masses, in particular higher $\mu$ are also possible. For lower
masses much smaller values of $\tan\beta$ can be sufficient.}
There are essentially no LHC
constraints on this scenario as long as the mass splitting between sleptons and the LSP
are sufficiently small. Dark matter direct detection enforces lower
limits on $\mu$, still leaving a wide parameter space in which the
WHL-contributions to $\amu^{\text{SUSY}}$ are dominant and large.
If slepton universality is assumed including staus, vacuum stability
imposes an upper limit on $\mu$; larger $\mu$ is possible if (at least
one) heavy stau is assumed and provides
further parameter space with large $\amu^{\text{SUSY}}$.

Again we provide a brief survey of model building efforts which lead
to constructions like the $(\tilde{W}\tilde{l})$-scenario
with a Wino-like LSP, light
sleptons and very large $\mu$.
 Ref.\ \cite{Yanagida:2020jzy} has
constructed an extreme variant of such a model
with Wino-like LSP and slepton masses
around 500 GeV  based on 
Higgs-anomaly mediated SUSY-breaking
\cite{Yanagida:2020jzy,Yin:2016shg,Evans:2013uza}; that construction
produces $\mu\gtrsim25$ TeV.
Ref.\ \cite{Cox:2018vsv} shows
that a similar scenario which involves both light Wino and Bino can follow from
gaugino+Higgs-mediated SUSY breaking. 
Ref.\ \cite{Endo:2019bcj} has also embedded a
$(\tilde{W}\tilde{l})$-like scenario in a UV-model
based on Higgs-mediated SUSY breaking. Such scenarios also had the
potential to explain not only $\damu$ but also the smaller deviation in the
electron magnetic moment $a_e$ \cite{Badziak:2019gaf,Endo:2019bcj}
(see however footnote \ref{aefootnote}).

\subsubsection{$(\tilde{H}\tilde{l})$-scenario with light sleptons and
  Higgsino }
\label{sec:HiggsinoSlepscenario}

The $(\tilde{H}\tilde{l})$-scenario is characterized by a Higgsino-like LSP and
sleptons significantly lighter than 700 GeV. Again, in view of current
LHC-constraints the slepton--LSP mass splitting cannot be much larger
than 100 GeV. The scenario is illustrated in
Fig.\ \ref{fig:WHsleptonscompressedb} in the $\mu$--$M_2$-plane. We again fix
the mass splitting $m_{\txl, \txr}=\mu+50$ GeV and we set the  Bino
mass to $M_1=2000$ GeV as reference values, although the considered observables are not very
sensitive to this choice (except that significantly lower $M_1$ can
lead to conflict with dark matter direct detection limits).

The behaviour of $\amu^{\text{SUSY}}$ shown by the red dashed lines and
  yellow/green coloured regions is quite different from the one in the previous
two scenarios. Since the Higgsino mass $\mu$ is small, only the
WHL-contributions of
Eqs.\ (\ref{eq:SUSY1Lnumerical},\ref{eq:SUSYMIapprox}) are
important. For this reason  the result shows the generic
$1/M_{\text{BSM}}^2$-behaviour explained in
Sec.\ \ref{sec:BSMoverview}  and 
drops quickly both with increasing $\mu$ or increasing
$M_2$, while the choice of $M_1$ has not much influence.
Still, e.g.\ if $\mu=300$ GeV the BNL deviation can be
explained at the 1$\sigma$ level even for $M_2=1$ TeV.

The constraints from LHC-recasting, shown in blue, are rather
strong. They originate from the chargino/slepton channel searches of
Eq.\ (\ref{Charginosleptonlimits}),
Refs.\ \cite{Aaboud:2018jiw,Sirunyan:2017lae}.
Compared to e.g.\ Fig.\ \ref{fig:ReproduceATLASplot} now $\mu$ instead of $M_1$
takes the role of the LSP-mass, and the recasting shows that the
resulting limits on the Wino-like chargino mass and 
thus on $M_2$ are  weaker than in
Fig.\ \ref{fig:ReproduceATLASplot}, extending only up to $M_2=700$ 
GeV in Fig.\ \ref{fig:WHsleptonscompressedb}.
The plot also shows additional cyan shaded parameter regions, which
are subject to
additional LHC-constraints implemented in a simplified way. Here,
both the slepton search of Ref.\ \cite{Aad:2019vnb} and the compressed-mass searches of
Ref.\ \cite{Aad:2019qnd} exclude small regions at small $M_2$ close to the $M_2 = \mu$ boundary and at
large $M_2\approx 2$ TeV, close to the bino mass. In both regions,
neutralino mixing happens to lead to
LSP--next-to-LSP mass splittings which are excluded.
Though not visible in the plot we
mention that the same compressed-mass searches also impose limits on
the Higgsino-like chargino/neutralino system and thereby exclude
parameter space with $\mu<200$ GeV.

It is well known that a
Higgsino-like LSP cannot
produce the observed relic density for such light Higgsinos as
considered here. Still there are relevant limits from dark matter
direct detection experiments \cite{Baer:2018rhs}. In the mass region
of interest for us, we find that the
Higgsino-LSP relic density is typically a factor 10 smaller than
the observed relic density.
This value is
higher than the relic density in the previous case of a Wino-like LSP. As a
result, stronger  dark matter direct detection limits are obtained on
$M_2$, shown as the red shaded band. In the plot they
require $M_2\gtrsim500\ldots1500$
GeV, depending on $\mu$.
As before, the dark matter direct detection limits apply only under
the assumption that the Higgsino-like LSP is stable.

In summary, the $(\tilde{H}\tilde{l})$-scenario is strongly
constrained by LHC chargino searches and by dark matter direct
detection constraints
(if the LSP is assumed to be stable). Still it allows values of $M_2$
as small as around $700$ GeV and $\mu$ around $200$ GeV which lead to 
$\amu^{\text{SUSY}}$ as large as $30\times10^{-10}$, but outside this small corner
of parameter space the values of $\amu^{\text{SUSY}}$ quickly drop.
\update{The average deviation can be explained at the $1\sigma$ level for LSP masses up
  to $350$ GeV for $\tan\beta=40$. At the $2\sigma$ level, much higher
LSP masses are possible.}

  In the model building literature, Ref.\ \cite{Han:2020exx} has considered a scenario
of the $(\tilde{H}\tilde{l})$-type with significant mass gap between
the Higgsino-LSP and the two gauginos, motivated within the context of a
model with seesaw mechanism and  SO(10) GUT constraints on the gaugino masses but
non-universal scalar masses. Although electroweak LHC and DMDD
constraints are not 
considered, this reference also finds only small viable contributions
to $\amu^{\text{SUSY}}$ as a result of model-specific
correlations to flavour-violating observables.

\subsubsection{$(\tilde{B}\tilde{W}\tilde{H})$-scenario with light charginos }
\begin{figure}[t]
  \begin{subfigure}[t]{0.405\textwidth}
    \centering\includegraphics[width=\textwidth]{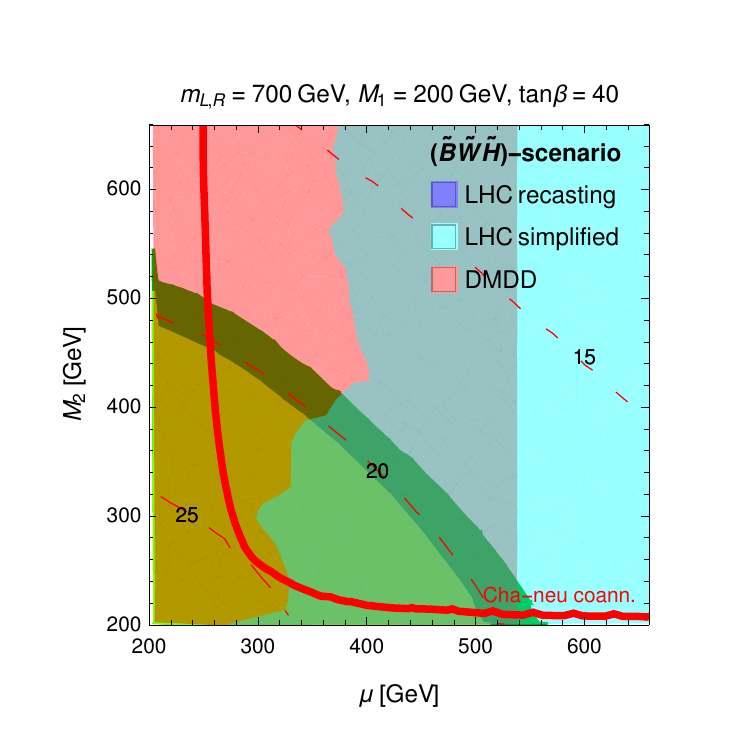}
    \caption{}
    \label{fig:chalightersleptonsa}
  \end{subfigure}
  \begin{subfigure}[t]{0.405\textwidth}
    \centering\includegraphics[width=\textwidth]{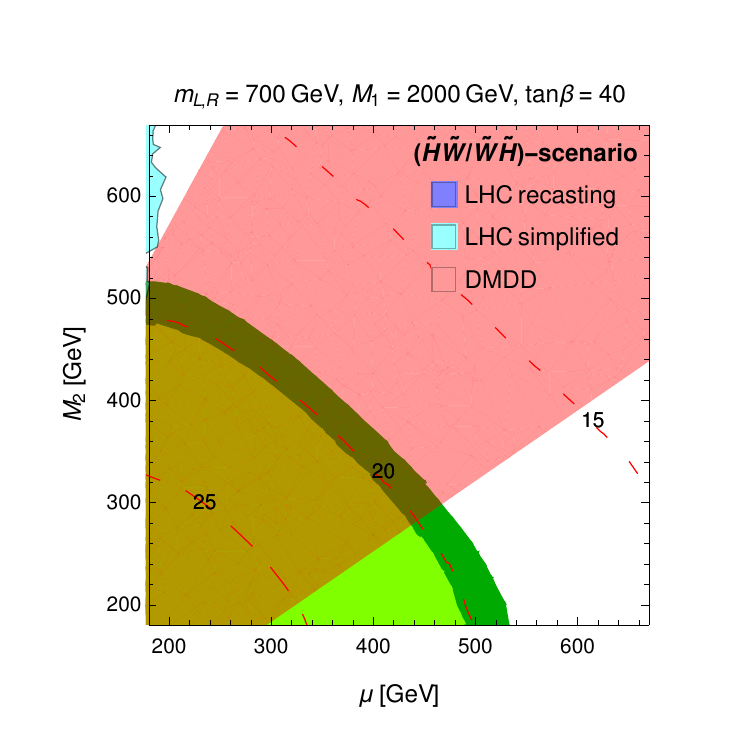}
    \caption{}
    \label{fig:chalightersleptonsb}
  \end{subfigure}
  \begin{subfigure}[t]{0.18\textwidth}
     \centering\includegraphics[scale=1.]{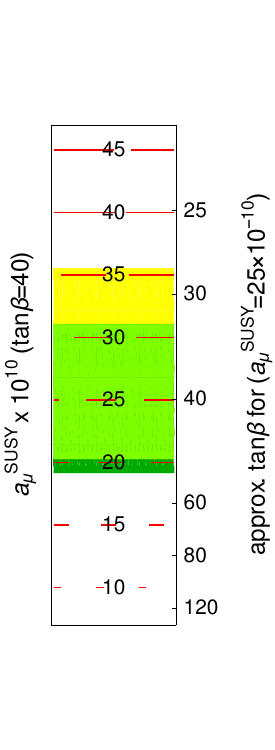}
  \end{subfigure} 
  \caption{\label{fig:chalightersleptons} (a)
    $(\tilde{B}\tilde{W}\tilde{H})$-scenario. (b)
    $(\tilde{H}\tilde{W}/\tilde{W}\tilde{H})$-scenario. For parameter values see the
    plots and the text.
    The red dashed contours correspond to values of
    $\amu^{\text{SUSY}}$ as indicated in the legend on the right; the yellow/green coloured
    regions correspond to the $1\sigma$    bands corresponding
    to the BNL deviation (\ref{eqn:BNLDiscrepancy}) and the new
    deviation including FNAL (\ref{eqn:avgDiscrepancy}), and their overlap. For the    
    $\tan\beta$-reinterpretation see caption of Fig.\ \ref{fig:briefsurveyplot}.
    The red shaded region is excluded by dark matter direct detection
    if the LSP is assumed stable; in the left plot this red region is
    the rectangle extending up to $\mu\approx540$ GeV. Both plots do not contain regions
    excluded by the LHC recasting. The cyan shaded
    regions correspond to the additional LHC limits implemented in a
    simplified way. In the left plot the cyan region is the large
    region extending from $\mu\gtrsim300\ldots400$ GeV to the right
    (it is partially overlaid with dark matter and $\amu^{\text{SUSY}}$-regions),
    it mainly arises
    from the stau-channel chargino search (\ref{eq:Chastaulimits}),
    Ref.\ \cite{Aaboud:2017nhr}; it is valid if light staus are assumed for
    stau-coannihilation. The red thick line in the left plot
    corresponds to the parameter strip 
    where chargino-neutralino coannihilation is possible; in this
    region the dark matter relic density can be correctly described
    without light staus and the LHC-constraint does not apply.
}
\end{figure}

The remaining scenarios differ from the previous ones in that they
involve two light charginos, while the 
sleptons are not assumed particularly light.
The $(\tilde{B}\tilde{W}\tilde{H})$-scenario discussed here
assumes both charginos to be lighter than the sleptons, but the
Bino-like neutralino to be the LSP. Hence $M_1<M_2,\mu$ with no
particular order between $M_2$ 
and $\mu$. The
scenario is illustrated in Fig.\ \ref{fig:chalightersleptonsa}.
In the figure we set $m_{\txl, \txr}=700$ GeV, safely but not too far above
the maximum LHC-limit of Ref.\ \cite{Aad:2019vnb}.
 We also fix $M_1=200$ GeV
and show $\amu^{\text{SUSY}}$ versus LHC and dark matter
constraints as a function of $M_2$ and $\mu$.

The behaviour of $\amu^{\text{SUSY}}$ is dominated by the WHL-contributions of
Eqs.\ (\ref{eq:SUSY1Lnumerical},\ref{eq:SUSYMIapprox}) which are
suppressed by the heavy slepton masses and further suppressed if $\mu$
and/or $M_2$ become heavy. The maximum  contribution to $\amu^{\text{SUSY}}$ in  Fig.\ \ref{fig:chalightersleptonsa} is around
$30\times10^{-10}$; if $\mu$ and $M_2$ are heavier than
$500$ GeV, $\amu^{\text{SUSY}}$ reaches at most  $15\times10^{-10}$
for $\tan\beta=40$ \update{(which
  is just within the $2\sigma$ region of the updated deviation
  $\damuNEW$}.
The chosen value of
$M_1=200$ GeV has almost no influence on $\amu^{\text{SUSY}}$. However the parameter
space of this $(\tilde{B}\tilde{W}\tilde{H})$-scenario is subject to
an interesting interplay of LHC and dark matter constraints.

First, since we assume the dark matter relic density to be correctly
explained, the limits from dark matter direct detection are
applicable, similarly to the plots in
Fig.\ \ref{fig:Binosleptonscompressed}. In the present case with
$M_1=200$ GeV, the region with $\mu\lesssim540$ GeV is excluded by
this constraint. The constraint is shown by the red shaded band (partially
overlaid with other coloured regions). As a result the entire region with
$\amu^{\text{SUSY}}>20\times10^{-10}$ is excluded, independently of other
details.

Second, in order to achieve the correct relic density for this case of
a Bino-like LSP in the given mass range, some coannihilation
mechanism must act. Since sleptons are heavy, the two options are either
chargino-coannihilation or
stau-coannihilation. Chargino-coannihilation requires a specific
chargino--LSP mass splitting. In Fig.\ \ref{fig:chalightersleptonsa}
the thick red line denotes the one-dimensional contour along which
chargino-coannihilation is possible due to appropriately small mass
splittings between the lightest chargino and the LSP. Along this red
contour we may assume 
staus to be heavy (e.g.\ degenerate with 1st and 2nd generation
sleptons or even heavier).
Anywhere outside the red line we must assume at least one or both staus
to be light and close to the LSP in mass for stau-coannihilation.

Finally, we can apply LHC-constraints. The constraints depend on the
coannihilation mechanism and the stau masses.
Along the red contour for chargino-coannihilation we assume the
staus to be heavy. Then the
LHC-limit from chargino searches
with $Wh$-channel \cite{Aad:2019vvf}, see
Eq.\ (\ref{eq:ChaWhlimits}), is relevant.
It turns out, however, that these limits 
do not exclude the chargino-coannihilation contour due to the small mass splittings.
The scenario with
Bino-like LSP and light charginos but heavier sleptons and staus has
also been investigated thoroughly in Ref.\ \cite{Athron:2018vxy}, where
these scenarios were found to be not constrained and in some cases
fitted excesses in the data. 

Everywhere outside the red contour, we need to assume
stau-coannihilation and light stau(s).
In this case, the constraint from chargino searches
with stau-channel \cite{Aaboud:2017nhr}, see
Eq.\ (\ref{eq:Chastaulimits}), applies.
If
we  apply this LHC-constraint as described in
Sec.\ \ref{sec:SUSYconstraints}, the cyan shaded region in the plot is 
excluded, which is essentially the entire region with $\mu\gtrsim350$ GeV.\footnote{As mentioned in the context of
  Eq.\ (\ref{eq:Chastaulimits})  the LHC-constraint of
Ref.\ \cite{Aaboud:2017nhr} is rather robust against
changes of the stau masses and mixings and against the Higgsino content
of the chargino. Hence we apply literally the constraints obtained in
the simplified-model interpretation of Ref.\ \cite{Aaboud:2017nhr} to
charginos with dominant decay into staus.}

As a result, the combination of dark matter and LHC-constraints
exclude the entire scenario with stau-coannihilation for $M_1=200$
GeV. Larger values of $M_1$ \update{above around $300$ GeV} would relax
LHC limits but lead to 
stronger dark matter direct detection limits (see footnote
\ref{footnotedarkmatterapprox}), thus 
leaving little room for large contributions to
$\amu$. 
\update{ If in the future the deviation decreases, this parameter region with significantly
  higher LSP masses $M_1$ and stau-coanihilation may become more promising.}

Ref.\ \cite{Hagiwara:2017lse} has considered $\amu$ versus LHC in the same
scenario as well, however evaluated for  $M_1\le50$ GeV. This smaller
value of $M_1$ leads to larger $\amu^{\text{SUSY}}$ and weaker dark matter constraints,
but also to
stronger LHC exclusion limits (assuming stau masses in between $M_1$
and the chargino masses) essentially excluding the entire
$(\mu,M_2)$-region of interest for $\amu$. Our larger value $M_1=200$
GeV reduces 
$\amu^{\text{SUSY}}$ but also leads to a parameter region with low chargino mass
allowed by LHC; however our parameter region is challenged by DMDD
constraints. This comparison highlights the complementarity between 
$\amu$, LHC and dark matter constraints.

In summary, the entire 
$(\tilde{B}\tilde{W}\tilde{H})$-scenario with light staus turns out to be strongly
under pressure. \update{A remaining possibility in this
$(\tilde{B}\tilde{W}\tilde{H})$-scenario is chargino-coannihilation
and heavier staus:
the small part of the red thick line at $\mu>540$ GeV is viable. Here
 $\damu$ reaches up to $20\times10^{-10}$ for
  $\tan\beta=40$, }\update{which is just sufficient to accommodate the
  deviation (\ref{eqn:avgDiscrepancy}) at the $1\sigma$
  level.}

The $(\tilde{B}\tilde{W}\tilde{H})$-scenario appears in an elaborate
model building construction of Ref.\ \cite{Altin:2017sxx} 
based on the Pati-Salam model with inverse seesaw mechanism for neutrino
masses. The benchmark points of that reference are very similar to the
region of  Fig.\ \ref{fig:chalightersleptonsa} with $\mu\sim450$
GeV, $M_2\sim300$ GeV; however that reference does not consider DMDD
constraints, which exclude such masses in our
Fig.\ \ref{fig:chalightersleptonsa}. In that model the right-handed
sneutrinos provide significant additional contributions to $\amu$,
enlarging the parameter space with large $\amu^{\text{SUSY}}$.
A variant of the scenario has been considered in
Ref.\ \cite{Pozzo:2018anw}. This reference investigates dark matter
generation via resonances, the so-called ``$Z$/$h$-funnel'' regions. It
shows that LHC-constraints allow the $h$-funnel region, i.e.\ an
LSP-mass around 60 GeV, together with large $\tan\beta\gtrsim20$ and
$\mu\gtrsim390$ GeV, which opens up additional parameter space similar
to the one of Fig.\ \ref{fig:chalightersleptonsa}.

\subsubsection{$(\tilde{H}\tilde{W}/\tilde{W}\tilde{H})$-scenarios
  with light charginos }
%

In the $(\tilde{H}\tilde{W}/\tilde{W}\tilde{H})$-scenarios again both charginos are lighter than the
sleptons, but now the Bino mass is also heavier, so the LSP is either Wino- or
Higgsino-like. Both scenarios are illustrated in
Fig.\ \ref{fig:chalightersleptonsb} in the $\mu$--$M_2$-plane. Like
in the previous case we set $m_{\txl, \txr}=700$ GeV, safely but not too far above
the LHC limit. We also set $M_1=2000$ GeV. This choice is not critical
--- it neither influences LHC-limits nor $\amu$, but it avoids limits
from dark matter direct detection (which would be similar to the
limits on $M_2$ discussed below).

The Figure shows that the values of $\amu^{\text{SUSY}}$ are very similar to the
previous $(\tilde{B}\tilde{W}\tilde{H})$-case. This is no surprise since $\amu^{\text{SUSY}}$ is dominated by the
WHL-contributions and the higher value of $M_1$ is inconsequential.
\update{Hence the new $\amu$ deviation can be generally
  explained if $\tan\beta$ is around $40$ or higher.}
 
In fact, the main difference
between the present scenarios and the previous
$(\tilde{B}\tilde{W}\tilde{H})$-scenario is the nature of the LSP and
the resulting very different dark matter and LHC constraints.

The LHC-constraints on the previous scenario were strong in case of
light staus. Since now in the $(\tilde{H}\tilde{W}/\tilde{W}\tilde{H})$-scenarios  there is no need to assume light
staus we assume the staus to be at least as heavy as the other
sleptons. As a result there are essentially no LHC constraints on the
scenarios, in line with the previous case and the findings of
Ref.\ \cite{Athron:2018vxy}. Only the regions with
very small $\mu$ or $M_2\lesssim200$
GeV are  subject to constraints from compressed-spectrum searches for
the Higgsino- or Wino-like chargino/neutralino system. In the plot
only a tiny cyan region at $\mu\approx200$ GeV is excluded in this
way, but the largest part of parameter space is allowed by LHC. 

The dark matter direct detection limits apply under the assumption
that the LSP is stable. As also mentioned in the context of
Fig.\ \ref{fig:WHsleptonscompressed} in Secs.\
\ref{sec:WinoSlepscenario} and \ref{sec:HiggsinoSlepscenario},
the LSP relic density is smaller than the
observed value in the displayed parameter region, hence only a
fraction of the observed dark matter can be explained. Nevertheless,
the LSP--nucleon cross sections are sufficiently high to imply
significant lower limits on the chargino masses. The limits are
stronger in case of a Higgsino-like LSP and exclude the largest part
of parameter space in the upper left part of the plot. The fact that
current direct detection constraints exclude a
    large fraction of the parameter space with Higgsino-like dark
    matter has already been observed in Ref.\ \cite{Baer:2018rhs},
    where also scenarios with non-thermal dark matter production and
    additional dark matter candidates such as axions were considered.
In the case of a
Wino-like LSP the dark matter relic density is smaller and the
constraints from direct detection are weaker, leaving open a larger
triangular region in the lower right part of the plot.

In summary, therefore, almost the entire Higgsino-like LSP region of the $(\tilde{H}\tilde{W}/\tilde{W}\tilde{H})$-scenario 
 is excluded by the combination of LHC
compressed-mass searches and dark matter direct detection.
Only a small region in the upper left part of the plot around 
$\mu\sim250$ GeV and
$M_2\sim600$ GeV remains viable. In this region, $\amu^{\text{SUSY}}$ is always
smaller than $20\times10^{-10}$ for $\tan\beta=40$ and \update{outside the
$1\sigma$ region of the new deviation $\damuNEW$.}
The Wino-like LSP region of the  $(\tilde{H}\tilde{W}/\tilde{W}\tilde{H})$-scenario 
in the triangular bottom
right region of the plot
\update{allows a larger viable parameter space, in which
a
$1\sigma$ explanation of $\damuNEW$ is possible for $\tan\beta=40$ and
even a full explanation $\amu^{\text{SUSY}}=25\times10^{-10}$ can be reached.
However for smaller $\tan\beta$, the current deviation is harder to explain.} In case
the dark
matter constraints are 
not applied (by assuming unstable LSP) both scenarios can accommodate
$\damu$ higher than $30\times10^{-10}$.

In the literature, several model-building efforts have led to
specific constructions with mass patterns of the 
$(\tilde{H}\tilde{W}/\tilde{W}\tilde{H})$ kind.
The scenario with either Higgsino- or Wino-like LSP can be motivated in
gauge-mediated SUSY breaking \cite{Bhattacharyya:2018inr}, extended by
non-minimal contributions to the soft-breaking parameters which allow
the Higgs, squark and slepton scalar mass parameters to be
non-universal.
In addition, in the context of gauge-mediated SUSY breaking
the lightest neutralino can decay into gravitinos; hence the DMDD
constraints of Fig.\ \ref{fig:chalightersleptonsb} do not apply and
the scenario of 
Ref.\ \cite{Bhattacharyya:2018inr} provides a
viable SUSY model explaining $\damu$.
The Higgsino-like LSP scenario can be motivated within
anomaly-mediated SUSY breaking \cite{Chowdhury:2015rja},
gaugino-mediated SUSY breaking \cite{Harigaya:2015kfa} or in the
context of electroweak finetuning considerations
\cite{Li:2016ucz,Padley:2015uma}. The scenario has also been
constructed in 
Ref.\ \cite{Harigaya:2015jba} as a
focus-point scenario (called FPNUS or FPHGM) based on gravity
mediation with
non-universal scalar masses or on Higgs-gaugino mediation.
All these references do
not apply DMDD constraints, which may be justified by assuming
R-parity violation \cite{Harigaya:2015jba}. Similar scenarios were
considered in a model with pseudo-Dirac gluino \cite{Li:2017fbg};
there, also DMDD constraints were investigated based only on the
weaker limits from LUX and PandaX, however also leading to significant 
constraints on parameter space.
Ref.\ \cite{Hussain:2017fbp} also considers the
$(\tilde{H}\tilde{W})$-scenario with Higgsino-like LSP, but here
universality between squarks and 
sleptons and quark flavour constraints imply only small contributions
to $\amu$.

\subsection{Summary}
\label{sec:SUSYSummary}

Here we briefly summarize our main results on SUSY explanations of
$\damuNEW$. Like for other BSM scenarios, negative results from LHC and
dark matter searches have significantly reduced the viable SUSY
parameter space. Simple traditionally considered cases such as the
Constrained MSSM are already excluded as explanations of
$\damu^{\text{BNL}}$ and now \update{also of $\damuNEW$}
\cite{Buchmueller:2013rsa,Bechtle:2015nua,Han:2016gvr,Athron:2017qdc}. 

In our detailed phenomenological analysis we focused on the general
MSSM without restrictions from GUT scale assumptions or specific SUSY
breaking mechanisms. The only
restrictions imposed by our analysis are a stable neutralino-like LSP
which constitutes (part or all of) dark matter and the
absence of flavour-violating soft SUSY-breaking parameters. For
simplicity we also consider equal masses of left- and right-handed
selectrons and smuons (called sleptons for short), while the
stau-masses are left arbitrary. In a series of footnotes
\ref{footnoteBHR}, \ref{footnoteMRSSM}, \ref{footnoteLFV}, \ref{footnoteTBMUlarge}  we
commented on alternative cases with ultra-high $\tan\beta$ or $\mu$,
enhancements via lepton-flavour violation, and the MRSSM without
$\tan\beta$ enhancement. The results of our analysis are as follows.

\begin{itemize}
  \item
\update{
  \underline{Scenario with heavy charginos and smuons}: the MSSM scenario with generally heavy masses,
  corresponding to the upper right quadrant of
  Fig.\ \ref{fig:briefsurveyplot} where LHC limits are trivially
  avoided is disfavoured
  as an explanation of $\damuNEW$. For such SUSY heavy masses, the
  current $\amu$ deviation can at most be
  explained if  $\tan\beta\gg40$ and/or $\mu\gg4$~TeV.}
\item
  \underline{$(\tilde{B}\tilde{l})$-scenario}:  \update{a
    promising} MSSM scenario is the
  $(\tilde{B}\tilde{l})$-scenario with Bino-like LSP and close-by
  sleptons to evade LHC limits. We identified three allowed  parameter
  regions particularly
  promising in view of $\damuNEW$: (1) Wino mass above LHC limits of
  around $900$ GeV (for LSP mass of $250$ GeV)
  and Higgsino mass $\mu$ of order
  $1$ TeV. Here all slepton and stau masses may be universal. (2) 
  Wino mass as before but $\mu$ in the multi-TeV region. Here at least one stau must be
  heavier to avoid vacuum stability constraints. (3) Light Wino with
  mass similar to the Bino and slepton masses. In all regions dark
  matter data implies a lower mass limit on $\mu$. The relic density
  can be generated via stau/slepton coannihilation; in region
  (3) also Wino coannihilation is possible.
\update{In all these cases
  the result for $\damuNEW$ after the FNAL measurement can be easily explained in a wide range
  of masses and $\tan\beta$ values.}

\item  
  \underline{ $(\tilde{W}\tilde{l})$- and $(\tilde{H}\tilde{l})$-scenarios}:
    the $(\tilde{W}\tilde{l})$- and $(\tilde{H}\tilde{l})$-scenarios are
  characterized by Wino- or Higgsino-like LSP; the sleptons are
  sufficiently close to the LSP to evade LHC limits. Specifically the
  $(\tilde{W}\tilde{l})$-scenario can lead to the largest $\amu^{\text{SUSY}}$
  of any MSSM scenario in a wide parameter space   via the WHL and
  BLR contributions: the DMDD constraints are weak
  and there are essentially no additional
  LHC constraints. \update{The new updated $\amu$ deviation can be accommodated
  in a wide range of masses and $\tan\beta$ values, see
  Fig.\ \ref{fig:WHsleptonscompresseda}. At the $1\sigma$ level and
  for $\tan\beta=40$ LSP masses above $400$ GeV are possible.}

  In the $(\tilde{H}\tilde{l})$-scenario $\amu^{\text{SUSY}}$ is
  dominated only by WHL contributions, and DMDD and
  LHC constrain the Wino mass to be rather heavy. Hence there is an
  upper limit to the possible values of $\amu^{\text{SUSY}}$,
  \update{but a $1\sigma$ explanation of the
    current $\amu$ deviation is possible for
    $\tan\beta=40$ and for LSP masses below $350$ GeV.}
  However, in both cases of Wino- or Higgsino-like LSP, the dark matter relic density cannot be
  fully accommodated simultaneously with $\damuNEW$, necessitating
  additional non-MSSM components 
  of dark matter such as gravitinos.

\item
  \underline{$(\tilde{B}\tilde{W}\tilde{H})$- and
        $(\tilde{H}\tilde{W}/\tilde{W}\tilde{H})$-scenarios}: 
  in the $(\tilde{B}\tilde{W}\tilde{H})$- and
        $(\tilde{H}\tilde{W}/\tilde{W}\tilde{H})$-scenarios both
  charginos are assumed to be lighter than the sleptons, which in turn are
  constrained by LHC data. 
  $\amu$ is rather limited in both scenarios. In the
  $(\tilde{B}\tilde{W}\tilde{H})$-scenario the Bino-like neutralino is
  the LSP and even lighter than both charginos. Outside the Bino-Wino
  coannihilation region this scenario is very
  strongly constrained by the combination of dark matter and LHC
  constraints. In the Bino-Wino coannihilation region, all 
  constraints 
  can be fulfilled for sufficiently large $\mu$, however in this
  region and  for
    $\tan\beta=40$
  \update{$\amu^{\text{SUSY}}$ is almost always at least
    $1\sigma$ lower than the observed deviation.}
  In the $(\tilde{H}\tilde{W}/\tilde{W}\tilde{H})$-scenarios either
  the Wino- or Higgsino-like neutralino is the LSP. Assuming staus as
  heavy as the other sleptons, there are no relevant LHC constraints.
  Although the full dark matter relic density is below the observed one
  (similarly to the $(\tilde{W}\tilde{l})$- and
  $(\tilde{H}\tilde{l})$-cases), direct detection
  constraints exist and require a mass splitting between the Higgsinos and
  Winos. The resulting $\amu^{\text{SUSY}}$ can be larger in case of a Wino-like
  LSP. \update{Such a scenario is well able to explain the
    observed deviation for $\tan\beta=40$ or slightly smaller
    $\tan\beta$ values.}

\end{itemize}
The discussion and the plots show that the general MSSM can explain
the current $\damuNEW$ in large parameter regions which will remain
hard to test at the LHC alone. More sensitive LHC measurements will
however sharpen the limits on SUSY particle masses; depending on the
scenario,  the mass of the LSP,  of the dark matter
coannihilation partner, or of the Wino-like chargino mass will
be further scrutinized. If the LHC search results remain negative,
this, together with theoretical constraints such as the stau vacuum
stability constraints, will particularly help to eliminate motivated high-scale
scenarios beyond the Constrained MSSM.

Future dark matter direct detection experiments are very promising in
view of MSSM explanations of $\damuNEW$. Both the
LUX-ZEPLIN (LZ) experiment \cite{LUX-ZEPLIN:2018poe} and the XENONnT
experiment \cite{XENON:2020kmp} have the potential to increase
the sensitivity to DMDD cross sections by more
than an order of magnitude, reaching close to the irreducible neutrino
background. These
experiments have the potential to discover evidence for SUSY dark
matter  if the MSSM explanation of $\damuNEW$ is correct, or to
significantly reduce the available parameter space. For the
longer-term future, $e^+e^-$ colliders offer great potential of
testing MSSM explanations of $\damuNEW$ more conclusively, and we
refer to Refs.\ \cite{Chakraborti:2021kkr,Chakraborti:2020vjp} for
more details. These references have shown that some part of the
dark matter coannihilation parameter space can be tested at an
$e^+e^-$ linear collider with $500$ GeV center-of-mass energy, while a
very large part of the parameter space can be tested at a multi-TeV
$e^+e^-$ collider such as CLIC \cite{CLICdp:2018cto}.

%% file: SectionSUSYAppendix.tex
\section{Details on LHC-constraints on SUSY parameter regions}
\label{app:SUSYLHC}

In section \ref{sec:SUSYconstraints}
we described our procedure for recasting LHC-constraints on the SUSY
parameter space; the procedure was then applied in section
\ref{sec:SUSYpheno} to investigate the impact of 
LHC-constraints on the SUSY parameter space. Here we provide further
details on the recasting.

\begin{figure}
  \null\hfill
  \includegraphics[scale=1]{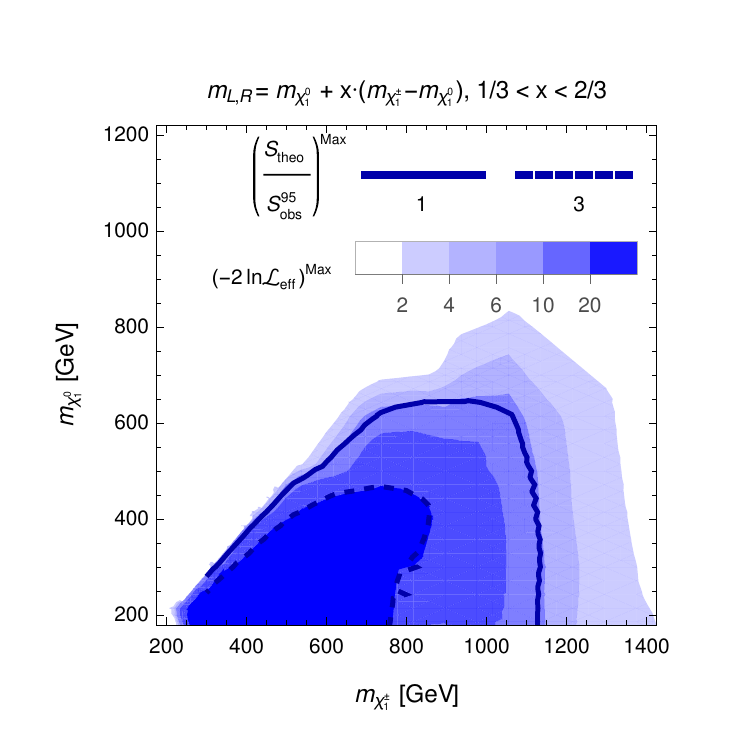}
  \hspace{0em}\hfill\null
  \caption{\label{fig:ReproduceATLASplot}
  Recasting of the exclusion limits corresponding to Ref.\ \cite{Aaboud:2018jiw},
    Fig.\ 8c. The generated MSSM parameter
    points have the mass hierarchy $M_1<m_{\txl, \txr}<M_2<\mu$, i.e.\ Bino-like
    LSP, intermediate sleptons of the 1st and 2nd generation, and a
    Wino-like pair of $\chi_2^0/\chi_1^\pm$.  The slepton
    masses satisfy
    $x=(m_{\txl, \txr}-m_{\chi_1^0})/(m_{\chi_1^\pm}-m_{\chi_1^0})$ with
    $1/3<x<2/3$. The thick solid blue contour can be directly compared to
    the 95\% C.L.\ exclusion contour of  Ref.\ \cite{Aaboud:2018jiw},
    Fig.\ 8c; the thick dashed blue contour corresponds to points
    where the predicted signal yield is three times higher than the
    respective ATLAS upper limit in at least one signal region. The
    blue coloured regions correspond to various values of the maximum
    effective          $(-2\ln{\cal
    L}_{\text{eff}})$ of any analysis implemented in
    \gambit/\colliderbit. The plots in section \ref{sec:SUSYpheno}
    show only the region
    corresponding to $(-2\ln{\cal
    L}_{\text{eff}})=6$, which is very close to the line where
    $\left(S_\text{theo}/S_{\text{obs}}^{95\%}\right)\ge1$ for at
    least one signal region; see text for more details.
    }
\end{figure}

    In Fig.\ \ref{fig:ReproduceATLASplot} we
    illustrate the recasting by reproducing the exclusion contour of
    the ATLAS chargino/neutralino search of Ref.\ \cite{Aaboud:2018jiw},
    Fig.\ 8c. Like in that reference, we have generated MSSM parameter
    points with mass hierarchy $M_1<m_{\txl, \txr}<M_2<\mu$, i.e.\ Bino-like
    LSP, intermediate sleptons of the 1st and 2nd generation, and a
    Wino-like pair of $\chi_2^0/\chi_1^\pm$. We allowed the slepton
    mass ratio parameter
    $x=(m_{\txl, \txr}-m_{\chi_1^0})/(m_{\chi_1^\pm}-m_{\chi_1^0})$ to be in
    the range $1/3<x<2/3$, while Ref.\ \cite{Aaboud:2018jiw} fixed
    $x=1/2$. The thick solid blue contour in our
    Fig.\ \ref{fig:ReproduceATLASplot} corresponds to points where the
    predicted signal yield for at least one signal region is equal to
    the respective ATLAS 95\% C.L.\ upper limit. It can be seen that
    this contour tracks the corresponding 95\% C.L.\ contour of
    Ref.\ \cite{Aaboud:2018jiw},      Fig.\ 8c, very well, i.e.\ to
    within 50 GeV. To illustrate the gradient we also show the
    thick dashed blue contour, corresponding to points where the
    predicted signal yield is three times higher than the respective
    ATLAS upper limit. Finally  we also show blue coloured regions corresponding
    to different values of the largest effective     $(-2\ln{\cal
    L}_{\text{eff}})$ of
    any implemented analysis (in this figure, this is always the 3-lepton channel
    analysis of Ref.\ \cite{Aaboud:2018jiw}).
    We find that numerically the contour with  $(-2\ln{\cal
    L}_{\text{eff}})^{\text{Max}}=6$ tracks very well the 95\%
    C.L.\ contour. We checked that the same would be true for most
    plots of our section \ref{sec:SUSYpheno} (the exceptions are very
    few cases where such a comparison is not possible because the
    relevant contribution to $(-2\ln{\cal
    L}_{\text{eff}})^{\text{Max}}$ comes from the CMS analysis of
    Ref.\ \cite{Sirunyan:2018iwl} for compressed spectra, which does
    not provide individual 95\% C.L.\ upper signal limits).

Fig.\ \ref{fig:ReproduceATLASplot} shows that the recasting
reproduces the corresponding original ATLAS exclusion contour very
well. On the other hand, several of the scenarios
presented in Sec.\ \ref{sec:SUSYpheno}, particularly 
Figs.\ \ref{fig:Binosleptonscompresseda},
and \ref{fig:chalightersleptons}
turned out to be entirely unconstrained by the LHC recasting. 
Hence we  present here quantitative results of our
recasting analysis, to confirm these statements and to expose further
details.\footnote{%
  Our results are compatible with related results of
  Ref.\ \cite{Athron:2018vxy} applying the  {\tt GAMBIT/ColliderBit}
  framework on chargino and neutralino searches with
  heavy sleptons, and we refer to that reference for further
  explanations of the weak LHC sensitivity to many realistic SUSY scenarios.}

We confirmed that out of all ATLAS and CMS analyses implemented in
\gambit/\colliderbit, the ATLAS electroweakino search of
Ref.\ \cite{Aaboud:2018jiw} is most sensitive in the parameter regions
in
Figs.\ \ref{fig:Binosleptonscompressed} and \ref{fig:chalightersleptons}. Within
this ATLAS 
search, the 3-lepton channel is most sensitive. The 3-lepton channel
in turn is divided into 11 signal regions. Hence we will mainly focus on the
results for these 11 signal regions in the following.

Tables \ref{tab:sampleLHCpointsunconstrainedscenarios} and \ref{tab:sampleLHCpointsunconstrainedscenariosresults} define
example SUSY parameter points and present detailed results. The first
four parameter
points represent the
$(\tilde{B}\tilde{l})$-, $(\tilde{B}\tilde{W}\tilde{H})$-, and
$(\tilde{H}\tilde{W})$-scenarios. They are examples obtained in a parameter scan
with particularly bad fit to experiment, i.e.\ particularly small
likelihood ratio (but they are still not excluded).
  The other two points represent the excluded region in
  Fig.\ \ref{fig:ReproduceATLASplot} with high or low chargino mass.

The  columns $\ln{\cal
      L}_{\text{eff}}^{\text{analysis}}$ of
Tab.\ \ref{tab:sampleLHCpointsunconstrainedscenarios}
  display  effective analysis-specific log-likelihood
  differences     $\ln{\cal
    L}_{\text{eff}}^{\text{analysis}}\equiv\ln{\cal
    L}_{\text{SR$^{\text{max}}$}}^{\text{analysis}}$, where
  $\text{SR}^{\text{max}}$ denotes the signal region of the respective
  analysis with the highest expected sensitivity, see 
  Sec.\ \ref{sec:SUSYconstraints} and 
  Ref.\ \cite{Balazs:2017moi}.
 The analyses are the ATLAS analyses
  of Ref.\ \cite{Aaboud:2018jiw} for the 3-lepton, 2-lepton
  0-jet and 2-lepton+jets channels, and to the CMS analyses of
  Ref.\ \cite{Sirunyan:2017lae} for the 2-same sign lepton and
  3-lepton channels.
The table shows that the ATLAS and CMS 3-lepton channels are most
sensitive for all parameter points.
  

Tab.\ \ref{tab:sampleLHCpointsunconstrainedscenariosresults} shows the
results obtained by \colliderbit for the 11 invidual signal regions of
the ATLAS 3-lepton analysis and
compares to the ATLAS results. The most important observation is that
all signal yield predictions of the first four parameter points are far smaller than the ATLAS signal
$95\%$ C.L.\ upper limits. The entries which come relatively closest,
i.e.\ which lead to the smallest negative log-likelihood difference,
are highlighted in boldface. A second observation is that these
entries are not always identical to the ones selected by \colliderbit
for evaluating the effective log-likelihood difference for the
next-to-last column of
Tab.\ \ref{tab:sampleLHCpointsunconstrainedscenarios}. The reason 
\cite{Balazs:2017moi} is that the selection is done assuming that the
observed counts match the background expectation; however for the
WZ-1Jc and slep-a signal regions this assumption is not true
($N_{\text{obs}}:N_{\text{exp}}=4:1.3\pm0.3$ and
$N_{\text{obs}}:N_{\text{exp}}=4:2.2\pm0.8$ respectively). 

On the other hand, for the two parameter points of the excluded region
in Fig.\ \ref{fig:ReproduceATLASplot} we observe that the high-mass
point is indeed excluded by the slep-e signal region, and the low-mass
point is excluded by a variety of signal regions by a large margin.

\begin{table}
  \centerline{
\scalebox{.8}{$\begin{array}{|c|c|c|c|c|c|c|c|c|c|c|c|}\hline
      \text{Point} & M_1 & M_2 & \mu & m_{L,R}
      & \ln{\cal
        L}_{\text{eff}}^{\text{ATLAS\_3Lep}}& \ln{\cal
        L}_{\text{eff}}^{\text{ATLAS\_2Lep0Jets}}&
       \ln{\cal
         L}_{\text{eff}}^{\text{ATLAS\_2LepPlusJets}}
       &
        \ln{\cal
          L}_{\text{eff}}^{\text{CMS\_2SSLep}}&
         \ln{\cal
           L}_{\text{eff}}^{\text{CMS\_3Lep}}
         \\\hline
          \text{$(\tilde{B}\tilde{l})_1$}
          & 200 & 1200 & 630 & 250 & -0.45 & 0.33 & 0.003&
          0.002 & 0.25  
          \\\hline
          \text{$(\tilde{B}\tilde{W}\tilde{H})_1$}
          & 200 & 296 & 346 & 700 & 0.50   & 0.07 &0.04& 0.05& -0.24
          \\\hline
          \text{$(\tilde{B}\tilde{W}\tilde{H})_2$}
          & 200 & 329 & 388 & 700 &1.05    &0.16& -0.37 & 0.05& -0.35
          \\\hline
          \text{$(\tilde{H}\tilde{W})_1$}
          & 2000 & 239 & 162 & 700 &-0.83  &0.17 &  -0.04    &  -0.04&
          -0.1   
          \\\hline
          \text{(Fig.\ \ref{fig:ReproduceATLASplot})$_1$}
          & 500 & 1039 & 2000 & 800 &-3.4  & -0.2& 0     &0 & -0.13
          \\\hline
          \text{(Fig.\ \ref{fig:ReproduceATLASplot})$_2$}
          & 204 & 350 & 2000 & 300 &-55.8  & -15.9& 0     & -6.3 &
          -56.9  
          \\\hline
\end{array}
$ }}
  \caption{\label{tab:sampleLHCpointsunconstrainedscenarios} Definitions
    and basic properties of sample parameter points. The first four
    points represent the $(\tilde{B}\tilde{l})$-,
    $(\tilde{B}\tilde{W}\tilde{H})$-, and
    $(\tilde{H}\tilde{W})$-scenarios and correspond to
  points in
  Figs.\ \ref{fig:Binosleptonscompressed} and \ref{fig:chalightersleptons}
  with particularly bad fit to experiment (though not excluded).
  The last two points represent the excluded region in
    Fig.\ \ref{fig:ReproduceATLASplot}.  The  columns $\ln{\cal
      L}_{\text{eff}}^{\text{analysis}}$
  display  effective analysis-specific log-likelihood
  differences obtained by \gambit/\colliderbit as
  described in Sec.\ \ref{sec:SUSYconstraints} and in
  Ref.\ \cite{Balazs:2017moi}. They correspond to the ATLAS analyses
  of Ref.\ \cite{Aaboud:2018jiw} for the 3-lepton, 2-lepton
  0-jet and 2-lepton+jets channels, and to the CMS analyses of
  Ref.\ \cite{Sirunyan:2017lae} for the 2-same sign lepton and 3-lepton
  channels.}
\end{table}

\begin{table}
  \centerline{
\scalebox{.82}{$
\begin{array}{|cccc|cccccc|}
  \hline
  \text{Region} & N_{\text{obs}} & N_{\text{exp}} & S^{95}_{\text{obs}} &
  \begin{array}{c}\text{$(\tilde{B}\tilde{l})_1$} \\ S\\
    \ln{\cal L}_{\text{SR}}
  \end{array}
 & \begin{array}{c}\text{$(\tilde{B}\tilde{W}\tilde{H})_1$} \\ S\\
    \ln{\cal L}_{\text{SR}}
   \end{array} &
  \begin{array}{c}\text{$(\tilde{B}\tilde{W}\tilde{H})_2$} \\ S\\
    \ln{\cal L}_{\text{SR}}
  \end{array} &
  \begin{array}{c}\text{$(\tilde{H}\tilde{W})_1$} \\ S\\
    \ln{\cal L}_{\text{SR}}
  \end{array}  
   &
  \begin{array}{c}\text{(Fig.\ \ref{fig:ReproduceATLASplot})$_1$} \\ S\\
    \ln{\cal L}_{\text{SR}}
  \end{array}  &
   \begin{array}{c}\text{(Fig.\ \ref{fig:ReproduceATLASplot})$_2$} \\ S\\
    \ln{\cal L}_{\text{SR}}
  \end{array}  \\\hline
 \left.
\begin{array}{c}
 \text{WZ-0Ja} \\
 \text{} \\
\end{array}
\right. & \left.
\begin{array}{c}
 21 \\
 \text{} \\
\end{array}
\right. & \left.
\begin{array}{c}
 21.70\pm 2.90 \\
 \text{} \\
\end{array}
\right. & \left.
\begin{array}{c}
 12.80 \\
 \text{} \\
\end{array}
\right. & \left.
\begin{array}{c}
 0.02 \\
0. \\
\end{array}
\right.  
& \left.
\begin{array}{c}
 0.41 \\
 -0. \\
\end{array}
\right. & \left.
\begin{array}{c}
 2.11 \\
 -0.11 \\
\end{array}
\right. & \left.
\begin{array}{c}
 1.31 \\
 -0.05 \\
\end{array}
\right. 
&\left.\begin{array}{c}     
  0\\0\\\end{array}\right.
  &\left.\begin{array}{c}8.98\\-1.33\\\end{array}\right.
\\
\hline
\left.
\begin{array}{c}
 \text{WZ-0Jb} \\
 \text{} \\
\end{array}
\right. & \left.
\begin{array}{c}
 1 \\
 \text{} \\
\end{array}
\right. & \left.
\begin{array}{c}
 2.70\pm 0.50 \\
 \text{} \\
\end{array}
\right. & \left.
\begin{array}{c}
 3.70 \\
 \text{} \\
\end{array}
\right. & \left.
\begin{array}{c}
 0.03 \\
 -0.02 \\
\end{array}
\right. 
& \left.
\begin{array}{c}
 0.37 \\
 -0.23 \\
\end{array}
\right. & \left.
\begin{array}{c}
 \mathbf{ 1.47} \\
 \mathbf{-1.0} \\
\end{array}
\right. & \left.
\begin{array}{c}
\mathbf{ 1.24} \\
\mathbf{ -0.83} \\
\end{array}
\right. 
&\left.\begin{array}{c}     
  0\\0\\\end{array}\right.
  &\left.\begin{array}{c}12.0\\-10.2\\\end{array}\right.
\\
\hline
\left.
\begin{array}{c}
 \text{WZ-0Jc} \\
 \text{} \\
\end{array}
\right. & \left.
\begin{array}{c}
 2 \\
 \text{} \\
\end{array}
\right. & \left.
\begin{array}{c}
 1.60\pm 0.30 \\
 \text{} \\
\end{array}
\right. & \left.
\begin{array}{c}
 4.80 \\
 \text{} \\
\end{array}
\right. & \left.
\begin{array}{c}
 0.12 \\
 0.03 \\
\end{array}
\right. 
& \left.
\begin{array}{c}
 0.26 \\
 0.04 \\
\end{array}
\right. & \left.
\begin{array}{c}
 0.78 \\
 0.02 \\
\end{array}
\right. & \left.
\begin{array}{c}
 0.53 \\
 0.05 \\
\end{array}
\right. 
&\left.\begin{array}{c}0.02\\ 0\\\end{array}\right.
  &\left.\begin{array}{c}10.9\\-6.69\\\end{array}\right.
  \\
\hline
\left.
\begin{array}{c}
 \text{WZ-1Ja} \\
 \text{} \\
\end{array}
\right. & \left.
\begin{array}{c}
 1 \\
 \text{} \\
\end{array}
\right. & \left.
\begin{array}{c}
 2.20\pm 0.50 \\
 \text{} \\
\end{array}
\right. & \left.
\begin{array}{c}
 3.20 \\
 \text{} \\
\end{array}
\right. & \left.
\begin{array}{c}
 0. \\
0. \\
\end{array}
\right. 
& \left.
\begin{array}{c}
 0.15 \\
 -0.08 \\
\end{array}
\right. & \left.
\begin{array}{c}
 0.33 \\
 -0.18 \\
\end{array}
\right. & \left.
\begin{array}{c}
 0.06 \\
 -0.03 \\
\end{array}
\right. 
&\left.\begin{array}{c} 0\\0\\\end{array}\right. &\left.\begin{array}{c}0.19\\-0.1\\\end{array}\right.\\
\hline
\left.
\begin{array}{c}
 \text{WZ-1Jb} \\
 \text{} \\
\end{array}
\right. & \left.
\begin{array}{c}
 3 \\
 \text{} \\
\end{array}
\right. & \left.
\begin{array}{c}
 1.80\pm 0.30 \\
 \text{} \\
\end{array}
\right. & \left.
\begin{array}{c}
 5.60 \\
 \text{} \\
\end{array}
\right. & \left.
\begin{array}{c}
 0.11 \\
 0.07 \\
\end{array}
\right. 
& \left.
\begin{array}{c}
 0.46 \\
 0.22 \\
\end{array}
\right. & \left.
\begin{array}{c}
 1.17 \\
 0.32 \\
\end{array}
\right. & \left.
\begin{array}{c}
 0.54 \\
 0.24 \\
\end{array}
\right. 
&\left.\begin{array}{c}0\\0\\\end{array}\right.
&\left.\begin{array}{c}3.55\\-0.28\\\end{array}\right.\\
\hline
\left.
\begin{array}{c}
 \text{WZ-1Jc} \\
 \text{} \\
\end{array}
\right. & \left.
\begin{array}{c}
 4 \\
 \text{} \\
\end{array}
\right. & \left.
\begin{array}{c}
 1.30\pm 0.30 \\
 \text{} \\
\end{array}
\right. & \left.
\begin{array}{c}
 7.20 \\
 \text{} \\
\end{array}
\right. & \left.
\begin{array}{c}
 0.13 \\
 0.21 \\
\end{array}
\right. 
& \left.
\begin{array}{c}
 0.33 \\
 0.49 \\
\end{array}
\right. & \left.
\begin{array}{c}
 \mathbf{0.87} \\
\mathbf{ 1.05} \\
\end{array}
\right. & \left.
\begin{array}{c}
 0.24 \\
 0.37 \\
\end{array}
\right. 
&\left.\begin{array}{c}0.03\\0.05\\\end{array}\right.
 &\left.\begin{array}{c}5.59\\0.94\\\end{array}\right. 
\\
\hline
\left.
\begin{array}{c}
 \text{slep-a} \\
 \text{} \\
\end{array}
\right. & \left.
\begin{array}{c}
 4 \\
 \text{} \\
\end{array}
\right. & \left.
\begin{array}{c}
 2.20\pm 0.80 \\
 \text{} \\
\end{array}
\right. & \left.
\begin{array}{c}
 6.80 \\
 \text{} \\
\end{array}
\right. & \left.
\begin{array}{c}
 0.12 \\
 0.07 \\
\end{array}
\right. 
& \left.
\begin{array}{c}
\mathbf{ 1.95} \\
\mathbf{ 0.50} \\
\end{array}
\right. & \left.
\begin{array}{c}
 0.60 \\
 0.29 \\
\end{array}
\right. & \left.
\begin{array}{c}
 0.44 \\
 0.22 \\
\end{array}
\right. 
&\left.\begin{array}{c}    0\\0\\\end{array}\right.
  &\left.\begin{array}{c}13.00\\-5.13\\\end{array}\right.
\\
\hline
\left.
\begin{array}{c}
 \text{slep-b} \\
 \text{} \\
\end{array}
\right. & \left.
\begin{array}{c}
 3 \\
 \text{} \\
\end{array}
\right. & \left.
\begin{array}{c}
 2.80\pm 0.40 \\
 \text{} \\
\end{array}
\right. & \left.
\begin{array}{c}
 5.20 \\
 \text{} \\
\end{array}
\right. & \left.
\begin{array}{c}
 0.11 \\
 0. \\
\end{array}
\right. 
& \left.
\begin{array}{c}
\mathbf{ 2.30} \\
\mathbf{ -0.48} \\
\end{array}
\right. & \left.
\begin{array}{c}
 0.94 \\
 -0.06 \\
\end{array}
\right. & \left.
\begin{array}{c}
 0.55 \\
 -0. \\
\end{array}
\right. 
&\left.\begin{array}{c}0.10\\ 0\\\end{array}\right.
&\left.\begin{array}{c}\mathbf{76.6}\\\mathbf{-66.5}\\\end{array}\right.\\
\hline
\left.
\begin{array}{c}
 \text{slep-c} \\
 \text{} \\
\end{array}
\right. & \left.
\begin{array}{c}
 9 \\
 \text{} \\
\end{array}
\right. & \left.
\begin{array}{c}
 5.40\pm 0.90 \\
 \text{} \\
\end{array}
\right. & \left.
\begin{array}{c}
 10.50 \\
 \text{} \\
\end{array}
\right. & \left.
\begin{array}{c}
 0.52 \\
 0.26 \\
\end{array}
\right. 
& \left.
\begin{array}{c}
 0.94 \\
 0.43 \\
\end{array}
\right. & \left.
\begin{array}{c}
 1.30 \\
 0.54 \\
\end{array}
\right. & \left.
\begin{array}{c}
 0.68 \\
 0.33 \\
\end{array}
\right. 
&\left.\begin{array}{c}0.14\\0.07\\\end{array}\right.
&\left.\begin{array}{c}\mathbf{81.0}\\\mathbf{-55.8}\\\end{array}\right.  \\
\hline
\left.
\begin{array}{c}
 \text{slep-d} \\
 \text{} \\
\end{array}
\right. & \left.
\begin{array}{c}
0 \\
 \text{} \\
\end{array}
\right. & \left.
\begin{array}{c}
 1.40\pm 0.40 \\
 \text{} \\
\end{array}
\right. & \left.
\begin{array}{c}
 3.00 \\
 \text{} \\
\end{array}
\right. & \left.
\begin{array}{c}
 0.30 \\
 -0.29 \\
\end{array}
\right. 
& \left.
\begin{array}{c}
{ 0.39} \\
{ -0.38} \\
\end{array}
\right. & \left.
\begin{array}{c}
{ 0.50} \\
{ -0.48} \\
\end{array}
\right. & \left.
\begin{array}{c}
 0.24 \\
 -0.23 \\
\end{array}
\right. 
&\left.\begin{array}{c}0.35\\-0.34\\\end{array}\right.
&\left.\begin{array}{c}42.2\\-42.1\\\end{array}\right.
\\
\hline
\left.
\begin{array}{c}
 \text{slep-e} \\
 \text{} \\
\end{array}
\right. & \left.
\begin{array}{c}
0 \\
 \text{} \\
\end{array}
\right. & \left.
\begin{array}{c}
 1.10\pm 0.20 \\
 \text{} \\
\end{array}
\right. & \left.
\begin{array}{c}
 3.30 \\
 \text{} \\
\end{array}
\right. & \left.
\begin{array}{c}
 \mathbf{0.45} \\
\mathbf{ -0.45} \\
\end{array}
\right. 
& \left.
\begin{array}{c}
 0.27 \\
 -0.27 \\
\end{array}
\right. & \left.
\begin{array}{c}
 0.34 \\
 -0.33 \\
\end{array}
\right. & \left.
\begin{array}{c}
 0.24 \\
 -0.23 \\
\end{array}
\right. 
&\left.\begin{array}{c}\mathbf{3.41}\\\mathbf{-3.40}\\\end{array}\right.
&\left.\begin{array}{c}12.5\\-12.5\\\end{array}\right.\\
\hline
\end{array}
$}
}
  \caption{\label{tab:sampleLHCpointsunconstrainedscenariosresults} Detailed
    results for the sample parameter points defined in
    Tab.\ \ref{tab:sampleLHCpointsunconstrainedscenarios}  versus the 3-lepton channel of
    the ATLAS analysis \cite{Aaboud:2018jiw}.
The first four columns are taken from Ref.\
\cite{Aaboud:2018jiw}, Table 15, and show the names of the signal
regions, the observed and expected background yields
$N_{\text{obs,exp}}$ as well as the model-independent signal upper
limits $S^{95}_{\text{obs}}$. The other columns show for each
parameter point the predicted signal yield $S$ and the resulting
log-likelihood difference as defined in Ref.\ \cite{Balazs:2017moi} (the signal
    uncertainties estimated by \colliderbit are at the level of $10\%$
    of the signal or less). Numbers  highlighted in boldface
    correspond to the entries with
    the highest significance and to the entries selected by
    \colliderbit for the overall $\ln{\cal
      L}_{\text{eff}}^{\text{ATLAS\_13TeV\_MultiLEP\_3Lep}}$ of
    Tab.\ \ref{tab:sampleLHCpointsunconstrainedscenarios} (selected
    ``on the basis of the signal region {\em expected} to give the
    strongest limit'' \cite{Balazs:2017moi}).}
\end{table}